\documentclass[11pt, a4paper]{article}
\usepackage[affil-it]{authblk}

\usepackage{amsmath}
\usepackage{amsthm}
\usepackage{amssymb}
\usepackage{graphicx}
\usepackage{subfigure}
\usepackage[colorlinks=false]{hyperref}
\usepackage[nodisplayskipstretch]{setspace}
\usepackage{color}
\usepackage{verbatim} 
\usepackage{setspace}
\usepackage{extsizes}
\usepackage{anysize} 
\usepackage{chngcntr}
\usepackage{lineno}
\usepackage{fancyhdr}
\marginsize{1.5cm}{1.5cm}{0.8cm}{0.8cm}

\counterwithin{equation}{section}

\newtheorem{proposition}{Proposition}[section]
\newtheorem{remark}{Remark}[section]
\newtheorem{theorem}{Theorem} 

\theoremstyle{plain} 
\newcommand{\thistheoremname}{}
\newtheorem{genericthm}[theorem]{\thistheoremname}

\begin{document}
\title{Robust Linear Estimation with Non-parametric Uncertainty:\\Average and Worst-case Performance (Full Version)}
\author{Gilberto O. Corr\^ea%
  \thanks{email: \texttt{gilberto\emph{{@}}lncc.br} - Corresponding author.}}
\affil{Laborat\'{o}rio Nacional de Computa\c{c}\~{a}o Cient\'{i}fica -- LNCC/MCTIC,\\
Petr\'{o}polis, Rio de Janeiro, Brazil - CEP: 25651-075}
{\author{Marlon M. L\'{o}pez-Flores%
  \thanks{email: \texttt{mmlf\emph{{@}}impa.br.}}}
\affil{Instituto Nacional de Matem\'{a}tica Pura e Aplicada -- IMPA,\\ Rio de Janeiro, Brazil, CEP: 22460-320}}

\date{}

\maketitle



\begin{abstract}
 In this paper, two types of linear estimators are considered for three related estimation problems involving set-theoretic uncertainty 
 pertaining to\ \ $\mathcal{H}_{2}$\ \  and\ \ $\mathcal{H}_{\infty}$\ \ balls of frequency-responses. The problems at stake correspond to 
 robust \ \ $\mathcal{H}_{2}$\ \  and\ \ $\mathcal{H}_{\infty}$\ \ estimation in the face of non-parametric ``channel-model' uncertainty and 
 to a nominal\ \ $\mathcal{H}_{\infty}$\ \  estimation problem. The estimators considered here are defined by minimizing the worst-case squared
 estimation error over the ``uncertainty set'' and by minimizing an average cost under the constraint that the worst-case error of any 
 admissible estimator does not exceed a prescribed value. The main point is to explore the derivation of estimators which may be viewed as
 less conservative alternatives to minimax estimators, or in other words, that allow for trade-offs between worst-case performance and better 
 performance over ``large'' subsets of the uncertainty set. The ``average costs'' over\ \ $\mathcal{H}_{2}-$signal balls are obtained as limits
 of averages over sets of finite impulse responses, as their length grows unbounded. The estimator design problems for the two types of 
 estimators and the three problems addressed here are recast as semi-definite programming problems (SDPs, for short). These SDPs are solved
 in the case of simple examples to illustrate the potential of the ``average cost/worst-case constraint'' estimators to mitigate the inherent
 conservatism of the minimax estimators.
\end{abstract}
\newpage

\section*{Notation}

\begin{itemize}
 \item[$\bullet$] $\mathbb{Z},\mathbb{R}$, $\mathbb{C}$, $\mathbb{Z}_{+}$, $\mathbb{R}_{+}$, and $\bar{\mathbb{R}}_{+}$ stand, respectively, for 
the sets of integers, real, complex,  positive integer, positive and non-negative real numbers.
 \item[$\bullet$]$\mathcal{S}^{p\times m}_{_{0}}-$ the set of all doubly-infinite sequences of matrices in\ \ \ $\mathbb{R}^{p\times m}$, \emph{i.e.},
$$\mathcal{S}^{p\times m}_{_{0}}=\{\mathbf{F}^s=\{\mathbf{F}_{_{k}}:k\in \mathbb{Z}\}:\mathbf{F}_{_{k}}\in \mathbb{R}^{p\times m}\}.$$
 \item[$\bullet$] $\mathcal{S}^{p\times m}_c-$ the set of all causal sequences, \emph{i.e.}, \
$\mathcal{S}^{p\times m}_c=\{\mathbf{F}^s\in \mathcal{S}^{p\times m}_{_{0}}:\forall k<0,\ \mathbf{F}_{_{k}}=0\}.$
 \item[$\bullet$] $r_c^{p\times m} - $ the set of all impulse responses of causal, stable and finite-dimensional systems, \emph{i.e.},
\begin{equation*}
\begin{split}
&r_c^{p\times m}= \{\mathbf{F}^s\in \mathcal{S}^{p\times m}_c: \mathbf{F}_{_{0}}=\mathbf{D}, \ \forall k\geq 1, \mathbf{F}_{_{k}}=\mathbf{C}_{_{\mathbf{F}}}\mathbf{A}^{k-1}_{_{\mathbf{F}}}\textbf{B}_{_{\mathbf{F}}} \ \text{for some}\ \textbf{A}_{_{\mathbf{F}}}\in \mathbb{R}^{n\times n},\\
&\ \ \ \ \ \ \ \ \ \ \ \  \  \ \ \ \ \ \ \ \ \ \ \ \ \ \ \ \ \ \ \ \ \ \ \ \ \ \ \ \ \ \ \  \mathbf{B}_{_{\mathbf{F}}}\in \mathbb{R}^{n\times m}, \mathbf{C}_{_{\mathbf{F}}}\in \mathbb{R}^{p\times n}\ \text{and}\ \mathbf{D} \in \mathbb{R}^{p\times m}\ \text{with}\ \rho (\mathbf{A}_{_{\mathbf{F}}})<1\}.
\end{split}
\end{equation*}
 \item[$\bullet$] $\mathcal{R}_c^{p\times m}-$ the set of frequency-responses corresponding to $r_c^{p\times m}$, \emph{i.e.},  $\mathcal{R}_c^{p\times m}={\cal F}(r^{p\times m}_c)$ where for\break $\mathbf{F}^s\in r^{p\times m}_c$, ${\cal F}(\mathbf{F}^s)$ denotes the Fourier transform of $\mathbf{F}^s$, \emph{i.e.},\
 $\mathbf{F}(e^{j\theta})= \displaystyle\sum^{\infty}_{k=0}\mathbf{F}_{_{k}}e^{-j\theta k}.
$
 \item[$\bullet$] $\mathcal{R}^{p\times m}-$ the set of all sums involving matrix functions in $\mathcal{R}_c^{p\times m}$ and the conjugate transposes thereof, \emph{i.e.}, $\mathcal{R}^{p\times m}=\{\textbf{G}=\textbf{E}_{_{1}}+\mathbf{E}_{_{2}}^{*}:\mathbf{E}_{_{i}}\in  \mathcal{R}_c^{p\times m}, i=1,2\},$\ where\ $\mathbf{E}^{\ast}(e^{j\theta})=\mathbf{E}(e^{j\theta})^{\ast}=\mathbf{E}(e^{-j\theta})^{^{\mathrm{T}}}.$
 \item[$\bullet$] $\mathcal{R}^{p\times m}_{_{h}}-$ matrix functions which are Hermitian on the unit circle, \emph{i.e.}, $$\mathcal{R}^{p\times m}_{_{h}}=\{\mathbf{G}=\mathbf{E}+\mathbf{E}^{\ast}:\mathbf{E}\in \mathcal{R}^{p\times m}_c\}.$$
 \item[$\bullet$] $\mathcal{R}^{p\times m}_{_{0}}-$ the subset of $\mathcal{R}^{p\times m}$ containing only matrix functions which are positive semidefinite on the unit
circle, \emph{i.e.}, $$\mathcal{R}^{p\times m}_{_{0}}=\{\mathbf{G}\in \mathcal{R}^{p\times m}_{_{h}}:\forall \theta \in[0,2\pi ]\ ,\ \mathbf{G}(e^{j\theta})\geq 0\}.$$
 \item[$\bullet$] $\{\mathbf{F}\}_{ca} - $  causal part of\ $\mathbf{F}\in \mathcal{R}^{p\times m}$ , \emph{i.e.}, for
 $$\mathbf{F}(e^{j\theta})=\displaystyle\sum^{\infty}_{k=-\infty}\mathbf{F}_{_{k}}e^{-j\theta k},\ \
\{\mathbf{F}\}_{ca}=\displaystyle\sum^{\infty}_{k=0}\mathbf{F}_{_{k}}e^{-j\theta k}.$$ 
 \item[$\bullet$] $\mathbf{M}^{\ast}$, $\rho (\mathbf{M})-$ the conjugate transpose and the spectral  radius of the matrix $\mathbf{M}$.
 \item[$\bullet$]$\langle \mathbf{F},\mathbf{G}\rangle \triangleq (2\pi )^{-1}\displaystyle\int^{2\pi}_{0}\operatorname{tr}\{\mathbf{F}(e^{j\theta})^{\ast}\mathbf{G}(e^{j\theta})\}d\theta, \|\mathbf{F}\|_{_{2}}^{^{2}}=\langle \mathbf{F},\mathbf{F}\rangle, \ \mathbf{F}\in \mathcal{R}^{p\times m},  \mathbf{G}\in \mathcal{R}^{p\times m}$;\break $\triangleq$ denotes equality by definition.
 \item[$\bullet$] $\langle \mathbf{A},\mathbf{B} \rangle_{_{F}} \triangleq \operatorname{tr}(\mathbf{A}^{*}\mathbf{B})$.
 \item[$\bullet$]$\{\mathbf{M}\}_{ij}$, $\mathbf{M}^{^{\mathrm{T}}}$ -- $ij$-th entry and transpose of the matrix $\mathbf{M}$.
 \item[$\bullet$]$\operatorname{tr}(\mathbf{A})$, $\|\mathbf{A}\|$, $\left\|\mathbf{A}\right\|_{_{F}}$, $\lambda_{min}(\mathbf{M})$ and $\lambda_{max}(\mathbf{M})$ denote  the trace, spectral norm and Frobenius norm of $\mathbf{A}\in\mathbb{C}^{m\times p}$ and the maximum and minimum eigenvalues of the Hermitian matrix $\mathbf{M}$.
 \item[$\bullet$] $\|\boldsymbol{x}\|_{_{E}}$ denotes the euclidean norm of $\boldsymbol{x}\in\mathbb{C}^{n}$.
 \item[$\bullet$] $\operatorname{diag} (\mathbf{A}_{_{1}},\ldots,\mathbf{A}_{_{n}})$ denotes a block diagonal matrix where $\mathbf{A}_{_{i}}$ is the $i-$th diagonal block.
 \item[$\bullet$] $\mathbf{A}\otimes \mathbf{B}$ denotes Kronecker product of the matrices $\mathbf{A}$ and $\mathbf{B}$, \emph{i.e.},\\
$
\mathbf{A}\otimes \mathbf{B}=\left[\begin{array}{ccc}
a_{_{11}}\mathbf{B}&\cdots &a_{_{1n}}\mathbf{B}\\
\vdots && \vdots\\
a_{_{m1}}\mathbf{B}& \cdots & a_{_{mn}}\mathbf{B}\\
\end{array}\right]$\ where $ \{a_{_{ij}}\}$ \ are the entries of the $m\times n$ matrix $\mathbf{A}$.

 \item[$\bullet$] $\operatorname{rvec} (\mathbf{F})$ denotes the column matrix $\{\boldsymbol{f}_{_{11}}\cdots \boldsymbol{f}_{_{1m}}\cdots \boldsymbol{f}_{_{p1}}\cdots \boldsymbol{f}_{_{pm}}\}^{^{\mathrm{T}}}$ where $\{\boldsymbol{f}_{_{ij}}\}$ are entries of the $p\times m$ matrix $\mathbf{F}$.
 \item[$\bullet$] $\operatorname{diag}(\{\mathbf{M}_{_{k}}\})$ denotes a block diagonal matrix where $\mathbf{M}_{_{k}}$ is the $k-$th diagonal block.
\end{itemize}

\newpage

\tableofcontents

\newpage

\section{Introduction}\label{sec:1}

Decision problems involving set-theoretic uncertainty on problem-data lead naturally to (robust) decision procedures based on minimizing,
with respect to the admissible procedures,the worst-case value of a given risk function over the problem data set involved -- for example,
to select an estimator (decision procedure) for signals corrupted by noise when several signal models (problem data) are envisaged, it is
natural to look for the estimator that minimizes the worst-case of the mean-square estimation error (risk function) over all signal
models considered. Indeed, such minimax procedures have been widely considered in control and estimation problems (see,
for example, [1] -- [13]). However, it has been acknowledged that minimax procedures tend to be ``overly conservative'' in the sense that
(roughly speaking) ``point-wise performance'' is compromised over much of the problem-data set in order to attain relatively small values
of the loss function at the  ``most unfavourable'' region of that set.

Attempts to overcome this drawback have been made on the basis of the so-called regret function associated to a given risk function -- namely, 
the difference between the risk function at a given pair (decision-rule, problem-data point) and its minimal value (over all decision rules) 
at the same problem-data point. In these works, robust decision procedures are sought which minimize the worst-case value of the regret function
(or approximation thereof) over the problem-data set (as done in [14] -- [16] in connection with estimation problems).

In spite of being an attractive concept, minimax regret problems are often difficult to solve, specially in problems involving dynamic
models, requiring either approximations of the regret function or restrictive assumptions on the problem set-up 
(see [17], for a brief discussion of this point). Thus, the motivation arises for pursuing other ways of obtaining less conservative robust decision procedures
(vis-\`a-vis a minimax procedure) which handle more directly the trade off between worst-case and ``point-wise'' performance. In this paper, this
theme is explored in connection with three closely-related discrete-time, linear-estimator design problems, namely, linear mean-squared error 
(MSE, for short) estimation involving non-parametric ``channel'' model
uncertainty ($\mathcal{H}_{2}-$balls of channel frequency-responses), nominal $\mathcal{H}_{\infty}$ estimation and robust $\mathcal{H}_{\infty}$ estimation with $\mathcal{H}_{\infty}$ model
uncertainty. In the case of nominal $\mathcal{H}_{\infty}$ estimation, the reduction of conservatism is connected with 
$\mathcal{H}_{2}-$balls of signals instead of ``channel'' frequency responses; whereas in the case of robust $\mathcal{H}_{\infty}$ 
estimation it pertains to both signals and channel models.

As the central issue here is to obtain, in a computationally-efficient way, minimax and other potentially  less  conservative  robust
estimators, for each of these problems the corresponding minimax  estimator -- design  problem is
cast as
a semi-definite programming problem (SDP, for short). Average cost  functionals are introduced for the MSE over $\mathcal{H}_{2}-$balls
of ``channel'' frequency-responses and for the (deterministic) estimation-error magnitudes over $\mathcal{H}_{2}-$balls of (information 
and noise) signals -- these cost functionals are derived as limits of averages taken over finite impulse-responses (FIRs, for short) of a given length, as their length grows 
unlimited. On the basis of the solutions to the minimax problems, classes of admissible linear estimators are defined to be used in the
formulation of ``average cost/worst-case constraint'' (a/w, for short) problems, -- \emph{i.e.}, minimization  with respect to admissible
estimators of an average cost functional under the constraint that the worst-case estimation error magnitude  (over the uncertain model 
or signal set) does not exceed a prescribed value. Invoking conditions obtained in the conversion of the minimax problems into SDPs, 
these a/w problems are also converted into SDPs. Simple numerical examples are then presented to illustrate the ``conservatism-reduction''
potential of the resulting estimators vis-\`{a}-vis the point-wise performance (over the uncertain sets) of the associated minimax 
estimators. 

It should be noted that robust estimation problems with $\mathcal{H}_{2}$ (MSE) and $\mathcal{H}_{\infty}$ estimation
criteria have attracted considerable attention (\emph{e.g.}, [5], [6], [8], [13], [22], [23], [25], [26], [29]--[31]) mainly in 
connection with parametric uncertainty. The nominal $\mathcal{H}_{\infty}$ estimation problem has also been successfully tackled 
(see, [19]--[21], [24] and its references).

These topics are revisited here mainly as set-ups for
exploring robust alternatives to minimax estimators. However, the robust estimation problems addressed here involve non-parametric
uncertainty (weighted $\mathcal{H}_{2}$ and $\mathcal{H}_{\infty}$ balls of frequency-responses). In addition, in the case of 
$\mathcal{H}_{\infty}$ filtering criteria considered here (as explained in greater
detail below), the radiuses of the ``information''
and noise signal balls are independently specified rather than being included (in a potentially conservative way) in a single ball of
larger radius. Finally, it is noted that the introduction of average costs over function balls in $\mathcal{H}_{2}$, the corresponding robust
estimator-design problems and their conversion to SDPs as well as the role of these estimators as less conservative alternatives to
minimax procedures have not been contemplated in the existing literature. 

This paper is organized as follows. In Section \ref{sec:2}, a linear estimation set-up is presented and the estimation problems treated here are explicitly formulated. In Section \ref{sec:3}, average performance criteria are
introduced in connection with $\mathcal{H}_{2}$ (MSE) robust estimation involving sets of possible ``channel'' models (frequency-responses) and with 
$\mathcal{H}_{\infty}$ nominal estimation involving $\mathcal{H}_{2}-$balls of signals. In Section \ref{sec:4}, minimax  problems for robust
$\mathcal{H}_{2}$ estimation and for nominal and robust $\mathcal{H}_{\infty}$ estimation are posed and converted into SDPs. In Section \ref{sec:5},
average cost/worst-case constraint problems are posed and also converted into SDPs. In Section \ref{sec:6}, the possible trade-offs between
worst-case and point-wise performance attainable with the a/w estimators are illustrated in simple examples. Unless otherwise stated, proofs are to be found in the Appendix.

\section{Background and Problem Formulation}\label{sec:2}

Consider the signal processing set-up of Figure 1 where the exogenous signals $\boldsymbol{y}$ and $\boldsymbol{v}$

pass through causal and stable, discrete-time, linear, multivariable filters 
$\mathcal{H}$ (channel), $\mathcal{H}_{_{\mathbf{I}}}$ (``reference'' filter, usually set to the identity or the $k$th--step delay) and $\mathcal{G}$ 
(estimator) with frequency-responses $\mathbf{H}\in \mathcal{R}_{c}^{m_{_{\boldsymbol{v}}}\times m_{_{\boldsymbol{y}}}}$, $\mathbf{H}_{_{\mathbf{I}}}\in \mathcal{R}_{c}^{m_{_{\boldsymbol{e}}}\times m_{_{\boldsymbol{y}}}}$ and
$\mathbf{G}\in \mathcal{R}_{c}^{m_{_{\boldsymbol{e}}}\times m_{_{\boldsymbol{v}}}}$.

\begin{figure}[h]
\begin{center}
\includegraphics[width=15cm]{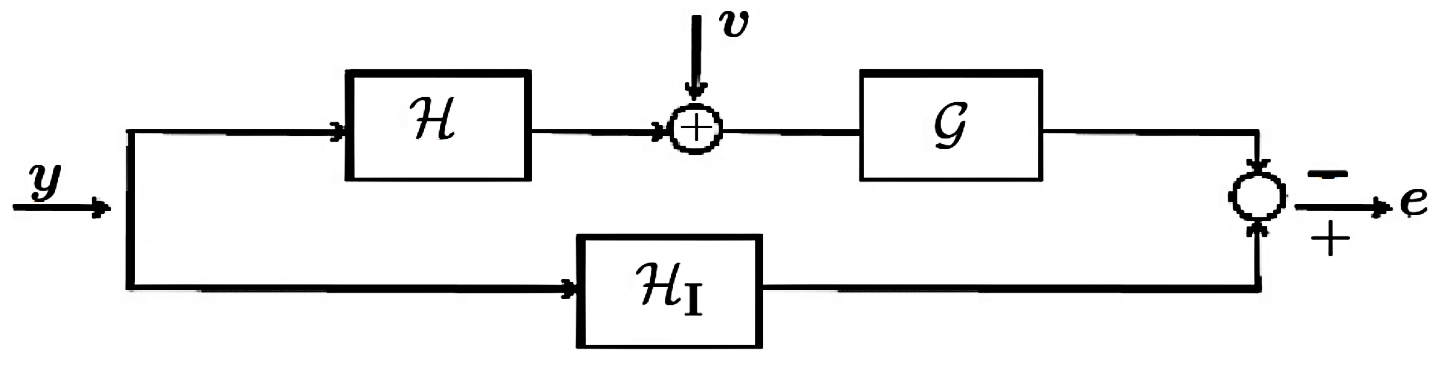}
\end{center}
\caption{Estimation set-up.}
\label{fig:01}
\end{figure}
\vspace{-1.0\baselineskip}

\subsection{$\mathcal{H}_{2}$ Estimation}\label{subsec:2.1}

In the first problem addressed in this paper, $\boldsymbol{y}$ and $\boldsymbol{v}$ are taken to be independent, wide-sense stationary,
discrete-time random processes with zero mean and power spectral densities\ 
$\boldsymbol{\Gamma}_{\boldsymbol{y}}\in \mathcal{R}_{_{0}}^{m_{_{\boldsymbol{y}}}\times m_{_{\boldsymbol{y}}}}$ and 
$\boldsymbol{\Gamma}_{\boldsymbol{v}}\in \mathcal{R}_{_{0}}^{m_{_{\boldsymbol{v}}}\times m_{_{\boldsymbol{v}}}}$. For a given set-up 
$(\mathbf{H}_{_{\mathbf{I}}}, \mathbf{H}, \boldsymbol{\Gamma}_{\boldsymbol{y}}, \boldsymbol{\Gamma}_{\boldsymbol{v}})$, the (nominal) performance of the
linear estimator defined by $\mathbf{G}$ is measured by the steady-state expected
value of $\boldsymbol{e}(t)^{^{\mathrm{T}}}\boldsymbol{e}(t)$, 
where $\boldsymbol{e}(t)$ denotes the estimation error signal, and is given by 
\begin{equation}\label{eq:01}
 \mathcal{J}(\mathbf{G}; \mathbf{H})\triangleq \left\langle(\mathbf{H}_{_{\mathbf{I}}}-\mathbf{G}\mathbf{H})\boldsymbol{\Gamma}_{\boldsymbol{y}},
 (\mathbf{H}_{_{\mathbf{I}}}-\mathbf{G}\mathbf{H})\right\rangle
 +\left\langle \mathbf{G}\boldsymbol{\Gamma}_{\boldsymbol{v}}, \mathbf{G} \right\rangle.
\end{equation}
(the connection between\ $\mathcal{J}(\mathbf{G}; \mathbf{H})$\ and\ $\boldsymbol{e}(t)$\ for finite\ $t$\ is briefly reviewed in the Appendix).

Accordingly, the estimation problem for a class $\mathcal{S}_{_{\mathbf{G}}}\subset \mathcal{R}_{c}^{m_{_{\boldsymbol{e}}}\times m_{_{\boldsymbol{v}}}}$ of
frequency-responses (defining the admissible estimators) is given by \ \ 
$\displaystyle\min_{\mathbf{G}\in \mathcal{S}_{_{\mathbf{G}}}}\mathcal{J}(\mathbf{G}; \mathbf{H})$. When channel model uncertainty is taken into account by
means of a set $\mathcal{S}_{_{\mathbf{H}}}\subset\mathcal{R}_{c}^{m_{_{\boldsymbol{v}}}\times m_{_{\boldsymbol{y}}}}$ of possible 
channel frequency-responses, the quality of a given $\mathbf{G}$ is usually assessed by its worst-case MSE over $\mathcal{S}_{_{\mathbf{H}}}$,
\emph{i.e.}, by\
$\bar{\mathcal{J}}(\mathbf{G}; \mathcal{S}_{_{\mathbf{H}}})\triangleq\sup\{\mathcal{J}(\mathbf{G}; \mathbf{H}): \mathbf{H}\in \mathcal{S}_{_{\mathbf{H}}}\}$\ \ and the corresponding
minimax estimator design problem is formulated as\
$\displaystyle\min_{\mathbf{G}\in \mathcal{S}_{_{\mathbf{G}}}}\bar{\mathcal{J}}(\mathbf{G}; \mathcal{S}_{_{\mathbf{H}}})$.

The major aim here is to introduce estimators which achieve better ``point-wise''\ performance (than that of
minimax estimators) over ``favourable'' subsets of $\mathcal{S}_{_{\mathbf{H}}}$ at the expense of a moderate increase in the resulting worst-case MSE
(over that of a minimax estimator). To this effect, estimation problems are posed in which a cost-functional is minimized under the constraint
that\ \ $\mathcal{J}(\mathbf{G}; \mathcal{S}_{_{\mathbf{H}}})$\ \ does not exceed a prescribed value, \emph{i.e.},
\begin{equation}\label{eq:01a0}
 \min_{\mathbf{G} \in \mathcal{S}_{_{\mathbf{G}}}}\ c(\mathbf{G}) \ \ \  \ \text{subject to}\ \ \ \ 
 \bar{\mathcal{J}}(\mathbf{G}; \mathcal{S}_{_{\mathbf{H}}})\leq (1+\alpha)\bar{\mathcal{J}}_{_{\mathrm{o}}},
\end{equation}
where\ \ $\bar{\mathcal{J}}_{_{\mathrm{o}}}= \inf\{\bar{\mathcal{J}}(\mathbf{G};\mathcal{S}_{_{\mathbf{H}}}):\mathbf{G} \in \mathcal{S}_{_{\mathbf{G}}})\}$, 
$\alpha>0$\ and \ $c(\cdot)$\  is a 
cost functional taking into account other properties of the MSE function
$\mathcal{J}(\mathbf{G}; \cdot ): \mathcal{S}_{_{\mathbf{H}}}\rightarrow \mathbb{R}$ other than its supremum.
 
The set $\mathcal{S}_{_{\mathbf{H}}}$ considered here is defined as a weighted $\mathcal{H}_{2}$--ball centered on the nominal, frequency-response\ \
$\mathbf{H}_{_{\boldsymbol{0}}}$, \emph{i.e.},
\begin{equation}\label{eq:01a}
\mathcal{S}_{_{\mathbf{H}}}\triangleq\left\{\mathbf{H}\in \mathcal{R}_{c}^{m_{_{\boldsymbol{v}}}\times m_{_{\boldsymbol{y}}}}: \|(\mathbf{H}-\mathbf{H}_{_{\boldsymbol{0}}})W\|_{_{2}}^{^{2}}\leq \gamma^{^{2}}\right\},
\end{equation}
where the weighting function\ \  $W\in \mathcal{R}_{c}^{m_{_{\boldsymbol{y}}}\times m_{_{\boldsymbol{y}}}}$ is such that \ \ $W^{^{-1}}\in \mathcal{R}_{c}^{m_{_{\boldsymbol{y}}}\times m_{_{\boldsymbol{y}}}}$.

\vspace*{3mm}

\begin{remark}
 $\mathcal{H}_{\infty}$--uncertainty on the channel frequency-response can also be cast (albeit conservatively) as $\mathcal{S}_{_{\mathbf{H}}}$ above with an 
 appropriate choice of the set-up data. Indeed, given
$$\check{\mathcal{S}}_{_{\mathbf{H}}}=\left\{\mathbf{H}\in \mathcal{R}_{c}^{m_{_{\boldsymbol{v}}}\times m_{_{\boldsymbol{y}}}}: \|(\mathbf{H}-\mathbf{H}_{_{\boldsymbol{0}}})\check{W}\|_{\infty}\leq \check{\gamma}\right\},$$
where\ \  $\check{W}$\ and \ $\check{W}^{^{-1}}\in \mathcal{R}_{c}^{m_{_{\boldsymbol{y}}}\times m_{_{\boldsymbol{y}}}}$, note that $\mathbf{H}\boldsymbol{\phi}_{_{\boldsymbol{y}}}=\mathbf{H}_{_{\boldsymbol{0}}}\boldsymbol{\phi}_{_{\boldsymbol{y}}}+(\mathbf{H}-\mathbf{H}_{_{\boldsymbol{0}}})\check{W}(\check{W}^{^{-1}}\boldsymbol{\phi}_{_{\boldsymbol{y}}})$ and, hence, 
$$\forall\ \mathbf{H}\in\check{\mathcal{S}}_{_{\mathbf{H}}}, \forall \alpha \in[0, 2\pi],\left\|\left[\left(\mathbf{H}-\mathbf{H}_{_{\boldsymbol{0}}}\right)\boldsymbol{\phi}_{_{\boldsymbol{y}}}\right](e^{j\alpha})\right\|_{_{F}}^{^{2}}\leq \check{\gamma}^{^{2}}\left\|\left[\check{W}^{^{-1}}\boldsymbol{\phi}_{_{\boldsymbol{y}}}\right](e^{j\alpha})\right\|_{_{F}}^{^{2}}=\check{\gamma}^{^{2}}|\boldsymbol{\phi}_{_{\boldsymbol{y}W}}(e^{j\alpha})|^{^{2}},$$
where\ \ $\boldsymbol{\phi}_{_{\boldsymbol{y}W}}(e^{j\alpha})^{*}\boldsymbol{\phi}_{_{\boldsymbol{y}W}}(e^{j\alpha})=\operatorname{tr}\left\{\left[\check{W}^{^{-1}}\boldsymbol{\phi}_{_{\boldsymbol{y}}}\right](e^{j\alpha})^{*}
\left[\check{W}^{^{-1}}\boldsymbol{\phi}_{_{\boldsymbol{y}}}\right](e^{j\alpha})\right\}$.

Thus, whenever the function in the right-hand side of the last equation does not have zeros on\break $\left\{e^{j\alpha}: \alpha\in[0,2\pi]\right\}$,\ $\boldsymbol{\phi}_{_{\boldsymbol{y}W}}$\  can be taken to be a spectral factor and, hence, $\forall\ \mathbf{H}\in\check{\mathcal{\alpha}}\in[0,2\pi]$
$$\left\|\left[\left(\mathbf{H}-\mathbf{H}_{_{\boldsymbol{0}}}\right)\boldsymbol{\phi}_{_{\boldsymbol{y}}}\boldsymbol{\phi}_{_{\boldsymbol{y}W}}^{^{-1}}\right](e^{j\alpha})\right\|_{_{F}}^{^{2}}\ \ \Rightarrow\ \ \left\|\left[\left(\mathbf{H}-\mathbf{H}_{_{\boldsymbol{0}}}\right)\boldsymbol{\phi}_{_{\boldsymbol{y}}}\boldsymbol{\phi}_{_{\boldsymbol{y}W}}^{^{-1}}\right](e^{j\alpha})\right\|_{_{2}}^{^{2}}.$$

Thus, given\  $\check{W}$\  and\  $\check{\gamma}$, taking a spectral factorization
$$W(e^{j\alpha})W(e^{j\alpha})^{*}=\left[\boldsymbol{\phi}_{_{\boldsymbol{y}}}\boldsymbol{\phi}_{_{\boldsymbol{y}W}}^{^{-1}}\right](e^{j\alpha})[\boldsymbol{\phi}_{_{\boldsymbol{y}}}\boldsymbol{\phi}_{_{\boldsymbol{y}W}}](e^{j\alpha})^{*}$$
and making $\gamma=\check{\gamma}$, it follows that $\check{\mathcal{S}}_{_{\mathbf{H}}}\subset
 \mathcal{S}_{_{\mathbf{H}}}$.\hfill$\nabla$
\end{remark}

\vspace*{3mm}

For $\mathcal{S}_{_{\mathbf{H}}}$ as in (\ref{eq:01a}), in addition to the minimax estimator defined by\ \underline{\emph{Prob.}\ $1$:} 
$\displaystyle\min_{\mathbf{G}\in \mathcal{S}_{_{\mathbf{G}}}}\ \bar{\mathcal{J}}(\mathbf{G}; \mathcal{S}_{_{\mathbf{H}}})$,\ where\ \ 
$\mathcal{S}_{_{\mathbf{G}}}\subset\mathcal{R}_{c}^{m_{_{\boldsymbol{v}}}\times m_{_{\boldsymbol{y}}}}$, \ another robust
estimator will be considered
which is defined as in (\ref{eq:01a0}) above with\ $c(\mathbf{G})$\ defined as a limit of the average values of \ $\mathcal{J}(\mathbf{G}; \cdot)$\ over classes of FIRs of increasing 
length -- the latter is derived in Subsection \ref{subsec:3.1}.

\subsection{Nominal $\mathcal{H}_{\infty}-$Filtering}\label{nom-H_inf-filtering}\label{subsec:2.2}
The second estimation problem considered here is a nominal, ``$\mathcal{H}_{\infty}-$filtering'' problem in which the class of admissible
estimators corresponds to linear systems with a prescribed maximum state-space dimension - $\mathcal{H}_{\infty}$ filtering problems have been
widely considered (see, for example, [19] -- [25] and references therein) and motivation for such problems vis-\`{a}-vis $\mathcal{H}_{2}$ 
filtering is briefly discussed in [25].

More specifically, consider the block-diagram of Figure \ref{fig:01} and, for $\boldsymbol{y}\in \mathcal{R}_{c}^{m_{_{\boldsymbol{y}}}}$ and 
$\boldsymbol{v}\in \mathcal{R}_{c}^{m_{_{\boldsymbol{v}}}}$, 
let 
$$\boldsymbol{e}(\boldsymbol{z}; \mathbf{G}, \mathbf{H}_{_{\boldsymbol{0}}})=\mathbf{H}_{_{\mathbf{I}}}\boldsymbol{y}-(\mathbf{G}\mathbf{H}_{_{\boldsymbol{0}}} \boldsymbol{y}+\mathbf{G}\boldsymbol{v})=(\mathbf{H}_{_{\mathbf{I}\boldsymbol{0}}}-\mathbf{G}\mathbf{H}_{_{\boldsymbol{0}\mathbf{I}}})\boldsymbol{z},$$
where\ $\mathbf{H}_{_{\boldsymbol{0}}}$\ is the ``nominal'' frequency-response,\  $\mathbf{H}_{_{\mathbf{I}\boldsymbol{0}}}=[\mathbf{H}_{_{\mathbf{I}}}\ \vdots\ \boldsymbol{0}_{m_{_{\boldsymbol{e}}}\times m_{_{\boldsymbol{v}}}}]$,\ $\mathbf{H}_{_{\boldsymbol{0}\mathbf{I}}}=[\mathbf{H}_{_{\boldsymbol{0}}}\ \vdots\ \mathbf{I}_{m_{_{\boldsymbol{v}}}}]$\ and\ 
$\boldsymbol{z}^{^{\mathrm{T}}}=[\boldsymbol{y}^{^{\mathrm{T}}}\ \vdots\ \boldsymbol{v}^{^{\mathrm{T}}}]$.

Let\ \ $S_{_{\boldsymbol{\alpha}}}=\{\boldsymbol{\alpha}\in \mathcal{R}_{c}^{m_{_{\boldsymbol{\alpha}}}}:
\|W_{_{\boldsymbol{\alpha}}}\boldsymbol{\alpha}\|_{_{2}}\leq \gamma_{_{\boldsymbol{\alpha}}}\}$, \ \
$\boldsymbol{\alpha}=\boldsymbol{y}, \boldsymbol{v}$, and the weighting function, \ $W_{_{\boldsymbol{\alpha}}}\in \mathcal{R}_{c}^{m_{\boldsymbol{\alpha}}\times m_{\boldsymbol{\alpha}}}$\ \ be such that\ \ $W_{_{\boldsymbol{\alpha}}}^{^{-1}}\in \mathcal{R}_{c}^{m_{\boldsymbol{\alpha}}\times m_{\boldsymbol{\alpha}}}$. The  ``$\mathcal{H}_{\infty}$''
filtering criterion $\mathcal{J}_{\infty}$ considered here is then defined as 
$$\mathcal{J}_{\infty}(\mathbf{G}; \mathbf{H}_{_{\boldsymbol{0}}})\triangleq\sup\{\|\boldsymbol{e}(\boldsymbol{z};
\mathbf{G}, \mathbf{H}_{_{\boldsymbol{0}}})\|_{_{2}}^{^{2}}: \boldsymbol{z}^{^{\mathrm{T}}}=[\boldsymbol{y}^{^{\mathrm{T}}}\ \vdots \ \boldsymbol{v}^{^{\mathrm{T}}}], 
\boldsymbol{y}\in S_{\boldsymbol{y}}, \boldsymbol{v}\in S_{\boldsymbol{v}}\}.$$

To simplify a little the derivation to follow, let\ $\bar{\boldsymbol{y}}=W_{_{\boldsymbol{y}}}\boldsymbol{y}$,\ 
$\bar{\boldsymbol{v}}=W_{_{\boldsymbol{v}}}\boldsymbol{y}$,\ $W_{_{\boldsymbol{z}}}=\operatorname{diag}(W_{_{\boldsymbol{y}}}, W_{_{\boldsymbol{v}}})$,\ 
$\bar{\boldsymbol{z}}=W_{_{\boldsymbol{z}}}\boldsymbol{z}$\ \ and\ \ 
$\bar{\boldsymbol{e}}(\bar{\boldsymbol{z}}; \mathbf{G}, \mathbf{H})=(\mathbf{H}_{_{\mathbf{I}\boldsymbol{y}}}-\mathbf{G}\mathbf{H}_{_{\boldsymbol{0}\boldsymbol{z}}})\bar{\boldsymbol{z}}$,\
where\ \ \ $\mathbf{H}_{_{\mathbf{I}\boldsymbol{y}}} \triangleq \mathbf{H}_{_{\mathbf{I}\boldsymbol{0}}}W_{\boldsymbol{z}}^{^{-1}}=[\mathbf{H}_{_{\mathbf{I}}}W_{\boldsymbol{y}}^{^{-1}}\ \vdots\ \boldsymbol{0}_{m_{_{\boldsymbol{e}}}\times m_{_{\boldsymbol{v}}}}]$\ \ \
and\ $\mathbf{H}_{_{\boldsymbol{0}\boldsymbol{z}}}\triangleq\mathbf{H}_{_{\boldsymbol{0}\mathbf{I}}}W_{\boldsymbol{z}}^{^{-1}}=[\mathbf{H}_{_{\boldsymbol{0}}}W_{\boldsymbol{y}}^{^{-1}}\ \vdots\ W_{\boldsymbol{v}}^{^{-1}}]$. Then, 
for
\ \ $\bar{S}_{_{\boldsymbol{\alpha}}}=\{\bar{\boldsymbol{\alpha}}\in \mathcal{R}_{c}^{m_{_{\boldsymbol{\alpha}}}}:
\|\bar{\boldsymbol{\alpha}}\|_{_{2}}\leq \gamma_{_{\boldsymbol{\alpha}}}\}$, \  $\mathcal{J}_{\infty}$\  can be rewritten as 
$$\mathcal{J}_{\infty}(\mathbf{G}; \mathbf{H}_{_{\boldsymbol{0}}})=\sup\{\|\bar{\boldsymbol{e}}(\bar{\boldsymbol{z}}; \mathbf{G}, \mathbf{H}_{_{\boldsymbol{0}}})\|_{_{2}}^{^{2}}:
\bar{\boldsymbol{z}}^{^{\mathrm{T}}}=[\bar{\boldsymbol{y}}^{^{\mathrm{T}}}\ \vdots\
\bar{\boldsymbol{v}}^{^{\mathrm{T}}}], \ \bar{\boldsymbol{y}} \in \bar{S}_{\boldsymbol{y}},\ \bar{\boldsymbol{v}} \in \bar{S}_{\boldsymbol{v}}\}.$$

\vspace*{3mm}

\begin{remark}
 Note that the $\mathcal{H}_{\infty}-$norm of the error system\ $(\mathbf{H}_{_{\mathbf{I}\boldsymbol{y}}}-\mathbf{G}\mathbf{H}_{_{\boldsymbol{0}\boldsymbol{z}}})$\ is given by\break
 $\sup\{\|\bar{\boldsymbol{e}}(\bar{\boldsymbol{z}}; \mathbf{G}, \mathbf{H})\|_{_{2}}^{^{2}}: 
 \bar{\boldsymbol{z}} \in \mathcal{R}_{c}^{m_{_{\boldsymbol{y}}} + m_{_{\boldsymbol{v}}}}, \ \|\bar{\boldsymbol{z}}\|_{_{2}}\leq 1\}$. Thus, the criterion \
 $\mathcal{J}_{\infty}$ \ introduced here differs from the usual\ \ $\mathcal{H}_{\infty}$  in that the\ $\mathcal{H}_{_{2}}$\ norms of the 
 ``information'' $(\boldsymbol{y})$ and noise $(\boldsymbol{v})$
signals are independently bounded, rather than having their squared sum subject to a single upper bound -- the alternative pursued here 
appears to be more natural in the signal processing set-up of Figure \ref{fig:01}. \hfill $\nabla$
\end{remark}
\vspace*{3mm}

For a given\ \ $\mathbf{H}\in \mathcal{R}_{c}^{m_{_{\boldsymbol{v}}}\times m_{_{\boldsymbol{y}}}}$\ \ and\ \
$\mathcal{S}_{_{\mathbf{G}}}^{^{\mathrm{o}}}\subset\mathcal{R}^{m_{_{\boldsymbol{e}}}\times m_{_{\boldsymbol{v}}}}$ the nominal 
``$\mathcal{H}_{\infty}-$estimation'' problem is then posed as follows
$$\text{\underline{\emph{Prob. $2$}}:}\
\displaystyle\min_{\mathbf{G}\in \mathcal{S}_{_{\mathbf{G}}}^{\mathrm{o}}}\ \mathcal{J}_{\infty}(\mathbf{G}; \mathbf{H}).$$

\subsection{Robust $\mathcal{H}_{\infty}$ Estimation}\label{subsec:2.3}
The third estimator-design problem tackled here is a robust ``$\mathcal{H}_{\infty}-$estimation'' problem in which the class of possible 
channel models is defined by weighted $\mathcal{H}_{\infty}-$balls of frequency responses. More specifically, let a nominal model \ \ 
$\mathbf{H}_{_{\boldsymbol{0}}}\in \mathcal{R}_{c}^{m_{_{\boldsymbol{v}}}\times m_{_{\boldsymbol{y}}}}$\ \ and the class $\mathcal{S}_{_{\mathbf{H}\infty}}$ be
given where
$$\mathcal{S}_{_{\mathbf{H}\infty}}\triangleq\{\mathbf{H}\in \mathcal{R}_{c}^{m_{_{\boldsymbol{v}}} \times m_{_{\boldsymbol{y}}}}: 
\|(\mathbf{H}-\mathbf{H}_{_{\boldsymbol{0}}})W_{_{\mathbf{H}}}\|_{\infty}\leq \gamma_{_{\mathbf{H}}}\},
\ W_{_{\mathbf{H}}}\in \mathcal{R}_{c}^{m_{_{\boldsymbol{y}}}\times m_{_{\boldsymbol{y}}}}\ \ \text{is such that}\ \
W_{_{\mathbf{H}}}^{^{-1}}\in \mathcal{R}_{c}^{m_{_{\boldsymbol{y}}}\times m_{_{\boldsymbol{y}}}}.$$


The worst-case \ ``$\mathcal{H}_{\infty}-$performance'' of a given estimator\ $\mathbf{G}$\ over\ $\mathcal{S}_{_{\mathbf{H}\infty}}$\ is
given by\break
$\sup\{\mathcal{J}_{\infty}(\mathbf{G}; \mathbf{H}): \mathbf{H}\in \mathcal{S}_{_{\mathbf{H}\infty}}\}$,\ \ or, equivalently, 
\begin{equation}\label{eq:32}
 \sup\left\{\|\bar{\boldsymbol{e}}(\bar{\boldsymbol{z}};  \mathbf{G}, \mathbf{H})\|_{_{2}}^{^{2}}:
 \bar{\boldsymbol{z}}^{^{\mathrm{T}}}= \begin{bmatrix} \bar{\boldsymbol{y}}^{^{\mathrm{T}}}\ \vdots \bar{\boldsymbol{v}}^{^{\mathrm{T}}} \end{bmatrix},\ 
 \bar{\boldsymbol{y}}\in \bar{\mathcal{S}}_{\boldsymbol{y}},\ \bar{\boldsymbol{v}}\in \bar{\mathcal{S}}_{\boldsymbol{v}},\
 \mathbf{H}\in \mathcal{S}_{_{\mathbf{H}\infty}}\right\}.
\end{equation}

Rewriting for\ \ $\mathbf{H}_{_{\boldsymbol{\delta}}}\triangleq(\mathbf{H}-\mathbf{H}_{_{\boldsymbol{0}}})$,\ \
$\bar{\boldsymbol{e}}(\bar{\boldsymbol{z}}; \mathbf{G}, \mathbf{H})
=(\mathbf{H}_{_{\mathbf{I}\boldsymbol{y}}}-\mathbf{G}\mathbf{H}_{\mathrm{o}\boldsymbol{z}})\bar{\boldsymbol{z}}
-\mathbf{G}\mathbf{H}_{_{\boldsymbol{\delta}}}W_{\boldsymbol{y}}^{^{-1}}\bar{\boldsymbol{y}}$\ \ and defining\ \ 
$\boldsymbol{w}\triangleq\mathbf{H}_{_{\boldsymbol{\delta}}}W_{\boldsymbol{y}}^{^{-1}}\bar{\boldsymbol{y}}$,\ \ 
$\bar{\boldsymbol{z}}_{a}^{^{\mathrm{T}}}=\begin{bmatrix}\bar{\boldsymbol{z}}^{^{\mathrm{T}}}\ \vdots\ \boldsymbol{w}^{^{\mathrm{T}}}\end{bmatrix}$,\ \
it follows that\ \
$\bar{\boldsymbol{e}}(\bar{\boldsymbol{z}}; \mathbf{G}, \mathbf{H})=
\boldsymbol{e}(\boldsymbol{z}_{a}; \mathbf{G})\triangleq(\mathbf{H}_{_{\mathbf{I}a}}-\mathbf{G}\mathbf{H}_{_{\mathrm{o}a}})\boldsymbol{z}_{a}$,\ \  where \ \ 
$\mathbf{H}_{_{\mathbf{I}a}}\triangleq
\begin{bmatrix}\mathbf{H}_{_{\mathbf{I}}}W_{\boldsymbol{y}}^{^{-1}}\ \vdots\ \boldsymbol{0}_{m_{_{\boldsymbol{e}}}\times m_{_{\boldsymbol{v}}}}\ \vdots\ 
\boldsymbol{0}_{m_{_{\boldsymbol{e}}}\times m_{_{\boldsymbol{v}}}}\end{bmatrix}$, \ \ and
$\mathbf{H}_{_{\mathrm{o}a}}\triangleq
\begin{bmatrix}\mathbf{H}_{_{\boldsymbol{0}}}W_{\boldsymbol{y}}^{^{-1}}\ \vdots\ W_{\boldsymbol{v}}^{^{-1}}\ \vdots\ \mathbf{I}_{m_{_{\boldsymbol{v}}}}\end{bmatrix}$, \ \ so that
(\ref{eq:32}) can be rewritten as \ \ 
\begin{equation}\label{eq:33}
\begin{split}
&\sup\left\{\|\boldsymbol{e}(\boldsymbol{z}_{a}; \mathbf{G})\|_{_{2}}^{^{2}}:
\boldsymbol{z}_{a}^{^{\mathrm{T}}}=\begin{bmatrix}\bar{\boldsymbol{y}}^{^{\mathrm{T}}}\ \vdots \bar{\boldsymbol{v}}^{^{\mathrm{T}}}\ \vdots\ \boldsymbol{w}\end{bmatrix}\right., \ 
\bar{\boldsymbol{v}} \in \bar{\mathcal{S}}_{\boldsymbol{v}}, \ (\bar{\boldsymbol{y}}, \boldsymbol{w})\ \ \text{is such that}\ \ 
\bar{\boldsymbol{y}}\in \bar{\mathcal{S}}_{\boldsymbol{y}}\\
& \ \ \ \ \ \ \ \  \ \ \ \text{and} \
\left.\exists \mathbf{H}_{_{\boldsymbol{\delta}}}\in \mathcal{R}_{c}^{m_{_{\boldsymbol{v}}}\times m_{_{\boldsymbol{y}}}}\ \ \text{such that} \ \ 
\|\mathbf{H}_{_{\boldsymbol{\delta}}}W_{_{\mathbf{H}}}\|_{\infty}\leq \gamma_{_{\mathbf{H}}}\ \ \text{and}\ \ 
\boldsymbol{w}=\mathbf{H}_{_{\boldsymbol{\delta}}}W_{\boldsymbol{y}}^{^{-1}}\bar{\boldsymbol{y}} \right\}.
\end{split}
\end{equation}

To obtain a more tractable optimization problem, the worst-case \ $\mathcal{H}_{\infty}-$performance index given by (\ref{eq:33}) will be replaced by an upper bound. To
this effect, note that\ $\boldsymbol{w}=\mathbf{H}_{_{\boldsymbol{\delta}}}W_{\boldsymbol{y}}^{^{-1}}\bar{\boldsymbol{y}}=
(\mathbf{H}_{_{\boldsymbol{\delta}}}W_{_{\mathbf{H}}})W_{_{\mathbf{H}\boldsymbol{y}}}\bar{\boldsymbol{y}}$, \ \ where\ \
$W_{_{\mathbf{H}\boldsymbol{y}}}\triangleq(W_{\boldsymbol{y}}W_{_{\mathbf{H}}})^{^{-1}}$, \ and, hence, for any pair\ $(\bar{\boldsymbol{y}}, \boldsymbol{w})$\ as in
(\ref{eq:33}), \ $\bar{\boldsymbol{y}}\in \bar{\mathcal{S}}_{y}$\ \ and \ \
$\|\boldsymbol{w}\|_{_{2}}^{^{2}}\leq \gamma_{_{\mathbf{H}}}^{^{2}}\|W_{_{\mathbf{H}\boldsymbol{y}}}\bar{\boldsymbol{y}}\|_{_{2}}^{^{2}}$. This observation leads to the following
upper bound on\ \  $\sup\{\mathcal{J}_{\infty}(\mathbf{G};\mathbf{H}): \mathbf{H}\in \mathcal{S}_{_{\mathbf{H}\infty}}\}$:
$$
\mathcal{J}_{\infty}^{a}(\mathbf{G})\triangleq\sup\left\{\|\boldsymbol{e}(\boldsymbol{z}_{a}; \mathbf{G})\|_{_{2}}^{^{2}}:
\boldsymbol{z}_{a}^{^{\mathrm{T}}}=\begin{bmatrix}\bar{\boldsymbol{y}}^{^{\mathrm{T}}}\ \vdots \bar{\boldsymbol{v}}^{^{\mathrm{T}}}\ \vdots\ \boldsymbol{w}\end{bmatrix}, \ 
\bar{\boldsymbol{v}} \in \bar{\mathcal{S}}_{\boldsymbol{v}}, \ 
\bar{\boldsymbol{y}}\in \bar{\mathcal{S}}_{\boldsymbol{y}}\ \text{and} \
\|\boldsymbol{w}\|_{_{2}}^{^{2}}\leq \gamma_{_{\mathbf{H}}}^{^{2}}\|W_{_{\mathbf{H}\boldsymbol{y}}}\bar{\boldsymbol{y}}\|_{_{2}}^{^{2}}
\right\}$$
and, for a given class\ $\mathcal{S}_{_{\mathbf{G}}}^{a} \subset \mathcal{R}^{m_{_{\boldsymbol{e}}}\times m_{_{\boldsymbol{v}}}}$ of admissible estimators, the corresponding robust estimator design problem is posed as
$$\text{\underline{\emph{Prob. $3$}}:}\
\displaystyle\min_{\mathbf{G}\in  \mathcal{S}_{_{\mathbf{G}}}^{a}}\ \mathcal{J}_{\infty}^{a}(\mathbf{G}).$$

\subsection{Average Cost/Worst-Case Constraint Problem}\label{subsec:2.4}

As \emph{Prob. $1-3$} are minimax problems, an alternative formulation for each of these problems is sought which allow for obtaining trade-offs between worst-case and 
point-wise performance in specific estimator-design exercises (leading to \emph{Prob.} $3-6$, respectively). To this effect, average criteria (say, $\boldsymbol{\eta}_{_{a\boldsymbol{v}}}$ and 
$\boldsymbol{\eta}_{_{a}}$) associated to\ $\mathcal{J}(\mathbf{G}; \cdot):\mathcal{S}_{_{\mathbf{H}}}\rightarrow \mathbb{R}$\ \ and\ \ 
$\|\bar{\boldsymbol{e}}(\cdot; \mathbf{G}, \mathbf{H})\|_{_{2}}^{^{2}}:
\bar{\mathcal{S}}_{\boldsymbol{y}}\times\bar{\mathcal{S}}_{\boldsymbol{v}}\rightarrow\mathbb{R}$ are derived in the next section. Then, 
on the basis of approximate solutions to the minimax problems, linear classes of frequency-responses for admissible estimators are defined and average cost/worst-case constraint  (``a/w'', for short) estimation problems are posed. More specifically, let\ \ $\bar{\mathcal{J}}_{_{\mathrm{o}}}$\ \ denote the 
optimal value of \emph{Prob. $1$}, \emph{i.e.},\ \
$\bar{\mathcal{J}}_{_{\mathrm{o}}}= \inf\{\bar{\mathcal{J}}(\mathbf{G}; \mathcal{S}_{_{\mathbf{H}}}):\mathbf{G}\in\mathcal{S}_{_{\mathbf{G}}}\}$\ \
and let\ \ $\mathbf{G}_{_{_{1}}}^{^{\mathrm{o}}}$\ \ denote an approximate solution of \emph{Prob. $1$}, \emph{i.e.},\ \
$\mathbf{G}_{_{_{1}}}^{^{\mathrm{o}}} \in \mathcal{S}_{_{\mathbf{G}}}$\ \ and\ \
$\bar{\mathcal{J}}(\mathbf{G}_{_{_{1}}}^{^{\mathrm{o}}}; \mathcal{S}_{\mathbf{H}})=(1+\varepsilon)\bar{\mathcal{J}}_{_{\mathrm{o}}}$\ \ for a
``small'' $\varepsilon>0$.\ \ For a minimal realization\ \ $(\mathbf{A}_{_{_{1}}}^{^{\mathrm{o}}}, \mathbf{B}_{_{_{1}}}^{^{\mathrm{o}}}, \mathbf{C}_{_{_{1}}}^{^{\mathrm{o}}},
\mathbf{D}_{_{_{1}}}^{^{\mathrm{o}}})$\ \ of\ \ $\mathbf{G}_{_{_{1}}}^{^{\mathrm{o}}}$\ \ let
$$\mathcal{S}_{_{\mathbf{G}\boldsymbol{\eta}}}=\{\mathbf{G}
=\mathbf{D}+\mathbf{C}\mathbf{Y}_{_{_{1}}}^{^{\mathrm{o}}}\mathbf{B}_{_{_{1}}}^{^{\mathrm{o}}}:
\mathbf{C}\in\mathbb{R}^{m_{_{\boldsymbol{e}}}\times n_{_{_{1}}}^{^{\mathrm{o}}}}\ \ \text{and}\ \ \mathbf{D}\in\mathbb{R}^{m_{_{\boldsymbol{e}}}\times m_{_{\boldsymbol{v}}}}\},$$
where\ \ $\mathbf{Y}_{_{_{1}}}^{^{\mathrm{o}}}(e^{j\phi})=(e^{j\phi}\mathbf{I}-\mathbf{A}_{_{_{1}}}^{^{\mathrm{o}}})^{^{-1}}$,\
$\mathbf{A}_{_{_{1}}}^{^{\mathbf{o}}}\in \mathbb{R}^{n_{_{_{1}}}^{^{\mathrm{o}}}\times n_{_{_{1}}}^{^{\mathrm{o}}}}$\ \ (note that\ 
$\mathbf{G}_{_{_{1}}}^{^{\mathrm{o}}}\in \mathcal{S}_{_{\mathbf{G}\boldsymbol{\eta}}}$).

A robust estimation problem is then posed as follows:
$$\text{\underline{\emph{Prob. $4$}}:}\ \min_{\mathbf{G}\in \mathcal{S}_{_{\mathbf{G}\boldsymbol{\eta}}}}\boldsymbol{\eta}(\mathbf{G})\ \ \
\text{subject to}\ \ \ \bar{\mathcal{J}}_{\infty}(\mathbf{G})\leq (1+\alpha)\bar{\mathcal{J}}_{_{\mathrm{o}}},\ \ $$
where $\alpha>\varepsilon$\ \ -- note that in \emph{Prob. $4$} the worst-case performance of an estimator given by\ \ $\mathbf{G}$\ \ (\emph{i.e.},\ \ 
$\bar{\mathcal{J}}_{\infty}(\mathbf{G})$)\ \ is allowed to be bigger than the minimum one ($\bar{\mathcal{J}}_{_{\mathrm{o}}}$) by at the
most\ \ $\alpha\bar{\mathcal{J}}_{_{\mathrm{o}}}$.

 Similar considerations are brought to bear on \emph{Prob. $2$} and $3$ thereby leading to corresponding ``a/w'' estimation problems 
 (\emph{Prob. $5$} and $6$). In Section \ref{sec:5}, \emph{Prob. $4 - 6$} are also converted into SDPs.

\section{Average Performance Criteria}\label{sec:3}

\subsection{Average MSE over $\mathcal{H}_{2}-$Balls of Frequency-Responses}\label{subsec:3.1}

In this case, the uncertain model class is\ \ $\mathcal{S}_{_{\mathbf{H}}}$ and the estimation criteria is\ \ 
$\mathcal{J}(\mathbf{G};\mathbf{H})$.

To simplify the derivation to follow, a simple change-of-variable is introduced replacing  $\mathbf{H}$ by $\mathbf{X}$, where
$\mathbf{X}\triangleq(\mathbf{H}-\mathbf{H}_{_{\boldsymbol{0}}})W$.\ Thus,
$\mathcal{J}(\mathbf{G}; \mathbf{H})=\mathcal{J}_{_{\mathbf{X}}}(\mathbf{G}; \mathbf{X})$\ \ and \ \ 
\begin{equation}\label{eq:02}
\bar{\mathcal{J}}(\mathbf{G}; \mathcal{S}_{_{\mathbf{H}}})=\bar{\mathcal{J}}_{_{\mathbf{X}}}(\mathbf{G}; \mathcal{S}_{_{\mathbf{X}}}),
\end{equation}
where
\begin{equation}\label{eq:03}
 \mathcal{J}_{_{\mathbf{X}}}(\mathbf{G}; \mathbf{X})= \left\langle \left(\mathbf{X}_{_{\boldsymbol{0}}}(\mathbf{G})-\mathbf{G}\mathbf{X}\right) \boldsymbol{\Gamma}_{_{\boldsymbol{y}1}}, \mathbf{X}_{_{\boldsymbol{0}}}(\mathbf{G})
 -\mathbf{G}\mathbf{X}\right\rangle+\left\langle \mathbf{G}\boldsymbol{\Gamma}_{\boldsymbol{v}}, \mathbf{G}\right\rangle,
\end{equation}
\begin{equation}\label{eq:04}
 \mathbf{X}_{_{\boldsymbol{0}}}(\mathbf{G})\triangleq(\mathbf{H}_{_{\mathbf{I}}}-\mathbf{G}\mathbf{H}_{_{\boldsymbol{0}}})W,\ \ \ 
 \bar{\mathcal{J}}_{_{\mathbf{X}}}(\mathbf{G}; \mathcal{S}_{_{\mathbf{X}}})\triangleq\sup\left\{\mathcal{J}_{_{\mathbf{X}}}(\mathbf{G}; \mathbf{X}): \mathbf{X} \in \mathcal{S}_{_{\mathbf{X}}}\right\},
 \end{equation}
\begin{equation}\label{eq:05}
 \mathcal{S}_{_{\mathbf{X}}}\triangleq\left\{\mathbf{X}: \mathcal{R}_{c}^{m_{_{\boldsymbol{v}}}\times m_{_{\boldsymbol{y}}}}: \|\mathbf{X}\|_{_{2}}\leq \gamma\right\}\ \ \ \text{and} \ \ \ 
 \boldsymbol{\Gamma}_{_{\boldsymbol{y}1}}=W^{^{-1}}\boldsymbol{\Gamma}_{\boldsymbol{y}}(W^{^{-1}})^{*}.
\end{equation}

To derive an average performance criterion to assess a given\ \ $\mathbf{G}$\ over \ $\mathcal{S}_{_{\mathbf{X}}}$\ \ (or, equivalently over \ \
$\mathcal{S}_{_{\mathbf{H}}}$) a 
family of subspaces \  $\mathcal{S}_{_{\mathbf{X}}}^{^{N}}\subset \mathcal{R}_{c}^{m_{_{\boldsymbol{v}}}\times m_{_{\boldsymbol{y}}}}$ is considered which is such that for any 
$\mathbf{X}\in \mathcal{R}_{c}^{m_{_{\boldsymbol{v}}}\times m_{_{\boldsymbol{y}}}}$,\ \ $\inf\left\{\|\mathbf{X}-\widehat{\mathbf{X}}\|_{_{2}}: \widehat{\mathbf{X}}\in \mathcal{S}_{_{\mathbf{X}}}^{^{N}}\right\}\rightarrow 0$\
\ 
as \ \ $N\rightarrow\infty$, \ \ namely, 
$$\mathcal{S}_{_{\mathbf{X}}}^{^{N}}\triangleq\left\{\mathbf{X}_{_{N}}(\boldsymbol{\beta})\in \mathcal{R}_{c}^{m_{_{\boldsymbol{v}}}
\times m_{_{\boldsymbol{y}}}}: \mathbf{X}_{_{N}}(\boldsymbol{\beta})=\boldsymbol{\beta}\check{\mathbf{Y}}_{_{\mathbf{X}}}^{^{N}},
\boldsymbol{\beta} \in \mathbb{R}^{m_{_{\boldsymbol{v}}}\times(n_{_{\mathbf{X}}}^{^{N}}+m_{_{\boldsymbol{y}}})}\right\},$$
where 
$\check{\mathbf{Y}}_{_{\mathbf{X}}}^{^{N}}\triangleq\begin{bmatrix}
                              \mathbf{Y}_{_{\mathbf{X}}}^{^{N}}\mathbf{B}_{_{N}}\\
                              \mathbf{I}_{m_{_{\boldsymbol{y}}}}
                             \end{bmatrix}$,\ $\mathbf{Y}_{_{\mathbf{X}}}^{^{N}}(e^{j\alpha})\triangleq(e^{j\alpha}\mathbf{I}-\mathbf{A}_{_{N}})^{^{-1}}$,\ \
                             $\rho(\mathbf{A}_{_{N}})<1$,\ \  
$(\mathbf{A}_{_{N}}, \mathbf{B}_{_{N}})$ is controllable and\break $\mathbf{A}_{_{N}}\in \mathbb{R}^{n_{_{\mathbf{X}}}^{^{N}}\times n_{_{\mathbf{X}}}^{^{N}}}$.

\vspace*{3mm}

\begin{remark}
 One such family of subspaces corresponds to $m_{_{\boldsymbol{v}}}\times m_{_{\boldsymbol{y}}}$ finite impulse responses (FIRs) of length $N$. This family will be
 exploited  in  the derivation to follow. \hfill $\nabla$
\end{remark} 

\vspace*{3mm}

 Let\ $\boldsymbol{\theta}=\operatorname{rvec}(\boldsymbol{\beta})$\ \ and (with a slight abuse of notation) write\ 
 $\mathbf{X}_{_{N}}(\boldsymbol{\beta})$\  as\ $\mathbf{X}_{_{N}}(\boldsymbol{\theta})$. Let\break
 $\mathcal{S}_{_{\boldsymbol{\theta}}}^{^{N}}\triangleq \left\{\boldsymbol{\theta} \in \mathbb{R}^{m_{\boldsymbol{\theta}}^{^{N}}}:
 \mathbf{X}_{_{N}}(\boldsymbol{\theta}) \in \mathcal{S}_{_{\mathbf{X}}}\right\}$, where 
 $m_{\boldsymbol{\theta}}^{^{N}}\triangleq m_{_{\boldsymbol{v}}}(n_{_{\mathbf{X}}}^{^{N}}+m_{_{\boldsymbol{y}}})$.  The average MSE attained with $\mathbf{G}$ over 
 $\mathcal{S}_{_{\mathbf{X}}}^{^{N}}\bigcap \mathcal{S}_{_{\mathbf{X}}}$ (with respect 
 to a uniform distribution on $\mathcal{S}_{_{\boldsymbol{\theta}}}^{^{N}}$) is given by
 $$\boldsymbol{\eta}_{_{N}}(\mathbf{G})=
 \left\{\int_{\mathcal{S}_{_{\boldsymbol{\theta}}}^{^{N}}}\mathcal{J}_{_{\mathbf{X}}}(\mathbf{G};
 \mathbf{X}_{_{N}}(\boldsymbol{\theta}))d \boldsymbol{\theta}\right\}\nu_{_{N}}^{^{-1}}, \ \ \text{where}\ \
 \nu_{_{N}}\triangleq\int_{\mathcal{S}_{_{\boldsymbol{\theta}}}^{^{N}}}d \boldsymbol{\theta} \ \ \text{and}\ \ d\boldsymbol{\theta}\ \
 \text{stands for}\ \ d\theta_{_{_{1}}}\mathellipsis d\theta_{_{m_{_{\boldsymbol{\theta}}}^{^{N}}}}.$$
 
Note that \ \ $\mathcal{S}_{_{\boldsymbol{\theta}}}^{^{N}}$\ \ can be written as
$$ \mathcal{S}_{_{\boldsymbol{\theta}}}^{^{N}}=\left\{\boldsymbol{\theta} \in \mathbb{R}^{m_{\boldsymbol{\theta}}^{^{N}}}: 
\left\langle\boldsymbol{\beta} P_{_{\boldsymbol{\beta}}}^{^{N}}, \boldsymbol{\beta}\right\rangle_{F}\leq \gamma^{^{2}}\right\},$$ 
where\ \ $P_{_{\boldsymbol{\beta}}}^{^{N}}\triangleq(1/2\pi)\displaystyle\int_{0}^{2\pi}\check{\mathbf{Y}}_{_{\mathbf{X}}}^{^{N}}(e^{j\alpha})\check{\mathbf{Y}}_{_{\mathbf{X}}}^{^{N}}(e^{j\alpha})^{*}d \alpha$, 
\ \ or equivalently, 
$$\mathcal{S}_{_{\boldsymbol{\theta}}}^{^{N}}=\left\{\boldsymbol{\theta}\in \mathbb{R}^{m_{\boldsymbol{\theta}}^{^{N}}}: 
\boldsymbol{\theta}^{^{\mathrm{T}}}P_{_{\boldsymbol{\theta}}}^{^{N}}\boldsymbol{\theta}\leq \gamma^{^{2}}\right\},
\ \ \text{where}\ \  P_{_{\boldsymbol{\theta}}}^{^{N}}\triangleq \mathbf{I}\otimes (P_{_{\boldsymbol{\beta}}}^{^{N}})^{^{\mathrm{T}}}.$$

Note also that (writing\ $\boldsymbol{\Gamma}_{_{\boldsymbol{y}1}}=\boldsymbol{\phi}_{_{\boldsymbol{y}1}}\boldsymbol{\phi}_{_{\boldsymbol{y}1}}^{*}$)\ 
$\mathbf{G}\mathbf{X}_{_{N}}(\boldsymbol{\theta})\boldsymbol{\phi}_{_{\boldsymbol{y}1}}=F_{_{N}}(\mathbf{G})\boldsymbol{\theta}$\ so that it follows from (\ref{eq:03}) 
that
$$\mathcal{J}_{_{\mathbf{X}}}(\mathbf{G}; \mathbf{X}_{_{N}}(\boldsymbol{\theta}))= \mathcal{J}(\mathbf{G};\mathbf{H}_{_{\boldsymbol{0}}})
-2 \left\langle Z_{y}(\mathbf{G}), F_{_{N}}(\mathbf{G})\boldsymbol{\theta}\right\rangle
+\left\langle F_{_{N}}(\mathbf{G})\boldsymbol{\theta}, F_{_{N}}(\mathbf{G})\boldsymbol{\theta}\right\rangle,$$
where \ \ $Z_{y}(\mathbf{G})\triangleq\operatorname{rvec}(\mathbf{X}_{_{\boldsymbol{0}}}(\mathbf{G})\boldsymbol{\phi}_{_{\boldsymbol{y}1}})$,\ \ $ F_{_{N}}(\mathbf{G})\triangleq
\mathbf{G}\otimes (\check{\mathbf{Y}}_{_{\mathbf{X}}}^{^{N}}\boldsymbol{\phi}_{_{\boldsymbol{y}1}})^{^{\mathrm{T}}}$.

As a result, \ \ $\boldsymbol{\eta}_{_{N}}(\mathbf{G})$\ \ can be written as
$$\boldsymbol{\eta}_{_{N}}(\mathbf{G})=\left\{\mathcal{J}(\mathbf{G}; \mathbf{H}_{_{\boldsymbol{0}}})\int_{\mathcal{S}_{_{\boldsymbol{\theta}}}^{^{N}}}d \boldsymbol{\theta} 
-2 \int_{\mathcal{S}_{_{\boldsymbol{\theta}}}^{^{N}}} \xi_{_{\mathcal{J}}}^{^{N}}(\mathbf{G})^{^{\mathrm{T}}}\boldsymbol{\theta} d \boldsymbol{\theta}
+\int_{\mathcal{S}_{_{\boldsymbol{\theta}}}^{^{N}}} \boldsymbol{\theta}^{^{\mathrm{T}}} P_{_{\mathcal{J}}}^{^{N}}(\mathbf{G})\boldsymbol{\theta}d \boldsymbol{\theta}\right\}\nu_{_{N}}^{^{-1}},$$
 where\ \
 $\xi_{_{\mathcal{J}}}^{^{N}}(\mathbf{G})^{^{\mathrm{T}}}
 =(1/2\pi)\displaystyle\int_{0}^{2\pi}\left\{Z_{y}(\mathbf{G})^{*}F_{_{N}}(\mathbf{G})\right\}(e^{j \alpha})d\alpha$ \ \ and\ \ 
$P_{_{\mathcal{J}}}^{^{N}}(\mathbf{G})\triangleq(1/2\pi)\displaystyle\int_{0}^{2\pi}
\left\{F_{_{N}}(\mathbf{G})^{*}F_{_{N}}(\mathbf{G})\right\}(e^{j \alpha})d\alpha.$\

It then follows from the fact that \ \ $\mathcal{S}_{_{\boldsymbol{\theta}}}^{^{N}}$\ \ is symmetric with respect to the origin that\
$\displaystyle\int_{\mathcal{S}_{_{\boldsymbol{\theta}}}^{^{N}}} \xi_{_{\mathcal{J}}}^{^{N}}(\mathbf{G})^{^{\mathrm{T}}}\boldsymbol{\theta} 
d \boldsymbol{\theta} =0$\ \ so that
$$\boldsymbol{\eta}_{_{N}}(\mathbf{G})=\mathcal{J}(\mathbf{G}; \mathbf{H}_{_{\boldsymbol{0}}})+
\nu_{_{N}}^{^{-1}}\int_{\mathcal{S}_{_{\boldsymbol{\theta}}}^{^{N}}}\boldsymbol{\theta}^{^{\mathrm{T}}}P_{_{\mathcal{J}}}^{^{N}}(\mathbf{G})\boldsymbol{\theta}
d \boldsymbol{\theta}.$$

Consider the following proposition (a proof is presented in the Appendix).
\begin{proposition}\label{prop:01}
 Let $n\triangleq m_{\boldsymbol{\theta}}^{^{N}}$\ \ and\ \
 $\widehat{P}_{_{\mathcal{J}}}^{^{N}}(\mathbf{G})\triangleq (P_{_{\boldsymbol{\theta}}}^{^{N}})^{^{-1/2}} 
 P_{_{\mathcal{J}}}^{^{N}}(\mathbf{G})(P_{_{\boldsymbol{\theta}}}^{^{N}})^{^{-1/2}}$:\\
\emph{\textbf{(a)}} 
$\nu_{_{N}}=\left|\operatorname{det}\left(P_{_{\boldsymbol{\theta}}}^{^{N}}\right)\right|^{^{-1/2}}\left(\dfrac{\gamma^{n}}{n}\right)\displaystyle\check{\prod}_{1, n-2}(2\pi)$,\ 
where\ \ $\displaystyle\check{\prod}_{k, K}\triangleq\displaystyle \prod_{i=k}^{K}\displaystyle\int_{0}^{\pi}\sin(\alpha_{i})^{n-1-i} d \alpha_{i}$,\ $K\geq k$,\ $K\leq n-2$,\\\\
\emph{\textbf{(b)}}  
\begin{eqnarray*}
\displaystyle\int_{\mathcal{S}_{_{\boldsymbol{\theta}}}^{^{N}}}\boldsymbol{\theta}^{^{\mathrm{T}}}P_{_{\mathcal{J}}}^{^{N}}(\mathbf{G})\boldsymbol{\theta} d\boldsymbol{\theta}
&=&\left\{\displaystyle\sum_{k=1}^{n}\left\{\widehat{P}_{_{\mathcal{J}}}^{^{N}}(\mathbf{G})\right\}_{kk}\right\}
\left|\operatorname{det}\left(P_{_{\boldsymbol{\theta}}}^{^{N}}\right)\right|^{^{-1/2}}\times\\
&&\ \ \ \ \ \ \ \ \ \ \ \ \ \ \ \ \ \ \ \ \ \ \  \ \ \times\left\{\left(\frac{\gamma^{n+2}}{n+2}\right)
\displaystyle\int_{0}^{\pi}\cos(\alpha_{_{_{1}}})^{^{2}}\sin(\alpha_{_{_{1}}})^{n-2}d \alpha_{_{_{1}}}\right\}\displaystyle\check{\prod}_{2, n-2}(2\pi),
\end{eqnarray*}
\emph{\textbf{(c)}} $\boldsymbol{\eta}_{_{N}}(\mathbf{G})=\mathcal{J}(\mathbf{G}; \mathbf{H}_{_{\boldsymbol{0}}})+ \dfrac{\gamma^{^{2}}}{n+2}
\displaystyle\sum_{k=1}^{n}\left\{\widehat{P}_{_{\mathcal{J}}}^{^{N}}(\mathbf{G})\right\}_{kk}.$\hfill $\nabla$
\end{proposition}

To obtain an explicit expression for $\boldsymbol{\eta}_{av}(\mathbf{G})=\displaystyle\lim_{N\rightarrow \infty} \boldsymbol{\eta}_{_{N}}(\mathbf{G})$, \
\ $\mathcal{S}_{_{\mathbf{X}}}^{^{N}}$\
is taken to be the set of frequency-responses corresponding to FIRs of length $N$ (\emph{i.e.}, impulse responses\ $\{\mathbf{F}_{k}:k=0,1, \mathellipsis\}$
such that $\forall k>N$, $\mathbf{F}_{k}=0$), with state-space realizations\ $(\mathbf{A}_{_{N}}, 
\mathbf{B}_{_{N}}, \mathbf{C}, \mathbf{D})$\ given by\
$\mathbf{A}_{_{N}}=\operatorname{diag}\left(\mathbf{A}_{_{c}}^{^{N}}, \mathellipsis, \mathbf{A}_{_{c}}^{^{N}}\right)$,\break 
$\mathbf{B}_{_{N}}=\operatorname{diag}\left(\boldsymbol{b}_{_{c}}^{^{N}}, \mathellipsis, \boldsymbol{b}_{_{c}}^{^{N}}\right) \in
\mathbb{R}^{(N-1)m_{_{\boldsymbol{y}}}\times m_{_{\boldsymbol{y}}}}$,\ $\newcommand*{\temp}{\multicolumn{1}{r|}{}}
\mathbf{A}_{_{c}}^{^{N}}=\left[\begin{array}{ccccc}
0 & \cdots & 0 &\temp & 0\\ \cline{1-5}
 && &\temp & 0\\
 &\mathbf{I}_{N-1}& &\temp & \vdots\\
 && &\temp & 0\\
\end{array}\right]
$, \ \ $\boldsymbol{b}_{_{c}}^{^{N}}=\mathbf{e}_{_{1}}(N)$, \ \ $n_{_{\mathbf{X}}}^{^{N}}=Nm_{_{\boldsymbol{y}}}$, \ \ 
$\mathbf{A}_{_{N}}\in \mathbb{R}^{n_{_{\mathbf{X}}}^{^{N}}\times n_{_{\mathbf{X}}}^{^{N}}}$,\
$\mathbf{C}=[\mathbf{F}_{_{1}}\mathbf{e}_{_{1}}(m)\cdots\mathbf{F}_{_{N}}\mathbf{e}_{_{1}}(m)\ \vdots \cdots \vdots 
\mathbf{F}_{_{1}}\mathbf{e}_{_{m}}(m)\cdots\mathbf{F}_{_{N}}\mathbf{e}_{_{m}}(m)]\in \mathbb{R}^{m_{_{\boldsymbol{v}}}\times n}$,\
$\mathbf{D}=\mathbf{F}_{_{0}}\in \mathbb{R}^{m_{_{\boldsymbol{v}}}\times m_{_{\boldsymbol{y}}}}$  \ \ and\ \ $n=m_{_{\boldsymbol{\theta}}}^{^{N}}=m_{_{\boldsymbol{v}}}m_{_{\boldsymbol{y}}}(N+1)$.

The desired limit is presented in the following proposition.

\begin{proposition}\label{prop:02}
 The average MSE criterion over\ \ $\mathcal{S}_{_{\mathbf{H}}}$\ \ defined by 
 $\boldsymbol{\eta}_{av}(\mathbf{G})\triangleq \displaystyle\lim_{N\rightarrow\infty}\boldsymbol{\eta}_{_{N}}(\mathbf{G})$\ \ is given by
 $$\boldsymbol{\eta}_{av}(\mathbf{G})=\mathcal{J}(\mathbf{G}; \mathbf{H}_{_{\boldsymbol{0}}})+\left(\gamma^{^{2}}/(m_{_{\boldsymbol{v}}}m_{_{\boldsymbol{y}}})\right)
 \left\langle \mathbf{G} \otimes \boldsymbol{\phi}_{_{\boldsymbol{y}1}}^{^{\mathrm{T}}}, \mathbf{G}\otimes \boldsymbol{\phi}_{_{\boldsymbol{y}1}}^{^{\mathrm{T}}}\right\rangle,$$
 where \ \ $\mathcal{S}_{_{\mathbf{H}}}$ \ \ is given by \emph {(\ref{eq:01a})} and\ \ $\boldsymbol{\phi}_{_{\boldsymbol{y}1}}\boldsymbol{\phi}_{_{\boldsymbol{y}1}}^{*}=W^{^{-1}}\boldsymbol{\Gamma}_{\boldsymbol{y}}(W^{^{-1}})^{*}$. \hfill$\nabla$    
 \end{proposition}

\vspace*{3mm} 
 
\begin{remark}
 The ``average'' criterion\ \ $\boldsymbol{\eta}_{av}(\mathbf{G})$\ \ consists of the nominal MSE\ \
 $(\mathcal{J}(\mathbf{G};\mathbf{H}_{_{\boldsymbol{0}}}))$\ \ supplemented by a ``channel-output noise'' term reflecting the ``effect'' of the 
 channel model perturbations on the input signal's power spectral density\ \ 
 $\boldsymbol{\Gamma}_{_{\boldsymbol{y}}}=\boldsymbol{\phi}_{_{\boldsymbol{y}}}\boldsymbol{\phi}_{_{\boldsymbol{y}}}^{*}$ -- indeed, in the SISO case, the additional term\ \
 $\gamma^{2}\langle\mathbf{G}\boldsymbol{\phi}_{_{\boldsymbol{y}1}}, \mathbf{G}\boldsymbol{\phi}_{_{\boldsymbol{y}1}}\rangle$\ \ is exactly like the observation noise term of\ \
 $\mathcal{J}(\mathbf{G};\mathbf{H}_{_{\boldsymbol{0}}})$\ \ (\emph{i.e.}, $\langle\mathbf{G}\boldsymbol{\Gamma}_{_{\boldsymbol{v}}}, \mathbf{G}\rangle$)\ \ with
 \ \ $\boldsymbol{\Gamma}_{_{\boldsymbol{v}}}$\ \ replaced by\ \ $\gamma^{^{2}}W^{^{-1}}\boldsymbol{\Gamma}_{_{\boldsymbol{y}}}(W^{^{-1}})^{*}.\hfill\nabla$
\end{remark}

\subsection{Average Estimation Error Over $\mathcal{H}_{2}$ Signal Balls}\label{subsec:3.2}

In this case, the estimation error magnitude for a pair of signals\ \ $(\bar{\boldsymbol{y}}, \bar{\boldsymbol{v}})$ \ \ is given by
$\|\bar{\boldsymbol{e}}(\bar{\boldsymbol{z}}; \mathbf{G},\mathbf{H})\|_{_{2}}^{^{2}}$\ where\ 
$\bar{\boldsymbol{z}}^{^{\mathrm{T}}}=[\bar{\boldsymbol{y}}^{^{\mathrm{T}}}\ \bar{\boldsymbol{v}}^{^{\mathrm{T}}}]$\ \ and the corresponding
average criterion is to be defined with respect to the set\ \ 
$\mathcal{S}_{_{\bar{\boldsymbol{z}}}}=\bar{\mathcal{S}}_{_{\boldsymbol{y}}}\times \bar{\mathcal{S}}_{_{\boldsymbol{v}}}$, where 
$\bar{\mathcal{S}}_{_{\boldsymbol{\alpha}}}$, $\boldsymbol{\alpha}=\bar{\boldsymbol{y}}, \bar{\boldsymbol{v}}$ is as defined in Section \ref{sec:2}. To this effect,
consider FIR subsets of 
$\bar{\mathcal{S}}_{_{\boldsymbol{\alpha}}}$, namely,
$$\{\boldsymbol{\alpha}_{_{N}}(\boldsymbol{\beta}_{_{\boldsymbol{\alpha}}})=
\mathbf{Y}_{_{\boldsymbol{\alpha}N}}^{a}\boldsymbol{\beta}_{_{\boldsymbol{\alpha}}},
\boldsymbol{\beta}_{_{\boldsymbol{\alpha}}}\in \mathcal{S}_{_{N}}^{^{\boldsymbol{\alpha}}}\}, \ \ 
\text{where}\ \
\mathcal{S}_{_{N}}^{^{\boldsymbol{\alpha}}}\triangleq\{\boldsymbol{\beta}_{_{\boldsymbol{\alpha}}}\in
\mathbb{R}^{n_{_{\boldsymbol{\alpha}N}}+m_{_{\boldsymbol{\alpha}}}}:
\|\boldsymbol{\alpha}_{_{N}}(\boldsymbol{\beta}_{_{\boldsymbol{\alpha}}})\|_{_{2}}
\leq \gamma_{_{\boldsymbol{\alpha}}}\},$$
$\mathbf{Y}_{_{\boldsymbol{\alpha}N}}^{a}\triangleq [\mathbf{C}_{_{\boldsymbol{\alpha}N}}\mathbf{Y}_{_{\boldsymbol{\alpha}N}}\ \vdots\ 
\mathbf{I}_{m_{_{\boldsymbol{\alpha}}}}],
\ \ \mathbf{Y}_{_{\boldsymbol{\alpha}N}}(e^{j \phi})=(e^{j\phi}\mathbf{I}-\mathbf{A}_{_{\boldsymbol{\alpha}N}})^{^{-1}},$
$\mathbf{A}_{_{\boldsymbol{\alpha}N}}=\check{\mathbf{A}}^{^{\mathrm{T}}}_{_{N\alpha}}\ \ \  \ \text{and}\ \  \ \ 
\mathbf{C}_{_{\boldsymbol{\alpha}N}}=\check{\mathbf{B}}^{^{\mathrm{T}}}_{_{N\alpha}}, \ \ \ \ \ n_{_{\boldsymbol{\alpha}N}}=Nm_{_{\alpha}},$
where\ \ $\check{\mathbf{A}}_{_{N\boldsymbol{\alpha}}}$\ \ and\ \ $\check{\mathbf{B}}_{_{N\boldsymbol{\alpha}}}$\ \ are given by\ \ $\mathbf{A}_{_{N}}$\ \ and\ \
$\mathbf{B}_{_{N}}$\ \ above with\ $m_{_{\boldsymbol{y}}}$\ replaced by\ $m_{_{\boldsymbol{\alpha}}}$.

The average value of the squared, estimation error over these signal sets is given by
$$\boldsymbol{\eta}_{_{N}}^{a}(\mathbf{G}; \mathbf{H})\triangleq\left\{\int_{\mathcal{S}_{_{N}}^{^{\boldsymbol{y}}}}\int_{\mathcal{S}_{_{N}}^{^{\boldsymbol{v}}}}
\left\|\bar{\boldsymbol{e}}(\bar{\boldsymbol{z}}_{_{N}}(\boldsymbol{\beta}_{\boldsymbol{y}}, 
\boldsymbol{\beta}_{\boldsymbol{v}});\mathbf{G}, \mathbf{H})\right\|_{_{2}}^{^{2}}d\boldsymbol{\beta}_{\boldsymbol{v}}d\boldsymbol{\beta}_{y}\right\}(\boldsymbol{\mu}_{_{N}}^{a})^{^{-1}},$$
where\  $\bar{\boldsymbol{z}}_{_{N}}(\boldsymbol{\beta}_{\boldsymbol{y}}, \boldsymbol{\beta}_{\boldsymbol{v}})^{^{\mathrm{T}}} \triangleq
[\bar{\boldsymbol{y}}_{_{N}}(\boldsymbol{\beta}_{\boldsymbol{y}})^{^{\mathrm{T}}}\ \vdots\ 
\bar{\boldsymbol{v}}_{_{N}}(\boldsymbol{\beta}_{\boldsymbol{v}})^{^{\mathrm{T}}}]$,\ 
$\boldsymbol{\mu}_{_{N}}^{a}\triangleq\displaystyle\int_{\mathcal{S}_{_{N}}^{^{\boldsymbol{y}}}}
\int_{\mathcal{S}_{_{N}}^{^{\boldsymbol{v}}}}d\boldsymbol{\beta}_{\boldsymbol{v}}d \boldsymbol{\beta}_{\boldsymbol{y}}$,\ and\
$d\boldsymbol{\beta}_{_{\boldsymbol{\alpha}}}\triangleq d\boldsymbol{\beta}_{1}, \mathellipsis,
d\boldsymbol{\beta}_{n_{_{\boldsymbol{\alpha}N}}+m_{_{\boldsymbol{\alpha}}}}$ or, equivalently,
\begin{eqnarray*}
 \boldsymbol{\mu}_{_{N}}^{a}\boldsymbol{\eta}_{_{N}}^{a}(\mathbf{G}; \mathbf{H})
 &=&\int_{\mathcal{S}_{_{N}}^{^{\boldsymbol{y}}}}\int_{\mathcal{S}_{_{N}}^{^{\boldsymbol{v}}}}
 \left\|(\mathbf{H}_{_{\mathbf{I}\boldsymbol{y}}}-\mathbf{G}\mathbf{H}_{\mathrm{o} \boldsymbol{z}})\begin{bmatrix}
                                                                                        \bar{\boldsymbol{y}}_{_{N}}(\boldsymbol{\beta}_{\boldsymbol{y}})\\
                                                                                        \bar{\boldsymbol{v}}_{_{N}}(\boldsymbol{\beta}_{\boldsymbol{v}})\\
                                                                                       \end{bmatrix}
\right\|_{_{2}}^{^{2}} d \boldsymbol{\beta}_{\boldsymbol{v}}d \boldsymbol{\beta}_{\boldsymbol{y}} \ \ \ \ \ \ \ \ \ \ \ \ \ \ \ \  \Leftrightarrow\\\\
 \boldsymbol{\mu}_{_{N}}^{a}\boldsymbol{\eta}_{_{N}}^{a}(\mathbf{G}; \mathbf{H})
 &=&\int_{\mathcal{S}_{_{N}}^{^{\boldsymbol{y}}}}\int_{\mathcal{S}_{_{N}}^{^{\boldsymbol{v}}}}
 \left\|(\mathbf{H}_{_{\mathbf{I}\boldsymbol{y}}}-\mathbf{G}\mathbf{H}_{\mathrm{o} \boldsymbol{z}})\begin{bmatrix}
                                                                                        \mathbf{Y}_{_{\boldsymbol{y}N}}^{a}& \boldsymbol{0}\\
                                                                                          \boldsymbol{0} & \mathbf{Y}_{_{\boldsymbol{v}N}}^{a}                                                                                     
                                                                                       \end{bmatrix}\begin{bmatrix}
                                                                                        \boldsymbol{\beta}_{\boldsymbol{y}}\\
                                                                                        \boldsymbol{\beta}_{\boldsymbol{v}}\\
                                                                                       \end{bmatrix}
\right\|_{_{2}}^{^{2}} d \boldsymbol{\beta}_{\boldsymbol{v}}d \boldsymbol{\beta}_{\boldsymbol{y}} \ \ \ \Leftrightarrow\\\\
\boldsymbol{\mu}_{_{N}}^{a}\boldsymbol{\eta}_{_{N}}^{a}(\mathbf{G}; \mathbf{H})
 &=&\int_{\mathcal{S}_{_{N}}^{^{\boldsymbol{y}}}}\int_{\mathcal{S}_{_{N}}^{^{\boldsymbol{v}}}}
\begin{bmatrix}
\boldsymbol{\beta}_{\boldsymbol{y}}^{^{\mathrm{T}}} &\vdots &\boldsymbol{\beta}_{\boldsymbol{v}}^{^{\mathrm{T}}}                                                                                      
\end{bmatrix}\boldsymbol{\Gamma}_{_{N}}^{\boldsymbol{e}}\begin{bmatrix}
                                        \boldsymbol{\beta}_{\boldsymbol{y}}\\
                                        \boldsymbol{\beta}_{\boldsymbol{v}}\\
                                        \end{bmatrix} d \boldsymbol{\beta}_{\boldsymbol{v}}d \boldsymbol{\beta}_{\boldsymbol{y}},
\end{eqnarray*}
where \ \ \ $\boldsymbol{\Gamma}_{_{N}}^{\boldsymbol{e}}\triangleq(1/2\pi)\displaystyle
\int_{0}^{2\pi}\mathbf{F}_{_{\mathbf{G}\mathbf{Y}}}(e^{j\phi})^{*}\mathbf{F}_{_{\mathbf{G}\mathbf{Y}}}(e^{j\phi})d\phi$\ \ \ and \ \ \ 
$\mathbf{F}_{_{\mathbf{G}\mathbf{Y}}}\triangleq(\mathbf{H}_{_{\mathbf{I}\boldsymbol{y}}}-\mathbf{G}\mathbf{H}_{\mathrm{o} \boldsymbol{z}})
\operatorname{diag}(\mathbf{Y}_{_{\boldsymbol{y}N}}^{a}, \mathbf{Y}_{_{\boldsymbol{v}N}}^{a})$.\

\noindent
Thus,\ \   for\ \ \  $\boldsymbol{\Gamma}_{_{N}}^{\boldsymbol{e}}\triangleq \begin{bmatrix}
                                                  \boldsymbol{\Gamma}_{_{N\boldsymbol{y}}}^{\boldsymbol{e}} & \boldsymbol{\Gamma}_{_{N1}}^{\boldsymbol{e}}\\
                                                  (\boldsymbol{\Gamma}_{_{N 1}}^{\boldsymbol{e}})^{^{\mathrm{T}}} & \boldsymbol{\Gamma}_{_{N\boldsymbol{v}}}^{\boldsymbol{e}}\\
                                                 \end{bmatrix}$,
\begin{eqnarray*}
 \boldsymbol{\mu}_{_{N}}^{a}\boldsymbol{\eta}_{_{N}}^{a}(\mathbf{G}; \mathbf{H})
 &=&\int_{\mathcal{S}_{_{N}}^{^{\boldsymbol{y}}}}\int_{\mathcal{S}_{_{N}}^{^{\boldsymbol{v}}}} 
 (\boldsymbol{\beta}_{\boldsymbol{y}}^{^{\mathrm{T}}}\boldsymbol{\Gamma}_{_{N\boldsymbol{y}}}^{\boldsymbol{e}}\boldsymbol{\beta}_{\boldsymbol{y}}+
 2\boldsymbol{\beta}_{\boldsymbol{y}}^{^{\mathrm{T}}}\boldsymbol{\Gamma}_{_{N1}}^{\boldsymbol{e}}\boldsymbol{\beta}_{\boldsymbol{v}}+
 \boldsymbol{\beta}_{\boldsymbol{v}}^{^{\mathrm{T}}}\boldsymbol{\Gamma}_{_{N\boldsymbol{v}}}^{\boldsymbol{e}}\boldsymbol{\beta}_{\boldsymbol{v}})
 d \boldsymbol{\beta}_{\boldsymbol{v}}d \boldsymbol{\beta}_{\boldsymbol{y}} \ \ \ \ \ \ \ \ \ \ \ \ \ \  \Leftrightarrow\\\\
 \boldsymbol{\mu}_{_{N}}^{a}\boldsymbol{\eta}_{_{N}}^{a}(\mathbf{G}; \mathbf{H})
 &=&\boldsymbol{\mu}_{_{\boldsymbol{v}N}}^{a}\int_{\mathcal{S}_{_{N}}^{^{\boldsymbol{y}}}}\boldsymbol{\beta}_{\boldsymbol{y}}^{^{\mathrm{T}}}\boldsymbol{\Gamma}_{_{N\boldsymbol{y}}}^{\boldsymbol{e}}\boldsymbol{\beta}_{\boldsymbol{y}}+ 
\boldsymbol{\mu}_{_{\boldsymbol{y}N}}^{a}\int_{\mathcal{S}_{_{N}}^{^{\boldsymbol{v}}}} \boldsymbol{\beta}_{\boldsymbol{v}}^{^{\mathrm{T}}}\boldsymbol{\Gamma}_{_{N\boldsymbol{v}}}^{\boldsymbol{e}}\boldsymbol{\beta}_{\boldsymbol{v}}
 d \boldsymbol{\beta}_{\boldsymbol{v}},
\end{eqnarray*}
where\ \ \ 
$\boldsymbol{\mu}_{_{\boldsymbol{\alpha}N}}^{a}\triangleq \displaystyle\int_{\mathcal{S}_{_{N}}^{^{\boldsymbol{\alpha}}}}d \boldsymbol{\beta}_{_{\boldsymbol{\alpha}}}$ \ \ \
(note that\ \  
$\displaystyle\int_{\mathcal{S}_{_{N}}^{^{\boldsymbol{y}}}}\int_{\mathcal{S}_{_{N}}^{^{\boldsymbol{v}}}}2 
\boldsymbol{\beta}_{\boldsymbol{y}}^{^{\mathrm{T}}}\boldsymbol{\Gamma}_{_{N1}}^{\boldsymbol{e}}\boldsymbol{\beta}_{\boldsymbol{v}}
d\boldsymbol{\beta}_{\boldsymbol{v}}d \boldsymbol{\beta}_{\boldsymbol{y}}=0$\ \ \ since\ \ \
$\boldsymbol{\beta}_{\boldsymbol{v}}\in \mathcal{S}_{_{N}}^{^{\boldsymbol{v}}}$\ \ $\Leftrightarrow$\ \
$-\boldsymbol{\beta}_{\boldsymbol{v}}\in \mathcal{S}_{_{N}}^{^{\boldsymbol{v}}}$).\\

\noindent
Thus, as \ \ $\boldsymbol{\mu}_{_{N}}^{a}=\boldsymbol{\mu}_{_{\boldsymbol{y}N}}^{a}\boldsymbol{\mu}_{_{\boldsymbol{v}N}}^{a}$, \ \ 
$\boldsymbol{\eta}_{_{N}}^{a}(\mathbf{G}, \mathbf{H})=\boldsymbol{\eta}_{_{\boldsymbol{y}N}}^{a}(\mathbf{G}; \mathbf{H})
+\boldsymbol{\eta}_{_{\boldsymbol{v}N}}^{a}(\mathbf{G}; \mathbf{H})$,\ \ \ where\ \ $\boldsymbol{\eta}_{_{\boldsymbol{y}N}}^{a}
=(\boldsymbol{\mu}_{_{\boldsymbol{y}N}}^{a})^{^{-1}}\displaystyle\int_{\mathcal{S}_{_{N}}^{^{\boldsymbol{y}}}}\boldsymbol{\beta}_{\boldsymbol{y}}^{^{\mathrm{T}}}\boldsymbol{\Gamma}_{_{N\boldsymbol{y}}}^{\boldsymbol{e}}\boldsymbol{\beta}_{\boldsymbol{y}}
d \boldsymbol{\beta}_{\boldsymbol{y}}$\ \ \ \ 
and\ \ \ \ $\boldsymbol{\eta}_{_{\boldsymbol{v}N}}^{a}
=(\boldsymbol{\mu}_{_{\boldsymbol{v}N}}^{a})^{^{-1}}\displaystyle\int_{\mathcal{S}_{_{N}}^{^{\boldsymbol{v}}}}
\boldsymbol{\beta}_{\boldsymbol{v}}^{^{\mathrm{T}}}\boldsymbol{\Gamma}_{_{N\boldsymbol{v}}}^{\boldsymbol{e}}\boldsymbol{\beta}_{\boldsymbol{v}}
d \boldsymbol{\beta}_{\boldsymbol{v}}$.

As a result, pursuing the path that led to $\boldsymbol{\eta}$ leads to the following proposition.

\begin{proposition}
$$\boldsymbol{\eta}_{_{\boldsymbol{\alpha}N}}^{a}=
\left\{\frac{\gamma_{_{\boldsymbol{\alpha}}}^{^{2}}}{m_{_{\boldsymbol{\alpha}N}}+2}\right\}(N+1)
\left\langle\bar{\mathcal{A}}_{_{\boldsymbol{\alpha}}}(\mathbf{G}; \mathbf{H}), \bar{\mathcal{A}}_{_{\boldsymbol{\alpha}}}(\mathbf{G}; \mathbf{H})\right\rangle,$$
where\ \ \ $m_{_{\boldsymbol{\alpha}N}}=(N+1)m_{_{\boldsymbol{\alpha}}}$, \ \ 
$\bar{\mathcal{A}}_{\boldsymbol{y}}(\mathbf{G}; \mathbf{H})=(\mathbf{H}_{_{\mathbf{I}}}-\mathbf{G}\mathbf{H})W_{\boldsymbol{y}}^{^{-1}}$\ \
and \ \ 
$\bar{\mathcal{A}}_{\boldsymbol{v}}(\mathbf{G}; \mathbf{H})=\mathbf{G}W_{\boldsymbol{v}}^{^{-1}}$.

Moreover, defining \ \ $\boldsymbol{\eta}^{a}(\mathbf{G}; \mathbf{H})=\displaystyle\lim_{N\rightarrow \infty}\boldsymbol{\eta}_{_{N}}^{a}(\mathbf{G}; 
\mathbf{H})$,\ \ it follows that
$$\boldsymbol{\eta}^{a}(\mathbf{G};  \mathbf{H})=
(\gamma_{_{\boldsymbol{y}}}^{^{2}}/m_{_{\boldsymbol{y}}})\left\langle\bar{\mathcal{A}}_{\boldsymbol{y}}(\mathbf{G}; \mathbf{H}), \bar{\mathcal{A}}_{\boldsymbol{y}}(\mathbf{G}; \mathbf{H})\right\rangle+
(\gamma_{_{\boldsymbol{v}}}^{^{2}}/m_{_{\boldsymbol{v}}})\left\langle\bar{\mathcal{A}}_{\boldsymbol{v}}(\mathbf{G}; \mathbf{H}), \bar{\mathcal{A}}_{\boldsymbol{v}}(\mathbf{G}; \mathbf{H})\right\rangle.$$
 \hfill$\nabla$
\end{proposition}

\vspace*{3mm}

\begin{remark}
 Note that the limit process yielding the ``average'' criterion\ \ $\boldsymbol{\eta}^{a}(\mathbf{G};\mathbf{H})$\ \ naturally led to the
 estimation error due to the deterministic signals in (the weighted $\mathcal{H}_{2}$--balls) $\mathcal{S}_{_{\boldsymbol{y}}}$\ \ and\ \ 
 $\mathcal{S}_{_{\boldsymbol{v}}}$\ \ being represented as the estimation MSE due to stochastic signals with power spectral densities\ \
 $(\gamma_{_{\boldsymbol{y}}}^{2}/m_{_{\boldsymbol{y}}})(W_{_{\boldsymbol{y}}}^{*}W_{_{\boldsymbol{y}}})^{^{-1}}$\  and\ \
 $(\gamma_{_{\boldsymbol{v}}}^{2}/m_{_{\boldsymbol{v}}})(W_{_{\boldsymbol{v}}}^{*}W_{_{\boldsymbol{v}}})^{^{-1}}$.\hfill$\nabla$
\end{remark}

\subsection{Average Criterion for the Robust $\mathcal{H}_{\infty}$ Problem}\label{subsec:3.3}

The results in Subsections \ref{subsec:3.1} and \ref{subsec:3.2} are now combined in a simple way to yield a cost-functional that takes into account the average 
estimation error over\ \ $\mathcal{H}_{2}$\ \ signals balls and the ``channel-model'' set
$$\mathcal{S}_{_{\mathbf{H}}\infty}=\{\mathbf{H}\in \mathcal{R}_{c}^{m_{_{\boldsymbol{v}}}\times m_{_{\boldsymbol{y}}}}:
\|(\mathbf{H}-\mathbf{H}_{_{\boldsymbol{0}}})W_{_{\mathbf{H}}}\|_{\infty}\leq \gamma_{_{\mathbf{H}}}\}$$
introduced in Subsection \ref{subsec:2.3}. This is done, in line with the derivation of\ \ $\boldsymbol{\eta}^{a}(\cdot)$\ \ and Remark 3.3, by viewing
the signal balls\ \ $\mathcal{S}_{\boldsymbol{y}}$\ \ and\ \ $\mathcal{S}_{\boldsymbol{v}}$\ \ as ``formally equivalent'' (for the purpose
of defining an average criterion) to stochastic signals with power spectral densities\ \ 
$\boldsymbol{\Gamma}_{_{\boldsymbol{y}}}^{a}=\boldsymbol{\phi}_{_{\boldsymbol{y}}}^{a}(\boldsymbol{\phi}_{_{\boldsymbol{y}}}^{a})^{*}$\ \
and \ \ 
$\boldsymbol{\Gamma}_{_{\boldsymbol{v}}}^{a}=\boldsymbol{\phi}_{_{\boldsymbol{v}}}^{a}(\boldsymbol{\phi}_{_{\boldsymbol{v}}}^{a})^{*}$,\ \
where\ \ $\boldsymbol{\phi}_{_{\boldsymbol{y}}}^{a}=(\gamma_{_{\boldsymbol{y}}}/\sqrt{m_{_{\boldsymbol{y}}}})W_{_{\boldsymbol{y}}}^{^{-1}}$,\ \
$\boldsymbol{\phi}_{_{\boldsymbol{v}}}^{a}=(\gamma_{_{\boldsymbol{v}}}/\sqrt{m_{_{\boldsymbol{v}}}})W_{_{\boldsymbol{v}}}^{^{-1}}$,\ \
and by (conservatively) taking into account the set\ \ $\mathcal{S}_{_{\mathbf{H}}\infty}$\ \ by means of a\ \ $\mathcal{H}_{2}-$ball of
frequency-responses, as described in Remark 2.1, namely,
$$\bar{\mathcal{S}}_{_{\mathbf{H}}\infty}=\{\mathbf{H}\in \mathcal{R}_{c}^{m_{_{\boldsymbol{v}}}\times m_{_{\boldsymbol{y}}}}:
\|(\mathbf{H}-\mathbf{H}_{_{\boldsymbol{0}}})\bar{W}\|_{_{2}}\leq \gamma_{_{\mathbf{H}}}\},$$
where\ \ $\bar{W}=\boldsymbol{\phi}_{_{\boldsymbol{y}}}^{a}\boldsymbol{\phi}_{_{\boldsymbol{y}W}}^{^{-1}}$\ \ and\ \ 
$\boldsymbol{\phi}_{_{\boldsymbol{y}W}}$\ \ is a spectral factor of
$$\boldsymbol{\phi}_{_{\boldsymbol{y}W}}(e^{j\alpha})\boldsymbol{\phi}_{_{\boldsymbol{y}W}}(e^{j\alpha})^{*}
=\operatorname{tr}\{[W_{_{\mathbf{H}}}^{^{-1}}\boldsymbol{\phi}_{_{\boldsymbol{y}}}^{a}](e^{j\alpha})^{*}[W_{_{\mathbf{H}}}^{^{-1}}\boldsymbol{\phi}_{_{\boldsymbol{y}}}^{a}](e^{j\alpha})\}.$$
 
Then, replacing\ \ $\gamma$,\ $\boldsymbol{\Gamma}_{_{\boldsymbol{y}}}$,\ $\boldsymbol{\Gamma}_{_{\boldsymbol{v}}}$,\ \ and \ \ 
$\boldsymbol{\phi}_{_{\boldsymbol{y}1}}$\ \ respectively by $\gamma_{_{\mathbf{H}}}$,\ $\boldsymbol{\Gamma}_{_{\boldsymbol{y}}}^{a}$,\ 
$\boldsymbol{\Gamma}_{_{\boldsymbol{v}}}^{a}$,\ \ and \ \ 
$\bar{\boldsymbol{\phi}}_{_{\boldsymbol{y}1}}=\bar{W}^{^{-1}}\boldsymbol{\phi}_{_{\boldsymbol{y}}}^{a}$\ \ in the expression of\ \ 
$\boldsymbol{\eta}_{_{a\boldsymbol{v}}}(\mathbf{G})$\ \ leads to (since 
$\bar{\boldsymbol{\phi}}_{_{\boldsymbol{y}1}}=\boldsymbol{\phi}_{_{\boldsymbol{y}W}}\mathbf{I}_{_{m_{_{\boldsymbol{y}}}}}$)
$$\boldsymbol{\eta}^{b}(\mathbf{G};\mathcal{S}_{_{\mathbf{H}}\infty})=\boldsymbol{\eta}^{a}(\mathbf{G};\mathbf{H}_{_{\boldsymbol{0}}})+
(\gamma_{_{\mathbf{H}}}^{^{2}}/m_{_{\boldsymbol{y}}}m_{_{\boldsymbol{v}}})
\left\langle\mathbf{G}\otimes(\boldsymbol{\phi}_{_{\boldsymbol{y}W}}\mathbf{I}_{_{m_{_{\boldsymbol{y}}}}}),
\mathbf{G}\otimes(\boldsymbol{\phi}_{_{\boldsymbol{y}W}}\mathbf{I}_{_{m_{_{\boldsymbol{y}}}}})\right\rangle$$
or, equivalently,\ \ \ $\boldsymbol{\eta}^{b}(\mathbf{G};\mathcal{S}_{_{\mathbf{H}}\infty})
=\boldsymbol{\eta}^{a}(\mathbf{G};\mathbf{H}_{_{\boldsymbol{0}}})+
(\gamma_{_{\mathbf{H}}}^{^{2}}/m_{_{\boldsymbol{v}}})
\left\langle\mathbf{G}\boldsymbol{\phi}_{_{\boldsymbol{y}W}},
\mathbf{G}\boldsymbol{\phi}_{_{\boldsymbol{y}W}}\right\rangle$.

Note that\ \ $\boldsymbol{\eta}^{b}(\cdot)$\ \ consists of the average criterion for the nominal\ \ $\mathcal{H}_{\infty}$\ \ problem plus an 
additive term which takes into account the weighting function\ \ $W_{_{\mathbf{H}}}$\ \ (by means of \
$\boldsymbol{\phi}_{_{\boldsymbol{y}W}}$) and the\ \  $\mathcal{H}_{\infty}-$uncertainty radius\ \ $\gamma_{_{\mathbf{H}}}$.

\section{Minimax Estimators}\label{sec:4}

In this section, the minimax problems \emph{Prob. $1 - 3$} are recast as SDPs.
\subsection{Minimax $\mathcal{H}_{2}$ Estimators with $\mathcal{H}_{2}$ Model Uncertainty}\label{subsec:4.1}

The first problem to be considered in this section is formulated as follows:
$$\underline{Prob.\ 1 :}\ \min_{\mathbf{G}\in \mathcal{S}_{_{\mathbf{G}}}}\bar{\mathcal{J}}(\mathbf{G}; \mathcal{S}_{_{\mathbf{H}}}),$$
where\ \  $\mathcal{S}_{_{\mathbf{H}}}\triangleq \left\{\mathbf{H} \in \mathcal{R}_{c}^{m_{_{\boldsymbol{v}}}\times m_{_{\boldsymbol{y}}}}: \|(\mathbf{H}-\mathbf{H}_{_{\boldsymbol{0}}})W\|_{_{2}}\leq \gamma\right\}$\ \  and\ \  
$\mathcal{S}_{_{\mathbf{G}}}$\ \ is a subset of\ \ $\mathcal{R}_{c}^{m_{_{\boldsymbol{e}}}\times m_{_{\boldsymbol{v}}}}$,\ \ or, equivalently 
(cf. (\ref{eq:02}) -- (\ref{eq:05}))
$$\underline{Prob.\ 1 :}\ \displaystyle\min_{\mathbf{G}\in \mathcal{S}_{_{\mathbf{G}}}}\bar{\mathcal{J}}_{_{\mathbf{X}}}(\mathbf{G}; \mathcal{S}_{_{\mathbf{X}}}),$$
where\ \  $\bar{\mathcal{J}}_{_{\mathbf{X}}}$\ \ and\ \ $\mathcal{S}_{_{\mathbf{X}}}$\ \ are defined by (\ref{eq:02}) -- (\ref{eq:05}).

The major aim of this Subsection is to recast \emph{Prob. $1$} as an SDP. This was also carried out in [18], but the SDP 
introduced here is simpler than the one previously obtained as one of the LMIs involved  in the latter was eliminated. This, together with
the fact that the simplified conditions are a part of the average cost/worst-case constraint problem below, provides motivation for 
presenting the modified SDP here.

Proceeding as in [18], the first step is to introduce the Lagrangian and dual functionals
\begin{equation}\label{eq:06}
 Lag(\mathbf{X}, \lambda; \mathbf{G})=\mathcal{J}_{_{\mathbf{X}}}(\mathbf{G}; \mathbf{X})-\lambda\left\{\|\mathbf{X}\|_{_{2}}^{^{2}}-\gamma^{^{2}}\right\}
\end{equation}
and
\begin{equation}\label{eq:07}
 \boldsymbol{\varphi}_{_{\mathbf{D}}}(\lambda; \mathbf{G})\triangleq\sup\left\{ Lag(\mathbf{X}, \lambda; \mathbf{G}): \mathbf{X}\in \mathcal{R}_{c}^{m_{_{\boldsymbol{v}}}\times m_{_{\boldsymbol{y}}}}\right\},
\end{equation}
so that Theorem 2 of [12] can be invoked to yield
\begin{equation}\label{eq:08}
 \bar{\mathcal{J}}_{_{\mathbf{X}}}(\mathbf{G}; \mathcal{S}_{_{\mathbf{X}}})=\inf\left\{\boldsymbol{\varphi}_{_{\mathbf{D}}}(\lambda; \mathbf{G}): \lambda>0\right\}.
\end{equation}
\\
To facilitate the derivation to follow,\ \  $\mathcal{J}_{_{\mathbf{X}}}(\mathbf{G}; \mathbf{X})$\ \ and\ \ $Lag(\mathbf{X}, \lambda; \mathbf{G})$\ \ are rewritten as 
\begin{equation*}
 \mathcal{J}_{_{\mathbf{X}}}(\mathbf{G}; \mathbf{X})=\|\mathbf{X}_{_{\boldsymbol{0}}}(\mathbf{G})\mathbf{F}_{\boldsymbol{y}}-\mathbf{G}\mathbf{X}\mathbf{F}_{\boldsymbol{y}}+\mathbf{G}\mathbf{F}_{\boldsymbol{v}}\|_{_{2}}^{^{2}}
\end{equation*}
and
\begin{equation}\label{eq:09}
 Lag(\mathbf{X}, \lambda; \mathbf{G})=\lambda \gamma^{^{2}}-L_{a}(\mathbf{X}, \lambda; \mathbf{G}),
\end{equation}
where\ \
$$L_{a}(\mathbf{X}, \lambda; \mathbf{G})=\left\langle\begin{bmatrix}
                                    \lambda \mathbf{I} & 0\\
                                    0 & -\mathbf{I}\\
                                   \end{bmatrix} \left(\begin{bmatrix}
                                    \mathbf{X}\mathbf{F}_{_{W}}\\
                                    \mathbf{G}\mathbf{X}\mathbf{F}_{\boldsymbol{y}}\\
                                   \end{bmatrix}- \mathbf{A}_{_{\boldsymbol{0}}}(\mathbf{G})\right), \left(\begin{bmatrix}
                                    \mathbf{X}\mathbf{F}_{_{W}}\\
                                    \mathbf{G}\mathbf{X}\mathbf{F}_{\boldsymbol{y}}\\
                                   \end{bmatrix}- \mathbf{A}_{_{\boldsymbol{0}}}(\mathbf{G})\right)\right\rangle,$$
$\mathbf{F}_{\boldsymbol{y}}\triangleq\begin{bmatrix}W^{^{-1}}\boldsymbol{\phi}_{_{\boldsymbol{y}}}& \vdots& \boldsymbol{0}_{m_{_{\boldsymbol{y}}}\times m_{_{\boldsymbol{v}}}}\end{bmatrix},\ 
\  \mathbf{F}_{\boldsymbol{v}}\triangleq\begin{bmatrix}\boldsymbol{0}_{m_{_{\boldsymbol{v}}}\times m_{_{\boldsymbol{y}}}}\ \vdots\ \phi_{_{\boldsymbol{v}}}\end{bmatrix}, \ \ 
\mathbf{F}_{_{W}}\triangleq\begin{bmatrix}\mathbf{I}_{m_{_{\boldsymbol{y}}}}\ \vdots\ \boldsymbol{0}_{m_{_{\boldsymbol{y}}}\times m_{_{\boldsymbol{v}}}}\end{bmatrix}$\
and\break
$\mathbf{A}_{_{\boldsymbol{0}}}(\mathbf{G})\triangleq \left[\begin{array}{c}
                                          \boldsymbol{0} \\ 
                                      \mathbf{X}_{_{\boldsymbol{0}}}(\mathbf{G})\mathbf{F}_{\boldsymbol{y}}+\mathbf{G}\mathbf{F}_{\boldsymbol{v}}
                                       \end{array}\right]$, or, equivalently, for $Z=\operatorname{rvec}(\mathbf{X})$
\begin{equation}\label{eq:10}
 L_{a}(Z, \lambda; \mathbf{G})= \left\langle\mathbf{M}(\lambda)\mathbf{F}Z-\boldsymbol{\mathcal{X}}_{_{\boldsymbol{0}}}(\mathbf{G})), 
 \mathbf{F}Z-\boldsymbol{\mathcal{X}}_{_{\boldsymbol{0}}}(\mathbf{G})\right\rangle,
\end{equation}
where\ \ $\mathbf{M}(\lambda)=\operatorname{diag}(\lambda\mathbf{I}_{m_{_{\boldsymbol{v}}}}, -\mathbf{I}_{m_{_{\boldsymbol{e}}}})\otimes \mathbf{I}_{(m_{_{\boldsymbol{y}}}+m_{_{\boldsymbol{v}}})}$,\ \ 
$\mathbf{F}=\begin{bmatrix}
             \mathbf{I}_{m_{_{\boldsymbol{v}}}}\otimes \mathbf{F}_{_{W}}^{^{\mathrm{T}}}\\
             \mathbf{G} \otimes \mathbf{F}_{\boldsymbol{y}}^{^{\mathrm{T}}}
            \end{bmatrix}$ \ \ and\ \ $\boldsymbol{\mathcal{X}}_{_{\boldsymbol{0}}}(\mathbf{G})=\operatorname{rvec}(\mathbf{A}_{_{\boldsymbol{0}}}(\mathbf{G}))$.
            
Note that it follows from (\ref{eq:07}), (\ref{eq:09}) and (\ref{eq:10}) that
\begin{equation}\label{eq:11}
 \boldsymbol{\varphi}_{_{\mathbf{D}}}(\lambda; \mathbf{G})= \lambda \gamma^{^{2}}- \inf\left\{L_{a}(Z, \lambda; \mathbf{G}): Z\in \mathcal{R}_{c}^{m_{_{\boldsymbol{v}}}m_{_{\boldsymbol{y}}}}\right\}.
\end{equation}

To proceed towards the conversion of \emph{Prob. $1$} into an SDP, the range of $\lambda$ in (\ref{eq:08}) is restricted to a set of values  ($\mathcal{S}_{_{\lambda}}$, say)
over which the $\inf$ of $L_{a}(\cdot)$ (see (\ref{eq:11})) can be recast as the maximum of a linear functional under a matrix inequality 
constraint (in the light of Lemma A1, [17]).  This is stated in the following proposition.

\begin{proposition}\label{prop:03}
 \emph{\textbf{(a)}} $\bar{\mathcal{J}}(\mathbf{G}; \mathcal{S}_{_{\mathbf{X}}})=\inf\{\boldsymbol{\varphi}_{_{\mathbf{D}}}(\lambda; \mathbf{G}): \lambda \in \mathcal{S}_{_{\lambda}}\}$,\ \ where\\
 $\mathcal{S}_{_{\lambda}}\triangleq\left\{\lambda>0: \forall\ \phi \in [0, 2\pi],\ \{\mathbf{F}^{*}\mathbf{M}(\lambda)\mathbf{F}\}(e^{j\alpha})>0\right\}.$\\
  \emph{\textbf{(b)}} For\ \ $\lambda\in \mathcal{S}_{_{\lambda}}$,\\
  $\inf\{L_{a}(Z, \lambda;\mathbf{G}): Z\in \mathcal{R}_{c}^{m_{_{\boldsymbol{v}}}m_{_{\boldsymbol{y}}}}\}=\sup\{\boldsymbol{x}_{_{\boldsymbol{0}}}^{T}\mathbf{P}\boldsymbol{x}_{_{\boldsymbol{0}}}: \mathbf{P}=\mathbf{P}^{^{\mathrm{T}}}\ \text{and}\ \ Q_{_{LQ}}(\mathbf{P}; \boldsymbol{\Sigma}_{a}, \mathbf{M}(\lambda))>0\},$\break
where\ \  $Q_{_{LQ}}\left(\mathbf{P}; \boldsymbol{\Sigma}_{a}, \mathbf{M}(\lambda)\right)=Q_{_{\mathcal{J}}}(\mathbf{P}; \mathbf{A}, \mathbf{B})
+\mathcal{S}\left(\boldsymbol{\Sigma}_{a}, \mathbf{M}(\lambda)\right)$,\ \  
$\boldsymbol{\Sigma}_{a}=\left(\mathbf{A}_{a}, \begin{bmatrix}\mathbf{B}_{a}\ \vdots\ \boldsymbol{b}_{a}\end{bmatrix}, \mathbf{C}_{a}, \begin{bmatrix}\mathbf{D}_{a}\ \vdots\ \boldsymbol{d}_{a}\end{bmatrix}\right)$ is a
realization of\ \  $\begin{bmatrix}\mathbf{F}& \vdots& -\boldsymbol{\mathcal{X}}_{_{\boldsymbol{0}}}(\mathbf{G})\end{bmatrix}$,\

 $\mathbf{A}=\begin{bmatrix}
                                                                                           \mathbf{A}_{a}& \boldsymbol{b}_{a}\\
                                                                                           \boldsymbol{0} & \boldsymbol{0}
                                                                                          \end{bmatrix}$, \ $\mathbf{B}=\begin{bmatrix}
                                                                                                                         \mathbf{B}_{a}\\
                                                                                                                         \boldsymbol{0}
                                                                                                                        \end{bmatrix}$,                                                                                                                         
                                                                                                                        $\boldsymbol{x}_{_{\boldsymbol{0}}}=\begin{bmatrix}
                                                                                                                                         0\\
                                                                                                                                         1
                                                                                                                                        \end{bmatrix}$, \ $\rho(\mathbf{A}_{_{a}})<1$,
                                                                                                                                        \
$\mathbf{R}=\begin{bmatrix}\mathbf{C}_{a}& \boldsymbol{d}_{a}& \mathbf{D}_{a}\end{bmatrix}$,\break
 $Q_{_{\mathcal{J}}}(\mathbf{P}; \mathbf{A}, \mathbf{B})\triangleq\begin{bmatrix}
                                           \mathbf{A}^{^{\mathrm{T}}}\mathbf{P}\mathbf{A}-\mathbf{P}& \mathbf{A}^{^{\mathrm{T}}}\mathbf{P}\mathbf{B}\\
                                           \mathbf{B}^{^{\mathrm{T}}}\mathbf{P}\mathbf{A} & \mathbf{B}^{^{\mathrm{T}}}\mathbf{P}\mathbf{B}
                                          \end{bmatrix}$\ \ and\ \ $\mathcal{S}(\boldsymbol{\Sigma}_{a}, \mathbf{M})\triangleq \mathbf{R}^{^{\mathrm{T}}}\mathbf{M}\mathbf{R}$.\\                                                                                                                                        
\emph{\textbf{(c)}} $\lambda \in \mathcal{S}_{_{\lambda}}$\ if and only if there exists\ \  $\mathbf{P}=\mathbf{P}^{^{\mathrm{T}}}$\ such that $Q_{_{LQ}}(\mathbf{P}; \boldsymbol{\Sigma}_{a}; \mathbf{M}(\lambda))>0$. 
\hfill$\nabla$                                                                                                                                              
\end{proposition}

In the light of (\ref{eq:07}) -- (\ref{eq:11}) and Proposition \ref{prop:03}, \ \ $\bar{\mathcal{J}}_{_{\mathbf{X}}}(\mathbf{G}; \mathcal{S}_{_{\mathbf{X}}})$ can be
written as
\begin{eqnarray}
 \bar{\mathcal{J}}_{_{\mathbf{X}}}(\mathbf{G}; \mathcal{S}_{_{\mathbf{X}}})&=& \inf\left\{\lambda\gamma^{^{2}}
 -\inf\left\{L_{a}(Z, \lambda, \mathbf{G}): Z\in \mathcal{R}_{c}^{m_{_{\boldsymbol{v}}}m_{_{\boldsymbol{y}}}}\right\}:\lambda \in \mathcal{S}_{_{\lambda}}\right\}\
 \ \ \ \ \ \ \ \ \ \ \ \ \ \ \ \ \ \ \ \ \ \ \  \Leftrightarrow\nonumber\\ 
 \bar{\mathcal{J}}_{_{\mathbf{X}}}(\mathbf{G}; \mathcal{S}_{_{\mathbf{X}}})&=& \inf\left\{\lambda\gamma^{^{2}}-\sup\left\{\boldsymbol{x}_{_{\boldsymbol{0}}}^{T}\mathbf{P}\boldsymbol{x}_{_{\boldsymbol{0}}}:
\mathbf{P}=\mathbf{P}^{^{\mathrm{T}}},\ Q_{_{LQ}}(\mathbf{P}; \boldsymbol{\Sigma}_{a}, \mathbf{M}(\lambda))>0 \right\}:\lambda \in \mathcal{S}_{_{\lambda}}\right\}\ \ \Leftrightarrow\nonumber\\ 
\bar{\mathcal{J}}_{_{\mathbf{X}}}(\mathbf{G}; \mathcal{S}_{_{\mathbf{X}}})&=& \inf\left\{\lambda\gamma^{^{2}}+\inf\left\{-\boldsymbol{x}_{_{\boldsymbol{0}}}^{T}\mathbf{P}\boldsymbol{x}_{_{\boldsymbol{0}}}:
\mathbf{P}=\mathbf{P}^{^{\mathrm{T}}},\ Q_{_{LQ}}(\mathbf{P}; \boldsymbol{\Sigma}_{a}, \mathbf{M}(\lambda))>0 \right\}:\lambda \in \mathcal{S}_{_{\lambda}}\right\}\ \Leftrightarrow\nonumber\\ 
\bar{\mathcal{J}}_{_{\mathbf{X}}}(\mathbf{G}; \mathcal{S}_{_{\mathbf{X}}})&=& \inf\left\{\lambda\gamma^{^{2}}+\boldsymbol{x}_{_{\boldsymbol{0}}}^{T}(-\mathbf{P})\boldsymbol{x}_{_{\boldsymbol{0}}}: \lambda>0,\
 \mathbf{P}=\mathbf{P}^{^{\mathrm{T}}},\ Q_{_{LQ}}(\mathbf{P}; \boldsymbol{\Sigma}_{a}, \mathbf{M}(\lambda))>0 \right\}.\label{eq:12}
 \end{eqnarray}

 The nonlinear term\ \  $\mathcal{S}(\boldsymbol{\Sigma}_{a}, \mathbf{M}(\lambda))$\ \ in the matrix inequality above can be eliminated on the basis
 of the
 Schur complement formula. To this effect, note that
 $$\begin{bmatrix}\mathbf{F}& \vdots& -\boldsymbol{\mathcal{X}}_{_{\boldsymbol{0}}}(\mathbf{G})\end{bmatrix}=
\left[\begin{array}{ccc}
 \mathbf{I}_{m_{_{\boldsymbol{v}}}}\otimes \mathbf{F}_{_{W}}^{^{\mathrm{T}}}  & \vdots&  \boldsymbol{0} \\ 
 \boldsymbol{0} &\vdots& \boldsymbol{0} \\
\end{array}\right]+ \begin{bmatrix}
                     \boldsymbol{0}\\
                     \mathbf{I}
                    \end{bmatrix}\mathbf{F}_{_{\boldsymbol{\mathcal{X}}}}(\mathbf{G}),$$
where $\mathbf{F}_{_{\boldsymbol{\mathcal{X}}}}(\mathbf{G})\triangleq\begin{bmatrix}\mathbf{G} \otimes \mathbf{F}_{\boldsymbol{y}}^{^{\mathrm{T}}}& \vdots & 
-\operatorname{rvec}(\mathbf{X}_{_{\boldsymbol{0}}}(\mathbf{G})\mathbf{F}_{\boldsymbol{y}}+\mathbf{G}\mathbf{F}_{\boldsymbol{v}})\end{bmatrix}$,\ so that, for a 
minimal realization\break
$\left(\mathbf{A}_{a}, \ \begin{bmatrix}\mathbf{B}_{a}& \vdots& \boldsymbol{b}_{a}\end{bmatrix},\ \widehat{\mathbf{C}}_{a},
\ \begin{bmatrix}\widehat{\mathbf{D}}_{a}\ \vdots\ \widehat{\boldsymbol{d}}_{a}\end{bmatrix}\right)$\ of \ $\mathbf{F}_{_{\boldsymbol{\mathcal{X}}}}(\mathbf{G})$,\ 
$\mathbf{C}_{a}= \left[\begin{array}{c}
                                          \boldsymbol{0} \\ \hline
                                      \mathbf{I}
                                       \end{array}\right]\widehat{\mathbf{C}}_{a}$,\ $\mathbf{D}_{a}= \left[\begin{array}{c}
                                          \mathbf{I}_{m_{_{\boldsymbol{v}}}}\otimes \mathbf{F}_{_{W}}^{^{\mathrm{T}}} \\ \hline
                                      \widehat{\mathbf{D}}_{a}
                                       \end{array}\right]$,\ $\boldsymbol{d}_{a}= \left[\begin{array}{c}
                                          \boldsymbol{0} \\ \hline
                                      \widehat{\boldsymbol{d}}_{a}
                                       \end{array}\right]$,\break $\mathbf{R}= \left[\begin{array}{c}
                                          \mathbf{R}_{_{_{1}}} \\ \hline
                                      \mathbf{R}_{_{2}}
                                       \end{array}\right]$,\
$\mathbf{R}_{_{_{1}}}=\begin{bmatrix}\boldsymbol{0}_{m_{_{\boldsymbol{v}}}(m_{_{\boldsymbol{y}}}+m_{_{\boldsymbol{v}}})\times n_{\mathbf{A}_{a}}}& \vdots& \boldsymbol{0}_{m_{_{\boldsymbol{v}}}(m_{_{\boldsymbol{y}}}+m_{_{\boldsymbol{v}}})\times 1} & \vdots& \mathbf{I}_{m_{_{\boldsymbol{v}}}}\otimes \mathbf{F}_{_{W}}^{^{\mathrm{T}}}\end{bmatrix}$,\  
                                       $\mathbf{R}_{_{2}}=\begin{bmatrix}\widehat{\mathbf{C}}_{a}& \vdots& \widehat{\boldsymbol{d}}_{a}& \vdots& \widehat{\mathbf{D}}_{a}\end{bmatrix}$\ and,\
                                       hence,\
$\mathcal{S}(\boldsymbol{\Sigma}_{a}, \mathbf{M}(\lambda))=\lambda\mathbf{R}_{_{_{1}}}^{^{\mathrm{T}}}\mathbf{R}_{_{_{1}}}-\mathbf{R}_{_{2}}^{^{\mathrm{T}}}\mathbf{R}_{_{2}}$. As a 
result,

$$Q_{_{LQ}}(P; \boldsymbol{\Sigma}_{a}, \mathbf{M}(\lambda))>0\ \ \Leftrightarrow\ \ \begin{bmatrix}
                                                                      Q_{_{\mathcal{J}}}(\mathbf{P}; \mathbf{A}, \mathbf{B})
                                                                      +\lambda\mathbf{R}_{_{_{1}}}^{^{\mathrm{T}}}\mathbf{R}_{_{_{1}}} & \mathbf{R}_{_{2}}^{^{\mathrm{T}}}&\\
                                                                      \mathbf{R}_{_{2}} & \mathbf{I}&
                                                                     \end{bmatrix}>0
$$
so that (\ref{eq:12}) can be rewritten ($\mathbf{Q}=-\mathbf{P}$) as
\begin{equation}\label{eq:13}
 \bar{\mathcal{J}}_{_{\mathbf{X}}}(\mathbf{G}; \mathcal{S}_{_{\mathbf{X}}})=\inf\left\{\lambda \gamma^{^{2}}+\boldsymbol{x}_{_{\boldsymbol{0}}}^{^{\mathrm{T}}}Q\boldsymbol{x}_{_{\boldsymbol{0}}}:
 \lambda>0, \mathbf{Q}=\mathbf{Q}^{^{\mathrm{T}}},\ Q_{_{\mathcal{J}a}}(\mathbf{Q}, \boldsymbol{\Sigma}_{a}, \mathbf{M}(\lambda))<0)\right\},
\end{equation}
where
\begin{equation}\label{eq:13a}
 Q_{_{\mathcal{J}a}}(\mathbf{Q}; \boldsymbol{\Sigma}_{a}, \mathbf{M}(\lambda))=\begin{bmatrix}
                                                                      Q_{_{\mathcal{J}}}(\mathbf{Q}; \mathbf{A}, \mathbf{B})
                                                                      -\lambda\mathbf{R}_{_{_{1}}}^{^{\mathrm{T}}}\mathbf{R}_{_{_{1}}} & \mathbf{R}_{_{2}}^{^{\mathrm{T}}}&\\
                                                                      \mathbf{R}_{_{2}} & -\mathbf{I}&
                                                                     \end{bmatrix}.
\end{equation}
                                                                     
Confining estimators' frequency-responses to a finite-dimensional subspace of 
$\mathcal{R}_{c}^{m_{_{\boldsymbol{e}}} \times m_{_{\boldsymbol{v}}}}$ (see Remark 4.2),\ $Prob.\ 1$
can be converted
to an SDP on the basis of (\ref{eq:13}). Indeed, let\ $\mathcal{S}_{_{\mathbf{G}}}$\ be defined as
\begin{equation}\label{eq:13b}
\mathcal{S}_{_{\mathbf{G}}}=\left\{\mathbf{G}=
\mathbf{D}+\mathbf{C}\mathbf{Y}_{_{\mathbf{G}}}\mathbf{B}_{_{\mathbf{G}}}: \mathbf{C}\in \mathbb{R}^{m_{_{\boldsymbol{e}}}\times n_{_{\mathbf{G}}} },\ 
\mathbf{D}\in \mathbb{R}^{m_{_{\boldsymbol{e}}}\times m_{_{\boldsymbol{v}}}}\right\},
\end{equation}
where\ \ $\mathbf{Y}_{_{\mathbf{G}}}(e^{j\phi})=(e^{j\phi}\mathbf{I}-\mathbf{A}_{_{\mathbf{G}}})^{^{-1}}$,\ 
$\mathbf{A}_{_{\mathbf{G}}}\in \mathbb{R}^{n_{_{\mathbf{G}}}\times n_{_{\mathbf{G}}}}$,\ $\rho(\mathbf{A}_{_{\mathbf{G}}})<1$,\ 
$\mathbf{B}_{_{\mathbf{G}}}\in \mathbb{R}^{n_{_{\mathbf{G}}}\times m_{_{\boldsymbol{v}}}}$,\ $(\mathbf{A}_{_{\mathbf{G}}}, \mathbf{B}_{_{\mathbf{G}}})$\ 
controllable, or equivalently,
$$\mathcal{S}_{_{\mathbf{G}}}=\left\{\mathbf{G}(\boldsymbol{\beta})=\boldsymbol{\boldsymbol{\beta}}\mathbf{Y}_{_{\mathbf{G}}}^{^{a}}:\ 
\boldsymbol{\boldsymbol{\beta}}=[\mathbf{C}\ \vdots\ \mathbf{D}]\in
\mathbb{R}^{m_{_{\boldsymbol{e}}}\times (n_{_{\mathbf{G}}}+m_{_{\boldsymbol{v}}})}\right\},$$
where\ \ $\mathbf{Y}_{_{\mathbf{G}}}^{^{a}}=\begin{bmatrix}
                              \mathbf{Y}_{_{\mathbf{G}}}\mathbf{B}_{_{\mathbf{G}}}\\
                              \mathbf{I}_{m_{_{\boldsymbol{v}}}}
                             \end{bmatrix}$. \ In this case, as
                             $\mathbf{X}_{_{\boldsymbol{0}}}(\mathbf{G})\mathbf{F}_{\boldsymbol{y}}-\mathbf{G}\mathbf{F}_{\boldsymbol{v}}
                             =(\mathbf{H}_{_{\mathbf{I}}}-\mathbf{G}\mathbf{H}_{_{\boldsymbol{0}}})W\mathbf{F}_{\boldsymbol{y}}-\mathbf{G}\mathbf{F}_{\boldsymbol{v}}$,\break
                            $\mathbf{F}_{_{\boldsymbol{\mathcal{X}}}}(\mathbf{G})=\begin{bmatrix}\left(\boldsymbol{\boldsymbol{\beta}}\otimes\mathbf{I}_{(m_{_{\boldsymbol{y}}}+m_{_{\boldsymbol{v}}})}\right)(\mathbf{Y}_{_{\mathbf{G}}}^{^{a}}\otimes \mathbf{F}_{\boldsymbol{y}}^{T})& \vdots& 
\boldsymbol{\mathcal{X}}_{_{\mathbf{I}W}}+\left(\boldsymbol{\boldsymbol{\beta}}\otimes \mathbf{I}_{(m_{_{\boldsymbol{y}}}+m_{_{\boldsymbol{v}}})}\right)\boldsymbol{\mathcal{X}}_{_{\boldsymbol{0}W}}\end{bmatrix}$,\ \
where\  $\boldsymbol{\mathcal{X}}_{_{\mathbf{I}W}}=-\operatorname{rvec}(\mathbf{H}_{_{\mathbf{I}}}W\mathbf{F}_{\boldsymbol{y}})$,\
$\boldsymbol{\mathcal{X}}_{_{\boldsymbol{0}W}}
=\operatorname{rvec}\left\{\mathbf{Y}_{_{\mathbf{G}}}^{^{a}}(\mathbf{H}_{_{\boldsymbol{0}W}}-\mathbf{F}_{\boldsymbol{v}})\right\}$\ \ 
and\ \ $\mathbf{H}_{_{\boldsymbol{0}W}}\triangleq \mathbf{H}_{_{\boldsymbol{0}}}W\mathbf{F}_{\boldsymbol{y}}$, or,
equivalently,\ \ 
$\mathbf{F}_{_{\boldsymbol{\mathcal{X}}}}(\mathbf{G})=\mathcal{A}_{_{F}}(\boldsymbol{\boldsymbol{\beta}})
\mathbf{F}_{_{\boldsymbol{\mathcal{X}}}}^{^{\mathrm{o}}}$,\ \ where
\begin{equation}\label{eq:14}
 \mathcal{A}_{_{F}}(\boldsymbol{\boldsymbol{\beta}})\triangleq\begin{bmatrix}\boldsymbol{\boldsymbol{\beta}}\otimes \mathbf{I}& \vdots& \mathbf{I}\end{bmatrix}\ \ \text{and}\ \
 \mathbf{F}_{_{\boldsymbol{\mathcal{X}}}}^{^{\mathrm{o}}}=
\left[\begin{array}{ccc}
\mathbf{Y}_{_{\mathbf{G}}}^{^{a}}\otimes \mathbf{F}_{\boldsymbol{y}}^{^{\mathrm{T}}}  & \vdots&  \boldsymbol{\mathcal{X}}_{_{\boldsymbol{0}W}} \\ 
 \boldsymbol{0} &\vdots& \boldsymbol{\mathcal{X}}_{_{\mathbf{I}W}} \\
\end{array}\right]
\end{equation}
so that, letting\ \  $\left(\mathbf{A}_{a},\ \begin{bmatrix}\mathbf{B}_{a}& \vdots& \boldsymbol{b}_{a}\end{bmatrix},\ \mathbf{C}_{_{\boldsymbol{\mathcal{X}}}},\
\begin{bmatrix}\mathbf{D}_{_{\boldsymbol{\mathcal{X}}}}& \vdots& \boldsymbol{d}_{_{\boldsymbol{\mathcal{X}}}}\end{bmatrix}\right)$\ \ 
denote a minimal realization
of\  $\mathbf{F}_{_{\boldsymbol{\mathcal{X}}}}^{^{\mathrm{o}}}$,
\begin{equation}\label{eq:15}
 \mathbf{R}_{_{2}}(\boldsymbol{\boldsymbol{\beta}})=\begin{bmatrix}\widehat{\mathbf{C}}_{a}& \vdots& \widehat{\boldsymbol{d}_{a}}& \vdots& \widehat{\mathbf{D}}_{a}\end{bmatrix}=
 \mathcal{A}_{_{F}}(\boldsymbol{\boldsymbol{\beta}})\begin{bmatrix}\mathbf{C}_{_{\boldsymbol{\mathcal{X}}}}& \vdots& \boldsymbol{d}_{_{\boldsymbol{\mathcal{X}}}}& \vdots& \mathbf{D}_{_{\boldsymbol{\mathcal{X}}}}\end{bmatrix}.
\end{equation}

\emph{Prob. $1$} can then be recast as an SDP as stated in the following proposition (it follows immediately from (\ref{eq:13}) -- (\ref{eq:15})).
\begin{proposition}\label{prop:04}
 \emph{Prob. $1$} can be recast as the following SDP:
 $$\min_{\boldsymbol{\boldsymbol{\beta}}, \mathbf{P}=\mathbf{P}^{^{\mathrm{T}}}, \lambda>0}\lambda \gamma^{^{2}}+\boldsymbol{x}_{_{\boldsymbol{0}}}^{^{\mathrm{T}}}\mathbf{P}\boldsymbol{x}_{_{\boldsymbol{0}}}\ \
 \text{subject to}\ \ Q_{_{\mathcal{J}\boldsymbol{\mathcal{X}}}}(\mathbf{P}, \lambda, \boldsymbol{\boldsymbol{\beta}})<0,$$
 where\ \ $Q_{_{\mathcal{J}\boldsymbol{\mathcal{X}}}}(\mathbf{P}, \lambda, \boldsymbol{\boldsymbol{\beta}})\triangleq\begin{bmatrix}
                                                                             Q_{_{\mathcal{J}}}(\mathbf{P}; \mathbf{A}, \mathbf{B})
                                                                             -\lambda \mathbf{R}_{_{_{1}}}^{^{\mathrm{T}}}\mathbf{R}_{_{_{1}}}& 
                                                                             \mathbf{R}_{_{2}}(\boldsymbol{\boldsymbol{\beta}})&\\
                                                                             \mathbf{R}_{_{2}}(\boldsymbol{\boldsymbol{\beta}})& -\mathbf{I}&
                                                                            \end{bmatrix}
$,\ \ $\mathbf{R}_{_{2}}(\boldsymbol{\boldsymbol{\beta}})$\ \ is an affine function of\ \ $\boldsymbol{\boldsymbol{\beta}}$,\break
$\mathbf{R}_{_{_{1}}}=[0\ \vdots\ 0\ \vdots\ \mathbf{I}_{m_{_{\boldsymbol{v}}}}\otimes \mathbf{F}_{_{W}}^{^{\mathrm{T}}}]$\ \ and\ \
$(\mathbf{A}_{a}, [\mathbf{B}_{a}\ \vdots\ \boldsymbol{b}_{a}], \mathbf{C}_{_{\boldsymbol{\mathcal{X}}}},
[\mathbf{D}_{_{\boldsymbol{\mathcal{X}}}}\ \vdots\ \boldsymbol{d}_{_{\boldsymbol{\mathcal{X}}}}])$\ \
is a minimal realization of\ \ $\mathbf{F}_{_{\boldsymbol{\mathcal{X}}}}^{^{\mathrm{o}}}$\ (given  by {(\ref{eq:14})}). \hfill $\nabla$
\end{proposition}

\vspace*{3mm}

\begin{remark}
 Once\ \ $\boldsymbol{\beta}_{_{\mathrm{o}}}$\  is obtained from a solution\ \ 
 $(\boldsymbol{\beta}_{_{\mathrm{o}}}, \mathbf{P}_{_{\mathrm{o}}}, \lambda_{_{\mathrm{o}}})$\ to the problem posed in Proposition \ref{prop:04}, the 
 corresponding minimax estimator has frequency-response given by\ \
 $\mathbf{G}(\boldsymbol{\beta}_{_{\mathrm{o}}})=\boldsymbol{\beta}_{_{\mathrm{o}}}\mathbf{Y}_{_{\mathbf{G}}}^{^{a}}$.
 \hfill$\nabla$ 
\end{remark}

\vspace*{3mm}

\begin{remark}
Taking\ $\mathcal{S}_{_{\mathbf{G}}}$\ to be a linear subspace of\ \ $\mathcal{R}_{c}^{m_{_{\boldsymbol{v}}}\times m_{_{\boldsymbol{y}}}}$\ \
is instrumental to the recasting of \emph{Prob. $1$} as a SDP. It would be quite natural and conceptually somewhat preferable to take\ \
$\mathcal{S}_{_{\mathbf{G}}}$\ \ to be the subset of\ \ $\mathcal{R}_{c}^{m_{_{\boldsymbol{v}}}\times m_{_{\boldsymbol{y}}}}$\  
corresponding to state-space realizations of a prescribed maximum dimension. This was indeed done in connection with robust estimator 
design problems involving the MSE and $\mathcal{H}_{\infty}$ criteria and parametric uncertainty classes (\emph{e.g.},[6]) without precluding their recasting as SDPs --
this is also the case with the nominal and robust\ $\mathcal{H}_{\infty}$\ estimation problems tackled in Subsections \ref{subsec:4.2} and \ref{subsec:4.3} 
below. However, in the case of the MSE criterion and non-parametric, $\mathcal{H}_{2}-$ball, channel-model uncertainty addressed here, the 
approach pursued in the subsequent sections to achieve the desired conversions into SDPs (hinging upon the so-called Elimination Lemma)
would not seem to be applicable beyond the SISO case.

Further justification for taking\ \ $\mathcal{S}_{_{\mathbf{G}}}$\ as in (\ref{eq:13b}) comes from the fact that any frequency-response in\ 
$\mathcal{R}_{c}^{m_{_{\boldsymbol{v}}}\times m_{_{\boldsymbol{y}}}}$\ can be approximated (as close as desired) with respect to the 
$\mathcal{H}_{2}-$norm in classes of FIRs (of sufficiently large length) and as mentioned above, such a class can be cast in the form of 
(\ref{eq:13b}). In addition, computing the optimal (over the whole of $\mathcal{R}_{c}^{m_{_{\boldsymbol{v}}}\times m_{_{\boldsymbol{y}}}}$)
nominal MSE estimator (say, $\mathbf{G}_{_{\mathrm{o}}}$\ with minimal realization\ \ $(\mathbf{A}_{_{\mathrm{o}}}$, $\mathbf{B}_{_{\mathrm{o}}}$, $\mathbf{C}_{_{\mathrm{o}}}$, $\mathbf{D}_{_{\mathrm{o}}})$) leads to a ``problem-specific'' and well-motivated class of admissible estimators (say, $\mathcal{S}_{_{\mathbf{G}}}^{^{\text{nom}}}$ as in (\ref{eq:13b}) with\ \ $\mathbf{A}_{_{\mathbf{G}}}=\mathbf{A}_{_{\mathrm{o}}}$,\  $\mathbf{B}_{_{\mathbf{G}}}=\mathbf{B}_{_{\mathrm{o}}}$ ), as done in the example presented in Subsection \ref{subsec:6.2}. 

Finally, it is noted that lower bounds on\
\ $\mu_{c}(\mathcal{S}_{_{\mathbf{X}}})=\inf\{\bar{\mathcal{J}}(\mathbf{G};\mathcal{S}_{_{\mathbf{X}}}):\mathbf{G}\in
\mathcal{R}_{c}^{m_{_{\boldsymbol{e}}}\times m_{_{\boldsymbol{v}}}}\}$\ can be obtained as optimal values of SDPs (see [18]) so that, 
in any given estimation exercise, upper bounds can be obtained on the increase of the minimax MSE brought about by confining 
estimators to\ \ $\mathcal{S}_{_{\mathbf{G}}}$ instead of optimizing over the whole of
\ $\mathcal{R}_{c}^{m_{_{\boldsymbol{e}}}\times m_{_{\boldsymbol{v}}}}$\ -- in fact, as illustrated in [18], the class
$\mathcal{S}_{_{\mathbf{G}}}^{^{\text{nom}}}$ mentioned above may lead to optimal minimax performance which is quite close to\ $\mu_{c}(\mathcal{S}_{_{\mathbf{X}}})$.\hfill$\nabla$ 
\end{remark}

\subsection{Nominal ``$\mathcal{H}_{\infty}-$Estimation''}\label{subsec:4.2}
Let $\mathbf{H}\in \mathcal{R}_{c}^{m_{_{\boldsymbol{v}}}\times m_{_{\boldsymbol{y}}}}$, 
$\mathcal{S}_{_{\mathbf{G}}}^{\mathrm{o}}\subset \mathcal{R}_{c}^{m_{_{\boldsymbol{e}}}\times m_{_{\boldsymbol{v}}}}$ and consider the ``$\mathcal{H}_{\infty}-$filtering''
problem
$$\underline{Prob.\ 2 :}\ \min_{\mathbf{G}\in \mathcal{S}_{_{\mathbf{G}}}^{\mathrm{o}}}\mathcal{J}_{\infty}(\mathbf{G}; \mathbf{H}).$$
The major aim of this section is to show that, for a given $n_{_{\mathbf{G}}}$ and
$$\mathcal{S}_{_{\mathbf{G}}}^{\mathrm{o}}=\{\mathbf{G}\in\mathcal{R}_{c}^{m_{_{\boldsymbol{e}}}\times m_{_{\boldsymbol{v}}}}: \mathbf{G}\ \text{has a realization}\ 
(\mathbf{A}_{_{\mathbf{G}}}, \mathbf{B}_{_{\mathbf{G}}}, \mathbf{C}_{_{\mathbf{G}}}, \mathbf{D}_{_{\mathbf{G}}})
\ \text{with}\ \mathbf{A}_{_{\mathbf{G}}}\in \mathbb{R}^{n_{_{\mathbf{G}}}\times n_{_{\mathbf{G}}}}, \rho(\mathbf{A}_{_{\mathbf{G}}})<1\},$$
a solution to \emph{Prob. $2$} can be obtained on the basis of SDPs.

To this effect and proceeding along the lines which led to Proposition \ref{prop:03}, let a Lagrangian and dual functional be given by
$$Lag_{\infty}(\bar{\boldsymbol{z}}, \boldsymbol{\sigma}; \mathbf{G}, \mathbf{H})\triangleq\|\bar{\boldsymbol{e}}(\bar{\boldsymbol{z}}; \mathbf{G}, \mathbf{H})\|_{_{2}}^{^{2}}
-\boldsymbol{\sigma}_{_{\boldsymbol{y}}}(\|\bar{\boldsymbol{y}}\|_{_{2}}^{^{2}}-\gamma_{\boldsymbol{y}}^{^{2}})-
\boldsymbol{\sigma}_{_{\boldsymbol{v}}}(\|\bar{\boldsymbol{v}}\|_{_{2}}^{^{2}}-\gamma_{\boldsymbol{v}}^{^{2}})$$
and\ \ $\boldsymbol{\varphi}_{_{\mathbf{D}\infty}}(\boldsymbol{\sigma}; \mathbf{G},
\mathbf{H}) \triangleq\sup\{Lag_{\infty}(\bar{\boldsymbol{z}}, \boldsymbol{\sigma}; \mathbf{G},
\mathbf{H}) : \bar{\boldsymbol{z}}\in \mathcal{R}_{c}^{m_{\boldsymbol{z}}}\}$, \ \ where\ \
$\boldsymbol{\sigma}=(\boldsymbol{\sigma}_{_{\boldsymbol{y}}}, \boldsymbol{\sigma}_{_{\boldsymbol{v}}})$ \ \ and\break
$m_{\boldsymbol{z}}=m_{_{\boldsymbol{y}}}+m_{_{\boldsymbol{v}}}$.

It then follows from Theorem 2 in [12] that
\begin{eqnarray*}
\mathcal{J}_{\infty}(\mathbf{G};\mathbf{H})=\bar{\boldsymbol{\varphi}}_{_{\mathbf{D}\infty}}(\mathbf{G};\mathbf{H})&\triangleq&\inf\{\boldsymbol{\varphi}_{_{\mathbf{D}\infty}}(\boldsymbol{\sigma};
\mathbf{G}, \mathbf{H}): \boldsymbol{\sigma}>0\}.
\end{eqnarray*}
Note now that
$$Lag_{\infty}(\bar{\boldsymbol{z}}, \boldsymbol{\sigma}; \mathbf{G}, \mathbf{H})
=\langle(\mathbf{H}_{_{\mathbf{I}\boldsymbol{y}}}-\mathbf{G}\mathbf{H}_{_{\boldsymbol{0}\boldsymbol{z}}})\bar{\boldsymbol{z}},
(\mathbf{H}_{_{\mathbf{I}\boldsymbol{y}}}-\mathbf{G}\mathbf{H}_{_{\boldsymbol{0}\boldsymbol{z}}})\bar{\boldsymbol{z}}\rangle 
- \langle\mathbf{M}_{_{\boldsymbol{\sigma}}}\bar{\boldsymbol{z}}, \bar{\boldsymbol{z}}\rangle 
+ \boldsymbol{\sigma}_{_{\boldsymbol{y}}}\gamma_{\boldsymbol{y}}^{^{2}}+ \boldsymbol{\sigma}_{_{\boldsymbol{v}}}\gamma_{\boldsymbol{v}}^{^{2}},$$
where\ \ $\mathbf{H}_{_{\mathbf{I}\boldsymbol{y}}}$\ \ and\ \ $\mathbf{H}_{_{\boldsymbol{0}}}$\ \ are defined in Subsection \ref{subsec:2.2} and\  
$\mathbf{M}_{_{\boldsymbol{\sigma}}}=
\operatorname{diag}(\boldsymbol{\sigma}_{_{\boldsymbol{y}}}\mathbf{I}_{m_{_{\boldsymbol{e}}}}, \boldsymbol{\sigma}_{_{\boldsymbol{v}}}
\mathbf{I}_{m_{_{\boldsymbol{v}}}})$. Thus, the dual 
functional \ $\varphi_{_{\mathbf{D\infty}}}$\
can be written as 
$$\boldsymbol{\varphi}_{_{\mathbf{D}\infty}}(\boldsymbol{\sigma}; \mathbf{G}, \mathbf{H})=\boldsymbol{\sigma}_{_{\boldsymbol{y}}}\gamma_{\boldsymbol{y}}^{^{2}}
+\boldsymbol{\sigma}_{_{\boldsymbol{v}}}\gamma_{\boldsymbol{v}}^{^{2}}-\check{\varphi}_{_{\mathbf{D}\infty}}(\boldsymbol{\sigma}; \mathbf{G}, \mathbf{H}),$$
where\ \ $\check{\varphi}_{_{\mathbf{D}\infty}}(\boldsymbol{\sigma}; \mathbf{G}, \mathbf{H})
=\inf\{\langle(\mathbf{M}_{_{\boldsymbol{\sigma}}}-\mathbf{F}_{_{\mathbf{G}}}^{*}\mathbf{F}_{_{\mathbf{G}}})\bar{\boldsymbol{z}},
\bar{\boldsymbol{z}}\rangle:\bar{\boldsymbol{z}}\in \mathcal{R}_{c}^{m_{\boldsymbol{yv}}}\}$\ \ and\ \ 
$\mathbf{F}_{_{\mathbf{G}}}\triangleq\mathbf{H}_{_{\mathbf{I}\boldsymbol{y}}}-\mathbf{G}\mathbf{H}_{_{\boldsymbol{0}\boldsymbol{z}}}$.

To proceed, consider the following proposition.\\

\begin{proposition}\label{prop:08a}
$\bar{\boldsymbol{\varphi}}_{_{\mathbf{D}\infty}}(\mathbf{G};\mathbf{H})$\ \  can be written as\\ 
$\bar{\boldsymbol{\varphi}}_{_{\mathbf{D}\infty}}(\mathbf{G};\mathbf{H})=\inf\{\boldsymbol{\sigma}_{_{\boldsymbol{y}}}\gamma_{_{\boldsymbol{y}}}^{^{2}}
+\boldsymbol{\sigma}_{_{\boldsymbol{v}}}\gamma_{\boldsymbol{v}}^{2}: \boldsymbol{\sigma}_{_{\boldsymbol{y}}}>0, \boldsymbol{\sigma}_{_{\boldsymbol{v}}}>0\ \ \
\text{and}\ \ \ \forall \phi \in [0, 2\pi],\ \ 
(\mathbf{M}_{_{\boldsymbol{\sigma}}}-\mathbf{F}_{_{\mathbf{G}}}^{*}\mathbf{F}_{_{\mathbf{G}}})(e^{j\phi})>0\}.$\hfill$\nabla$
\end{proposition}

In the light of Proposition \ref{prop:08a}, the next proposition is an immediate consequence of the so-called (discrete-time) bounded-real lemma ([27]).

\begin{proposition}\label{prop:10}
 \begin{equation*}\label{eq:22-a}
 \mathcal{J}_{\infty}(\mathbf{G}, \mathbf{H})=
 \inf\{\boldsymbol{\sigma}_{_{\boldsymbol{y}}}\gamma_{\boldsymbol{y}}^{^{2}}
 +\boldsymbol{\sigma}_{_{\boldsymbol{v}}}\gamma_{\boldsymbol{v}}^{^{2}}:\boldsymbol{\sigma}_{_{\boldsymbol{y}}}>0, 
 \boldsymbol{\sigma}_{_{\boldsymbol{v}}}>0,\mathbf{P}=\mathbf{P}^{^{\mathrm{T}}}>0\ \text{and}\ 
 \ Q_{_{\mathbf{B}\mathbf{R}}}(\mathbf{P}; \boldsymbol{\Sigma}_{_{\mathbf{F}\mathbf{G}}}, 
\mathbf{M}_{_{\boldsymbol{\sigma}}})<0\} 
 \end{equation*}
 where
 \begin{equation*}
Q_{_{\mathbf{B}\mathbf{R}}}(\mathbf{P}; \boldsymbol{\Sigma}_{_{\mathbf{F}\mathbf{G}}}, 
\mathbf{M}_{_{\boldsymbol{\sigma}}})\triangleq\begin{bmatrix}
                       \mathbf{A}_{_{\mathbf{F}\mathbf{G}}}^{^{\mathrm{T}}}\mathbf{P}\mathbf{A}_{_{\mathbf{F}\mathbf{G}}}-\mathbf{P}& \vdots &
                       \mathbf{A}_{_{\mathbf{F}\mathbf{G}}}^{^{\mathrm{T}}}\mathbf{P}\mathbf{B}_{_{\mathbf{F}\mathbf{G}}}\\
                       \mathbf{B}_{_{\mathbf{F}\mathbf{G}}}^{^{\mathrm{T}}}\mathbf{P}\mathbf{A}_{_{\mathbf{F}\mathbf{G}}}& \vdots &
                       \mathbf{B}_{_{\mathbf{F}\mathbf{G}}}^{^{\mathrm{T}}}\mathbf{P}\mathbf{B}_{_{\mathbf{F}\mathbf{G}}}
                       \end{bmatrix}
                      + \begin{bmatrix}
                                       \mathbf{C}_{_{\mathbf{F}\mathbf{G}}}^{^{\mathrm{T}}}\\
                                       \mathbf{D}_{_{\mathbf{F}\mathbf{G}}}^{^{\mathrm{T}}}
                                       \end{bmatrix} [\mathbf{C}_{_{\mathbf{F}\mathbf{G}}}\ \mathbf{D}_{_{\mathbf{F}\mathbf{G}}}]                               -\begin{bmatrix}
                                         \boldsymbol{0}& \boldsymbol{0}\\
                                         \boldsymbol{0}& \mathbf{M}_{_{\boldsymbol{\sigma}}}
                                        \end{bmatrix} 
\end{equation*}
and \ \ \ $\boldsymbol{\Sigma}_{_{\mathbf{F}\mathbf{G}}}=(\mathbf{A}_{_{\mathbf{F} \mathbf{G}}}, 
                              \mathbf{B}_{_{\mathbf{F} \mathbf{G}}}, \mathbf{C}_{_{\mathbf{F} \mathbf{G}}}, \mathbf{D}_{_{\mathbf{F} \mathbf{G}}})$\ \ \ is a 
                              realization of\ \ \ $\mathbf{F}_{_{\mathbf{G}}}$.
 \hfill$\nabla$
 \end{proposition}
 
As a result, for such $\mathcal{S}_{_{\mathbf{G}}}^{\mathrm{o}}$ and in the light of Proposition \ref{prop:10}, \emph{Prob. $2$} can be 
stated as
$$\underline{Prob.\ 2a :}\ \displaystyle\min_{\begin{smallmatrix}\boldsymbol{\theta},\ \rho(\mathbf{A}_{_{\mathbf{G}}})<1,\\
\boldsymbol{\sigma}_{_{\boldsymbol{y}}}>0,\ \boldsymbol{\sigma}_{_{\boldsymbol{v}}}>0 \\ 
\mathbf{P}=\mathbf{P}^{^{\mathrm{T}}}>0
 \end{smallmatrix}}
 \boldsymbol{\sigma}_{_{\boldsymbol{y}}}\gamma_{\boldsymbol{y}}^{^{2}}+\boldsymbol{\sigma}_{_{\boldsymbol{v}}}\gamma_{\boldsymbol{v}}^{^{2}}
\ \ \ \ \ \text{subject to}\ \ \ \
 Q_{_{\mathbf{B}\mathbf{R}}}(\mathbf{P}, \boldsymbol{\Sigma}_{_{\mathbf{F}\mathbf{G}}}(\boldsymbol{\theta}), \mathbf{M}_{_{\boldsymbol{\sigma}}})<0, 
$$
where\ \ $\boldsymbol{\theta}=\begin{bmatrix}
                              \mathbf{A}_{_{\mathbf{G}}}&\mathbf{B}_{_{\mathbf{G}}}\\
                              \mathbf{C}_{_{\mathbf{G}}}&\mathbf{D}_{_{\mathbf{G}}}
                              \end{bmatrix}$.\\
                              
To convert $Prob.\ 2a$ into a SDP, the approach to the\ \ $\mathcal{H}_{\infty}$\ control problem pursued in [27]   is followed here. Its first 
step is to separate the estimator parameter $\boldsymbol{\theta}$ from $(\mathbf{P}, \boldsymbol{\sigma})$ in the constraint
above so that the so-called Elimination Lemma (see [27]) can be invoked. This is done rewriting the condition\ ``$Q_{_{\mathbf{B}\mathbf{R}}}(\cdot)<0$''\ and exploiting tha fact that\ \ 
$\mathbf{A}_{_{\mathbf{F}\mathbf{G}}}= \mathbf{A}_{_{\mathrm{o}}}+\mathbf{A}_{L}(\boldsymbol{\theta})$,\ \ 
$\mathbf{B}_{_{\mathbf{F}\mathbf{G}}}= \mathbf{B}_{_{\mathrm{o}}}+\mathbf{B}_{L}(\boldsymbol{\theta})$,\ \
$\mathbf{C}_{_{\mathbf{F}\mathbf{G}}}= \mathbf{C}_{_{\mathrm{o}}}+\mathbf{C}_{L}(\boldsymbol{\theta})$,\ \
$\mathbf{D}_{_{\mathbf{F}\mathbf{G}}}= \mathbf{D}_{_{\mathrm{o}}}+\mathbf{D}_{L}(\boldsymbol{\theta})$, \ \ where $\mathbf{A}_{L}$,\ $\mathbf{B}_{L}$, $\mathbf{C}_{L}$,\
$\mathbf{D}_{L}$\ \ are linear functions of\ $\boldsymbol{\theta}$\  and\ $(\mathbf{A}_{_{\mathrm{o}}}, \mathbf{B}_{_{\mathrm{o}}}, \mathbf{C}_{_{\mathrm{o}}}, \mathbf{D}_{_{\mathrm{o}}})$\  are given matrices.
This leads to the following proposition.

\begin{proposition}\label{prop:11}
 \emph{\textbf{(a)}}\ \ $Q_{_{\mathbf{B}\mathbf{R}}}(\mathbf{P}, \boldsymbol{\Sigma}_{_{\mathbf{F}\mathbf{G}}}(\boldsymbol{\theta}), \mathbf{M}(\boldsymbol{\sigma}))<0 \ \ \ \Leftrightarrow\ \ \ \boldsymbol{\psi}(\mathbf{P},\boldsymbol{\sigma}, \boldsymbol{\theta})>0$,\\
where
 $$\boldsymbol{\psi}(\mathbf{P}, \boldsymbol{\sigma}, \boldsymbol{\theta})\triangleq
\begin{bmatrix}
\mathbf{P}^{^{-1}}&\mathbf{A}_{_{\mathbf{F}\mathbf{G}}}&\mathbf{B}_{_{\mathbf{F}\mathbf{G}}}&\boldsymbol{0}_{n_{_{\mathbf{F}\mathbf{G}}}\times m_{_{\boldsymbol{e}}}}\\
 \mathbf{A}_{_{\mathbf{F}\mathbf{G}}}^{^{\mathrm{T}}}&\mathbf{P}&\boldsymbol{0}_{n_{_{\mathbf{FG}}}\times m_{\boldsymbol{yv}}} & \mathbf{C}_{_{\mathbf{F}\mathbf{G}}}^{^{\mathrm{T}}}\\
 \mathbf{B}_{_{\mathbf{F}\mathbf{G}}}^{^{\mathrm{T}}}&\boldsymbol{0}_{m_{\boldsymbol{yv}}\times n_{_{\mathbf{FG}}}}&\mathbf{M}_{_{\boldsymbol{\sigma}}}& \mathbf{D}_{_{\mathbf{FG}}}^{^{\mathrm{T}}}\\
\boldsymbol{0}_{m_{_{\boldsymbol{e}}}\times n_{_{\mathbf{FG}}}}&\mathbf{C}_{_{\mathbf{FG}}}&\mathbf{D}_{_{\mathbf{FG}}}&\mathbf{I}_{m_{_{\boldsymbol{e}}}}
 \end{bmatrix}.$$\\

 \noindent
 \emph{\textbf{(b)}}\ \ $\boldsymbol{\psi}(\mathbf{P}, \boldsymbol{\sigma},\boldsymbol{\theta})
 =\boldsymbol{\psi}_{_{\mathrm{o}}}(\mathbf{P}, \boldsymbol{\sigma})+ \mathbf{T}_{_{a}}^{^{\mathrm{T}}}\boldsymbol{\theta}\mathbf{T}_{_{b}}
 +(\mathbf{T}_{_{a}}^{^{\mathrm{T}}}\boldsymbol{\theta}\mathbf{T}_{_{b}})^{^{\mathrm{T}}}>0,$\\
 where
 $$\boldsymbol{\psi}_{_{\mathrm{o}}}(\mathbf{P}, \boldsymbol{\sigma})\triangleq
\begin{bmatrix}
\mathbf{P}^{^{-1}}& \vdots&\mathbf{A}_{_{\mathrm{o}}}&\vdots&\mathbf{B}_{_{\mathrm{o}}}&\vdots&\boldsymbol{0}_{n_{_{\mathbf{FG}}}\times m_{_{\boldsymbol{e}}}}\\
 \mathbf{A}_{_{\mathrm{o}}}^{^{\mathrm{T}}}&\vdots&\mathbf{P}&\vdots&\boldsymbol{0}_{n_{_{\mathbf{FG}}}\times m_{\boldsymbol{yv}}}&\vdots & \mathbf{C}_{_{\mathrm{o}}}^{^{\mathrm{T}}}\\
 \mathbf{B}_{_{\mathrm{o}}}^{^{\mathrm{T}}}&\vdots&\boldsymbol{0}_{m_{\boldsymbol{yv}}\times n_{_{\mathbf{FG}}}}&\vdots&\mathbf{M}_{_{\boldsymbol{\sigma}}}& \vdots&\mathbf{D}_{_{\mathbf{I}\boldsymbol{y}}}^{^{\mathrm{T}}}\\
\boldsymbol{0}_{m_{_{\boldsymbol{e}}}\times n_{_{\mathbf{FG}}}}&\vdots&\mathbf{C}_{_{\mathrm{o}}}&\vdots&\mathbf{D}_{_{\mathbf{I}\boldsymbol{y}}}&\vdots&\mathbf{I}_{m_{_{\boldsymbol{e}}}}
 \end{bmatrix},
$$

\bigskip

$$\mathbf{T}_{_{a}}=\begin{bmatrix}
                  \boldsymbol{0}&\mathbf{I}&\vdots&\boldsymbol{0}& \boldsymbol{0}&\vdots& \boldsymbol{0}& \vdots& \boldsymbol{0}\\
                 \boldsymbol{0}&\boldsymbol{0}&\vdots&\boldsymbol{0}& \boldsymbol{0}&\vdots& \boldsymbol{0}&\vdots& -\mathbf{I}
                  \end{bmatrix},\ \ \ \ \ \
\mathbf{T}_{_{b}}=\begin{bmatrix}
                  \boldsymbol{0}&\boldsymbol{0}&\vdots&\boldsymbol{0}& \mathbf{I}&\vdots& \boldsymbol{0}&\vdots& \boldsymbol{0}\\
                 \boldsymbol{0}&\boldsymbol{0}&\vdots&\mathbf{C}_{_{a\boldsymbol{z}}}& \boldsymbol{0}&\vdots& \mathbf{D}_{_{\mathrm{o}\boldsymbol{z}}}&\vdots& \boldsymbol{0}
                  \end{bmatrix},
$$
$\mathbf{A}_{_{\mathrm{o}}}$,\ $\mathbf{B}_{_{\mathrm{o}}}$,\ $\mathbf{C}_{_{\mathrm{o}}}$,\ $\mathbf{D}_{_{\mathbf{I}\boldsymbol{y}}}$,\ $\mathbf{C}_{_{a\boldsymbol{z}}}$\ and \
$\mathbf{D}_{_{\mathrm{o}\boldsymbol{z}}}$\ are given matrices (see Appendix).\hfill$\nabla$
\end{proposition}

\bigskip

Now, it follows from the Elimination Lemma that\

\noindent
\textbf{(\emph{i})}\ If\ \ $(\mathbf{P}, \boldsymbol{\sigma},\boldsymbol{\theta})$\ \  is such that\ \ 
$\boldsymbol{\psi}(\mathbf{P}, \boldsymbol{\sigma},\boldsymbol{\theta})>0$\ \ \ \ \ \ $\Rightarrow$
\begin{equation}\label{eq:25}
 \mathbf{W}_{_{a}}^{^{\mathrm{T}}}\boldsymbol{\psi}_{_{\mathrm{o}}}(\mathbf{P}, \boldsymbol{\sigma})\mathbf{W}_{_{a}}>0\ \ \ \ \ \ \ \ \text{and}\ \ \ \ \ \ \ \ 
 \mathbf{W}_{_{b}}^{^{\mathrm{T}}}\boldsymbol{\psi}_{_{\mathrm{o}}}(\mathbf{P}, \boldsymbol{\sigma})\mathbf{W}_{_{b}}>0,
\end{equation}
where the columns of\ \ $\mathbf{W}_{_{a}}$\ \ and\ \ $\mathbf{W}_{_{b}}$\ \ respectively constitute bases for the null spaces of\ \ 
$\mathbf{T}_{_{a}}$\ \ and\ \ $\mathbf{T}_{_{b}}$;\

\noindent
\textbf{(\emph{ii})}\ If\ $\mathbf{P}$\ \ and \ \ $\boldsymbol{\sigma}$\ \ are such that (\ref{eq:25}) hold, 
$\exists\ \boldsymbol{\theta}$\ \ such that\ \ $\boldsymbol{\psi}(\boldsymbol{\sigma},\mathbf{P}, \boldsymbol{\theta})>0$.

Thus, it follows from \textbf{(\emph{i})}, \textbf{(\emph{ii})} and Proposition \ref{prop:11} that \emph{Prob. $2$} can be converted into two
problems to be solved in sequence as stated in the following proposition.\

\begin{proposition}\label{prop:12}
 \emph{\textbf{(a)}}\ The optimal value\  $\mathcal{J}_{\infty}^{\mathrm{o}}(\mathbf{H})$\ of \emph{Prob. $2$} is given by 
 \begin{eqnarray*}
 \mathcal{J}_{\infty}^{\mathrm{o}}(\mathbf{H})&=&\inf\{\boldsymbol{\sigma}_{_{\boldsymbol{y}}}\gamma_{\boldsymbol{y}}^{^{2}}
 +\boldsymbol{\sigma}_{_{\boldsymbol{v}}}\gamma_{\boldsymbol{v}}^{^{2}}: \boldsymbol{\sigma}_{_{\boldsymbol{y}}}>0, \ \boldsymbol{\sigma}_{_{\boldsymbol{v}}}>0,\ 
\mathbf{P}=\mathbf{P}^{^{\mathrm{T}}}>0,\
 \boldsymbol{\theta}\in \mathbb{R}^{(n_{_{\mathbf{G}}}+m_{_{\boldsymbol{e}}})\times(n_{_{\mathbf{G}}}+m_{_{\boldsymbol{v}}})}\\\
 &&\ \ \ \ \ \ \ \ \ \ \ \ \ \ \ \ \ \ \ \ \ \ \ \ \ \ \ \ \ \ 
 \ \ \ \ \ \ \ \ \ \ \ \ \ \ \ \ \  \ \ \ \ \ \ \ \ \ \ \ \ \ \ \ \ \ \ \ \ \ \ \ \   \text{are such that}\ \ \boldsymbol{\psi}(\boldsymbol{\sigma},\mathbf{P},\boldsymbol{\theta})>0\}
 \end{eqnarray*}
is also given by
$$
 \mathcal{J}_{\infty}^{\mathrm{o}}(\mathbf{H})=\inf\{\boldsymbol{\sigma}_{_{\boldsymbol{y}}}\gamma_{\boldsymbol{y}}^{^{2}}
 +\boldsymbol{\sigma}_{_{\boldsymbol{v}}}\gamma_{\boldsymbol{v}}^{^{2}}: \boldsymbol{\sigma}_{_{\boldsymbol{y}}}>0, \ \boldsymbol{\sigma}_{_{\boldsymbol{v}}}>0,\ 
\mathbf{P}=\mathbf{P}^{^{\mathrm{T}}}>0\ \text{are such that}\ \ (\ref{eq:25})\ \text{holds}.\}
$$
\emph{\textbf{(b)}}\ If\ \ $\mathbf{P}^{^{\mathrm{o}}}$\ and\ $\boldsymbol{\sigma}^{^{\mathrm{o}}}$\ are such that \emph{(\ref{eq:25})} holds there exists
 \ $\boldsymbol{\theta}$\ such that
\begin{equation}\label{eq:26}
 \boldsymbol{\psi}_{_{\mathrm{o}}}(\mathbf{P}^{^{\mathrm{o}}}, \boldsymbol{\sigma}^{^{\mathrm{o}}})+\mathbf{T}_{_{a}}^{^{\mathrm{T}}}\boldsymbol{\theta}\mathbf{T}_{_{b}}
 +(\mathbf{T}_{_{a}}^{^{\mathrm{T}}}\boldsymbol{\theta}\mathbf{T}_{_{b}})^{^{\mathrm{T}}}>0
\end{equation}
and\ $\rho(\mathbf{A}_{_{\mathbf{G}}})<1$.

Moreover, if\ \ $\boldsymbol{\sigma}_{_{\boldsymbol{y}}}^{^{\mathrm{o}}}\gamma_{\boldsymbol{y}}^{^{2}}
 +\boldsymbol{\sigma}_{_{\boldsymbol{v}}}^{^{\mathrm{o}}}\gamma_{\boldsymbol{v}}^{^{2}}=\mathcal{J}_{\infty}^{\mathrm{o}}(\mathbf{H})+\varepsilon$\ \ then any\ \
  $\boldsymbol{\theta}$\ for which  holds is such that\break 
  $\mathcal{J}_{\infty}(\mathbf{G}(\boldsymbol{\theta}); \mathbf{H})\leq \mathcal{J}_{\infty}^{\mathrm{o}}(\mathbf{H})+\varepsilon$.
\hfill$\nabla$ 
\end{proposition}

\begin{remark}
 In words, an approximate solution, say\ $\widehat{\mathbf{G}}$,\ to \emph{Prob. $2$} can be obtained on the basis of an approximate solution\ 
 $(\mathbf{P}^{^{\mathrm{o}}}, \boldsymbol{\sigma}^{^{\mathrm{o}}})$\ of \emph{Prob. 2b} in the following way:\\
find a solution \ $\widehat{\boldsymbol{\theta}}$\ of the LMI given in \emph{(\ref{eq:26})} (which is guaranteed to exist) and  take \ $\widehat{\mathbf{G}}$\ with 
realization \ $(\widehat{\mathbf{A}}_{_{\mathbf{G}}},
\widehat{\mathbf{B}}_{_{\mathbf{G}}}, \widehat{\mathbf{C}}_{_{\mathbf{G}}}, \widehat{\mathbf{D}}_{_{\mathbf{G}}})$, where\ \
$\widehat{\boldsymbol{\theta}}=\begin{bmatrix}
                               \widehat{\mathbf{A}}_{_{\mathbf{G}}}&\widehat{\mathbf{B}}_{_{\mathbf{G}}}\\
                               \widehat{\mathbf{C}}_{_{\mathbf{G}}}&\widehat{\mathbf{D}}_{_{\mathbf{G}}}\\
                               \end{bmatrix}$.\hfill$\nabla$
\end{remark}

Note that whereas, for a given pair\ $(\mathbf{P}^{^{\mathrm{o}}}, \boldsymbol{\sigma}^{^{\mathrm{o}}})$\ which satisfies (\ref{eq:25}),
(\ref{eq:26})
is a LMI on \ $\boldsymbol{\theta}$ (with non-empty solution set), the constraint (\ref{eq:25}) of \emph{Prob. $2\emph{c}$} is non-linear
on\ $\mathbf{P}$\ as it is 
affine on\ $(\mathbf{P},\mathbf{P}^{^{-1}})$.\ However, exploiting the ``zero-structure''\ of \ $W_{a}$\ and\ $W_{b}$\ as well an specific 
parametrization of\ $\mathbf{P}$ (as done in [27]), \ (\ref{eq:25}) can be converted into
two LMIs (on the ``free''\ parameters of\ $\mathbf{P}$), as is 
now stated in detail.

To this effect, $\mathbf{P}$\ is written as\ $\mathbf{P}=\begin{bmatrix}
                                                                \mathbf{S}&\mathbf{N}\\
                                                                \mathbf{N}^{^{\mathrm{T}}}&\mathbf{X}\\
                                                               \end{bmatrix}$ \ \ \ and \ \ \  $\mathbf{P}^{^{-1}}=\begin{bmatrix}
                                                                \mathbf{R}&\mathbf{M}\\
                                                                \mathbf{M}^{^{\mathrm{T}}}&\mathbf{Z}\\
                                                               \end{bmatrix}$, \ \ where the dimensions of \ $\mathbf{S}$,\ $\mathbf{X}$,
                                                               $\mathbf{R}$\ \ and\ \ $\mathbf{Z}$ are equal, \emph{i.e.},\
                                                               $n_{_{\mathbf{G}}}$\ is taken to be equal to\ 
                                                               $n_{_{\mathrm{o}\boldsymbol{z}}}+n_{_{\mathbf{I}\boldsymbol{y}}}$ (see Appendix). Then, as
                                                               $\mathbf{R}=(\mathbf{S}-\mathbf{N}\mathbf{X}^{^{-1}}\mathbf{N}^{^{\mathbf{T}}})^{^{-1}}$,\ $\mathbf{P}$\
                                                               is parametrized by\ $\mathbf{S}$,\ $\mathbf{X}$\ and\  $\mathbf{R}$ (rather than\
                                                               $\mathbf{S}$,\ $\mathbf{X}$\ and\ $\mathbf{N}$), $\mathbf{S}=\mathbf{S}^{^{\mathrm{T}}}>0$, $\mathbf{R}=\mathbf{R}^{^{\mathrm{T}}}>0$, $\mathbf{X}=\mathbf{X}^{^{\mathrm{T}}}>0$ with $\mathbf{S}\geq \mathbf{R}^{^{-1}}$ \ 
                                                               ($\Leftrightarrow$\ \ $\begin{bmatrix}
                                                                \mathbf{S}&\mathbf{I}\\
                                                                \mathbf{I}&\mathbf{R}\\
                                                               \end{bmatrix}\geq 0$),\ $\mathbf{N}$\ is given by 
                                                               $\mathbf{N}\mathbf{X}^{^{-1}}\mathbf{N}^{^{\mathrm{T}}}=\mathbf{S}-\mathbf{R}^{^{-1}}$.
Condition  (\ref{eq:25}) on\ $\mathbf{P}$\ is then converted into LMIs on\ $\mathbf{S}$\ and\ $\mathbf{R}$, as it is now precisely stated:

\begin{proposition}\label{prop:13}
 Let\ $\mathbf{P}=\begin{bmatrix}
                                \mathbf{S}&\mathbf{N}\\
                                \mathbf{N}^{^{\mathrm{T}}}&\mathbf{X}\\
                               \end{bmatrix}$, \ $\mathbf{S}=\mathbf{S}^{^{\mathrm{T}}}\in \mathbb{R}^{n_{_{a\boldsymbol{z}}}\times n_{_{a\boldsymbol{z}}}}$\ and \
                               $\mathbf{X}=\mathbf{X}^{^{\mathrm{T}}}\in \mathbb{R}^{n_{_{\mathbf{G}}}\times n_{_{\mathbf{G}}}}$,\ 
                               $n_{_{a\boldsymbol{z}}}=n_{_{\mathbf{I}\boldsymbol{y}}}+n_{_{\mathrm{o}\boldsymbol{z}}}$,\ $n_{_{\mathbf{G}}}=n_{_{a\boldsymbol{z}}}$ and write\ 
                               $\mathbf{P}^{^{-1}}=\begin{bmatrix}
                                                              \mathbf{R}&\mathbf{M}\\
                                                              \mathbf{M}^{^{\mathrm{T}}}& \mathbf{Z}\\
                                                             \end{bmatrix}$.\\
                                                             
\noindent
\emph{\textbf{(a)}} Let\ \ \ $(\mathbf{A}_{_{\mathbf{I}\boldsymbol{y}}}, \mathbf{B}_{_{\mathbf{I}\boldsymbol{y}}}, \mathbf{C}_{_{\mathbf{I}\boldsymbol{y}}}, \mathbf{D}_{_{\mathbf{I}\boldsymbol{y}}})$\ \ \ and\ \ \ $(\mathbf{A}_{_{\mathrm{o}\boldsymbol{z}}}, \mathbf{B}_{_{\mathrm{o}\boldsymbol{z}}}, \mathbf{C}_{_{\mathrm{o}\boldsymbol{z}}}, \mathbf{D}_{_{\mathrm{o}\boldsymbol{z}}})$\ \ be minimal realizations of\ \ \
$\mathbf{H}_{_{\mathbf{I}\boldsymbol{y}}}$\ \ \ and\ \ \ $\mathbf{H}_{_{\mathrm{o}\boldsymbol{z}}}$. 
\begin{equation}\label{eq:27}
\mathbf{W}_{_{a}}^{^{\mathrm{T}}}\boldsymbol{\psi}_{_{\mathrm{o}}}(\mathbf{P}, \boldsymbol{\sigma})\mathbf{W}_{_{a}}>0 \ \ \ \Leftrightarrow\ \ 
\mathbf{P}>0\ \ \text{and}\ \ Q_{a}(\mathbf{R})\triangleq\begin{bmatrix}
                              \mathbf{R}-\mathbf{A}_{_{a\boldsymbol{z}}}\mathbf{R}\mathbf{A}_{_{a\boldsymbol{z}}}^{\mathbf{T}}& \mathbf{B}_{_{a\boldsymbol{z}}}\\
                              \mathbf{B}_{_{a\boldsymbol{z}}}^{^{\mathrm{T}}}& \mathbf{M}_{_{\boldsymbol{\sigma}}}
                             \end{bmatrix}>0,
\end{equation}
where\ \  $\mathbf{A}_{_{a\boldsymbol{z}}}\triangleq\operatorname{diag}(\mathbf{A}_{_{\mathbf{I}\boldsymbol{y}}}, \mathbf{A}_{_{\mathrm{o}\boldsymbol{z}}})$\ \ \ and\ \ \ 
$\mathbf{B}_{_{a\boldsymbol{z}}}\triangleq \begin{bmatrix}
                                           \mathbf{B}_{_{\mathbf{I}\boldsymbol{y}}}\\
                                           \mathbf{B}_{_{\mathrm{o}\boldsymbol{z}}} 
                                           \end{bmatrix}$.\\
                                           
\noindent
\emph{\textbf{(b)}}\ $\mathbf{W}_{_{b}}^{^{\mathrm{T}}}\boldsymbol{\psi}_{_{\mathrm{o}}}(\mathbf{P}, \boldsymbol{\sigma})\mathbf{W}_{_{b}}>0$\ \ 
$\Leftrightarrow$\ \ \ $\mathbf{P}>0\ \ \ \text{and}\ \ \ Q_{_{b}}(\mathbf{S},\boldsymbol{\sigma})>0$,\\ where\ \ \
$Q_{_{b}}(\mathbf{S})=Q_{_{b1}}(\mathbf{S},\boldsymbol{\sigma})-\mathbf{E}_{_{b\mathrm{o}}}^{^{\mathrm{T}}}\mathbf{E}_{_{b\mathrm{o}}}$,\
$Q_{_{b1}}(\mathbf{S}, \boldsymbol{\sigma})=\operatorname{diag}(\mathbf{S},
\boldsymbol{\sigma}_{_{\boldsymbol{y}}}\mathbf{I}_{m_{_{\boldsymbol{y}}}})
+\boldsymbol{\sigma}_{_{\boldsymbol{v}}}\mathbf{E}_{_{b}}^{^{\mathrm{T}}}\mathbf{E}_{_{b}}
-\boldsymbol{\psi}_{_{b}}^{^{\mathrm{T}}}\mathbf{S}\boldsymbol{\psi}_{_{b}}$, $\mathbf{E}_{_{b\mathrm{o}}}$, $\mathbf{E}_{_{b}}$\ and\
$\boldsymbol{\psi}_{_{b}}$\ are given in the Appendix.  \hfill$\nabla$
\end{proposition}

Combining Propositions (\ref{prop:12}) and (\ref{prop:13}) lead to the conversion of \emph{Prob. $2$}\ into two SDPs, as follows.

\begin{proposition}\label{prop:14}
 \emph{\textbf{(a)}}\ The optimal value of \emph{Prob. $2$} equals the optimal value of
$$\text{\underline{\emph{Prob. 2d}}:}\
\displaystyle\min_{\begin{smallmatrix}
                    \boldsymbol{\sigma}>0,\\
                    \mathbf{S}=\mathbf{S}^{^{\mathrm{T}}}>0,\\
                    \mathbf{R}=\mathbf{R}^{^{\mathrm{T}}}>0\\
                    \end{smallmatrix}}\ 
\boldsymbol{\sigma}_{_{\boldsymbol{y}}}\gamma_{\boldsymbol{y}}^{^{2}}
 +\boldsymbol{\sigma}_{_{\boldsymbol{v}}}\gamma_{\boldsymbol{v}}^{^{2}}\ \ \ \text{subject to}\ \ \ Q_{a}(\mathbf{R}, \boldsymbol{\sigma})>0,
 \ Q_{b}(\mathbf{S}, \boldsymbol{\sigma})>0\ \ \text{and}\ \ 
 \begin{bmatrix}
  \mathbf{S}&\mathbf{I}\\
  \mathbf{I}&\mathbf{R}\\
 \end{bmatrix}\geq0.
$$ 
\noindent
 \emph{\textbf{(b)}}\ If \ $(\boldsymbol{\sigma}^{^{\mathrm{o}}},\mathbf{S}_{_{\mathrm{o}}}, \mathbf{R}_{_{\mathrm{o}}})$\ is a feasible solution of
 \emph{Prob. 2d} and\ 
 $\boldsymbol{\sigma}_{_{\boldsymbol{y}}}^{^{\mathrm{o}}}\gamma_{\boldsymbol{y}}^{^{2}}
 +\boldsymbol{\sigma}_{_{\boldsymbol{v}}}^{^{\mathrm{o}}}\gamma_{\boldsymbol{v}}^{^{2}}=\mathcal{J}_{\infty}^{\mathrm{o}}(\mathbf{H})+\varepsilon$,\ then,
 for any\ \
 $\mathbf{X}=\mathbf{X}^{^{\mathrm{T}}}>\boldsymbol{0}$\ and any unitary matrix\ \ $\mathbf{V}$,\ \ defining\ \
 $\mathbf{P}^{^{\mathrm{o}}}=\begin{bmatrix}
                              \mathbf{S}_{_{\mathrm{o}}}&\mathbf{Q}_{_{\mathbf{SR}}}\mathbf{V}\mathbf{X}^{^{1/2}}\\
                              (\mathbf{Q}_{_{\mathbf{SR}}}\mathbf{V}\mathbf{X}^{^{1/2}})^{^{\mathrm{T}}}&\mathbf{X}\\
                              \end{bmatrix}$,\ \ where\ \ 
                              $\mathbf{Q}_{_{\mathbf{SR}}}=(\mathbf{S}_{_{\mathrm{o}}}-\mathbf{R}_{_{\mathrm{o}}}^{^{-1}})^{^{1/2}}$,\ it follows that\
                              $\mathbf{P}^{^{\mathrm{o}}}>0$\ and there exists
                              \ $\boldsymbol{\theta}$\ 
                              such that\ \ $\boldsymbol{\psi}_{_{\mathrm{o}}}(\mathbf{P}^{^{\mathrm{o}}},
                              \boldsymbol{\sigma}^{^{\mathrm{o}}})
+\mathbf{T}_{_{a}}^{^{\mathrm{T}}}\boldsymbol{\theta}\mathbf{T}_{_{b}}
+(\mathbf{T}_{_{a}}\boldsymbol{\theta}\mathbf{T}_{_{b}})^{^{\mathrm{T}}}>0$\ and\ 
$\rho(\mathbf{A}_{_{\mathbf{G}}})<1$. Moreover, for any such\ $\boldsymbol{\theta}$,\
$\mathcal{J}_{\infty}(\mathbf{G}(\boldsymbol{\theta}), \mathbf{H})\leq \mathcal{J}_{\infty}^{\mathrm{o}}(\mathbf{H})+\varepsilon$. \hfill$\nabla$
\end{proposition}

\begin{remark}
 An approximate solution\ \ $\mathbf{G}_{_{\mathbf{M}}}$\  to \emph{Prob.} $2$ can be obtained on the basis of Proposition \emph{\ref{prop:14}} in 
 the following way: solve \emph{Prob.} $2$\emph{d} to get\ \ $(\mathbf{S}_{_{\mathrm{o}}}, \mathbf{R}_{_{\mathrm{o}}}, 
 \boldsymbol{\sigma}_{_{\boldsymbol{y}}}^{^{\mathrm{o}}}, \boldsymbol{\sigma}_{_{\boldsymbol{v}}}^{^{\mathrm{o}}})$; for\ \ 
 $\mathbf{P}_{_{\mathrm{o}}}$\ \ as defined in Proposition \emph{\ref{prop:14}} \emph{\textbf{(b)}} obtain a solution\ \ 
 $\boldsymbol{\theta}_{_{\mathrm{o}}}$\ \
 to the LMI above involving\ \ $\boldsymbol{\psi}_{_{\mathrm{o}}}(\mathbf{P}^{^{\mathrm{o}}}, \boldsymbol{\sigma}^{^{\mathrm{o}}})$; obtain
 a realization\ \ $(\mathbf{A}_{_{\mathrm{o}}}, \mathbf{B}_{_{\mathrm{o}}}, \mathbf{C}_{_{\mathrm{o}}}, \mathbf{D}_{_{\mathrm{o}}})$\ \ for\ \
 $\mathbf{G}_{_{\mathbf{M}}}$,\ \ where\ \ $\boldsymbol{\theta}_{_{\mathrm{o}}}=\begin{bmatrix}
                                                                   \mathbf{A}_{_{\mathrm{o}}}&\mathbf{B}_{_{\mathrm{o}}}\\
                                                                   \mathbf{C}_{_{\mathrm{o}}}&\mathbf{D}_{_{\mathrm{o}}}\\
                                                                  \end{bmatrix}$.\hfill$\nabla$
\end{remark}

\subsection{Minimax $\mathcal{H}_{\infty}$ Estimation for $\mathcal{H}_{\infty}-$Balls of Uncertain Models}\label{subsec:4.3}

\emph{Prob. $3$} (see Subsection \ref{subsec:2.3}) is now approached exactly along the lines pursued in connection with \emph{Prob. $2$}, with the class of admissible estimators given by
\begin{eqnarray*}
 \mathcal{S}_{_{\mathbf{G}}}^{a}\triangleq&&\{\mathbf{G}\ \text{has a realization}\ (\mathbf{A}_{_{\mathbf{G}}},\mathbf{B}_{_{\mathbf{G}}},\mathbf{C}_{_{\mathbf{G}}},\mathbf{D}_{_{\mathbf{G}}}):\\
 &&\mathbf{A}_{_{\mathbf{G}}}\in\mathbb{R}^{n_{_{\mathbf{G}}}\times m_{_{\boldsymbol{y}}}},\mathbf{C}_{_{\mathbf{G}}}\in\mathbb{R}^{m_{_{\boldsymbol{e}}}\times n_{_{\mathbf{G}}}^{a}},\mathbf{D}_{_{\mathbf{G}}}\in\mathbb{R}^{m_{_{\boldsymbol{e}}}\times m_{_{\boldsymbol{v}}}},\rho(\mathbf{A}_{_{\mathbf{G}}})<1\}.
\end{eqnarray*}

The first step of this approach is to give a characterization of\ \ $\mathcal{J}_{\infty}^{a}$\ \ involving a matrix inequality constraint.

To this effect, recall that $\boldsymbol{e}(\boldsymbol{z}_{a};\mathbf{G})=\mathbf{F}_{_{\mathbf{G}a}}\boldsymbol{z}_{a}$,\ where\ \
$\boldsymbol{z}_{_{a}}^{^{\mathrm{T}}}\triangleq 
\begin{bmatrix}\bar{\boldsymbol{y}}^{^{\mathrm{T}}}&\bar{\boldsymbol{v}}^{^{\mathrm{T}}}&\bar{\boldsymbol{w}}^{^{\mathrm{T}}}\end{bmatrix}$,\
$\mathbf{F}_{_{\mathbf{G}a}}\triangleq\mathbf{H}_{_{\mathbf{I}a}}-\mathbf{G}\mathbf{H}_{_{\mathrm{o}a}}$,\ and consider the Lagrangian and 
dual functionals\ $Lag_{\infty}^{a}(\cdot)$\ and\ $\boldsymbol{\varphi}_{_{\mathbf{D}\infty}}^{a}(\cdot)$\ 
\begin{eqnarray*}
 Lag_{\infty}^{a}(\boldsymbol{z}_{a}; \boldsymbol{\sigma}, \mathbf{G})&=&\|\boldsymbol{e}(\boldsymbol{z}_{a}; \mathbf{G})\|_{_{2}}^{^{2}}
 - \boldsymbol{\sigma}_{_{\boldsymbol{y}}}(\|\bar{\boldsymbol{y}}\|_{_{2}}^{^{2}}-\gamma_{\boldsymbol{y}}^{^{2}})
 - \boldsymbol{\sigma}_{_{\boldsymbol{v}}}(\|\bar{\boldsymbol{v}}\|_{_{2}}^{^{2}}-\gamma_{\boldsymbol{v}}^{^{2}})  -\boldsymbol{\sigma}_{_{\boldsymbol{w}}}(\|\boldsymbol{w}\|_{_{2}}^{^{2}}
-\gamma_{_{\mathbf{H}}}^{^{2}}\|W_{_{\mathbf{H}\mathbf{Y}}}\bar{\boldsymbol{y}}\|_{_{2}}^{^{2}}),\\
\boldsymbol{\varphi}_{_{\mathbf{D}\infty}}^{a}(\boldsymbol{\sigma}; \mathbf{G})&=&\sup\{Lag_{\infty}^{a}(\boldsymbol{z}_{a}; \boldsymbol{\sigma}, \mathbf{G}):
\boldsymbol{z}_{a}\in \mathcal{R}_{c}^{m_{_{\boldsymbol{y}}}+2m_{_{\boldsymbol{v}}}}\},\ 
\ \text{where}\ \ \boldsymbol{\sigma}=(\boldsymbol{\sigma}_{_{\boldsymbol{y}}}, \boldsymbol{\sigma}_{_{\boldsymbol{v}}},
\boldsymbol{\sigma}_{_{\boldsymbol{w}}}).
\end{eqnarray*}
It then follows from Theorem 2 in [12] that
$$\mathcal{J}_{\infty}^{a}(\mathbf{G})=\bar{\boldsymbol{\varphi}}_{_{\mathbf{D}\infty}}^{a}(\mathbf{G})\triangleq\inf\{\boldsymbol{\varphi}_{_{\mathbf{D}\infty}}^{a}(\boldsymbol{\sigma}; \mathbf{G}):
\boldsymbol{\sigma}=(\boldsymbol{\sigma}_{_{\boldsymbol{y}}}, \boldsymbol{\sigma}_{_{\boldsymbol{v}}}, \boldsymbol{\sigma}_{_{\boldsymbol{w}}}), 
\boldsymbol{\sigma}_{_{\boldsymbol{y}}}>0, \boldsymbol{\sigma}_{_{\boldsymbol{v}}}>0, \boldsymbol{\sigma}_{_{\boldsymbol{w}}}>0\}.$$

Note now that
$$Lag_{\infty}^{a}(\boldsymbol{z}_{a}; \boldsymbol{\sigma}, \mathbf{G})
=\left\langle\mathbf{F}_{_{\mathbf{G}a}}\boldsymbol{z}_{a},\mathbf{F}_{_{\mathbf{G}a}}\boldsymbol{z}_{a}\right\rangle
-\left\langle\mathbf{M}_{_{\boldsymbol{\sigma}}}^{a}\boldsymbol{z}_{a}, \boldsymbol{z}_{a}\right\rangle
+\boldsymbol{\sigma}_{_{\boldsymbol{w}}}\gamma_{_{\mathbf{H}}}^{^{2}}\left\langle W_{_{\mathbf{H}\boldsymbol{y}}}\bar{\boldsymbol{y}}, W_{_{\mathbf{H}\boldsymbol{y}}}\bar{\boldsymbol{y}}\right\rangle
+\boldsymbol{\sigma}_{_{\boldsymbol{y}}}\gamma_{\boldsymbol{y}}^{^{2}}+\boldsymbol{\sigma}_{_{\boldsymbol{v}}}\gamma_{\boldsymbol{v}}^{^{2}}$$
or, equivalently,
$$Lag(\boldsymbol{z}_{a}; \boldsymbol{\sigma}, \mathbf{G})
=\left\langle(\mathbf{F}_{_{\mathbf{G}W}}^{*}\mathbf{F}_{_{\mathbf{G}W}}-\mathbf{M}_{_{\boldsymbol{\sigma}}}^{a})\boldsymbol{z}_{a}, \boldsymbol{z}_{a}\right\rangle
+\boldsymbol{\sigma}_{_{\boldsymbol{y}}}\gamma_{\boldsymbol{y}}^{^{2}} + \boldsymbol{\sigma}_{_{\boldsymbol{v}}}\gamma_{\boldsymbol{v}}^{^{2}},$$
where\\ $\mathbf{M}_{_{\boldsymbol{\sigma}}}^{a}\triangleq
\operatorname{diag}(\boldsymbol{\sigma}_{_{\boldsymbol{y}}}\mathbf{I}_{m_{_{\boldsymbol{y}}}},
\boldsymbol{\sigma}_{_{\boldsymbol{v}}}\mathbf{I}_{m_{_{\boldsymbol{v}}}},
\boldsymbol{\sigma}_{_{\boldsymbol{w}}}\mathbf{I}_{m_{_{\boldsymbol{v}}}})$,\ \
$\mathbf{F}_{_{\mathbf{G}W}}\triangleq\begin{bmatrix}
                                    \boldsymbol{\sigma}_{_{\boldsymbol{w}}}^{^{1/2}}\gamma_{_{\mathbf{H}}}W_{_{\mathbf{H}\boldsymbol{y}}}^{a}\\
                                    \mathbf{F}_{_{\mathbf{G}a}}
                                   \end{bmatrix}$,\ \
$W_{_{\mathbf{H}\boldsymbol{y}}}^{a}\triangleq\begin{bmatrix}W_{_{\mathbf{H}\boldsymbol{y}}}\ \vdots\ \boldsymbol{0}_{m_{_{\boldsymbol{y}}}\times m_{_{\boldsymbol{v}}}}\
\vdots\ \boldsymbol{0}_{m_{_{\boldsymbol{y}}}\times m_{_{\boldsymbol{v}}}} \end{bmatrix}$.\\

Proceeding as in Subsection \ref{subsec:4.2}, it follows that
\begin{eqnarray}
\mathcal{J}_{\infty}^{a}(\mathbf{G})
&=&\inf\{\boldsymbol{\sigma}_{_{\boldsymbol{y}}}\gamma_{\boldsymbol{y}}^{^{2}}+\boldsymbol{\sigma}_{_{\boldsymbol{v}}}\gamma_{\boldsymbol{v}}^{^{2}}: 
\boldsymbol{\sigma}_{_{\boldsymbol{y}}}>0, \boldsymbol{\sigma}_{_{\boldsymbol{v}}}>0, \boldsymbol{\sigma}_{_{\boldsymbol{w}}}>0\ \ \text{and}\ \
\forall \phi \in [0, 2\pi], \mathbf{F}_{_{\mathbf{G}W}}(e^{j\phi})^{*}\mathbf{F}_{_{\mathbf{G}W}}(e^{j\phi})-\mathbf{M}_{_{\boldsymbol{\sigma}}}^{a}<0\}\label{eq:28a}.\nonumber\\
\end{eqnarray}
Noting now that
$$\mathbf{F}_{_{\mathbf{G}W}}(e^{j\phi})^{*}\mathbf{F}_{_{\mathbf{G}W}}(e^{j\phi})-\mathbf{M}_{_{\boldsymbol{\sigma}}}^{a}<0\ \ \ \Leftrightarrow\ \ \
\mathbf{I}-(\mathbf{M}_{_{\boldsymbol{\sigma}}}^{a})^{^{-1/2}}\mathbf{F}_{_{\mathbf{G}W}}(e^{j\phi})^{*}\mathbf{F}_{_{\mathbf{G}W}}(e^{j\phi})(\mathbf{M}_{_{\boldsymbol{\sigma}}}^{a})^{^{-1/2}}>0,$$
it follows that
$$\forall \phi \in [0, 2\pi], \ \mathbf{F}_{_{\mathbf{G}W}}(e^{j\phi})^{*}\mathbf{F}_{_{\mathbf{G}W}}(e^{j\phi})-\mathbf{M}_{_{\boldsymbol{\sigma}}}^{a}<0
 \ \ \Leftrightarrow\ \ \ \|\mathbf{F}_{_{\mathbf{G}W}}(\mathbf{M}_{_{\boldsymbol{\sigma}}}^{a})^{^{-1/2}}\|_{\infty}<1 $$
or, equivalently, taking a realization \ \ 
$\displaystyle\boldsymbol{\Sigma}_{_{\mathbf{G}W}}=(\mathbf{A}_{_{\mathbf{G}W}}, \mathbf{B}_{_{\mathbf{G}W}},
\mathbf{C}_{_{\mathbf{G}W}}, \mathbf{D}_{_{\mathbf{G}W}})$ \ \ of \ \ 
$\mathbf{F}_{_{\mathbf{G}W}}$,\ $\rho(\mathbf{A}_{_{\mathbf{G}W}})<1$,\ \ and invoking the discrete-time, bounded-real lemma (as done in the nominal\ \ $\mathcal{H}_{\infty}$\ \ derivation 
above)\break $\exists\ \mathbf{P}=\mathbf{P}^{^{\mathrm{T}}}>0$\ \ such that\ \
$Q_{_{\mathbf{B}\mathbf{R}}}(\mathbf{P}; \boldsymbol{\Sigma}_{_{\mathbf{G}W}}, \mathbf{M}_{_{\boldsymbol{\sigma}}}^{a})<0$.

Thus,
\begin{eqnarray}
 \mathcal{J}_{\infty}^{a}(\mathbf{G})&=&\inf\{ \boldsymbol{\sigma}_{_{\boldsymbol{y}}}\gamma_{\boldsymbol{y}}^{^{2}}
 +\boldsymbol{\sigma}_{_{\boldsymbol{v}}}\gamma_{\boldsymbol{v}}^{^{2}}: 
 \boldsymbol{\sigma}=(\boldsymbol{\sigma}_{_{\boldsymbol{y}}},\boldsymbol{\sigma}_{_{\boldsymbol{v}}}, \boldsymbol{\sigma}_{_{\boldsymbol{w}}}), \boldsymbol{\sigma}_{_{\boldsymbol{y}}}>0,
 \boldsymbol{\sigma}_{_{\boldsymbol{v}}}>0, \boldsymbol{\sigma}_{_{\boldsymbol{w}}}>0, \mathbf{P}=\mathbf{P}^{^{\mathrm{T}}}>0\nonumber\\
 &&\  \ \ \ \ \  \ \ \ \  \ \ \  \ \ \  \ \ \   \ \ \ \ \ \ \ \ \ \ \ \ \ \ \ \ \ \ \ \ \ \ \ \ \ \ \ \ \ \ \ \ \ \ \ \ \ \ \ \ \ \ \ \ \ 
 \text{and}\ \  Q_{_{\mathbf{B}\mathbf{R}}}(\mathbf{P}; \boldsymbol{\Sigma}_{_{\mathbf{G}W}}, \mathbf{M}_{_{\boldsymbol{\sigma}}}^{a})<0 \}.\label{eq:34}
\end{eqnarray}
Now, as shown in Section \ref{subsec:4.2}, 
$$Q_{_{\mathbf{B}\mathbf{R}}}(\mathbf{P}; \boldsymbol{\Sigma}_{_{\mathbf{G}W}}, \mathbf{M}_{_{\boldsymbol{\sigma}}}^{a})<0\ \ \ \Leftrightarrow\ \ \ \boldsymbol{\psi}_{a}(\mathbf{P}, \boldsymbol{\sigma}, \boldsymbol{\theta})>0,$$
where\ \ \ $\boldsymbol{\theta}=\begin{bmatrix}
                                  \mathbf{A}_{_{\mathbf{G}}}&\mathbf{B}_{_{\mathbf{G}}}\\
                                  \mathbf{C}_{_{\mathbf{G}}}&\mathbf{D}_{_{\mathbf{G}}}\\
                                \end{bmatrix}$, \ \ $(\mathbf{A}_{_{\mathbf{G}}}, \mathbf{B}_{_{\mathbf{G}}}, \mathbf{C}_{_{\mathbf{G}}}, \mathbf{D}_{_{\mathbf{G}}})$ is
a realization of\ $\mathbf{G}$\  and\                              
$$\boldsymbol{\psi}_{a}(\mathbf{P}, \boldsymbol{\sigma}, \boldsymbol{\theta})
\triangleq\begin{bmatrix}
\mathbf{P}^{^{-1}}&\mathbf{A}_{_{\mathbf{G}W}}&\mathbf{B}_{_{\mathbf{G}W}}&\boldsymbol{0}\\
\mathbf{A}_{_{\mathbf{G}W}}^{^{\mathrm{T}}}&\mathbf{P}&\boldsymbol{0}&\mathbf{C}_{_{\mathbf{G}W}}^{^{\mathrm{T}}}\\
\mathbf{B}_{_{\mathbf{G}W}}^{^{\mathrm{T}}}&\boldsymbol{0}&\mathbf{M}_{_{\boldsymbol{\sigma}}}^{a}&\mathbf{D}_{_{\mathbf{G}W}}^{^{\mathrm{T}}}\\
\boldsymbol{0}&\mathbf{C}_{_{\mathbf{G}W}}&\mathbf{D}_{_{\mathbf{G}W}}&\mathbf{I}\\
\end{bmatrix}
$$
or, equivalently, separating the estimator parameter\ $\boldsymbol{\theta}$\  from \ $(\mathbf{P}, \boldsymbol{\sigma})$,
\begin{equation}\label{eq:35}
 Q_{_{\mathbf{B}\mathbf{R}}}(\mathbf{P}; \boldsymbol{\Sigma}_{_{\mathbf{G}W}}, \mathbf{M}_{_{\boldsymbol{\sigma}}}^{a})<0\ \ \ \Leftrightarrow\ \ \ 
 \boldsymbol{\psi}_{a}^{\mathrm{o}}(\mathbf{P}, \boldsymbol{\sigma})+\mathbf{T}_{_{_{1}}}^{^{\mathrm{T}}}\boldsymbol{\theta}\mathbf{T}_{_{_{2}}}
+ \mathbf{T}_{_{_{2}}}^{^{\mathrm{T}}}\boldsymbol{\theta}^{^{\mathrm{T}}}\mathbf{T}_{_{_{1}}}>0,
 \end{equation}
where
$$\boldsymbol{\psi}_{a}^{\mathrm{o}}(\mathbf{P}, \boldsymbol{\sigma})
=\begin{bmatrix}
\mathbf{P}^{^{-1}}&\mathbf{A}_{a}&\mathbf{B}_{a}&\boldsymbol{0}\\
\mathbf{A}_{a}^{^{\mathrm{T}}}&\mathbf{P}&\boldsymbol{0}&\mathbf{C}_{a}^{^{\mathrm{T}}}\\
\mathbf{B}_{a}^{^{\mathrm{T}}}&\boldsymbol{0}&\mathbf{M}_{_{\boldsymbol{\sigma}}}^{a}&\mathbf{D}_{a}^{^{\mathrm{T}}}\\
\boldsymbol{0}&\mathbf{C}_{a}&\mathbf{D}_{a}&\mathbf{I}\\
\end{bmatrix},
$$
where\  $\mathbf{A}_{_{\mathbf{G}W}}=\mathbf{A}_{a}+\mathbf{A}_{_{L}}(\boldsymbol{\theta})$,\ 
$\mathbf{B}_{_{\mathbf{G}W}}=\mathbf{B}_{a}+\mathbf{B}_{_{L}}(\boldsymbol{\theta})$,\ 
$\mathbf{C}_{_{\mathbf{G}W}}=\mathbf{C}_{a}+\mathbf{C}_{_{L}}(\boldsymbol{\theta})$,\ $\mathbf{D}_{_{\mathbf{G}W}}=\mathbf{D}_{a}+\mathbf{D}_{_{L}}(\boldsymbol{\theta})$,\ \ 
$\mathbf{A}_{_{L}}(\cdot)$,\ \ $\mathbf{B}_{_{L}}(\cdot)$,\ \ $\mathbf{C}_{_{L}}(\cdot)$\ \ and\ \ $\mathbf{D}_{_{L}}(\cdot)$ \ are linear functions 
(of $\boldsymbol{\theta}$)\ given in the Appendix,\ $$\mathbf{T}_{_{_{1}}}=\begin{bmatrix}
                  \boldsymbol{0} & \mathbf{I}     & \vdots & \boldsymbol{0} & \boldsymbol{0} & \vdots & \boldsymbol{0} & \boldsymbol{0} & \vdots & \boldsymbol{0} & \boldsymbol{0}\\
                  \boldsymbol{0} & \boldsymbol{0} & \vdots & \boldsymbol{0} & \boldsymbol{0} & \vdots & \boldsymbol{0} & \boldsymbol{0} & \vdots & \boldsymbol{0} & -\mathbf{I}
                 \end{bmatrix},\ \ \mathbf{T}_{_{2}}=\begin{bmatrix}
                  \boldsymbol{0} & \boldsymbol{0} & \vdots & \boldsymbol{0} & \mathbf{I} & \vdots & \boldsymbol{0} & \boldsymbol{0} & \vdots & \boldsymbol{0} & \boldsymbol{0}\\
                  \boldsymbol{0} & \boldsymbol{0} & \vdots & \widehat{\mathbf{C}}_{_{\mathrm{o}a}} & \boldsymbol{0} & \vdots & \mathbf{D}_{_{\mathrm{o}a}}^{^{\boldsymbol{z}}} & \mathbf{I} & \vdots & \boldsymbol{0} & \boldsymbol{0} \end{bmatrix},$$
$\widehat{\mathbf{C}}_{_{\mathrm{o}a}}=\begin{bmatrix}\boldsymbol{0}_{m_{_{\boldsymbol{v}}} \times n_{_{Wa}}}\ \vdots\ \mathbf{C}_{_{\mathrm{o}a}}\end{bmatrix}$\ \ and
\ $\mathbf{D}_{_{\mathrm{o}a}}^{^{\boldsymbol{z}}}=\begin{bmatrix}\mathbf{D}_{_{\mathbf{H}\mathrm{o}}}\mathbf{D}_{_{W\boldsymbol{y}}}^{^{-1}}& \vdots
& \mathbf{D}_{_{W\boldsymbol{v}}}^{^{-1}}\end{bmatrix}$,\ 
$\mathbf{A}_{_{a}}=\operatorname{diag}(\mathbf{A}_{_{a1}},\boldsymbol{0}_{n_{_{\mathbf{G}}}\times n_{_{\mathbf{G}}}})$,\
$n_{_{\mathbf{G}}}$
is the dimension of $\mathbf{A}_{_{a1}}$,\ $\mathbf{A}_{_{a1}}$,\ $\mathbf{B}_{_{a}}$,\ $\mathbf{C}_{_{a}}$\ and\  $\mathbf{D}_{_{a}}$\ 
are also given in the Appendix.

For \ $\boldsymbol{\psi}_{a}^{\mathrm{o}}(\mathbf{P}; \boldsymbol{\sigma})$, \ $\mathbf{T}_{_{_{1}}}$\ and\ $\mathbf{T}_{_{_{2}}}$\ so defined, it follows
from (\ref{eq:34}) and 
(\ref{eq:35}) that
\begin{eqnarray*}
\mathcal{J}_{\infty}^{a}(\mathbf{G}(\boldsymbol{\theta}))&=&\inf\{\boldsymbol{\sigma}_{_{\boldsymbol{y}}}\gamma_{\boldsymbol{y}}^{^{2}}
+\boldsymbol{\sigma}_{_{\boldsymbol{v}}}\gamma_{\boldsymbol{v}}^{^{2}}:
\boldsymbol{\sigma}_{_{\boldsymbol{y}}}>0, \boldsymbol{\sigma}_{_{\boldsymbol{v}}}>0, \boldsymbol{\sigma}_{_{\boldsymbol{w}}}>0, \mathbf{P}=\mathbf{P}^{^{\mathrm{T}}}>0\ \ \text{and} \\
&&\ \ \ \ \ \ \  \ \ \ \ \ \ \ \ \ \ \ \ \ \ \ \ \ \ \ \ \ \ \  \ \ \ \ \  \  \ \ \ \ \ \ \ \ \ \ \  \ \ \ \ \boldsymbol{\psi}_{a}^{\mathrm{o}}(\mathbf{P}, \boldsymbol{\sigma})+\mathbf{T}_{_{_{1}}}^{^{\mathrm{T}}}\boldsymbol{\theta}\mathbf{T}_{_{_{2}}}
+\mathbf{T}_{_{_{2}}}^{^{\mathrm{T}}}\boldsymbol{\theta}^{^{\mathrm{T}}}\mathbf{T}_{_{_{1}}}>0\}. 
\end{eqnarray*}

Bringing in matrices\  $W_{_{_{1}}}$\ and\ $W_{_{_{2}}}$\ whose columns respectively constitute bases for the null spaces of\ $\mathbf{T}_{_{_{1}}}$\ and\ $\mathbf{T}_{_{_{2}}}$ together with 
the corresponding constraints
\begin{equation}\label{eq:36}
 W_{_{_{1}}}^{^{\mathrm{T}}}\boldsymbol{\psi}_{a}^{\mathrm{o}}(\mathbf{P}, \boldsymbol{\sigma})W_{_{_{1}}}>0\ \ \ \ \text{and}\ \ \ \  W_{_{_{2}}}^{^{\mathrm{T}}}\boldsymbol{\psi}_{a}^{\mathrm{o}}(\mathbf{P}, \boldsymbol{\sigma})W_{_{_{2}}}>0
\end{equation}
and invoking the Elimination Lemma, a statement entirely similar to Proposition \ref{prop:12} can be seen to hold as follows.

\begin{proposition}\label{prop:16}
Let
\begin{eqnarray*}
 \mathcal{S}_{_{\mathbf{G}}}^{a}&\triangleq&\{ \mathbf{G}\ \text{has a realization}\ 
(\mathbf{A}_{_{\mathbf{G}}},\mathbf{B}_{_{\mathbf{G}}}, \mathbf{C}_{_{\mathbf{G}}}, \mathbf{D}_{_{\mathbf{G}}}):\\ 
&&\mathbf{A}_{_{\mathbf{G}}}\in \mathbb{R}^{n_{_{\mathbf{G}}}^{a}\times n_{_{\mathbf{G}}}^{a}}, 
\mathbf{B}_{_{\mathbf{G}}}\in \mathbb{R}^{n_{_{\mathbf{G}}}^{a}\times m_{_{\boldsymbol{v}}}}, \mathbf{C}_{_{\mathbf{G}}}\in
\mathbb{R}^{m_{_{\boldsymbol{e}}}\times n_{_{\mathbf{G}}}^{a}},
\mathbf{D}_{_{\mathbf{G}}}\in \mathbb{R}^{m_{_{\boldsymbol{e}}}\times m_{_{\boldsymbol{v}}}}, \rho(\mathbf{A}_{_{\mathbf{G}}})<1\},
\end{eqnarray*}
and  \ \  $\mathcal{J}_{\mathrm{o}\infty}^{a}\triangleq\inf\{\mathcal{J}_{\infty}^{a}(\mathbf{G}):\mathbf{G}\in  \mathcal{S}_{_{\mathbf{G}}}^{a}\}$.
Then \emph{(\textbf{a})}\ \ $\forall 
\mathbf{G}\in  \mathcal{S}_{_{\mathbf{G}}}^{a}$
\begin{eqnarray*}
\mathcal{J}_{\infty}^{a}(\mathbf{G})&\triangleq&\inf\{\boldsymbol{\sigma}_{_{\boldsymbol{y}}}\gamma_{\boldsymbol{y}}^{^{2}}
+\boldsymbol{\sigma}_{_{\boldsymbol{v}}}\gamma_{\boldsymbol{v}}^{^{2}}:
\boldsymbol{\sigma}_{_{\boldsymbol{y}}}>0, \boldsymbol{\sigma}_{_{\boldsymbol{v}}}>0, \boldsymbol{\sigma}_{_{\boldsymbol{w}}}>0, \mathbf{P}
=\mathbf{P}^{^{\mathrm{T}}}>0,\\
&&\ \ \ \ \ \ \  \ \ \ \ \ \ \ \  \ \ \ \ \ \ \ 
\boldsymbol{\theta}\in \mathbb{R}^{(n_{_{\mathbf{G}}}^{a} + m_{_{\boldsymbol{e}}})\times(n_{_{\mathbf{G}}}^{a}+m_{_{\boldsymbol{v}}})} \ \ \text{and}  \ \  \boldsymbol{\psi}_{a}^{\mathrm{o}}(\mathbf{P}, \boldsymbol{\sigma})+\mathbf{T}_{_{_{1}}}^{^{\mathrm{T}}}\boldsymbol{\theta}\mathbf{T}_{_{_{2}}}
+\mathbf{T}_{_{_{2}}}^{^{\mathrm{T}}}\boldsymbol{\theta}^{^{\mathrm{T}}}\mathbf{T}_{_{_{1}}}>0\}, 
\end{eqnarray*}
where\ \ $\boldsymbol{\theta}=\begin{bmatrix}
                               \mathbf{A}_{_{\mathbf{G}}}&\mathbf{B}_{_{\mathbf{G}}}\\
                               \mathbf{C}_{_{\mathbf{G}}}&\mathbf{D}_{_{\mathbf{G}}}\\
                               \end{bmatrix}.$\\ \\
\emph{(\textbf{b})} $\mathcal{J}_{\mathrm{o}\infty}^{a}$\ is given by
$$\mathcal{J}_{\mathrm{o}\infty}^{a}=\inf\{\boldsymbol{\sigma}_{_{\boldsymbol{y}}}\gamma_{\boldsymbol{y}}^{^{2}}
+\boldsymbol{\sigma}_{_{\boldsymbol{v}}}\gamma_{\boldsymbol{v}}^{^{2}}:
\boldsymbol{\sigma}_{_{\boldsymbol{y}}}>0, \boldsymbol{\sigma}_{_{\boldsymbol{v}}}>0, \boldsymbol{\sigma}_{_{\boldsymbol{w}}}>0, \mathbf{P}=\mathbf{P}^{^{\mathrm{T}}}>0
\ \text{and\  \emph{(\ref{eq:36})}\ holds}\}.$$

Moreover, if \ $\mathbf{P}^{^{\mathrm{o}}}$\ and \ $\boldsymbol{\sigma}^{^{\mathrm{o}}}=(\boldsymbol{\sigma}_{_{\boldsymbol{y}}}^{^{\mathrm{o}}}, \boldsymbol{\sigma}_{_{\boldsymbol{v}}}^{^{\mathrm{o}}},
\boldsymbol{\sigma}_{_{\boldsymbol{w}}}^{^{\mathrm{o}}})$\ satisfy the constraints above and\
$\boldsymbol{\sigma}_{_{\boldsymbol{y}}}^{^{\mathrm{o}}}\gamma_{\boldsymbol{y}}^{^{2}}
+\boldsymbol{\sigma}_{_{\boldsymbol{v}}}^{^{\mathrm{o}}}\gamma_{\boldsymbol{v}}^{^{2}}=\mathcal{J}_{\mathrm{o} \infty}^{a}+\varepsilon$,\ then there exists\ $\boldsymbol{\theta}$\ such that
$$\boldsymbol{\psi}_{a}^{\mathrm{o}}(\mathbf{P}^{^{\mathrm{o}}}, \boldsymbol{\sigma}^{^{\mathrm{o}}})+\mathbf{T}_{_{_{1}}}^{^{\mathrm{T}}}\boldsymbol{\theta}\mathbf{T}_{_{_{2}}}
+\mathbf{T}_{_{_{2}}}^{^{\mathrm{T}}}\boldsymbol{\theta}^{^{\mathrm{T}}}\mathbf{T}_{_{_{1}}}>0$$
and, for any such $\boldsymbol{\theta}$,\ \ $\mathcal{J}_{\infty}^{a}(\mathbf{G}(\boldsymbol{\theta}))\leq \mathcal{J}_{\mathrm{o} \infty}^{a}+\varepsilon$. \hfill$\nabla$
\end{proposition}

Proceeding along the lines that led to Propositions \ref{prop:13} and \ref{prop:14},  (\ref{eq:36}) is converted into LMIs on\ \ $\mathbf{R}$\ 
and\ $\mathbf{S}$, where\ $\mathbf{P}=\begin{bmatrix}
                                        \mathbf{S}&\mathbf{N}\\
                                        \mathbf{N}^{^{\mathrm{T}}}& \mathbf{X}\\
                                       \end{bmatrix}$\ and\ $\mathbf{P}^{^{-1}}=\begin{bmatrix}
                                                                                                \mathbf{R}&\mathbf{M}^{^{\mathrm{T}}}\\
                                                                                                \mathbf{M}^{^{\mathrm{T}}}&\mathbf{Z}\\
                                                                                                 \end{bmatrix}$, as it is now stated.
\bigskip
                                                                                                 
\begin{proposition}\label{prop:16a}
\emph{\textbf{(a)}} $W_{_{_{1}}}^{^{\mathrm{T}}}\boldsymbol{\psi}_{a}^{\mathrm{o}}(\mathbf{P}, \boldsymbol{\sigma})W_{_{_{1}}}>0$\ \ $\Leftrightarrow$\ \
$\boldsymbol{\sigma}_{_{\boldsymbol{w}}}>0$,\ $\mathbf{P}>0$,\ and\ \ $Q_{_{1}}(\mathbf{R},\boldsymbol{\sigma})>0$,                                    
\noindent                                                             
where
\begin{eqnarray*}
Q_{_{_{1}}}(\mathbf{R}, \boldsymbol{\sigma})\triangleq\left[\begin{smallmatrix}
                                                  \mathbf{R}&\mathbf{B}_{_{a1}}^{^{\boldsymbol{z}}}&\boldsymbol{0}\\
                                                  (\mathbf{B}_{_{a1}}^{^{\boldsymbol{z}}})^{^{\mathrm{T}}}& \mathbf{M}_{_{\boldsymbol{z}}}^{^{\boldsymbol{\sigma}}}&(\mathbf{D}_{_{a\boldsymbol{z}}}^{^{W}})^{^{\mathrm{T}}}\\
                                                  \boldsymbol{0}&\mathbf{D}_{_{a\boldsymbol{z}}}^{^{W}}&\boldsymbol{\sigma}_{_{\boldsymbol{w}}}^{_{-1}}\mathbf{I}_{m_{_{\boldsymbol{y}}}}
                                                  \end{smallmatrix}\right]-\left[\begin{smallmatrix}
                                                                \mathbf{A}_{_{a1}}\\
                                                                \boldsymbol{0}\\
                                                                \mathbf{C}_{_{a1}}^{^{W}}\\
                                                                \end{smallmatrix}\right]\mathbf{R}\left[\begin{smallmatrix}\mathbf{A}_{_{a1}}^{^{\mathrm{T}}}&\boldsymbol{0}&(\mathbf{C}_{_{a1}}^{^{W}})^{^{\mathrm{T}}}\end{smallmatrix}\right],
\end{eqnarray*}                                                      
$\mathbf{C}_{_{a1}}^{^{W}}
=[\gamma_{_{\mathbf{H}}}\mathbf{C}_{_{W\mathbf{H}\boldsymbol{y}}}\ \vdots\
\boldsymbol{0}_{m_{_{\boldsymbol{y}}}\times n_{_{\mathbf{A}\mathbf{H}\mathbf{I}a}}}\ \vdots \boldsymbol{0}_{m_{_{\boldsymbol{y}}}\times
n_{_{\mathbf{A}\mathrm{o}a}}}]$\ \ and\ \  $\mathbf{D}_{_{a\boldsymbol{z}}}^{^{W}}
=[\gamma_{_{\mathbf{H}}}\mathbf{D}_{_{W\mathbf{H}\boldsymbol{y}}}\ \vdots\ 
\boldsymbol{0}_{m_{_{\boldsymbol{y}}}\times m_{_{\boldsymbol{v}}}}]$,\ \
$\mathbf{M}_{_{\boldsymbol{\sigma}}}^{^{\boldsymbol{z}}}
=\operatorname{diag}(\boldsymbol{\sigma}_{_{\boldsymbol{y}}}\mathbf{I}_{m_{_{\boldsymbol{y}}}}, \boldsymbol{\sigma}_{_{\boldsymbol{v}}}\mathbf{I}_{m_{_{\boldsymbol{v}}}})$,\ \
$\mathbf{A}_{_{a1}}$\ \ and\ \ $\mathbf{B}_{_{a1}}^{^{\boldsymbol{z}}}$\ are given in the Appendix.\\

\noindent
\emph{\textbf{(b)}} $W_{_{_{2}}}^{^{\mathrm{T}}}\boldsymbol{\psi}_{_{a}}^{\mathrm{o}}(\mathbf{P}, \boldsymbol{\sigma})W_{_{_{2}}}>0\ \ \ \Leftrightarrow \ \ \ 
\mathbf{P}>0\ \ \text{and}\ \ Q_{_{2}}(\mathbf{S},\boldsymbol{\sigma})>0$,\  where
\begin{eqnarray*}
Q_{_{_{2}}}(\mathbf{S}, \boldsymbol{\sigma})&\triangleq&\check{Q}_{_{2}}(\mathbf{S}, \boldsymbol{\sigma})-\mathbf{E}_{_{\mathrm{o}}}^{^{\mathrm{T}}}\mathbf{E}_{_{\mathrm{o}}}-\mathbf{E}_{_{\boldsymbol{s}}}^{^{\mathrm{T}}}\mathbf{S}\mathbf{E}_{_{\boldsymbol{s}}},\\\\\
\check{Q}_{_{2}}(\mathbf{S}, \boldsymbol{\sigma})&\triangleq&\left[\begin{smallmatrix}
                                                                             \mathbf{S}&\boldsymbol{0}\\
                                                                             \boldsymbol{0}& \mathbf{M}_{_{\boldsymbol{\sigma}}}^{^{\boldsymbol{z}}}\\
                                                                             \end{smallmatrix}\right]+\boldsymbol{\sigma}_{_{\boldsymbol{w}}}\left[\begin{smallmatrix} 
                                                                                           \widehat{\mathbf{C}}_{_{\mathrm{o}a}}^{^{\mathrm{T}}}\\
                                                                                           (\mathbf{D}_{_{\mathrm{o}a}}^{^{\boldsymbol{z}}})^{^{\mathrm{T}}}\\
                                                                                           \end{smallmatrix}\right]
                                                                                           [\widehat{\mathbf{C}}_{_{\mathrm{o}a}}\ \ \mathbf{D}_{_{\mathrm{o}a}}^{^{\boldsymbol{z}}}]
                                                                                           -\boldsymbol{\sigma}_{_{\boldsymbol{w}}}\left[\begin{smallmatrix}
                                       (\mathbf{C}_{_{a1}}^{^{W}})^{^{\mathrm{T}}}\\
                                       (\mathbf{D}_{_{a\boldsymbol{z}}}^{^{W}})
                                       \end{smallmatrix}\right][\mathbf{C}_{_{a1}}^{^{W}}\ \vdots\ \mathbf{D}_{_{a\boldsymbol{z}}}^{^{W}}],
\end{eqnarray*}
$\mathbf{E}_{_{\mathrm{o}}}$\ \ and\ \ $\mathbf{E}_{_{\boldsymbol{s}}}$ are given in the Appendix.\hfill$\nabla$                                                                                                                                                                                                                                                                                                                                                                                                                                                                                                                                                                                         
\end{proposition}

Finally, combining Proposition \ref{prop:16} and Proposition \ref{prop:16a} leads to the counterpart of Proposition \ref{prop:14} which is now stated (the proofs of ``$\mathbf{P}^{^{\mathrm{o}}}>0$''
and ``$\rho(\mathbf{A}_{_{\mathbf{G}}})<1$'' follow exactly the same argument invoked in the 
proof of Proposition \ref{prop:14}).

\begin{proposition}\label{prop:17}
 \emph{\textbf{(a)}} The optimal value of \ \ $\mathcal{J}_{\mathrm{o}\infty}^{a}$\ \ of \emph{Prob. $3$} equals to the optimal value of the following problem
  \begin{eqnarray*}
&&\underline{Prob.\ 3a}\ \ \displaystyle\min_{\begin{smallmatrix}\boldsymbol{\sigma}_{_{\boldsymbol{y}}}>0,\ \boldsymbol{\sigma}_{_{\boldsymbol{v}}}>0,\ \boldsymbol{\sigma}_{_{\boldsymbol{w}}}>0,\\
\mathbf{S}=\mathbf{S}^{^{\mathrm{T}}}>0,\ \mathbf{R}=\mathbf{R}^{^{\mathrm{T}}}>0
 \end{smallmatrix}} \ \boldsymbol{\sigma}_{_{\boldsymbol{y}}}\gamma_{\boldsymbol{y}}^{^{2}}+\boldsymbol{\sigma}_{_{\boldsymbol{v}}}\gamma_{\boldsymbol{v}}^{^{2}}\\ 
&&\text{subject to}\ \ \  \ Q_{_{_{1}}}(\mathbf{R}, \boldsymbol{\sigma})>0, \ Q_{_{_{2}}}(\mathbf{S}, \boldsymbol{\sigma})>0, \ \left[\begin{smallmatrix}
                                                                                                                  \mathbf{S} & \mathbf{I}\\
                                                                                                                  \mathbf{I} & \mathbf{R}
                                                                                                                 \end{smallmatrix}\right]>0,
\end{eqnarray*}
where\ \ $Q_{_{1}}(\cdot)$\ is affine on\ \ $(\mathbf{R},\boldsymbol{\sigma}_{_{\boldsymbol{y}}},\boldsymbol{\sigma}_{_{\boldsymbol{v}}})$\ \ and\ \ $Q_{_{2}}(\cdot)$\ is affine (see Proposition \emph{\ref{prop:16a}}).\medskip

\noindent
\emph{\textbf{(b)}} If\ $(\boldsymbol{\sigma}_{_{\boldsymbol{y}}}^{^{\mathrm{o}}},\ \boldsymbol{\sigma}_{_{\boldsymbol{v}}}^{^{\mathrm{o}}},\
\boldsymbol{\sigma}_{_{\boldsymbol{w}}}^{^{\mathrm{o}}},\ \mathbf{S}^{^{\mathrm{o}}},\ \mathbf{R}^{^{\mathrm{o}}})$\ is a feasible solution
of \emph{Prob. 3a} \ and 
$\boldsymbol{\sigma}_{_{\boldsymbol{y}}}^{^{\mathrm{o}}}\gamma_{\boldsymbol{y}}^{^{2}}+\boldsymbol{\sigma}_{_{\boldsymbol{v}}}^{^{\mathrm{o}}}\gamma_{\boldsymbol{v}}^{^{2}}
=\mathcal{J}_{\mathrm{o}\infty}^{a}+\varepsilon$,\ then for any\ $\mathbf{X}_{a}=\mathbf{X}_{a}^{^{\mathrm{T}}}>0$\ and any unitary matrix\ \ $\mathbf{V}$,\ defining \
$\mathbf{P}^{^{\mathrm{o}}}=\left[\begin{smallmatrix}
                \mathbf{S}^{^{\mathrm{o}}}&Q_{_{\mathbf{SR}}}^{a}\mathbf{V}\mathbf{X}_{a}^{^{1/2}}\\
                (Q_{_{\mathbf{SR}}}^{a}\mathbf{V}\mathbf{X}_{a}^{^{1/2}})^{^{\mathrm{T}}}&\mathbf{X}\\
                \end{smallmatrix}\right]$ \ where\ $Q_{_{\mathbf{SR}}}^{a}=[\mathbf{S}^{^{\mathrm{o}}}-(\mathbf{R}^{^{\mathrm{o}}})^{^{-1}}]^{^{1/2}}$, it follows that
                \ $\mathbf{P}^{^{\mathrm{o}}}>0$\ and there exists\ 
                $\boldsymbol{\theta}$\ such that
$$\boldsymbol{\psi}_{a}^{\mathrm{o}}(\mathbf{P}^{^{\mathrm{o}}}, \boldsymbol{\sigma}^{^{\mathrm{o}}})+\mathbf{T}_{_{_{1}}}^{^{\mathrm{T}}}\boldsymbol{\theta}\mathbf{T}_{_{_{2}}}
+\mathbf{T}_{_{_{2}}}^{^{\mathrm{T}}}\boldsymbol{\theta}^{^{\mathrm{T}}}\mathbf{T}_{_{_{1}}}>0$$
and\ \ $\rho(\mathbf{A}_{_{\mathbf{G}}})<1$. Moreover, for any such\ 
$\boldsymbol{\theta}$,\ $\mathcal{J}_{\infty}^{a}(\mathbf{G}(\boldsymbol{\theta}))\leq \mathcal{J}_{\mathrm{o}\infty}^{a}+\varepsilon$.\hfill$\nabla$
\end{proposition}

\vspace*{3mm}

\begin{remark}
Note that\ \ $Q_{_{1}}(\mathbf{R},\boldsymbol{\sigma})$\ \ is not affine on\ \ $\boldsymbol{\sigma}_{_{\boldsymbol{w}}}$. However, if a value is assigned to\ $\boldsymbol{\sigma}_{_{\boldsymbol{w}}}$\ so that it is no longer a decision variable of \emph{Prob. $3a$}, the resulting problem is a
SDP on the remaining ones\ \ $(\mathbf{S}, \mathbf{R}, \boldsymbol{\sigma}_{_{\boldsymbol{y}}}, \boldsymbol{\sigma}_{_{\boldsymbol{v}}})$.\ Thus, a natural way
of tackling \emph{Prob. $3a$} is by means of a line search with respect to\ $\boldsymbol{\sigma}_{_{\boldsymbol{w}}}$\ at each step of which a  SDP is solved\ (if\ \ $W_{_{\mathbf{H}\boldsymbol{y}}}$\  is constant\ ``$Q_{_{1}}(\mathbf{R}, \boldsymbol{\sigma})>0$'' can be rewritten as a LMI). Note, in addition, that for a
given solution\ $(\mathbf{P}^{^{\mathrm{o}}}, \boldsymbol{\sigma}^{^{\mathrm{o}}}_{_{\boldsymbol{y}}}, \boldsymbol{\sigma}^{^{\mathrm{o}}}_{_{\boldsymbol{v}}}, \boldsymbol{\sigma}^{^{\mathrm{o}}}_{_{W}})$\ of $Prob.\ 3a$,  $\boldsymbol{\theta}$\ is obtained from the LMI in
Proposition \emph{\ref{prop:17}(b)}.\hfill$\nabla$ 
\end{remark}

\section{Robust Estimation Based on an Average Criteria}\label{robust-estimation-based-average-criterion}\label{sec:5}

The major aim of this section is to formulate estimation problems ($Prob.\ 4 - 6$ below) based on the
average cost-functionals\ $\boldsymbol{\eta}_{a\boldsymbol{v}}$,\ $\boldsymbol{\eta}^{a}$\ and\ $\boldsymbol{\eta}^{b}$  and convert them  into SDPs, with the purpose of enabling trade-offs to be achieved between worst-case and ``pointwise'' performance. $Prob.\ 4 - 6$ can be respectively cast as
$$ \min_{\mathbf{G}\in\mathcal{S}_{_{\mathbf{G}}}^{^{i}}}\boldsymbol{\eta}^{^{i}}(\mathbf{G})\ \ \ \text{subject to}\ \ \
\mathcal{J}^{^{i}}(\mathbf{G})\leq (1+\alpha)\mathcal{J}_{_{\mathrm{o}}}^{^{i}},\ \ \alpha>0$$
where\ \ $(\boldsymbol{\eta}^{^{1}},\mathcal{J}^{^{1}})=(\boldsymbol{\eta}_{_{a\boldsymbol{v}}},\bar{\mathcal{J}}_{_{\mathbf{X}}})$,\ $(\boldsymbol{\eta}^{^{2}},\mathcal{J}^{^{2}})=(\boldsymbol{\eta}^{^{a}},\mathcal{J}_{\infty})$,\  $(\boldsymbol{\eta}^{^{3}},\mathcal{J}^{^{3}})=(\boldsymbol{\eta}^{^{b}},\mathcal{J}_{\infty}^{a})$,\  and\  $\mathcal{J}_{_{\mathrm{o}}}^{^{i}}$,\ $i=1,2,3$\ are respectively the optimal values of the minimax problems $Prob.\ 1 - 3$. As the main point here is to generate ``less conservative'' alternatives to the corresponding minimax estimators, in each case\ $\mathcal{S}_{_{\mathbf{G}}}^{^{i}}$\ is taken to be a linear class
$$\mathcal{S}_{_{\mathbf{G}}}^{^{i}}\triangleq\left\{\mathbf{G}(\boldsymbol{\beta})=\boldsymbol{\beta}\mathbf{Y}_{a}^{^{i}}: \boldsymbol{\beta}=\begin{bmatrix}\mathbf{C}_{_{\mathbf{G}}} & \mathbf{D}_{_{\mathbf{G}}}\end{bmatrix}, \mathbf{C}_{_{\mathbf{G}}}\in\mathbb{R}^{m_{_{\boldsymbol{e}}}\times n_{_{\mathbf{G}}}^{^{i}}},\ \mathbf{D}_{_{\mathbf{G}}}\in \mathbb{R}^{m_{_{\boldsymbol{e}}}\times m_{_{\boldsymbol{y}}}}\right\},$$
where\  $\mathbf{Y}_{a}^{^{i}}=\begin{bmatrix}
                                  \mathbf{Y}_{_{\mathbf{G}}}^{^{i}}\mathbf{B}_{_{\mathbf{G}}}^{^{i}}\\
                                  \mathbf{I}_{m_{_{\boldsymbol{y}}}}
                                  \end{bmatrix}$,\ $\mathbf{Y}_{_{\mathbf{G}}}^{^{i}}(e^{j\phi})=(e^{j\phi}\mathbf{I}-\mathbf{A}_{_{\mathbf{G}}}^{^{i}})^{-1}$\ and\ $(\mathbf{A}_{_{\mathbf{G}}}^{^{i}},\mathbf{B}_{_{\mathbf{G}}}^{^{i}},\mathbf{C}_{_{\mathbf{G}}}^{^{i}}, \mathbf{D}_{_{\mathbf{G}}}^{^{i}})$\  is
                                  a minimal realization of an $\varepsilon-$approximate solution\ \ $\mathbf{G}^{^{i}}$ to the associated minimax 
                                  problem ($\mathbf{G}^{^{i}}$\ is a feasible solution of \emph{Prob.\ $i$} and\
                                  $\mathcal{J}^{^{i}}(\mathbf{G})\leq \mathcal{J}_{_{\mathrm{o}}}^{^{i}}+\varepsilon$) and $\alpha>\varepsilon$. The three derivations to convert $Prob.\ 4 - 6$ into 
                                  SDPs are entirely analogous and go as follows: \textbf{(\emph{a})} the results of Section \ref{sec:4} for problems $1-3$ provide the constraints that impose upperbounds on worst-case performance of any admissible estimator in each linear class\ $\mathcal{S}_{_{\mathbf{G}}}^{^{i}}$; \textbf{(\emph{b})}  then each average, quadratic cost functional\ \ $\boldsymbol{\eta}^{^{i}}$\ is converted into a linear one with the introduction of additional LMI constraints.
                                  
With respect to \textbf{(\emph{b})}, note that the cost-functionals\ $\boldsymbol{\eta}_{av}$,\  $\boldsymbol{\eta}^{a}$\ and\ $\boldsymbol{\eta}^{b}$\ can be written as
\begin{eqnarray*}
 \boldsymbol{\eta}_{av}(\mathbf{G})&=&C(\mathbf{G};\boldsymbol{\Gamma}_{_{\boldsymbol{y}}}, \boldsymbol{\Gamma}_{_{\boldsymbol{v}}})+ \left(\gamma^{2}/(m_{_{\boldsymbol{y}}}m_{_{\boldsymbol{v}}})\right)\left\langle\mathbf{G}\otimes \boldsymbol{\phi}_{_{\boldsymbol{y}1}}^{^{\mathrm{T}}},\mathbf{G}\otimes \boldsymbol{\phi}_{_{\boldsymbol{y}1}}^{^{\mathrm{T}}}\right\rangle\\
 \boldsymbol{\eta}_{a}(\mathbf{G})&=&C(\mathbf{G};\boldsymbol{\Gamma}_{_{\boldsymbol{y}}}^{a},\boldsymbol{\Gamma}_{_{\boldsymbol{v}}}^{a})\ \ \ \ \ \ \ \text{and}\ \ \ \ \ \ \ \ \boldsymbol{\eta}_{b}(\mathbf{G})=C(\mathbf{G};\boldsymbol{\Gamma}_{_{\boldsymbol{y}}}^{a},\boldsymbol{\Gamma}_{_{\boldsymbol{v}}}^{b}),
\end{eqnarray*}
where\ \ $C(\mathbf{G};\widehat{\boldsymbol{\Gamma}}_{_{\boldsymbol{y}}}, \widehat{\boldsymbol{\Gamma}}_{_{\boldsymbol{v}}})\triangleq\left\langle(\mathbf{H}_{_{\mathbf{I}}}-\mathbf{G}\mathbf{H}_{_{\boldsymbol{0}}})\widehat{\boldsymbol{\Gamma}}_{_{\boldsymbol{y}}},(\mathbf{H}_{_{\mathbf{I}}}-\mathbf{G}\mathbf{H}_{_{\boldsymbol{0}}})\right\rangle+\left\langle\mathbf{G}\widehat{\boldsymbol{\Gamma}}_{_{\boldsymbol{v}}},\mathbf{G}\right\rangle$,\\
$\boldsymbol{\Gamma}_{_{\boldsymbol{\alpha}}}^{a}=(\gamma_{_{\boldsymbol{\alpha}}}^{^{2}}/m_{_{\boldsymbol{\alpha}}})(W_{_{\boldsymbol{\alpha}}}^{*}W_{_{\boldsymbol{\alpha}}})^{-1}$,\ $\boldsymbol{\alpha}=\boldsymbol{y}, \boldsymbol{v}$\ \ \ \ \ \ \ and\ \ \ \ \ \ \  $\boldsymbol{\Gamma}_{_{\boldsymbol{v}}}^{b}=\boldsymbol{\Gamma}_{_{\boldsymbol{v}}}^{a}+(\gamma_{_{\mathbf{H}}}^{^{2}}/m_{_{\boldsymbol{v}}})\boldsymbol{\phi}_{_{\boldsymbol{y}W}}\boldsymbol{\phi}_{_{\boldsymbol{y}W}}^{*}\mathbf{I}_{m_{_{\boldsymbol{v}}}}$. 

As a result, the replacement of\ $\boldsymbol{\eta}_{av}$,\ $\boldsymbol{\eta}^{a}$\ and\ $\boldsymbol{\eta}$\ by linear cost-functionals will be based on the following equalities: for\ \ $\mathbf{G}(\boldsymbol{\beta})\in \mathbf{S}_{_{\mathbf{G}}}^{^{i}}$,
\begin{equation}\label{eq:sec5-eq01}
 C\left(\mathbf{G}(\boldsymbol{\beta});\widehat{\boldsymbol{\Gamma}}_{_{\boldsymbol{y}}},\widehat{\boldsymbol{\Gamma}}_{_{\boldsymbol{v}}}\right)=\inf\left\{\operatorname{tr}(\mathbf{P}_{c}):\mathbf{P}_{c}=\mathbf{P}_{c}^{^{\mathrm{T}}}\ \ \text{and}\ \ Q_{_{\mathcal{J}q}}(\mathbf{P}_{c}, [\mathbf{I}_{m_{_{\boldsymbol{e}}}}\ \vdots\ \boldsymbol{\beta}];Q_{c})\geq 0 \right\},
\end{equation}
\begin{equation}\label{eq:sec5-eq02}
 \left\langle\mathbf{G}(\boldsymbol{\beta})\otimes\boldsymbol{\phi}_{_{\boldsymbol{y}1}}^{^{\mathrm{T}}},\mathbf{G}(\boldsymbol{\beta})\otimes\boldsymbol{\phi}_{_{\boldsymbol{y}1}}^{^{\mathrm{T}}}\right\rangle=\inf\left\{\operatorname{tr}(\mathbf{R}_{c}):\mathbf{R}_{c}=\mathbf{R}_{c}^{^{\mathrm{T}}}\ \ \text{and}\ \ Q_{_{\mathcal{J}q}}(\mathbf{R}_{c}, \boldsymbol{\beta}\otimes\mathbf{I};Q_{_{\mathbf{G}c}})\geq 0 \right\},
\end{equation}
where\ \ $Q_{_{\mathcal{J}q}}(\mathbf{P}, \mathbf{M};Q)\triangleq\begin{bmatrix}
                                                                  \mathbf{P}&\mathbf{M}Q^{^{1/2}}\\
                                                                  Q^{^{1/2}}\mathbf{M}^{^{\mathrm{T}}}&\mathbf{I}
                                                                 \end{bmatrix},
$
\begin{equation}\label{eq:sec5-eq03}
 Q_{c}\triangleq(1/2\pi)\int_{0}^{2\pi}\left\{\begin{bmatrix}
                                               \mathbf{H}_{_{\mathbf{I}}}\\
                                               -\mathbf{Y}_{a}^{^{i}}\mathbf{H}_{_{\boldsymbol{0}}}
                                              \end{bmatrix}\widehat{\boldsymbol{\Gamma}}_{_{\boldsymbol{y}}}\begin{bmatrix}
                                               \mathbf{H}_{_{\mathbf{I}}}\\
                                               -\mathbf{Y}_{a}^{^{i}}\mathbf{H}_{_{\boldsymbol{0}}}
                                              \end{bmatrix}^{*}+\begin{bmatrix}
                                                                \boldsymbol{0}\\
                                                                \mathbf{Y}_{a}^{^{i}}
                                                                \end{bmatrix} \widehat{\boldsymbol{\Gamma}}_{_{\boldsymbol{v}}}\begin{bmatrix}
                                                                                    \boldsymbol{0}\\
                                                                                    \mathbf{Y}_{a}^{^{i}}
                                                                                    \end{bmatrix}^{*}
                                                                                    \right\}(e^{j\phi})d\phi,
\end{equation}
\medskip
\begin{equation}\label{eq:sec5-eq04}
 Q_{_{\mathbf{G}c}}\triangleq(1/2\pi)\int_{0}^{2\pi}\left(\mathbf{Y}_{a}^{^{i}}\otimes\boldsymbol{\phi}_{_{\boldsymbol{y}1}}^{^{\mathrm{T}}}\right)\left(\mathbf{Y}_{a}^{^{i}}\otimes\boldsymbol{\phi}_{_{\boldsymbol{y}1}}^{^{\mathrm{T}}}\right)^{*}(e^{j\phi})d\phi.
\end{equation}

\subsection{MSE Estimation With $\mathcal{H}_{2}$ Model Uncertainty}\label{subsec:5.1}

A robust estimation problem is now considered in which the cost functional $\boldsymbol{\eta}_{av}$ (cf. Proposition \ref{prop:02}) is minimized with respect
to\ $\mathbf{G}$\ under the constraint that the worst-case MSE of a given estimator does not exceed a prescribed value, \emph{i.e.},
$$\underline{Prob.\ 4:}\ \min_{\mathbf{G}\in \mathcal{S}_{_{\mathbf{G}}}^{^{1}}}\boldsymbol{\eta}_{a\boldsymbol{v}}(\mathbf{G})\ \ \ \text{subject to}\ \ \
\bar{\mathcal{J}}_{_{\boldsymbol{\mathcal{X}}}}(\mathbf{G}: \mathcal{S}_{_{\boldsymbol{\mathbf{X}}}})\leq (1+\alpha) \mathcal{J}_{_{\mathrm{o}}}^{^{1}},$$
where\ \ $\mathcal{J}_{_{\mathrm{o}}}^{^{1}}\triangleq\inf\left\{\bar{\mathcal{J}}_{_{\boldsymbol{\mathcal{X}}}}(\mathbf{G};
\mathcal{S}_{_{\boldsymbol{\mathbf{X}}}}): \mathbf{G}\in \mathcal{S}_{_{\mathbf{G}}}\right\}$.

To convert $Prob.\ 4$ into an SDP the constraint above is recast in the light of (\ref{eq:13}) and (\ref{eq:13a}) as follows.

\begin{proposition}\label{prop:05}
 $\mathbf{G}\in\mathcal{S}_{_{\mathbf{G}}}^{^{1}}$\ satisfies the constraint \ 
 ``$\bar{\mathcal{J}}_{_{\boldsymbol{\mathcal{X}}}}(\mathbf{G}; \boldsymbol{\mathcal{X}})\leq (1+\alpha)\mathcal{J}_{_{\mathrm{o}}}^{^{1}}$''\
 in
 $Prob.\ 4$ if

 ``$\exists \lambda >0 \ \ \text{and}\ \ \mathbf{P}=\mathbf{P}^{^{\mathrm{T}}}>0$ \ \ such that\ \
\emph{\textbf{(\emph{i})}}\ $\lambda \gamma^{^{2}}+\boldsymbol{x}_{_{\boldsymbol{0}}}^{^{\mathrm{T}}}\mathbf{P}\boldsymbol{x}_{_{\boldsymbol{0}}}\leq (1+\alpha)\mathcal{J}_{_{\mathrm{o}}}^{^{1}}$\ \ and\\ 
 \emph{\textbf{(\emph{ii})}}\ $Q_{_{\mathcal{J}a}}(\mathbf{P}, \Sigma_{a}(\mathbf{G}), \mathbf{M}(\lambda))<0$". Moreover, the optimal value of \emph{Prob.\ $4$} equals the optimal value of the following problem
 $$\min_{\mathbf{G}\in \mathcal{S}_{_{\mathbf{G}}}^{^{1}}, \mathbf{P}=\mathbf{P}^{^{\mathrm{T}}}>0, \lambda>0}\boldsymbol{\eta}_{av}(\mathbf{G})\ \ \text{subject to}\ \ \emph{\text{\textbf{(\emph{i})}}}\ \ and\ \ \emph{\text{\textbf{(\emph{ii})}}}
,$$
 where $Q_{_{\mathcal{J}a}}(\cdot)$ is given by \emph{(\ref{eq:13a})}.\hfill$\nabla$
\end{proposition}

Thus, noting that\ \ $Q_{_{\mathcal{J}a}}(\mathbf{P}, \boldsymbol{\Sigma}_{a}(\mathbf{G}(\boldsymbol{\beta})), \mathbf{M}(\lambda))<0$ if and 
only if\ \ $Q_{_{\mathcal{J}\mathbf{X}}}(\mathbf{P}, \lambda, \boldsymbol{\beta})<0$, 
it follows from equations (\ref{eq:sec5-eq01}) and (\ref{eq:sec5-eq02})
(replacing\ $(\widehat{\boldsymbol{\Gamma}}_{_{\boldsymbol{y}}},\widehat{\boldsymbol{\Gamma}}_{_{\boldsymbol{v}}},
\mathbf{Y}_{a}^{^{i}}$)\ by\ $(\boldsymbol{\Gamma}_{_{\boldsymbol{y}}},\boldsymbol{\Gamma}_{_{\boldsymbol{v}}},\mathbf{Y}_{a}^{^{1}}$))\ and  
Proposition \ref{prop:05} that \emph{Prob. $4$} can be converted into a SDP as stated in the following proposition.

\begin{proposition}\label{prop:06}
 $Prob.\ 4$ can be recast as 
  \begin{eqnarray*}
&&\displaystyle\min_{\begin{smallmatrix}\boldsymbol{\boldsymbol{\beta}},\ \mathbf{P}_{_{\boldsymbol{\eta}}}=\mathbf{P}_{_{\boldsymbol{\eta}}}^{^{\mathrm{T}}},\ \mathbf{P}_{_{\mathcal{J}}}=\mathbf{P}_{_{\mathcal{J}}}^{^{\mathrm{T}}}\\ 
 \mathbf{P}=\mathbf{P}^{^{\mathrm{T}}},\ \lambda>0\end{smallmatrix}}\operatorname{tr}(\mathbf{P}_{_{\mathcal{J}}})+\operatorname{tr}(\mathbf{P}_{_{\boldsymbol{\eta}}})\\
&&\text{subject to}\\
&&\ \ \ \ \ Q_{_{\mathcal{J}q}}(\mathbf{P}_{_{\mathcal{J}}}, [\mathbf{I}_{m_{_{\boldsymbol{e}}}}\ \vdots\ \boldsymbol{\beta}]; \mathbf{Q}_{_{\mathrm{o}\mathbf{G}}})\geq0, \ \ Q_{_{\mathcal{J}q}}(\mathbf{P}_{_{\boldsymbol{\eta}}}, \boldsymbol{\boldsymbol{\beta}}\otimes \mathbf{I}; \mathbf{Q}_{_{_{\mathbf{G}\boldsymbol{y}}}})\geq0,\\
&&\ \ \ \ \ \lambda \gamma^{^{2}}+\boldsymbol{x}_{_{\boldsymbol{0}}}^{^{\mathrm{T}}}\mathbf{P}\boldsymbol{x}_{_{\boldsymbol{0}}}\leq (1+\alpha)\mathcal{J}_{_{\mathrm{o}}}^{^{1}},\ \ \ \ \text{and}\ \ \ \ 
Q_{_{\mathcal{J}X}}(\mathbf{P}, \lambda, \boldsymbol{\beta})<0,
\end{eqnarray*}
where\ \ $\boldsymbol{x}_{_{\boldsymbol{0}}}$\ \ and\ \ $Q_{_{\mathcal{J}X}}(\mathbf{P}, \lambda, \boldsymbol{\beta})$\ \ are as in Proposition \emph{\ref{prop:04}},\ $Q_{_{\mathrm{o}\mathbf{G}}}$\ and\ $Q_{_{\mathbf{G}\boldsymbol{y}}}$\ are defined as\ $Q_{c}$\ and\ $Q_{_{\mathbf{G}c}}$ \emph{(}in \emph{(\ref{eq:sec5-eq03})}, \emph{(\ref{eq:sec5-eq04}))}\  replacing\ \ $(\widehat{\boldsymbol{\Gamma}}_{_{\boldsymbol{y}}},\widehat{\boldsymbol{\Gamma}}_{_{\boldsymbol{v}}}, \mathbf{Y}_{a}^{^{i}})$\ \ by\ \ $(\boldsymbol{\Gamma}_{_{\boldsymbol{y}}},\boldsymbol{\Gamma}_{_{\boldsymbol{v}}},\mathbf{Y}_{a}^{^{1}})$. \hfill$\nabla$
\end{proposition}

\subsection{A Nominal $\mathcal{H}_{\infty}$ Estimator Based on an Average Cost}\label{subsec:5.2}

In this subsection, a linear filter is sought in a class of admissible ones which minimizes the average cost\ \ 
$\boldsymbol{\eta}^{a}(\cdot; \mathbf{H}_{_{\boldsymbol{0}}})$\  
(for a given ``nominal'' $\mathbf{H}_{_{\boldsymbol{0}}}$) under the constraint that its worst-case estimation error over \ \
$\mathcal{S}_{\boldsymbol{y}}\times \mathcal{S}_{\boldsymbol{v}}$\ \ does not exceed  a prescribed value.

To this effect, consider the optimization problem
$$\text{\underline{\emph{Prob. $5$}}:}\
\displaystyle\min_{\mathbf{G}\in \mathcal{S}_{_{\mathbf{G}}}^{^{2}}}\ \boldsymbol{\eta}^{a}(\mathbf{G};\mathbf{H}_{_{\boldsymbol{0}}})\ \ \
\text{subject to}\ \ \ 
\mathcal{J}_{\infty}(\mathbf{G}; \mathbf{H}_{_{\boldsymbol{0}}})\leq (1+\alpha)\mathcal{J}_{_{\mathrm{o}}}^{^{2}},$$ 
where\ \ \ $\mathcal{J}_{_{\mathrm{o}}}^{^{2}}=\inf\{\mathcal{J}_{\infty}(\mathbf{G};\mathbf{H}_{_{\boldsymbol{0}}}): \mathbf{G}\in  \mathcal{S}_{_{\mathbf{G}}}^{^{2}}\}$.

In the light of (\ref{eq:22-a}), \emph{Prob. $5$} can be recast as
\begin{equation}\label{eq:29}
 \displaystyle\min_{\begin{smallmatrix}
                    \boldsymbol{\beta}>0,\\
                    \boldsymbol{\sigma}_{_{\boldsymbol{y}}}>0,\boldsymbol{\sigma}_{_{\boldsymbol{v}}}>0,\\
                   \mathbf{P}=\mathbf{P}^{^{\mathrm{T}}}>0\\
                    \end{smallmatrix}}\ \boldsymbol{\eta}^{a}(\mathbf{G}(\boldsymbol{\beta});\mathbf{H}) \ \ \ \  \ \ \ 
                    \text{subject to}\ \ \ \ \  \
                    \left\{
\begin{array}{rl}
 & \boldsymbol{\sigma}_{_{\boldsymbol{y}}}\gamma_{\boldsymbol{y}}^{^{2}}+\boldsymbol{\sigma}_{_{\boldsymbol{v}}}\gamma_{\boldsymbol{v}}^{^{2}}\leq (1+\alpha)\mathcal{J}_{_{\mathrm{o}}}^{^{2}},\\
&Q_{_{\mathbf{B}\mathbf{R}}}\left(\mathbf{P}; \displaystyle\Sigma_{_{\mathbf{F}\mathbf{G}}}^{^{\mathrm{o}}}(\mathbf{G}(\boldsymbol{\beta})),\mathbf{M}_{_{\boldsymbol{\sigma}}}\right)<0
 \end{array} \right.,
\end{equation}
where\ \ $\displaystyle\Sigma_{_{\mathbf{F}\mathbf{G}}}^{^{\mathrm{o}}}(\mathbf{G}(\boldsymbol{\beta}))=\left(\mathbf{A}_{_{\mathbf{F}\mathbf{G}}}^{^{\mathrm{o}}}, 
\mathbf{B}_{_{\mathbf{F}\mathbf{G}}}^{^{\mathrm{o}}}, \mathbf{C}_{_{\mathbf{F}\mathbf{G}}}(\boldsymbol{\beta}), 
\mathbf{D}_{_{\mathbf{F}\mathbf{G}}}(\boldsymbol{\beta})\right)$\ \ is a realization of\\ 
$\mathbf{F}_{_{\mathbf{G}}}(\boldsymbol{\beta})=\mathbf{H}_{_{\mathbf{I}\boldsymbol{y}}}
-\boldsymbol{\beta}\mathbf{Y}_{_{\mathbf{G}}}^{a}\mathbf{H}_{_{\mathrm{o}\boldsymbol{z}}}$,\ \ obtained from a minimal realization
$(\mathbf{A}_{_{\mathbf{F}\mathbf{G}}}^{^{\mathrm{o}}},\mathbf{B}_{_{\mathbf{F}\mathbf{G}}}^{^{\mathrm{o}}}, \widehat{\mathbf{C}}_{_{\mathbf{F}\mathbf{G}}},\widehat{\mathbf{D}}_{_{\mathbf{F}\mathbf{G}}})$\ \ of\ \ $\left[\begin{smallmatrix}
                              \mathbf{H}_{_{\mathbf{I}\boldsymbol{y}}}\\
                              -\mathbf{Y}_{_{\mathbf{G}}}^{a}\mathbf{H}_{_{\mathrm{o}\boldsymbol{z}}}
                              \end{smallmatrix}\right]$,\ and\ $\begin{bmatrix}\mathbf{C}_{_{\mathbf{F}\mathbf{G}}}(\boldsymbol{\beta})&\vdots& \mathbf{D}_{_{\mathbf{F}\mathbf{G}}}(\boldsymbol{\beta})\end{bmatrix}=\begin{bmatrix}\mathbf{I}_{m_{_{\boldsymbol{e}}}}&\vdots&\boldsymbol{\beta}\end{bmatrix}\begin{bmatrix}\widehat{\mathbf{C}}_{_{\mathbf{F}\mathbf{G}}}&\vdots&\widehat{\mathbf{D}}_{_{\mathbf{F}\mathbf{G}}}\end{bmatrix}$.\\    
                              
The second constraint in (\ref{eq:29}) can be rewritten as\ \ $\check{Q}_{_{a}}(\mathbf{P},\boldsymbol{\sigma},\boldsymbol{\beta})>0$

\noindent
where \ 
$
 \check{Q}_{_{a}}(\mathbf{P},\boldsymbol{\sigma},\boldsymbol{\beta})\triangleq\newcommand*{\temp}{\multicolumn{1}{r|}{}}
\left[\begin{array}{ccc}
 \check{Q}_{_{a1}}(\mathbf{P}, \boldsymbol{\sigma})& \temp& \begin{array}{c}
                             \mathbf{C}_{_{\mathbf{F}\mathbf{G}}}(\boldsymbol{\beta})^{^{\mathrm{T}}}\\
                              \mathbf{D}_{_{\mathbf{F}\mathbf{G}}}(\boldsymbol{\beta})^{^{\mathrm{T}}}
                              \end{array}               \\ \cline{1-3}
\begin{array}{cc}\mathbf{C}_{_{\mathbf{F}\mathbf{G}}}(\boldsymbol{\beta}) & \mathbf{D}_{_{\mathbf{F}\mathbf{G}}}(\boldsymbol{\beta})\end{array} &\temp& \mathbf{I}_{m_{_{\boldsymbol{e}}}} \\
\end{array}\right]$,\\
$\check{Q}_{_{a1}}(\mathbf{P}, \boldsymbol{\sigma})=\begin{bmatrix}
                    \mathbf{P}& \boldsymbol{0}\\
                     \boldsymbol{0}& \mathbf{M}_{_{\boldsymbol{\sigma}}}\\
                    \end{bmatrix}-\begin{bmatrix}
                                   \mathbf{A}_{_{\mathbf{F}\mathbf{G}}}^{^{\mathrm{o\ T}}}\\
                                   \mathbf{B}_{_{\mathbf{F}\mathbf{G}}}^{^{\mathrm{o\ T}}}\\
                                   \end{bmatrix}\mathbf{P}\begin{bmatrix} 
                                                                        \mathbf{A}_{_{\mathbf{F}\mathbf{G}}}^{^{\mathrm{o}}} & \mathbf{B}_{_{\mathbf{F}\mathbf{G}}}^{^{\mathrm{o}}}
                                                                        \end{bmatrix}
$.

It then follows from equation (\ref{eq:sec5-eq01}) (replacing\ 
$(\widehat{\boldsymbol{\Gamma}}_{_{\boldsymbol{y}}},\widehat{\boldsymbol{\Gamma}}_{_{\boldsymbol{v}}},\mathbf{Y}_{a}^{^{i}})$\ \  by\ \
$(\boldsymbol{\Gamma}_{_{\boldsymbol{y}}}^{a},\boldsymbol{\Gamma}_{_{\boldsymbol{v}}}^{a}, \mathbf{Y}_{a}^{^{2}})$)\  and  (\ref{eq:sec5-eq04}) -- (\ref{eq:29}) that \emph{Prob. $5$} can be recast as an SPD as follows.

\begin{proposition}\label{prop:15}
 \emph{Prob. $5$} can be recast as\\ 
 \noindent
 $ \displaystyle\min_{\begin{smallmatrix}
                     \boldsymbol{\beta},\ \boldsymbol{\sigma}_{_{\boldsymbol{y}}}>0, \ \boldsymbol{\sigma}_{_{\boldsymbol{v}}}>0\\ 
                    \mathbf{P}=\mathbf{P}^{^{\mathrm{T}}}\\
                    \mathbf{R}_{a}=\mathbf{R}_{a}^{^{\mathrm{T}}}\\
                    \end{smallmatrix}}\ \operatorname{tr}\{\mathbf{R}_{a}\} \ \ \  \text{subject to}\ \  \ 
                    \left\{
\begin{array}{rl}
 Q_{_{\mathcal{J}q}}(\mathbf{R}_{a}, [\mathbf{I}_{m_{_{\boldsymbol{e}}}}\ \vdots\ \boldsymbol{\beta}];\boldsymbol{\Gamma}_{_{\boldsymbol{\eta}}})\geq 0,\ \  \boldsymbol{\sigma}_{_{\boldsymbol{y}}}\gamma_{\boldsymbol{y}}^{^{2}}+\boldsymbol{\sigma}_{_{\boldsymbol{v}}}\gamma_{\boldsymbol{v}}^{^{2}}\leq (1+\alpha)\mathcal{J}_{_{\mathrm{o}}}^{^{2}},\\
\text{and}\ \ \check{Q}_{_{a}}\left(\mathbf{P},\boldsymbol{\sigma},\boldsymbol{\beta}\right)>0, \end{array} \right.$,\\
where\ $\boldsymbol{\Gamma}_{_{\boldsymbol{\eta}}}$\ is defined as\  $Q_{c}$ \emph{(}in \emph{(\ref{eq:sec5-eq03}))}  replacing\ \
$(\widehat{\boldsymbol{\Gamma}}_{_{\boldsymbol{y}}},\widehat{\boldsymbol{\Gamma}}_{_{\boldsymbol{v}}}, \mathbf{Y}_{a}^{^{i}})$\ \ by\ \
$(\boldsymbol{\Gamma}_{_{\boldsymbol{y}}}^{a},\boldsymbol{\Gamma}_{_{\boldsymbol{v}}}^{a},\mathbf{Y}_{a}^{^{2}})$.\hfill$\nabla$
 \end{proposition}
\vspace*{3mm}

\begin{remark}\label{remark:5.2}
The frequency-response of the ``average-cost'', $\mathcal{H}_{\infty}$ estimator is obtained from a solution\ \ 
$\left(\boldsymbol{\beta}_{_{\mathrm{o}}}, \mathbf{P}_{_{\mathrm{o}}}, \mathbf{R}_{_{a}}^{^{\mathrm{o}}},
\boldsymbol{\sigma}_{_{\boldsymbol{y}}}^{^{\mathrm{o}}}, \boldsymbol{\sigma}_{_{\boldsymbol{v}}}^{^{\mathrm{o}}}\right)$\ \ of the SDP introduced 
in 
Proposition \emph{\ref{prop:15}} and is given by\ $\mathbf{G}(\boldsymbol{\beta}_{_{\mathrm{o}}})
=\boldsymbol{\beta}_{_{\mathrm{o}}}\mathbf{Y}_{a}^{^{2}}$.\hfill$\nabla$    
\end{remark}

\subsection{A Robust $\mathcal{H}_{\infty}$ Estimator Based on an Average Cost}\label{subsec:5.3}

In this subsection, a robust estimator is introduced for the set-up of Subsection \ref{subsec:4.3}, with the aim
of enabling trade-offs to be achieved between worst-case and ``point-wise'' performance over the $\mathcal{H}_{\infty}-$balls of channel
frequency-responses and the $\mathcal{H}_{2}-$balls of exogenous signals.

More specifically, consider the following optimization problem
$$\text{\underline{\emph{Prob. $6$}}:}\
\displaystyle\min_{\mathbf{G}\in  \mathcal{S}_{_{\mathbf{G}}}^{a}}\ \boldsymbol{\eta}^{b}(\mathbf{G})\ \ \ \text{subject to}\ \ \ 
\mathcal{J}_{\infty}^{a}(\mathbf{G})\leq (1+\alpha)\mathcal{J}_{_{\mathrm{o}}}^{^{3}},$$ 
where\ \ \ $\mathcal{J}_{_{\mathrm{o}}}^{^{3}}=\inf\{\mathcal{J}_{\infty}^{a}(\mathbf{G}): \mathbf{G}\in  \mathcal{S}_{_{\mathbf{G}}}^{^{3}}\}$.

It then follows from (\ref{eq:34}) (in the same way $Prob.\ 5$ was recast as (\ref{eq:29})) that \emph{Prob. $6$}  can be stated as                            
\begin{equation}
\displaystyle\min_{\begin{smallmatrix}
                    \boldsymbol{\sigma}_{_{\boldsymbol{y}}}>0,\boldsymbol{\sigma}_{_{\boldsymbol{v}}}>0, \boldsymbol{\sigma}_{_{\boldsymbol{w}}}>0 \\
                    \boldsymbol{\beta},\ \mathbf{P}=\mathbf{P}^{^{\mathrm{T}}}>0\\
                    \end{smallmatrix}}\ \boldsymbol{\eta}^{b}(\mathbf{G};\mathbf{H}_{_{\boldsymbol{0}}})
\ \ \text{subject to}\ \
\boldsymbol{\sigma}_{_{\boldsymbol{y}}}\gamma_{\boldsymbol{y}}^{^{2}}
+\boldsymbol{\sigma}_{_{\boldsymbol{v}}}\gamma_{\boldsymbol{v}}^{^{2}}\leq (1+\alpha)\mathcal{J}_{_{\mathrm{o}}}^{^{3}}, \ Q_{_{\mathbf{B}\mathbf{R}}}(\mathbf{P}; \boldsymbol{\Sigma}_{_{\mathbf{G}W}}^{^{b}}(\boldsymbol{\sigma}_{_{\boldsymbol{w}}},\boldsymbol{\beta}), \mathbf{M}_{_{\boldsymbol{\sigma}}}^{a})<0,\label{eq:sec5-eq07}
\end{equation}
where\ $\boldsymbol{\beta}=\begin{bmatrix}\mathbf{C}_{_{\mathbf{G}}}&\vdots& \mathbf{D}_{_{\mathbf{G}}}\end{bmatrix}$, \ \ 
$\displaystyle\boldsymbol{\Sigma}_{_{\mathbf{G}W}}^{^{b}}(\boldsymbol{\sigma}_{_{\boldsymbol{w}}},\boldsymbol{\beta})
=(\mathbf{A}_{_{\mathbf{G}W}}^{^{b}},\ \mathbf{B}_{_{\mathbf{G}W}}^{^{b}},\ 
\mathbf{C}_{_{\mathbf{G}W}}(\boldsymbol{\sigma}_{_{\boldsymbol{w}}},\boldsymbol{\beta}),\ 
\mathbf{D}_{_{\mathbf{G}W}}(\boldsymbol{\sigma}_{_{\boldsymbol{w}}},\boldsymbol{\beta}))$\  is a realization of\
$\mathbf{F}_{_{\mathbf{G}W}}(\boldsymbol{\beta})$ (see Subsection \ref{subsec:4.3})) obtained from a minimal
realization\ \ $(\mathbf{A}_{_{\mathbf{G}W}}^{^{b}},\mathbf{B}_{_{\mathbf{G}W}}^{^{b}},\widehat{\mathbf{C}}_{_{\mathbf{G}W}},
\widehat{\mathbf{D}}_{_{\mathbf{G}W}})$\ of\
$[(W_{_{\mathbf{H}\boldsymbol{y}}}^{a})^{^{\mathrm{T}}}\ \vdots\ \mathbf{H}_{_{\mathbf{I}a}}^{^{\mathrm{T}}}\ \vdots\ 
(\mathbf{Y}_{_{a}}^{^{3}}\mathbf{H}_{_{\mathrm{o}a}})^{^{\mathrm{T}}}]^{^{\mathrm{T}}}$\ letting\ \
$[\mathbf{T}_{_{\boldsymbol{y}}}^{^{\mathrm{T}}}\ \vdots\ 
\mathbf{T}_{_{\boldsymbol{e}}}(\boldsymbol{\beta})^{^{\mathrm{T}}}]^{^{\mathrm{T}}}\triangleq
\operatorname{diag}(\mathbf{I}_{m_{_{\boldsymbol{y}}}}, [\mathbf{I}_{m_{_{\boldsymbol{e}}}}\ \vdots\ 
\boldsymbol{\beta}])[\widehat{\mathbf{C}}_{_{\mathbf{G}W}}\ \vdots\ \widehat{D}_{_{\mathbf{G}W}}]$,\
$\mathbf{M}_{_{\boldsymbol{\sigma}}}^{a}$\ as in Subsection \ref{subsec:4.3}\ and\ \
$[\mathbf{C}_{_{\mathbf{G}W}}(\boldsymbol{\sigma}_{_{\boldsymbol{w}}},\boldsymbol{\beta})\ \vdots\ 
\mathbf{D}_{_{\mathbf{G}W}}(\boldsymbol{\sigma}_{_{\boldsymbol{w}}},\boldsymbol{\beta})]=
\operatorname{diag}(\boldsymbol{\sigma}_{_{\boldsymbol{w}}}^{^{1/2}}\gamma_{_{\mathbf{H}}}\mathbf{I}_{m_{_{\boldsymbol{y}}}},
\mathbf{I}_{m_{_{\boldsymbol{e}}}})[\mathbf{T}_{_{\boldsymbol{y}}}^{^{\mathrm{T}}}\ \vdots\
\mathbf{T}_{_{\boldsymbol{e}}}(\boldsymbol{\beta})^{^{\mathrm{T}}}]^{^{\mathrm{T}}}$.
                                                                                               
It can then be shown that
\begin{equation}\label{eq:sec5-eq08}
Q_{_{\mathbf{B}\mathbf{R}}}(\mathbf{P}; \boldsymbol{\Sigma}_{_{\mathbf{G}W}}^{^{b}}(\boldsymbol{\sigma}_{_{\boldsymbol{w}}},\boldsymbol{\beta}), \mathbf{M}_{_{\boldsymbol{\sigma}}}^{a})<0\ \ \ \Leftrightarrow\ \ \ \check{Q}_{_{b}}(\mathbf{P}, \boldsymbol{\sigma}, \boldsymbol{\beta})>0,
\end{equation}
where
\begin{equation}\label{eq:34b}
 \check{Q}_{_{b}}(\mathbf{P}, \boldsymbol{\sigma}, \boldsymbol{\beta})\triangleq
 \newcommand*{\temp}{\multicolumn{1}{r|}{}}\left[\begin{array}{ccc}
 \check{Q}_{_{b1}}(\mathbf{P}, \boldsymbol{\sigma}) - 
 \boldsymbol{\sigma}_{w}\gamma_{_{\mathbf{H}}}^{^{2}}(\mathbf{T}_{_{\boldsymbol{y}}})^{^{\mathrm{T}}}\mathbf{T}_{_{\boldsymbol{y}}}
 &\temp& \mathbf{T}_{_{\boldsymbol{e}}}(\boldsymbol{\beta})^{^{\mathrm{T}}}\\ \cline{1-3}
\mathbf{T}_{_{\boldsymbol{e}}}(\boldsymbol{\beta})&\temp& \mathbf{I}_{m_{_{\boldsymbol{e}}}} \\
\end{array}\right],
\end{equation}
$\check{Q}_{_{b1}}(\mathbf{P}, \boldsymbol{\sigma})\triangleq\begin{bmatrix}
                                                    \mathbf{P}&\boldsymbol{0}\\
                                                    \boldsymbol{0}& \mathbf{M}_{_{\boldsymbol{\sigma}}}^{a}\\
                                                   \end{bmatrix}-\begin{bmatrix}
                                                                 (\mathbf{A}_{_{\mathbf{G}W}}^{^{b}})^{^{\mathrm{T}}}\\
                                                                 (\mathbf{B}_{_{\mathbf{G}W}}^{^{b}})^{^{\mathrm{T}}}
                                                                \end{bmatrix}\mathbf{P}\begin{bmatrix} \mathbf{A}_{_{\mathbf{G}W}}^{^{b}} & \mathbf{B}_{_{\mathbf{G}W}}^{^{b}}\end{bmatrix}$.\\

In the light of (\ref{eq:sec5-eq07}), (\ref{eq:sec5-eq08}) and equation (\ref{eq:sec5-eq01}) (replacing\ \ $(\widehat{\boldsymbol{\Gamma}}_{_{\boldsymbol{y}}},\widehat{\boldsymbol{\Gamma}}_{_{\boldsymbol{v}}}, \mathbf{Y}_{a}^{^{i}})$\ \ by\ \ $(\boldsymbol{\Gamma}_{_{\boldsymbol{y}}},\widehat{\boldsymbol{\Gamma}}_{_{\boldsymbol{v}}}, \mathbf{Y}_{a}^{^{3}})$), \emph{Prob. $6$} can be recast as a SDP as it is now stated in detail.
\begin{proposition}\label{prop:18}
 \emph{Prob. $6$} can be recast as the following SDP
 \begin{eqnarray*}
&&\displaystyle\min_{\begin{smallmatrix}
                    \boldsymbol{\sigma}_{_{\boldsymbol{y}}}>0,\boldsymbol{\sigma}_{_{\boldsymbol{v}}}>0, \boldsymbol{\sigma}_{_{\boldsymbol{w}}}>0 \\
                    \boldsymbol{\beta},\ \mathbf{P}=\mathbf{P}^{^{\mathrm{T}}},\ \mathbf{R}=\mathbf{R}^{^{\mathrm{T}}}\\
                    \end{smallmatrix}}\ \operatorname{tr}(\mathbf{R})\\
&&\ \ \ \text{subject to}\ \ \ Q_{_{\mathcal{J}q}}(\mathbf{R},[\mathbf{I}_{m_{_{\boldsymbol{e}}}}\ \vdots\ \boldsymbol{\beta}];
\boldsymbol{\Gamma}_{_{\boldsymbol{\eta}}}^{b})\geq0,\  
\boldsymbol{\sigma}_{_{\boldsymbol{y}}}\gamma_{\boldsymbol{y}}^{^{2}}
+\boldsymbol{\sigma}_{_{\boldsymbol{v}}}\gamma_{\boldsymbol{v}}^{^{2}}\leq (1+\alpha)\mathcal{J}_{_{\mathrm{o}}}^{^{3}},\\ 
&&\ \ \ \ \ \ \ \ \ \ \  \ \ \ \ \ \ \ \text{and}\ \  \check{Q}_{_{b}}(\mathbf{P},\boldsymbol{\sigma}, \boldsymbol{\beta})>0,
\end{eqnarray*}
where\ \ $\check{Q}_{_{b}}(\mathbf{P}, \boldsymbol{\sigma}, \boldsymbol{\beta})$\ \ is given by \emph{(\ref{eq:34b})},\ \ 
$\boldsymbol{\sigma}=(\boldsymbol{\sigma}_{_{\boldsymbol{y}}},\ \boldsymbol{\sigma}_{_{\boldsymbol{v}}},
\boldsymbol{\sigma}_{_{\boldsymbol{w}}})$,\ \ and\ \ $\boldsymbol{\Gamma}_{_{\boldsymbol{\eta}}}^{b}$\ is defined as\ \
$Q_{c}$\ \emph{(}in \emph{(\ref{eq:sec5-eq03}))} replacing\ \ 
$(\widehat{\boldsymbol{\Gamma}}_{_{\boldsymbol{y}}},\widehat{\boldsymbol{\Gamma}}_{_{\boldsymbol{v}}},
\mathbf{Y}_{_{a}}^{^{i}})$\ \ by\
$(\boldsymbol{\Gamma}_{_{\boldsymbol{y}}}^{a},\boldsymbol{\Gamma}_{_{\boldsymbol{v}}}^{b},\mathbf{Y}_{_{a}}^{^{3}})$.    \hfill$\nabla$ 
\end{proposition}

\begin{remark}
 In exactly the same way described in Remark \emph{\ref{remark:5.2}}, the frequency-response of the ``average-cost'' estimators for the robust 
 $\mathcal{H}_{\infty}$ problem defined by \emph{Prob. $6$} is obtained from a solution $(\boldsymbol{\beta}_{_{\mathrm{o}}}, 
 \mathbf{P}_{_{\mathrm{o}}}, \mathbf{R}_{_{\mathrm{o}}}, \boldsymbol{\sigma}_{_{\boldsymbol{y}}}^{^{\mathrm{o}}},
 \boldsymbol{\sigma}_{_{\boldsymbol{v}}}^{^{\mathrm{o}}}, \boldsymbol{\sigma}_{_{\boldsymbol{w}}}^{^{\mathrm{o}}})$\ \ of the SDP introduced in 
 Proposition \emph{\ref{prop:18}} as
 $\mathbf{G}(\boldsymbol{\beta}_{_{\mathrm{o}}})
 =\boldsymbol{\beta}_{_{\mathrm{o}}}\mathbf{Y}_{_{a}}^{^{3}}$.\hfill$\nabla$
\end{remark}

\section{Comparing Robust Estimators}\label{sec:6}
In this section, the average cost/worst-case constraint estimators introduced in Section \ref{sec:5} are compared with the corresponding minimax ones. 
The main issues addressed here are first discussed in connection with robust $\mathcal{H}_{2}$\ estimation.

\subsection{Robust $\mathcal{H}_{2}$ Estimators}\label{subsec:6.1}

The assessment of a linear MSE estimator given by $\mathbf{G}\in \mathcal{R}_{c}^{m_{_{\boldsymbol{e}}}\times m_{_{\boldsymbol{v}}}}$ in connection with a model set\
$\mathcal{S}_{_{\mathbf{H}}}$\ may be carried out in terms of various features of (functionals computed on) the ``MSE function''\
$\mathcal{J}(\mathbf{G};\cdot): \mathcal{S}_{_{\mathbf{H}}}\rightarrow \mathbb{R}$. As pointed out in Section \ref{sec:1}, the supremum of\ 
$\mathcal{J}(\mathbf{G}; \cdot)$\ over\ 
$\mathcal{S}_{_{\mathbf{H}}}$\ (denoted by\ $\bar{\mathcal{J}}(\mathbf{G}; \mathcal{S}_{_{\mathbf{H}}})$), \emph{i.e.}, the worst-case MSE 
over\ $\mathcal{S}_{_{\mathbf{H}}}$,\ 
has been extensively used as the main assessment feature in the case of set-theoretic, model uncertainty, leading to minimax estimators 
(denoted below by $\mathbf{G}_{_{\mathbf{M}}}$). To mitigate the conservatism of such estimators other features may be considered together with\
$\bar{\mathcal{J}}(\mathbf{G}; \mathcal{S}_{_{\mathbf{H}}})$\ such as the ``average ($L_{1}-$norm)'' of $\mathcal{J}(\mathbf{G}; \cdot)$\
over $\mathcal{S}_{_{\mathbf{H}}}$\ or the nominal MSE,\ $\mathcal{J}(\mathbf{G};\mathbf{H}_{_{\boldsymbol{0}}})$,\ which lead to the
estimator defined by $Prob.\ 4$ above with\ \ $\boldsymbol{\eta}(\mathbf{G})=\boldsymbol{\eta}_{av}(\mathbf{G})$  -- this estimator will be denoted by 
$\mathbf{G}_{_{a\boldsymbol{v}}}$.

When comparing this estimator with the minimax estimator the questions naturally arise as to whether the improvement brought about 
at the expense of the increase in the worst-case MSE is significant (at some points of $\mathcal{S}_{_{\mathbf{H}}}$) or whether\ $\mathbf{G}_{_{a\boldsymbol{v}}}$\ outperforms\ $\mathbf{G}_{_{\mathbf{M}}}$\ on a ``sizable part''\ of $\mathcal{S}_{_{\mathbf{H}}}$.

In the first case (range of ``point-wise improvement''), a suitable additional feature of\ $\mathcal{J}(\mathbf{G}; \cdot)$\break
(for $\mathbf{G}=\mathbf{G}_{_{a\boldsymbol{v}}}$) 
could be defined by 
$$\boldsymbol{\eta}_{_{PW}}(\mathbf{G})\triangleq\sup\left\{\mathcal{J}(\mathbf{G}_{_{\mathbf{M}}}; \mathbf{H})-\mathcal{J}(\mathbf{G};
\mathbf{H}): \mathbf{H}\in \mathcal{S}_{_{\mathbf{H}}}\right\},$$
or, in relative terms, by 
$$\boldsymbol{\eta}_{_{\mathbf{R}W}}(\mathbf{G})=\inf\left\{\mathcal{J}(\mathbf{G}; 
\mathbf{H})/\mathcal{J}(\mathbf{G}_{_{\mathbf{M}}};\mathbf{H}): \mathbf{H}\in \mathcal{S}_{_{\mathbf{H}}}\right\}.$$

\vspace*{3mm}

\begin{remark}
It should be noted that for a given $\mathbf{G}\in \mathcal{S}_{_{\mathbf{G}}}$, $\boldsymbol{\eta}_{_{PW}}(\mathbf{G})$ can also be characterized as the optimal 
value of an SDP along the lines which led to Proposition \emph{\ref{prop:03}} and equation \emph{(\ref{eq:12})}.\hfill$\nabla$
\end{remark}

\vspace*{3mm}

 In the second case (``relative size'' of the subset of\ $\mathcal{S}_{_{\mathbf{H}}}$\ over which point-wise performance was improved), the 
 ``improvement set''\ $\mathcal{S}_{_{I}}$\ could be defined as\ 
$\mathcal{S}_{_{I}}(\mathbf{G}; \mathbf{G}_{_{\mathbf{M}}})=\left\{\mathbf{H}\in \mathcal{S}_{_{\mathbf{H}}}: 
\mathcal{J}(\mathbf{G}; \mathbf{H})< \mathcal{J}(\mathbf{G}_{_{\mathbf{M}}}; \mathbf{H})\right\}$\ or, equivalently,\
$$\mathcal{S}_{_{I}}(\mathbf{G}; \mathbf{G}_{_{\mathbf{M}}})=\left\{\mathbf{X}\in \mathcal{S}_{_{\mathbf{X}}}: \delta_{_{\mathcal{J}}}(\mathbf{X}; \mathbf{G})< 0\right\},$$
where\ 
$\delta_{_{\mathcal{J}}}(\mathbf{X};\mathbf{G})\triangleq\mathcal{J}_{_{\mathbf{X}}}(\mathbf{G}; \mathbf{X})- \mathcal{J}_{_{\mathbf{X}}}(\mathbf{G}_{_{\mathbf{M}}}; X)$, \emph{i.e.},\  
$\delta_{_{\mathcal{J}}}(\mathbf{X}; \mathbf{G})= \delta_{_{\mathcal{J}\mathrm{o}}}-2\delta_{_{\ell}}(\mathbf{X}; \mathbf{G})+\delta_{_{q}}(\mathbf{X}; \mathbf{G})$,\break
$\delta_{_{\mathcal{J}\mathrm{o}}}=\mathcal{J}(\mathbf{G};\mathbf{H}_{_{\boldsymbol{0}}})-\mathcal{J}(\mathbf{G}_{_{\mathbf{M}}}; \mathbf{H}_{_{\boldsymbol{0}}})$, \
$\delta_{_{\ell}}(\mathbf{X}; \mathbf{G})=\ell_{_{\mathbf{X}}}(\mathbf{X}; \mathbf{G})-\ell_{_{\mathbf{X}}}(\mathbf{X}; \mathbf{G}_{_{\mathbf{M}}})$,\break
$\delta_{_{q}}(\mathbf{X}; \mathbf{G})=q_{_{\mathbf{X}}}(\mathbf{X}; \mathbf{G})
- q_{_{\mathbf{X}}}(\mathbf{X};\mathbf{G}_{_{\mathbf{M}}})$,\ 
$\ell_{_{\mathbf{X}}}(\mathbf{X}; \mathbf{G})= \left\langle \mathbf{G}\mathbf{X}\boldsymbol{\Gamma}_{_{\boldsymbol{y}1}}, \mathbf{X}_{_{\boldsymbol{0}}}(\mathbf{G})\right\rangle$\ and\break
$q_{_{\mathbf{X}}}(\mathbf{X}; \mathbf{G})= \left\langle \mathbf{G} \mathbf{X}\boldsymbol{\Gamma}_{_{\boldsymbol{y}1}}, \mathbf{G}\mathbf{X} \right\rangle$. 

Due to the difficulty in comparing the ``volume''\ \ (in $\mathcal{H}_{2}$)\ \ of\ \ 
$\mathcal{S}_{_{I}}(\mathbf{G}; \mathbf{G}_{_{\mathbf{M}}})$\ \ with that of\ \ 
$\mathcal{S}_{_{\mathbf{X}}}$,\ \ ``projections'' of\ \ $\mathcal{S}_{_{I}}(\mathbf{G}; \mathbf{G}_{_{\mathbf{M}}})$\ \ along radial line segments in \ $\mathcal{S}_{_{\mathbf{X}}}$\  
are considered in the search for conditions which indicate that\ $\mathbf{G}$\ improves on\ $\mathbf{G}_{_{\mathbf{M}}}$\ over a ``sizeable part''\ of 
$\mathcal{S}_{_{\mathbf{X}}}$. To this effect, consider radial line segments of model perturbations\ $\mathbf{X}$\ (the nominal model 
corresponds to $\mathbf{X}=0$)\ \
$\mathcal{S}_{_{\boldsymbol{\beta}}}(\widehat{\mathbf{X}})=\left\{\mathbf{X}=\boldsymbol{\beta}\widehat{\mathbf{X}}: \boldsymbol{\beta}\in [-1, 1]\right\}$,\ where\
$\widehat{\mathbf{X}}\in \mathcal{R}_{c}^{m_{_{\boldsymbol{v}}}\times m_{_{\boldsymbol{y}}}}$\ is such that\ $\|\widehat{\mathbf{X}}\|_{_{2}}=1$\ \ and\ \
$\delta_{_{\ell}}(\widehat{\mathbf{X}})\leq 0$ -- note that no loss of generality is incurred by the condition\ 
``$\delta_{_{\ell}}(\widehat{\mathbf{X}})\leq 0$''\ \ since\ \
$\mathcal{S}_{_{\boldsymbol{\beta}}}(\widehat{\mathbf{X}})$\ \ and\ \ $\mathcal{S}_{_{\boldsymbol{\beta}}}(-\widehat{\mathbf{X}})$\ \ define the same line segment. Along one such segment 
 the projection of\ $\mathcal{S}_{_{I}}(\cdot)$\ (say, $\mathcal{S}_{_{I\boldsymbol{\beta}}}(\widehat{\mathbf{X}};\mathbf{G})$) is given by
$$\mathcal{S}_{_{I \boldsymbol{\beta}}}(\widehat{\mathbf{X}}; \mathbf{G})=\left\{\boldsymbol{\beta}\in [-1, 1]:
\delta_{_{\mathcal{J}\mathrm{o}}}+2\boldsymbol{\beta}|\delta_{_{\ell}}(\widehat{\mathbf{X}}; \mathbf{G})|+\boldsymbol{\beta}^{^{2}}\delta_{_{q}}(\widehat{\mathbf{X}}; \mathbf{G})<0\right\}$$
so that, if the length (Lebesgue measure)\ $ \mu_{_{I}}(\widehat{\mathbf{X}}; \mathbf{G})$\ of\ $\mathcal{S}_{_{I\boldsymbol{\beta}}}(\widehat{\mathbf{X}}; \mathbf{G})$\ is greater 
than\ $\alpha\in[0, 2]$,\ it can be said that\ ``$\mathbf{G}$\ improves on\ $\mathbf{G}_{_{\mathbf{G}}}$\ on a fraction of the segment defined
by\ $\widehat{\mathbf{X}}$ which is larger than $\alpha/2$''.

If this holds for any radial linear segment, it can be said that\ $\mathbf{G}$\ improves on\ $\mathbf{G}_{_{\mathbf{M}}}$\ (in a ``radial'' sense)
 \ on ``more than $\alpha/2$''\
of the perturbed model set.

A lower bound on how much\ $\mathbf{G}$\ improves on\ $\mathbf{G}_{_{\mathbf{M}}}$\ in this sense is provided by the next proposition.

\begin{proposition}\label{prop:07}
 Let $\bar{\delta}_{_{\ell}}\triangleq\gamma\left\|\left\{\left(\mathbf{G}^{*}\mathbf{X}_{_{\mathrm{o}}}(\mathbf{G})
 -\mathbf{G}_{_{\mathbf{M}}}^{*}\mathbf{X}_{_{\mathrm{o}}}(\mathbf{G}_{_{\mathbf{M}}})\right)\boldsymbol{\Gamma}_{_{\boldsymbol{y}1}}\right\}_{ca}\right\|_{_{2}}$\ \
 and\\
 $\bar{\delta}_{_{q}}\triangleq\bar{\lambda}_{\infty}\left(\mathbf{G}^{*}\mathbf{G}
 -\mathbf{G}_{_{\mathbf{M}}}^{*}\mathbf{G}_{_{\mathbf{M}}}\right)\|\boldsymbol{\phi}_{_{\boldsymbol{y}1}}\|_{\infty}^{^{2}}\gamma^{^{2}}$,\ \ 
 $\boldsymbol{\Gamma}_{_{\boldsymbol{y}1}}=\boldsymbol{\phi}_{_{\boldsymbol{y}1}}\boldsymbol{\phi}_{_{\boldsymbol{y}1}}^{*}$, \ \ where for\ \ $\mathbf{E}(e^{j\phi})
 =\mathbf{E}_{a}(e^{j\phi})+\mathbf{E}_{a}(e^{j\phi})^{*}$\ \ with\break
 $\mathbf{E}_{a}\in \mathcal{R}_{c}^{m\times m}$,\ \
 $\bar{\lambda}_{\infty}(\mathbf{E})\triangleq\sup\left\{\lambda_{\max}(\mathbf{E}(e^{j\phi})): \phi \in [0, 2\pi]\right\}$\ \ and\ \ $\lambda_{\max}(\mathbf{P})$\ \ 
 denotes the maximum eigenvalue of the Hermitian matrix $\mathbf{P}$.  Let\
 $\nu_{a}\triangleq(1/2)\left|\delta_{_{\mathcal{J}\mathrm{o}}}\right|/\bar{\delta}_{_{\ell}}$,\
 $\nu_{c}\triangleq|\delta_{_{\mathcal{J}\mathrm{o}}}|/\bar{\delta}_{_{q}}$,\
 $\nu_{_{\beta}}\triangleq(1/2)\nu_{c}\left/\left\{\left(\bar{\delta}_{_{\ell}}/\bar{\delta}_{_{q}}\right)^{^{2}}+\nu_{c}\right\}^{^{1/2}}\right.$\ \ and
 define\ \ $\mu_{_{I}}^{^{\mathrm{o}}}(\mathbf{G})=\min\left\{2, 1+\nu_{a}, 1+\nu_{b}, 2\nu_{c}^{^{1/2}}\right\}$. Let\ $\mathbf{G}$\ be such that \ 
 $\delta_{_{\mathcal{J}\mathrm{o}}}(\mathbf{G})<0$. Then\ \  $\forall\ \widehat{\mathbf{X}}\in \mathcal{R}_{c}^{m_{_{\boldsymbol{v}}}\times m_{_{\boldsymbol{y}}}}$\ \ 
 such that\ \ $\|\widehat{\mathbf{X}}\|_{_{2}}=\gamma^{^{2}}$,\ $ \mu_{_{I}}(\widehat{\mathbf{X}}; \mathbf{G})\geq \mu_{_{I}}^{^{\mathrm{o}}}(\mathbf{G})$.\hfill$\nabla$
\end{proposition}

In the next section, three numerical examples will be presented in which a given robust estimator\ $\mathbf{G}$\ will be assessed on the basis of its 
worst-case performance ($\bar{\mathcal{J}}_{_{\mathbf{X}}}(\mathbf{G})$), ``range of point-wise improvement'' ($\eta_{_{PW}}(\mathbf{G})$\ or\ $\boldsymbol{\eta_{_{\mathbf{R}W}}}(\mathbf{G})$) and of
estimates of 
the ``relative size''\ of the ``improvement set''\ $\mathcal{S}_{_{I}}(\mathbf{G}; \mathbf{G}_{_{\mathbf{M}}})$. The first two
performance indexes
will be computed with SDPs whereas the third feature will be (conservatively) assessed by means of Proposition \ref{prop:07} and by 
estimates of the Lebesgue measure of\ $\mathcal{S}_{_{I}}(\mathbf{G}; \mathbf{G}_{_{\mathbf{M}}})\bigcap \mathcal{S}_{_{\mathbf{X}}}^{^{N}}$\ 
obtained from pseudo-random samples.

\subsection{Numerical Examples}\label{subsec:6.2}

A simple SISO numerical example is now presented to illustrate the potential of the estimator given by $\mathbf{G}_{a v}$ (obtained on the
basis of $Prob.\ 4$) 
in the search of trade-offs between worst-case and ``point-wise''~performance over the perturbed model set. The comparison between this estimator and the
minimax estimator given by $\mathbf{G}_{_{\mathbf{M}}}$ (obtained on the basis of $Prob.\ 1$) will be based on the estimate
$\mu_{_{I}}^{^{\mathrm{o}}}(\mathbf{G}_{_{a\boldsymbol{v}}})/2$ of the ``fraction of the model set''~upon which $\mathbf{G}_{_{a\boldsymbol{v}}}$ improves on 
$\mathbf{G}_{_{\mathbf{M}}}$ and on the largest (over the ``channel'' model set) relative decrease of the estimator error brought about by $\mathbf{G}_{_{a\boldsymbol{v}}}$,\ \emph{i.e.},\ $1-\boldsymbol{\eta}_{_{\mathbf{R}\mathbf{M}}}(\mathbf{G}_{_{av}})$,\ \ where
$$\boldsymbol{\eta}_{_{\mathbf{R}\mathbf{M}}}(\mathbf{G}_{_{a\boldsymbol{v}}})
=\inf\{\mathcal{J}(\mathbf{G}_{_{a\boldsymbol{v}}}; \mathbf{H})/\mathcal{J}(\mathbf{G}_{_{\mathbf{M}}}; \mathbf{H}):
\mathbf{H} \in \mathcal{S}_{_{\mathbf{H}}}\}.$$

Additionally, for randomly-generated samples of SISO, FIR model perturbations\ $\boldsymbol{X}$\ of a pre-specified length $L$ (say $\{\boldsymbol{X}_{1}, \mathellipsis, \boldsymbol{X}_{N_{_{\boldsymbol{x}}}}\}$,\ \ with\ \
$\boldsymbol{X}$\  uniformly distributed on\
$\left\{\boldsymbol{\theta}\in\mathbb{R}^{L+1}:\|\boldsymbol{\theta}\|_{_{E}}\leq\gamma\right\}$), the following statistics where computed \ \ $i_{_{fN}}=i_{_{N}}/N_{_{\boldsymbol{x}}}$,\ where\ $i_{_{N}}$\ is the number
of instances\ \ $\boldsymbol{X}_{_{i}}$\ \ such that\ $\mathcal{J}(\mathbf{G}_{_{a\boldsymbol{v}}}; \mathbf{H}_{_{\boldsymbol{0}}}+\boldsymbol{X}_{_{i}})< \mathcal{J}(\mathbf{G}_{_{\mathbf{M}}}; \mathbf{H}_{_{\boldsymbol{0}}}+X_{i})$, \ \
and\ \ $\min\{\mathcal{J}(\mathbf{G}_{_{a\boldsymbol{v}}}; \mathbf{H}_{_{\boldsymbol{0}}}+\boldsymbol{X}_{_{i}})/ \mathcal{J}(\mathbf{G}_{_{\mathbf{M}}}, \mathbf{H}_{_{\boldsymbol{0}}}+\boldsymbol{X}_{_{i}}): i=1, \mathellipsis,  N_{_{\boldsymbol{x}}}\}$ 
-- note that\ \ $i_{_{fN}}$\ \ is a consistent estimator for the ratio between the ``volumes'' of the ``improvement set'' (say $v_{_{IL}}$) and that of\ \ 
$\mathcal{S}_{_{\boldsymbol{X}^{^{N}}}}$ (say $v_{_{\boldsymbol{X}L}}$) when\ \ $\boldsymbol{X}$\ \ is confined to the set of FIRs of length $L$.

The possible point-wise improvements were to be obtained at the expenses of a pre-specified, maximum allowed increase in the worst-case 
MSE vis-\`a-vis that attained with the minimax estimator.

To generate the numerical results, the following sequence of computations is required for a given set-up specified by\ \ 
$(\boldsymbol{\Gamma}_{\boldsymbol{y}}, \boldsymbol{\Gamma}_{\boldsymbol{v}}, \mathbf{H}_{_{\boldsymbol{0}}}, \mathbf{H}_{_{\mathbf{I}}}, W, \gamma)$:\medskip

\noindent
\textbf{({\emph{i})}} Compute (an approximation to) the solution\ $\mathbf{G}_{_{\mathrm{o}}}$\ (with minimal realization\
($\mathbf{A}_{_{\mathrm{o}}}, \mathbf{B}_{_{\mathrm{o}}}, \mathbf{C}_{_{\mathrm{o}}}, \mathbf{D}_{_{\mathrm{o}}}$)) of the nominal MSE 
problem\ \ 
$\displaystyle\min_{\mathbf{G}\in \mathcal{R}^{m_{_{\boldsymbol{e}}}\times m_{_{\boldsymbol{v}}}}}\mathcal{J}(\mathbf{G};
\mathbf{H}_{_{\boldsymbol{0}}})$ --
note that\
$\mathbf{G}_{_{\mathrm{o}}}=
\{\mathbf{H}_{_{\mathbf{I}}}\boldsymbol{\Gamma}_{\boldsymbol{y}}\mathbf{H}_{_{\boldsymbol{0}}}^{*}(\boldsymbol{\psi}_{_{\mathrm{o}}}^{*})^{^{-1}}\}_{ca}\boldsymbol{\psi}_{_{\mathrm{o}}}^{^{-1}}$\ \
where\ \ $\boldsymbol{\psi}_{_{\mathrm{o}}}$\ \ is a spectral factor of\ \
$\boldsymbol{\Gamma}_{\boldsymbol{v}}+\mathbf{H}_{_{\boldsymbol{0}}}\boldsymbol{\Gamma}_{\boldsymbol{y}}\mathbf{H}_{_{\boldsymbol{0}}}^{*}= 
\boldsymbol{\psi}_{_{\mathrm{o}}}\boldsymbol{\psi}_{_{\mathrm{o}}}^{*}$.\medskip

\noindent
\textbf{({\emph{ii})}} Compute the frequency-response\ \ $\mathbf{G}_{_{\mathbf{M}}}$\ \ of the minimax MSE estimator in the class\\ 
$\mathcal{S}_{_{\mathbf{G}}}=\mathcal{S}_{_{\mathbf{G}}}^{nom}\triangleq\left\{\mathbf{G}(\boldsymbol{\beta})=\boldsymbol{\beta}\mathbf{Y}_{_{\mathrm{o}}}^{^{a}}:
\boldsymbol{\beta}\in\mathbb{R}^{m_{_{\boldsymbol{e}}}\times (n_{_{\mathrm{o}}}+m_{_{\boldsymbol{v}}})}\right\}$,\ \ where\ \
$\mathbf{Y}_{_{\mathrm{o}}}^{^{a}}=\begin{bmatrix}
                     \mathbf{Y}_{_{\mathrm{o}}}\mathbf{B}_{_{\mathrm{o}}}\\
                     \mathbf{I}_{m_{_{\boldsymbol{v}}}}\\
                    \end{bmatrix}$\ \ and\ \ $\mathbf{Y}_{_{\mathrm{o}}}(e^{j\phi})=(e^{j\phi}\mathbf{I}-\mathbf{A}_{_{\mathrm{o}}})^{^{-1}}$,\ \ and the 
corresponding worst-case error\ \ $\bar{\mathcal{J}}_{_{\mathbf{X}}}(\mathcal{S}_{_{\mathbf{G}}})$,\ solving the SDP of Proposition \ref{prop:04}.\medskip

\noindent
\textbf{({\emph{iii})}} Choose an upper bound for worst-case performance\ \ 
$\boldsymbol{\eta}_{_{\mathcal{J}}}=(1+\alpha)\bar{\mathcal{J}}_{_{\mathbf{X}}}(\mathcal{S}_{_{\mathbf{G}}})$, \ $\alpha\in (0,1)$\  and obtain\
$\mathbf{G}_{_{a\boldsymbol{v}}}$\ solving the SDP of Proposition \ref{prop:06}.\medskip

\noindent
\textbf{({\emph{iv})}} Compute\ \ $ \mu_{_{I}}^{^{\mathrm{o}}}(\mathbf{G}_{_{a\boldsymbol{v}}})$\ \ as defined in Proposition \ref{prop:07}.\medskip

\noindent
\textbf{({\emph{v})}} Compute\ \
$\boldsymbol{\eta}_{_{\mathbf{R}W}}(\mathbf{G}_{_{a\boldsymbol{v}}})=
\inf\left\{\mathcal{J}(\mathbf{G}_{_{a\boldsymbol{v}}};\mathbf{H})-\lambda\mathcal{J}(\mathbf{G}_{_{\mathbf{M}}}; \mathbf{H}): 
\mathbf{H}\in\mathcal{S}_{_{\mathbf{H}}},\ \lambda\in \mathbb{R}_{+}\right\}$,\ \emph{i.e.},\\
$\boldsymbol{\eta}_{_{\mathbf{R}W}}(\mathbf{G}_{_{a\boldsymbol{v}}})=\inf\left\{\lambda>0: f_{_{\mathbf{R}W}}(\lambda)\leq 0\right\}$,\ \
$f_{_{\mathbf{R}W}}(\lambda)=\inf\{\mathcal{J}(\mathbf{G}_{_{a\boldsymbol{v}}};\mathbf{H})-\lambda \mathcal{J}(\mathbf{G}_{_{\mathbf{M}}}; \mathbf{H}): 
\mathbf{H}\in \mathcal{S}_{_{\mathbf{H}}}\}$\ \ by means of a line search with respect to \ $\lambda$\ with\ $f_{_{\mathbf{R}W}}(\lambda)$ ( 
for a given value of\ $\lambda$) computed by means of a SDP.\medskip

To ensure that tight confidence intervals for\ \ $v_{_{IL}}/v_{_{\boldsymbol{X}L}}$\ \ can be constructed from a sample\ 
$\{\boldsymbol{X}_{1},\mathellipsis, \boldsymbol{X}_{N_{_{\boldsymbol{x}}}}\}$,\ a lower bound on\ \ $N_{_{\boldsymbol{x}}}$\ \ is enforced to ensure that a double-sided 
confidence interval of length\ \ $2\varepsilon$\ \ around \ \ $i_{_{fN}}$\ \  has confidence level of\ $(1-\delta)$,  namely, \ 
$N_{_{\boldsymbol{x}}}\geq\dfrac{1}{2\varepsilon^{^{2}}}\log(2/\delta)$\ ([28]) -- thus, with\  $\varepsilon=10^{-2}$\ and\ 
$\delta=10^{-2}$,\ $N_{_{\boldsymbol{x}}}\geq16.505$.

A simple, SISO numerical experiment was carried out with the following data:\\
$m_{_{\boldsymbol{y}}}=m_{_{\boldsymbol{v}}}=m_{_{\boldsymbol{e}}}=1$,\ 
$\boldsymbol{\Gamma}_{\boldsymbol{y}}=\boldsymbol{\sigma}_{_{\boldsymbol{y}}}^{^{2}}$,\ 
$\boldsymbol{\Gamma}_{\boldsymbol{v}}=\boldsymbol{\sigma}_{_{\boldsymbol{v}}}^{^{2}}$,\ 
$\boldsymbol{\sigma}_{_{\boldsymbol{v}}}=0.1\boldsymbol{\sigma}_{_{\boldsymbol{y}}}$,\ $\boldsymbol{\sigma}_{_{\boldsymbol{y}}}=5$,\ $\mathbf{H}_{_{\mathbf{I}}}=1$,\ 
$\mathbf{H}_{_{\boldsymbol{0}}}(e^{j\phi})=\displaystyle\sum_{k=0}^{5}\mathbf{M}_{k}e^{-j\phi k}$,\ 
$\begin{bmatrix}\mathbf{M}_{0}& \mathbf{M}_{1}& \cdots& \mathbf{M}_{5}\end{bmatrix}=2 
\times \begin{bmatrix}1.0 & -1.3963 & 0.9638 & -0.8713 & 0.5593 & -0.1389\end{bmatrix}$,\ $W=1$,\
$\alpha=0.15$\ (see \textbf{(\emph{iii})} above),\ $\gamma=0.3\|\mathbf{H}_{_{\boldsymbol{0}}}\|_{_{2}}$.

The results obtained were as follows:\\
$\delta_{_{\mathcal{J}\mathrm{o}}}=-6.7966$,\ $\mu_{_{I}}^{^{\mathrm{o}}}(\mathbf{G}_{_{a\boldsymbol{v}}})=1.4684$\ 
($\mu_{_{I}}^{^{\mathrm{o}}}(\mathbf{G}_{_{\mathbf{B}}})=1.4703$) indicating that \ $\mathbf{G}_{_{a\boldsymbol{v}}}$\ improved on\  $\mathbf{G}_{_{\mathbf{M}}}$\ 
over at least\ $0.7342\%$\ of the perturbed model set\ $\mathcal{S}_{_{\mathbf{H}}}$;\ 
$\boldsymbol{\eta}_{_{\mathbf{R}W}}(\mathbf{G}_{_{a\boldsymbol{v}}})\approx0.2500$\ showing that at certain points of the perturbed model set\ 
$\mathbf{G}_{_{a\boldsymbol{v}}}$\ yields a MSE value which is close to a $75\%$ reduction of the one obtained with $\mathbf{G}_{_{\mathbf{M}}}$ -- the maximum 
point-wise improvement observed in the sample in the case of\ $L=6$,\ achieved at a frequency-response\ $\mathbf{H}_{a}$ (say) corresponds 
to\ \ $\mathcal{J}(\mathbf{G}_{_{a\boldsymbol{v}}}; \mathbf{H}_{a})= 3.0202$\ \ and\ \ $\mathcal{J}(\mathbf{G}_{_{\mathbf{M}}}; \mathbf{H}_{a})=12.5158$.

The role of\ \ $\mu_{_{I}}^{^{\mathrm{o}}}$\ \ as a conservative estimate of the relative size of the ``improvement set'' is borne out by the values of\ \ 
$i_{_{fN}}$\ \ obtained in three Monte Carlo experiments, each one with\ \ $N_{_{\boldsymbol{x}}}=65\ 000$ samples with FIR model perturbations
of length\ $L=6, 9$ and $13$. The obtained values were as follows:\ \
for\ \ $L=6$,\ \ $i_{_{fN}}=0.9673$ for $L=9$, $i_{_{fN}}=0.9822$ and for $L=13$, $i_{_{fN}}=0.9923$.

Summing up, for\ \ a $15\%$\ \ increase in the worst-case MSE over its minimum value, the robust estimator given
by\ \ $\mathbf{G}_{_{a\boldsymbol{v}}}$\ \ is guaranteed in this example to improve on\ \ $\mathbf{G}_{_{\mathbf{M}}}$\ \ over at least ``0.73 of the 
perturbed model set\ \ $\mathcal{S}_{_{\mathbf{H}}}$,'' \ bringing about a decrease in the MSE error (vis-\`{a}-vis that of the minimax
estimator) by up to\ $75\%$\ in some points of\ \ $\mathcal{S}_{_{\mathbf{H}}}$ -- in fact, the ``relative size'' of the improvement set may 
be expected to be substantially bigger than $0.73$ as Monte Carlo results with FIR model perturbations of length \ 6, 9 and 13 led to
lower confidence bounds (with 0.99 confidence level) for this ratio respectively greater than 0.94,\ 0.97\ and\ 0.98 (\emph{i.e}, the ``improvement set'' in each these cases is
at least $94\%$,\  $97\%$\ and\ $98\%$ of the uncertain model set).

The numerical results for this SISO example are summarized in Tables (recall that\  $\bar{\mathcal{J}}$,\ $\boldsymbol{\eta}^{a}$\ and\ 
$\boldsymbol{\eta}_{_{\mathbf{R}W}}$ are the worst-case, average smallest relative
MSE on $\mathcal{S}_{_{\mathbf{H}}}$)
\begin{table}[h!]
\begin{center}
\begin{tabular}{ r|c|c|c|c|}
\multicolumn{1}{r}{}
 &  \multicolumn{1}{c|}{$\bar{\mathcal{J}}(\mathbf{G})$}
 & \multicolumn{1}{c|}{$\boldsymbol{\eta}^{a}(\mathbf{G})$} & 
 \multicolumn{1}{c|}{$\mathcal{J}(\mathbf{G};\mathbf{H}_{_{\boldsymbol{0}}})$}  & 
 \multicolumn{1}{c}{$\boldsymbol{\eta}_{_{\mathbf{R}W}}(\mathbf{G})$} \\
\cline{2-5}
$\mathbf{G}_{_{a\boldsymbol{v}}}$ & 26.8604 & 12.6040 &10.1015&$\approx 0.25$ \\
\cline{1-5}
$\mathbf{G}_{_{\mathbf{M}}}$ & 23.3569 & 17.4249 &16.9366& 1  \\
\cline{2-5}
\end{tabular}
\end{center}
\caption{Performance on\ $\mathcal{S}_{_{\mathbf{H}}}$.}
\end{table}

Table 2 exhibits the relative frequencies of the improvement set $i_{_{fN}}$ in three experiments involving\ 65,000 samples of FIR perturbations of length\
$L_{_{\text{FIR}}}=$ 6, 9\ and\ 13, respectively
\begin{table}[h!]
\begin{center}
 \begin{tabular}{ r|c|c|c|}
\cline{2-4}
\multicolumn{1}{r}{$L_{_{\text{FIR}}}$}
 &  \multicolumn{1}{|c|}{6}
 & \multicolumn{1}{c|}{9} & 
 \multicolumn{1}{c|}{13}  \\
\cline{1-4}
$i_{_{fN}}$ & 0.9673 & 0.9822 & 0.9923 \\
\cline{2-4}
\end{tabular}
\end{center}
\caption{Relative frequency of the improvement set.}
\end{table}

In addition, similar numerical experiments were performed with two, $2 \times 2$ MIMO examples.
The first one involves a FIR of length 4 as nominal channel model $(\mathbf{H}_{_{\boldsymbol{0}}})$, a 
``coloured'' signal\ $\boldsymbol{y}$\ (filtered white-noise), with\ $\boldsymbol{\phi}_{_{\boldsymbol{y}}}$\ as the frequency-response of the shaping filter ($\mathbf{H}_{_{\boldsymbol{0}}}$\ and\ $\boldsymbol{\phi}_{_{\boldsymbol{y}}}$ are presented in the Appendix), white observation noise with covariance matrix given by
$\boldsymbol{\sigma}_{_{\boldsymbol{v}}}^{^{2}}\mathbf{I}$ and signal-to-noise ratio (at the channel output)\  
$\boldsymbol{\sigma}_{_{\boldsymbol{v}}}/\|\mathbf{H}_{_{\boldsymbol{0}}}
\boldsymbol{\phi}_{_{\boldsymbol{y}}}\|_{_{2}}=0.2$, \ perturbations radius\ 
$\gamma=0.2\times \|\mathbf{H}_{_{\boldsymbol{0}}}\boldsymbol{\phi}_{_{\boldsymbol{y}}}\|_{_{2}}$\ and average MSE minimization allowing for $0.1$ increase on worst-case MSE above its minimum value. The numerical examples for the first MIMO example are presented in Tables 3 and 4.

\begin{table}[h!]
\begin{center}
 \begin{tabular}{ r|c|c|c|c|}
\multicolumn{1}{r}{}
 &  \multicolumn{1}{c|}{$\mathcal{J}(\mathbf{G})$}
 & \multicolumn{1}{c|}{$\boldsymbol{\eta}^{a}(\mathbf{G})$} & 
 \multicolumn{1}{c|}{$\mathcal{J}(\mathbf{G};\mathbf{H}_{_{\boldsymbol{0}}})$}& \multicolumn{1}{c}{$\check{\boldsymbol{\eta}}_{_{\mathbf{R}W}}(\mathbf{G})$}  \\
\cline{2-5}
$\mathbf{G}_{_{a\boldsymbol{v}}}$ & 3.8090  & 1.961 & 1.8027 & $\approx 0.45$ \\
\cline{1-5}
$\mathbf{G}_{_{\mathbf{M}}}$ & 3.4628 & 2.5947 & 2.5289 & 1 \\
\cline{2-5}
\end{tabular}
\end{center}
\caption{MIMO example 1: Performance on\ $\mathcal{S}_{_{\mathbf{H}}}$.}
\end{table}

The relative frequencies of the improvement set in three experiments involving FIR perturbations of length $L_{_{\text{FIR}}}=$ 4, 8 and 10 are displayed in Table 3.

\begin{table}[h!]
\begin{center}
 \begin{tabular}{ r|c|c|c|}
 \cline{2-4}
\multicolumn{1}{r}{$L_{_{\text{FIR}}}$}
 &  \multicolumn{1}{|c|}{4}
 & \multicolumn{1}{c|}{8} & 
 \multicolumn{1}{c|}{10}  \\
\cline{1-4}
$i_{_{fN}}$ & 0.9987 & 0.9991 & 0.9999 \\
\cline{2-4}
\end{tabular}
\end{center}
\caption{MIMO example 1: Relative frequency of the improvement set.}
\end{table}

Analogously to what happened in the case of the SISO example presented above, with small allowed increase on worst-case performance ($10\%$ in this example) it is possible to achieve significantly better ``point wise'' performance with\ 
$\mathbf{G}_{_{a\boldsymbol{v}}}$, including reduction up to 
$55\%$ on the MSE of the minimum estimator at specific channel 
models and very high frequency of improvement over 
$\mathbf{G}_{_{\mathbf{M}}}$\ on Monte-Carlo experiments.

To further illustrate the ``pointwise'' improvements brought about by the a/w estimator, a plot is presented below which exhibits the variation of the error estimation criterion attained by the minimax and the a/w estimators over a path in the ``uncertain'' model set (with FIRs of length 4). This path goes from the most favourable to the most unfavourable model for the a/w estimator, passing through the nominal model.

\begin{figure}[h!]
\begin{center}
\includegraphics[width=10.5cm]{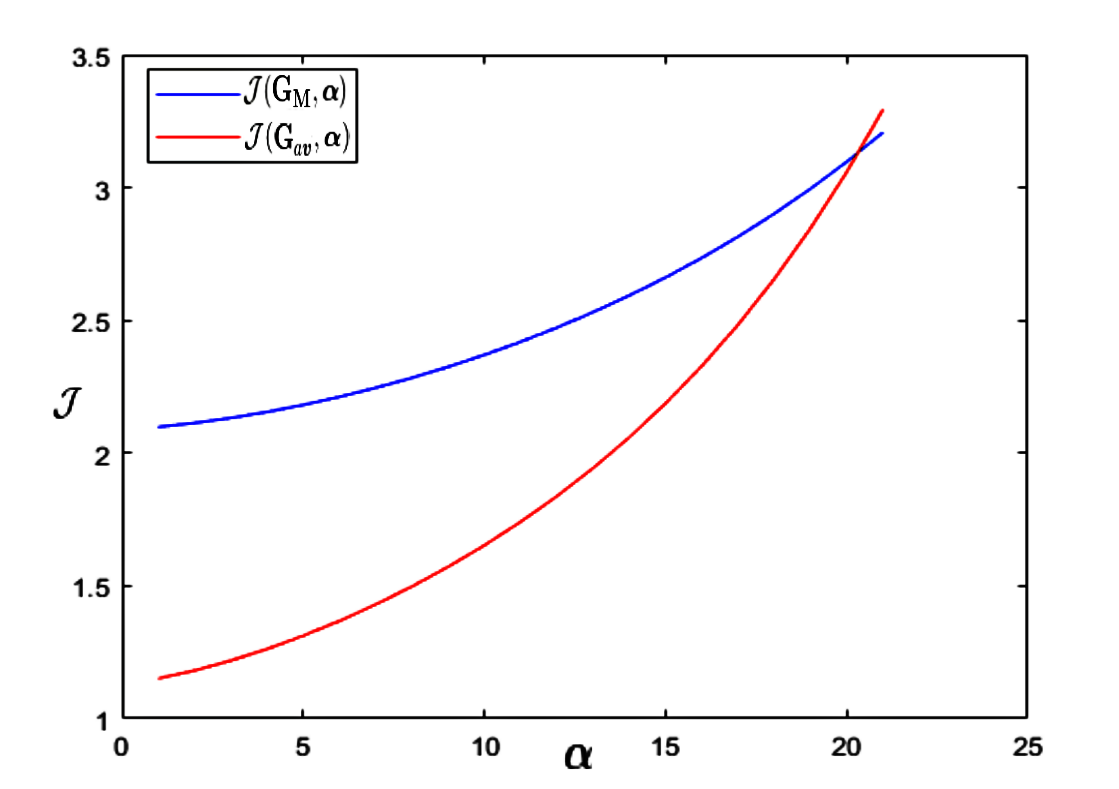}
\end{center}
\caption{Pointwise performance of $\mathbf{G}_{_{a\boldsymbol{v}}}$ and $\mathbf{G}_{_{\mathbf{M}}}$.}
\label{fig:02}
\end{figure}

 It can be seen that, at the most unfavourable model for the a/w
estimator (corresponding to the value of $21$ in the horizontal axis), its steady-state MSE exceeds that of the minimax estimator by about 10\%; whereas at the most favourable model (value of 1 in the horizontal axis) its steady-state MSE is about 0.54 of the MSE of the minimax estimator (the nominal model corresponds to the value 11 in the horizontal axis). In
addition, in about ``90\% of the path'' the pointwise performance of $\mathbf{G}_{_{a\boldsymbol{v}}}$ is
superior to that of $\mathbf{G}_{_{\mathbf{M}}}$.

In the second MIMO example, $\mathbf{H}_{_{\boldsymbol{0}}}$\ is given by
a fifth-order state-space model with non-zero eigenvalues (see the Appendix), 
$\gamma=0.3\times\|\mathbf{H}_{_{\boldsymbol{0}}}\boldsymbol{\phi}_{_{\boldsymbol{y}}}\|_{_{2}}$\ and the remaining data is as in the first MIMO example. The numerical results obtained are given in Tables 5 and 6.

\begin{table}[h!]
\begin{center}
 \begin{tabular}{ r|c|c|c|c|}
\multicolumn{1}{r}{}
 &  \multicolumn{1}{c|}{$\mathcal{J}(\mathbf{G})$}
 & \multicolumn{1}{c|}{$\boldsymbol{\eta}^{a}(\mathbf{G})$} & 
 \multicolumn{1}{c|}{$\mathcal{J}(\mathbf{G};\mathbf{H}_{_{\boldsymbol{0}}})$}& \multicolumn{1}{c}{$\check{\boldsymbol{\eta}}_{_{\mathbf{R}W}}(\mathbf{G})$}  \\
\cline{2-5}
$\mathbf{G}_{_{a\boldsymbol{v}}}$ & 3.2680 & 1.8904 & 1.5001 & $\approx 0.6$ \\
\cline{1-5}
$\mathbf{G}_{_{\mathbf{M}}}$ & 3.0293 & 2.0494 & 1.6340 & 1 \\
\cline{2-5}
\end{tabular}
\end{center}
\caption{MIMO example 2: Performance on\ $\mathcal{S}_{_{\mathbf{H}}}$.}
\end{table}

\begin{table}[h!]
\begin{center}
 \begin{tabular}{ r|c|c|c|}
 \cline{2-4}
\multicolumn{1}{r}{$L_{_{\text{FIR}}}$}
 &  \multicolumn{1}{|c|}{4}
 & \multicolumn{1}{c|}{8} & 
 \multicolumn{1}{c|}{10}  \\
\cline{1-4}
$i_{_{fN}}$ & 0.9104 & 0.9632 & 0.9747 \\
\cline{2-4}
\end{tabular}
\end{center}
\caption{MIMO example 2: Relative frequency of the improvement set.}
\end{table}

In this case, with worst-case MSE only $8\%$ larger than that of
$\mathbf{G}_{_{\mathbf{M}}}$, $\mathbf{G}_{_{a\boldsymbol{v}}}$\ achieves 
MSE reduction of up to $40\%$ at specific points of the channel model set and improves on\ $\mathbf{G}_{_{\mathbf{M}}}$\ over most of the Monte-Carlo samples examined.

\subsection{Nominal $\mathcal{H}_{\infty}$ Estimation}\label{subsec:6.3}

In this subsection, possible ways of comparing the performance of the linear estimators defined by $Prob.\ 2$ and $Prob.\ 5$ (with 
frequency-response\ $\mathbf{G}_{_{\mathbf{M}}}$\ and\ $\mathbf{G}$, respectively) are discussed along the lines pursued in Subsection \ref{subsec:6.1}.

In this case, the largest improvement in pointwise performance (by $\mathbf{G}$ over $\mathbf{G}_{_{\mathbf{M}}}$) is defined by
$$\boldsymbol{\eta}_{_{\mathbf{P}}}^{\infty}(\mathbf{G})=
\sup\left\{\left\langle\boldsymbol{\Gamma}_{_{\boldsymbol{\delta}}}\bar{\boldsymbol{z}},\bar{\boldsymbol{z}}\right\rangle:
\bar{\boldsymbol{z}}\in \mathcal{S}_{_{\bar{\boldsymbol{z}}}}\right\},\ \ \text{where}\ \ 
\boldsymbol{\Gamma}_{_{\boldsymbol{\delta}}}=\boldsymbol{\Gamma}_{_{\boldsymbol{e}0}}-\boldsymbol{\Gamma}_{_{\boldsymbol{e}1}},$$
$\boldsymbol{\Gamma}_{_{\boldsymbol{e}0}}=\mathbf{H}_{_{\boldsymbol{ez}}}(\mathbf{G}_{_{\mathbf{M}}})^{*}
\mathbf{H}_{_{\boldsymbol{ez}}}(\mathbf{G}_{_{\mathbf{M}}})$,\ \ $\boldsymbol{\Gamma}_{_{\boldsymbol{e}1}}
=\mathbf{H}_{_{\boldsymbol{ez}}}(\mathbf{G})^{*}\mathbf{H}_{_{\boldsymbol{ez}}}(\mathbf{G})$,\ \ 
$\mathbf{H}_{_{\boldsymbol{ez}}}=\mathbf{H}_{_{\mathbf{I}\boldsymbol{y}}}-\mathbf{G}\mathbf{H}_{_{\mathrm{o}\boldsymbol{z}}}$\ \ and\break
$\mathcal{S}_{_{\bar{\boldsymbol{z}}}}=\left\{\bar{\boldsymbol{z}}\in \mathcal{R}_{c}^{m_{_{\boldsymbol{y}}}+m_{_{\boldsymbol{v}}}}:
\bar{\boldsymbol{z}}=\begin{bmatrix}\bar{\boldsymbol{y}}^{^{\mathrm{T}}}&\vdots&\bar{\boldsymbol{v}}^{^{\mathrm{T}}}\end{bmatrix},\ 
\|\bar{\boldsymbol{y}}\|_{_{2}}\leq \gamma_{_{\boldsymbol{y}}},\ \|\bar{\boldsymbol{v}}\|_{_{2}}\leq \gamma_{_{\boldsymbol{v}}}\right\}$.\medskip

It may also be of interest to look at the ``largest'' relative improvement in pointwise performance brought about by\ \ $\mathbf{G}$, 
\emph{i.e.},\ $1-\boldsymbol{\eta}_{_{\mathbf{R}}}^{\infty}(\mathbf{G})$, for $\mathbf{G}$ such that $\boldsymbol{\eta}_{_{\mathbf{R}}}^{\infty}\leq1$, where
$$\boldsymbol{\eta}_{_{\mathbf{R}}}^{\infty}(\mathbf{G})
\triangleq\inf\left\{\left\langle\boldsymbol{\Gamma}_{_{\boldsymbol{e}1}}\bar{\boldsymbol{z}}, \bar{\boldsymbol{z}}\right\rangle/
\left\langle\boldsymbol{\Gamma}_{_{\boldsymbol{e}0}}\bar{\boldsymbol{z}}, \bar{\boldsymbol{z}}\right\rangle:
\bar{\boldsymbol{z}}\in\bar{\mathcal{S}}_{_{\boldsymbol{z}}}, \left\langle\boldsymbol{\Gamma}_{_{\boldsymbol{e}0}}\bar{\boldsymbol{z}},
\bar{\boldsymbol{z}}\right\rangle \neq0\right\}.$$

A lower bound on\ \ $\boldsymbol{\eta}_{_{\mathbf{P}}}^{\infty}(\mathbf{G})$\ and an upper bound on\ \ 
$\boldsymbol{\eta}_{_{\mathbf{R}}}^{\infty}(\mathbf{G})$\ can be approximately computed on the basis of the following statement.

\begin{proposition}\label{prop:08}
 For\ \  $\boldsymbol{\Gamma}=\boldsymbol{E}_{_{\boldsymbol{\Gamma}}}+\boldsymbol{E}_{_{\boldsymbol{\Gamma}}}^{*}$,\ where\ \
 $\boldsymbol{E}_{_{\boldsymbol{\Gamma}}} \in \mathcal{R}_{c}^{m\times m}$, let\\
 $\bar{\lambda}_{\infty}\left(\boldsymbol{\Gamma}\right)
 \triangleq\sup\left\{\bar{\lambda}(\boldsymbol{E}_{_{\boldsymbol{\Gamma}}}(e^{j\phi})): \phi \in [0, 2\pi]\right\}$, \ where\
 $\bar{\lambda}(\mathbf{R})$ is the maximum eigenvalue of the Hermitian matrix $\mathbf{R}$. Then,\\
 \noindent
 \emph{\textbf{(a)}} $\boldsymbol{\eta}_{_{\mathbf{P}}}^{\infty}(\mathbf{G})\geq 
 \bar{\lambda}_{\infty}\left(\mathbf{M}_{_{\gamma}}\boldsymbol{\Gamma}_{_{\boldsymbol{\delta}}}\mathbf{M}_{_{\gamma}}\right)$,\ \ where\ \ 
 $\mathbf{M}_{_{\gamma}}=\operatorname{diag}\left(\gamma_{_{\boldsymbol{y}}}\mathbf{I}_{m_{_{\boldsymbol{y}}}},
 \gamma_{_{\boldsymbol{v}}}\mathbf{I}_{m_{_{\boldsymbol{v}}}}\right)$.
 
 \noindent
 \emph{\textbf{(b)}} If\ \ $\mu>0$\ \ is such that\ \ $\bar{\lambda}_{\infty}
 \left\{\mathbf{M}_{_{\gamma}}(\mu\boldsymbol{\Gamma}_{_{\boldsymbol{e}0}}-\boldsymbol{\Gamma_{_{\boldsymbol{e}1}}})\mathbf{M}_{_{\gamma}}\right\}\geq0$,\ \
 $\boldsymbol{\eta}_{_{\mathbf{R}}}^{\infty}(\mathbf{G})\leq \mu$.\hfill$\nabla$
 \end{proposition}

 A simple SISO numerical example is now presented to illustrate the possible trade-off between worst-case and ``pointwise'' performance enabled
 by an  average cost/worst-case constraint estimator. The nominal model in this case is $(1/2)\mathbf{H}_{_{0}}$,\ where\ 
 $\mathbf{H}_{_{0}}$\ is as in Subsection \ref{subsec:6.1}. The remaining elements 
 of the estimation set-up are $\mathbf{H}_{_{\mathbf{I}}}=1$,\ $\gamma_{_{\boldsymbol{y}}}=5$,\ $\gamma_{_{\boldsymbol{v}}}=0.5$,\ $m_{_{\boldsymbol{y}}}=1$,\ $m_{_{\boldsymbol{v}}}=m_{_{\boldsymbol{e}}}=1$,\   $W_{_{\boldsymbol{y}}}=1$,\ and\ $W_{_{\boldsymbol{v}}}=1$. The estimators obtained from the approximate solution of $Prob.\ 2$ 
 ($\mathbf{G}_{_{\mathbf{M}}}$, see Proposition \ref{prop:14} and the subsequent ``Remark'') and $Prob.\ 5$ ($\mathbf{G}_{_{a\boldsymbol{v}}}$, see Proposition
 \ref{prop:15} and Remark \ref{remark:5.2}) give rise, respectively, to 
 worst-case, squared estimation errors smaller than $18.3903$ and $19.7594$ (an increase of less than $10\%$ in the minimum value of the 
 worst-case performance index). Using Proposition \ref{prop:08} and approximately computing (taking a grid on $[0, 2\pi]$) 
 $\bar{\lambda}_{\infty}(\mathbf{M}_{_{\gamma}} \boldsymbol{\Gamma}_{_{\boldsymbol{\delta}}}\mathbf{M}_{_{\gamma}})$\ \ and\ \
 $\bar{\lambda}_{\infty}\left\{\mathbf{M}_{_{\gamma}}\left(\mu\boldsymbol{\Gamma}_{_{\boldsymbol{e}0}}
 -\boldsymbol{\Gamma}_{_{\boldsymbol{e}1}}\right)\mathbf{M}_{_{\gamma}}\right\}$\ \ for several values of $\mu$, the following results were obtained: 
 $\boldsymbol{\eta}_{_{\mathbf{P}}}^{\infty}(\mathbf{G}_{_{a\boldsymbol{v}}})\geq 8.5866$\ \ and\ \
 $\boldsymbol{\eta}_{_{\mathbf{R}}}^{\infty}(\mathbf{G}_{_{a\boldsymbol{v}}})\leq 0.15$ -- in words, at some points in the disturbance set
 \ \ $\mathcal{S}_{_{\bar{\boldsymbol{z}}}}$,\ $\mathbf{G}_{_{av}}$\ \ diminishes the squared estimation error attained by 
 $\mathbf{G}_{_{\mathrm{o}}}$  by at least $8.6$ (in the range of $(0,\ 18.3903)$) and brings it down to $15\%$ of its value.
 
 In addition, Monte Carlo experiments involving\ \ $\bar{\boldsymbol{y}} \in \bar{\mathcal{S}}_{_{\boldsymbol{y}}}$\ \ and\ \ 
 $\bar{\boldsymbol{v}}\in \bar{\mathcal{S}}_{\boldsymbol{v}}$ defined by FIRs of prescribed length (say,\ $L$) were carried out to estimate the ``relative size''
 (Lebesgue measure) of the set of signal pairs on which $\mathbf{G}_{_{a \boldsymbol{v}}}$ yields a small squared estimation error than
 $\mathbf{G}_{_{\mathbf{M}}}$. Such an estimate is obtained from a pseudo-random sample of $N_{_{s}}=30\ 000$\ \ pairs\ \
 $\bar{\boldsymbol{z}}_{_{k}}^{^{\mathrm{T}}}=\left(\bar{\boldsymbol{y}_{_{k}}}^{^{\mathrm{T}}}, \bar{\boldsymbol{v}}_{_{k}}^{^{\mathrm{T}}}\right)$\ \ 
 computing the relative frequency
$\frac{1}{N_{_{s}}}\sum_{i=1}^{N_{_{s}}}i_{_{I}}\left(\bar{\boldsymbol{z}}_{_{k}}; \mathbf{G}_{_{\boldsymbol{\eta}}}\right)$,\ \ where\
  $i_{_{I}}\left(\bar{\boldsymbol{z}}_{_{k}}; \mathbf{G}_{_{\boldsymbol{\eta}}}\right)=1$,\ \ if\ \
  $\|\boldsymbol{e}\left(\bar{\boldsymbol{z}}_{_{k}};\mathbf{G}_{_{\boldsymbol{\eta}}}\right)\|_{_{2}}
  <\left\|\boldsymbol{e}\left(\bar{\boldsymbol{z}}_{_{k}}; \mathbf{G}_{_{\mathbf{M}}}\right)\right\|_{_{2}}$\ \ and\ \
  $i_{_{I}}\left(\bar{\boldsymbol{z}}_{_{k}}; \mathbf{G}_{_{\boldsymbol{\eta}}}\right)=0$\ \  otherwise. The sample\  $\{\bar{\boldsymbol{z}}_{_{k}}:k=1,\mathellipsis,N_{_{\mathcal{S}}}\}$, in turn, is obtained from\ $N_{_{\mathcal{S}}}$\ independent samples of FIRs of length\ $L$,\ $\boldsymbol{F}_{_{\boldsymbol{\alpha}}}=(\boldsymbol{F}_{_{\boldsymbol{\alpha}1}},\mathellipsis, \boldsymbol{F}_{_{\boldsymbol{\alpha}(L+1)}})$,\ $\boldsymbol{\alpha} = \boldsymbol{y},\boldsymbol{v}$,\  with\ $\boldsymbol{F}_{_{\boldsymbol{\alpha}}}$\  uniformly distributed on\ $\{\boldsymbol{F}\in\mathbb{R}^{L+1}:\|\boldsymbol{F}\|_{_{E}}^{^{2}}\leq \gamma_{_{\boldsymbol{\alpha}}}^{^{2}}\}$.    
  
  The relative frequency obtained in an experiment with FIRs of length $30$ was $0.7114$ so that with the sample size equal to $30\ 000$, a confidence
  interval for the desired probability was obtained with lower limit equal to $0.70$ and confidence level equal to $0.99$ (cf. Subsection \ref{subsec:6.1}). In other similar
  experiments, results were obtained to the effect that the corresponding probabilities grow with the length of the FIRs involved.
  
  Summing up, this simple example illustrates the possibility of improving on\ \ $\mathbf{G}_{_{\mathbf{M}}}$\ \ over a ``large portion'' of the disturbance
  set, with significant absolute and relative decreases in the squared estimation error at some points of the set, if a relatively small increase is
  allowed in the worst-case performance index over its achievable minimum. Similar numerical results were obtained in the case of robust\ $\mathcal{H}_{\infty}$\ estimation.

\section{Concluding Remarks} 
In this paper, three basic linear estimation problems involving set-theoretical uncertainty
were revisited with the major aim of designing estimators which may be viewed as alterna-
tives to minimax estimators. The problems addressed were robust\ $\mathcal{H}_{2}$\ and\ $\mathcal{H}_{\infty}$ estimation in the face of 
non-parametric ``channel- model'' uncertainty ($\mathcal{H}_{2}$\ and\ $\mathcal{H}_{\infty}$ balls of frequency-responses) and a nominal
Hinf problem (in this case, set-theoretical uncertainty pertains to\ $\mathcal{H}_{2}$ balls of ``information'' and noise signals). To 
provide trade-offs between worst-case and ``pointwise'' performance over the uncertainty set, in each case, average criteria on 
$\mathcal{H}_{2}$ balls were derived as limits of averages over sets of FIRs of a given length (say, $L$) as\ $L$\ grows unbounded. Linear
estimation design problems were then formulated as minimization of an average cost under the constraint that worst-case performance of any 
admissible estimator  does not exceed a prescribed value. The corresponding minimax and average cost/worst-case constraint problems were all
recast as SDPs. A brief discussion was presented on how to compare such estimators considering the ``size'' of the part of the uncertainty set
on which a ``constrained-average'' estimator improves on the corresponding minimax estimator, as well as on how much absolute or relative
improvement is brought about by the former at some points of the uncertainty set. The SDPs involved were solved in the case of simple examples
and the numerical results obtained indicate the potential of this approach to provide attractive alternatives to minimax estimators.

\newpage
\section{References}\label{sec:7}

\begin{itemize} \itemsep0pt

\item[{[1]}] H. V. Poor, ``On robust Wiener filtering," \emph{IEEE Trans. Automat. Contr.}, vol. AC-25, pp. 521-526, June 1980.
\item[{[2]}] S. Verd\'{u} and H. V. Poor, ``On minimax robustness: A general approach and applications", \emph{IEEE Trans. Inform. Theory}, vol. IT-30, pp. 328-340, Mar. 1984.
\item[{[3]}] S. A. Kassam and H. V. Poor, ``Robust techniques for signal processing: A survey", \emph{Proc. IEEE}, vol. 73, pp. 433-481, Mar. 1985.
\item[{[4]}] G. V. Moustakides  \&  S.A. Kassam,  ``Minimax equalization for random signals", \emph{IEEE Trans. Commun.}, vol. 33, no. 8, pp. 820-825, 1985.
\item[{[5]}] J. C. Geromel, ``Optimal linear filtering under parameter uncertainty'', \emph{IEEE Trans.  Signal Processing},  vol. 47, no. 1, pp. 168-175, 1999.
\item[{[6]}] J. C. Geromel, J. Bernussou, G. Garcia,  \& M. C. de Oliveira, `` $H_2$ and $/H_\infty$ robust filtering of discrete-time linear systems", \emph{SIAM Journal on Control Optimization}, vol. 35, no. 5, pp. 1353-1368, 2000.
\item[{[7]}] A. H. Sayed,  ``A framework for state-space estimation with uncertain models", \emph{IEEE Trans. Aut. Control}, vol. 46,  no. 7, pp. 998-1013, 2001.
\item[{[8]}] Y. Guo \& B. C. Levy,  ``Robust MSE equalizer design for MIMO communication systems in the presence of model uncertainties", \emph{IEEE Trans. Signal Process.}, vol. 54, no. 5, pp. 1840-1852, 2006.
\item[{[9]}] V. A. Ugrinovski \& I. R. Petersen. ``Robust ISI equalization for uncertain channels via minimax optimal filtering", \emph{Int. Journal. Adapt. Cont. and Signal Process.}, vol. 20, no. 3, pp. 99-122, 2006.
\item[{[10]}] N. Vucic \& H. Boche, ``Robust minimax equalization of imperfectly-known, frequency-selective MIMO channels", \emph{Proc. of the ACSSC (Asilomar Conf. on Signals, Systems and Computers)}, pp. 1611-1615, 2007.
\item[{[11]}] M. D. Nisar  \&  W. Utschick,  ``Minimax robust a priori information aware channel equalization", \emph{IEEE Trans. Signal Process.}, vol. 59,  no. 4, pp. 1734-1740, 2011.
\item[{[12]}] V. A. Yakubovich, ``Nonconvex optimization problems: the infinite-horizon, linear-quadratic control problem with quadratic contraints'', \emph{Syst. Control Lett.}, vol. 19, pp. 13-22, 1992.
\item[{[13]}] T. Basar, ``Optimum performance levels for minimax filters, predictors and smoothers", \emph{Syst. Contr. Lett.}, vol. 16, no. 5, pp. 309-317, 1991.
\item[{[14]}] Y. Eldar and N. Mehrav,  ``A competitive minimax aproach to robust estimation of random parameters", \emph{IEEE Trans. Signal Process.}, vol. 52,  no. 7, pp. 1931-1946, 2004.
\pagebreak
\item[{[15]}] Y. C. Eldar, A. Ben-Tal \& A. Nemirovski, ``Linear minimax regret estimation of deterministic parameters with bounded data uncertainties'',  \emph{IEEE Trans. Signal Process.}, vol. 52, No. 8, pp. 2177-2188, August 2004.
\item[{[16]}] S. S. Kozat  \&  A. T. Erdogan,  ``Competitive linear estimation under model uncertainties", \emph{IEEE Trans. Signal Process.}, vol. 58,  no. 4, pp. 2388-2393, 2010.
\item[{[17]}] G. O. Corr\^{e}a \& A. Talavera, ``Competitive robust estimation for uncertain linear dynamic models'', \emph{IEEE Trans. Signal Process.}, vol. 65, no. 18, pp. 4847- 4861, 2017. 
\item[{[18]}] G. O. Corr\^ea , ``Robust MIMO equalization for non-parametric channel model uncertainty", IEEE Trans. Signal Process., vol. 62, no. 6, pp. 1335-1347, 2014. 
\item[{[19]}] M. J. Grimble \& A. E. Sayed, ``Solution of the $H_{\infty}$ optimal linear filtering problem for discrete-time systems'', \emph{IEEE Trans. Acoust., Speech, Signal Process.}, vol. 38, pp. 1092-1104, July 1990.
\item[{[20]}] K. M. Nagpal \& P. P. Khargonekar, ``Filtering and smoothing in an $H_{\infty}$ setting'', \emph{IEEE Trans. Automat. Contr.}, vol. 36, pp. 152-166, 1991.
\item[{[21]}] U. Shaked, ``$H_{\infty}$ minimum error state estimation of linear stationary processes'', \emph{IEEE Trans. Automat. Contr.}, vol. 35, pp. 554-558, 1990.
\item[{[22]}] Y. Theodor, U. Shaked \& C. E. de Souza, ``A game theory approach to robust discrete-time $H_{\infty}$-estimation'', \emph{IEEE Trans. Signal Process.}, vol. 42, pp. 1486-1495, June 1994.
\item[{[23]}] L. Xie, C. E. de Souza \& M. Fu, ``$H_{\infty}$ estimation for linear discrete-time uncertain systems'', Int. J. Robust Nonlinear Contr., vol. 1, pp. 111-123, 1991.
\item[{[24]}] U. Shaked \& Y. Theodor, ``$H_{\infty}$-optimal estimation: a tutorial'', \emph{Proceedings 31st IEEE Conference on Decision and Control}, pp. 2278-2286, Tucson, Arizona, Dec. 1992.
\item[{[25]}] H. Li, M. Fu.,`` A linear matrix inequality approach to robust $H_{\infty}$ filtering'', \emph{IEEE Trans. on Signal Process.}, vol. 45, no. 9, pp. 2338-2349, 1997.
\item[{[26]}] Y. Theodor \& U. Shaked  ``Robust discrete-time, minimum-variance filtering", \emph{IEEE Trans. Signal Process.}, vol. 44, no. 2, pp. 181-189, 1996.
\item[{[27]}] P. Gahinet \& P. Apkarian, ``A linear matrix inequality approach to $H_{\infty}$ control'', \emph{Int. J. Robust Nonlinear Contr.}, vol. 4, pp. 421-428, 1994.
\item[{[28]}] R. Tempo \& H. Ishii, ``Monte Carlo and Las Vegas randomized algorithms for systems and control'', \emph{European Journal of Control}, vol. 13, no. 2-3, pp. 189-203, 2007.
\item[{[29]}]  I. R. Petersen \& D. C. McFarlane, ``Optimal guaranteed cost filtering for uncertain discrete-time linear systems'', Int. J. Robust Nonlinear Control, vol. 6, no. 4, pp. 267-280, 1996.
\item[{[30]}]  L. Xie, Y. C. Soh \& C. E. de Souza, ``Robust Kalman Filtering for uncertain discrete-time systems'', IEEE Trans. Autom. Control, vol. 39, no. 6, pp. 1310-1314, Jun. 1994.
\item[{[31]}] C. E. de Souza \& A. Trofino. ``A linear matrix inequality approach to the design of robust $\mathcal{H}_{2}$ filters''.
In, ``Advances in Linear Matrix Inequality Methods in Control'' (Advances in Design and Control), L. El Ghaoui  \& S. Niculescu (eds.),
SIAM, Philadelphia, PA, 2000.
\end{itemize}

\newpage
\section{Appendix}

\subsection*{Notation for Problem Data}\label{subsec:A}

 For a string of symbols $\boldsymbol{s}$, a state-space realization\
 $(\mathbf{A}_{_{\boldsymbol{s}}},\mathbf{B}_{_{\boldsymbol{s}}},\mathbf{C}_{_{\boldsymbol{s}}},\mathbf{D}_{_{\boldsymbol{s}}})$\ is
 denoted by\ $\boldsymbol{\Sigma}_{_{\boldsymbol{s}}}$. 
 
\subsubsection*{Estimation Set-up and $\mathcal{H}_{2}$ Problem}\label{subsubsec:A.1}
$\mathbf{H}_{_{\mathbf{I}}}$, $\boldsymbol{\Gamma}_{_{\boldsymbol{y}}}=\boldsymbol{\phi}_{_{\boldsymbol{y}}}\boldsymbol{\phi}_{_{\boldsymbol{y}}}^{*}$, $\boldsymbol{\Gamma}_{_{\boldsymbol{v}}}=\boldsymbol{\phi}_{_{\boldsymbol{v}}}\boldsymbol{\phi}_{_{\boldsymbol{v}}}^{*}$ (Subsection \ref{subsec:2.1}).\\
$\mathbf{H}_{_{\boldsymbol{0}}}:$ nominal ``channel'' frequency-response with minimal realization (mr, for short)\
$\boldsymbol{\Sigma}_{_{\mathbf{H}\boldsymbol{0}}}$.\\
$W$: weighting function for $\mathcal{H}_{2}-$uncertainty (eq. (\ref{eq:01a})).

\subsubsection*{Nominal $\mathcal{H}_{\infty}$ Problem}\label{subsubsec:A.2}

$W_{_{\boldsymbol{\alpha}}}$, $\boldsymbol{\alpha}=\boldsymbol{y}, \boldsymbol{v}$: weighting function for the $\mathcal{H}_{2}-$signal ball\ $\mathcal{S}_{_{\boldsymbol{\alpha}}}$\ with mr\ $\boldsymbol{\Sigma}_{_{W\boldsymbol{\alpha}}}$.\\
$\mathbf{H}_{_{\mathbf{I}\boldsymbol{y}}}\triangleq[\mathbf{H}_{_{\mathbf{I}}}W_{_{\boldsymbol{y}}}^{-1}\ \vdots\ \boldsymbol{0}_{m_{_{\boldsymbol{e}}}\times m_{_{\boldsymbol{v}}}}]$\ and\ $\mathbf{H}_{_{\mathrm{o}\boldsymbol{z}}}\triangleq[\mathbf{H}_{_{\mathrm{o}}}W_{_{\boldsymbol{y}}}^{-1}\ \vdots\ W_{_{\boldsymbol{v}}}^{-1}]$\ with mrs\ $\boldsymbol{\Sigma}_{_{\mathbf{I}\boldsymbol{y}}}$\ and\ $\boldsymbol{\Sigma}_{_{\mathrm{o}\boldsymbol{z}}}$.
\subsubsection*{Robust $\mathcal{H}_{\infty}$ Problem}\label{subsubsec:A.3}
$\mathbf{H}_{_{\mathbf{I}a}}=[\mathbf{H}_{_{\mathbf{I}\boldsymbol{y}}}\ \vdots\ \boldsymbol{0}_{m_{_{\boldsymbol{e}}}\times m_{_{\boldsymbol{v}}}}]$\ and\ $\mathbf{H}_{_{\mathrm{o}a}}=[\mathbf{H}_{_{\mathrm{o}\boldsymbol{z}}}\ \vdots\ \mathbf{I}_{m_{\boldsymbol{v}}}]$\ with mrs\ $\boldsymbol{\Sigma}_{_{\mathbf{H}\mathbf{I}a}}$\ and\ $\boldsymbol{\Sigma}_{_{\mathrm{o}a}}$.\\
$W_{_{\mathbf{H}}}:$ weighting function for $\mathcal{H}_{\infty}-$uncertainty.\\
$W_{_{\mathbf{H}\boldsymbol{y}}}\triangleq(W_{_{\boldsymbol{y}}}W_{_{\mathbf{H}}})^{-1}$\ with mr\ $\boldsymbol{\Sigma}_{_{W\mathbf{H}\boldsymbol{y}}}$, $W_{_{\mathbf{H}\boldsymbol{y}}}^{^{a}}=W_{_{\mathbf{H}\boldsymbol{y}}}[\mathbf{I}_{m_{_{\boldsymbol{y}}}}\ \vdots\ \boldsymbol{0}_{m_{_{\boldsymbol{y}}}\times m_{_{\boldsymbol{v}}}}\ \vdots\ \boldsymbol{0}_{m_{_{\boldsymbol{y}}}\times m_{_{\boldsymbol{v}}}}]$.

\subsubsection*{Matrix definitions for Proposition \ref{prop:11} and \ref{prop:13}}

Given\ $\boldsymbol{\Sigma}_{_{\mathbf{I}\boldsymbol{y}}}$, $\boldsymbol{\Sigma}_{_{\mathrm{o}\boldsymbol{z}}}$\ as
in as in the definitions for the \textbf{Nominal $\mathcal{H}_{\infty}$ Problem}  above above,\\ $\mathbf{A}_{_{\mathbf{I}\boldsymbol{y}}}\in\mathbb{R}^{n_{_{\mathbf{I}\boldsymbol{y}}}\times
n_{_{\mathbf{I}\boldsymbol{y}}}}$, $\mathbf{A}_{_{\mathrm{o}\boldsymbol{z}}}\in\mathbb{R}^{n_{_{\mathrm{o}\boldsymbol{z}}}\times
n_{_{\mathrm{o}\boldsymbol{z}}}}$, $n_{_{\mathbf{G}}}=n_{_{\mathbf{I}\boldsymbol{y}}}+n_{_{\mathrm{o}\boldsymbol{z}}}$, 
$\mathbf{A}_{_{\mathrm{o}}}=\operatorname{diag}(\mathbf{A}_{_{\mathbf{I}\boldsymbol{y}}}, \mathbf{A}_{_{\mathrm{o}\boldsymbol{z}}},
\boldsymbol{0}_{n_{_{\mathbf{G}}}\times n_{_{\mathbf{G}}}})$.\\
$\mathbf{B}_{_{\mathrm{o}}}^{^{\mathrm{T}}}=[\mathbf{B}_{_{\mathbf{I}\boldsymbol{y}}}^{^{\mathrm{T}}}\ \vdots\
\mathbf{B}_{_{\mathrm{o}\boldsymbol{z}}}^{^{\mathrm{T}}}\ \vdots\ \boldsymbol{0}_{m_{_{\boldsymbol{z}}}\times n_{_{\mathbf{G}}}}]$, 
$m_{_{\boldsymbol{z}}}=m_{_{\boldsymbol{y}}}+m_{_{\boldsymbol{v}}}$.\\
$\mathbf{C}=[\mathbf{C}_{_{\mathbf{I}\boldsymbol{y}}}\ \vdots\
\boldsymbol{0}_{m_{_{\boldsymbol{e}}}\times(n_{_{\mathrm{o}\boldsymbol{z}}}+n_{_{\mathbf{G}}})}]$, 
$\mathbf{C}_{_{a\boldsymbol{z}}}=[\boldsymbol{0}_{m_{_{\boldsymbol{v}}}\times n_{_{\mathbf{I}\boldsymbol{y}}}}\ \vdots\
\mathbf{C}_{_{\mathrm{o}\boldsymbol{z}}}]$, 
$\widehat{\mathbf{C}}_{_{\mathrm{o}}}=[\mathbf{C}_{_{\mathbf{I}\boldsymbol{y}}}\ \vdots\
\boldsymbol{0}_{m_{_{\boldsymbol{e}}}\times n_{_{\mathrm{o}\boldsymbol{z}}}}]$.\\
$\mathbf{D}_{_{\mathbf{I}\boldsymbol{y}1}}=\mathbf{D}_{_{\mathbf{I}\boldsymbol{y}}}([\mathbf{I}_{m_{_{\boldsymbol{y}}}}\ \vdots\
\boldsymbol{0}_{m_{_{\boldsymbol{y}}}\times m_{_{\boldsymbol{v}}}}])^{^{\mathrm{T}}}$.\\ 
$\mathbf{E}_{_{b}}\triangleq-\mathbf{D}_{_{W\boldsymbol{v}}}[\mathbf{C}_{_{a\boldsymbol{z}}}\ \vdots\
\mathbf{D}_{_{\mathbf{H}\mathrm{o}}}\mathbf{D}_{_{W\boldsymbol{y}}}^{^{-1}}]$,\ $\mathbf{E}_{_{b\mathrm{o}}}\triangleq[\widehat{\mathbf{C}}_{_{\mathrm{o}}}\ \vdots\ \mathbf{D}_{_{\mathbf{I}\boldsymbol{y}1}}]$.\\ 
$\boldsymbol{\psi}_{_{b}}\triangleq[\mathbf{A}_{_{a\boldsymbol{z}}}\ \vdots\ \mathbf{B}_{_{a\boldsymbol{z}1}}]
+\mathbf{B}_{_{a\boldsymbol{z}2}}\mathbf{E}_{_{b}}$,
$\mathbf{A}_{_{a\boldsymbol{z}}}=\operatorname{diag}(\mathbf{A}_{_{\mathbf{I}\boldsymbol{y}}}, \mathbf{A}_{_{\mathrm{o}\boldsymbol{z}}})$,
$\mathbf{B}_{_{a\boldsymbol{z}}}=[\mathbf{B}_{_{a\boldsymbol{z}1}}\ \vdots\ \mathbf{B}_{_{a\boldsymbol{z}2}}]
\triangleq[\mathbf{B}_{_{\mathbf{I}\boldsymbol{y}}}^{^{\mathrm{T}}}\ \vdots\ \mathbf{B}_{_{\mathrm{o}\boldsymbol{z}}}^{^{\mathrm{T}}}]^{^{\mathrm{T}}}$,\\ $\mathbf{B}_{_{a\boldsymbol{z}1}}=\mathbf{B}_{_{a\boldsymbol{z}}}([\mathbf{I}_{m_{_{\boldsymbol{y}}}}\ \vdots\ \boldsymbol{0}_{m_{_{\boldsymbol{y}}}\times m_{\boldsymbol{v}}}])^{^{\mathrm{T}}}$.

\subsubsection*{Matrix Definitions for Subsection \ref{subsec:4.3}}

Given\ $\boldsymbol{\Sigma}_{_{\mathrm{o}a}}$, $\boldsymbol{\Sigma}_{_{\mathbf{H}\mathbf{I}a}}$\ and\ 
$\boldsymbol{\Sigma}_{_{W\mathbf{H}\boldsymbol{y}}}$ as in the definitions for the \textbf{Robust $\mathcal{H}_{\infty}$ Problem}  above,\break
$\mathbf{A}_{_{a1}}=\operatorname{diag}(\mathbf{A}_{_{W\mathbf{I}\boldsymbol{y}}}, \mathbf{A}_{_{\mathbf{H}\mathbf{I}a}}, \mathbf{A}_{_{\mathrm{o}a}})$, $n_{_{\mathbf{G}}}^{a}$ denotes the dimension of $\mathbf{A}_{_{a1}}$.\\
$\mathbf{B}_{_{a}}^{^{\mathrm{T}}}=[\mathbf{B}_{_{a1}}^{^{\mathrm{T}}}\ \vdots\ \boldsymbol{0}_{(m_{_{\boldsymbol{y}}}+2m_{_{\boldsymbol{v}}})\times n_{_{\mathbf{G}}}^{a}}]$, $\mathbf{B}_{_{a1}}^{^{\mathrm{T}}}=[\mathbf{B}_{_{W\mathbf{H}\boldsymbol{y}}}^{^{\mathrm{T}}}\ \vdots\ \mathbf{B}_{_{\mathbf{H}\mathbf{I}a}}^{^{\mathrm{T}}}\ \vdots\ \mathbf{B}_{_{\mathrm{o}a}}^{^{\mathrm{T}}}]$,\ $\mathbf{B}_{_{W\mathbf{H}\boldsymbol{y}}}^{a}=\mathbf{B}_{_{W\mathbf{H}\boldsymbol{y}}}[\mathbf{I}_{m_{_{\boldsymbol{y}}}}\ \vdots\ \boldsymbol{0}_{m_{_{\boldsymbol{y}}}\times 2m_{_{\boldsymbol{v}}}}]$,\\ $\mathbf{B}_{_{a1}}^{^{\boldsymbol{z}}}=\mathbf{B}_{_{a1}}([\mathbf{I}_{m_{\boldsymbol{z}}}\ \vdots\ \boldsymbol{0}_{m_{_{\boldsymbol{z}}}\times m_{_{\boldsymbol{v}}}}])^{^{\mathrm{T}}}$.\\
$\mathbf{C}_{_{a}}=\operatorname{diag}(\boldsymbol{\sigma}_{_{\boldsymbol{w}}}^{^{1/2}}\mathbf{I}_{m_{\boldsymbol{y}}}, \mathbf{I}_{m_{_{\boldsymbol{e}}}})[\check{\mathbf{C}}_{_{Wa}}\ \vdots\ \boldsymbol{0}_{(m_{_{\boldsymbol{y}}}+m_{_{\boldsymbol{e}}})\times(n_{_{\mathrm{o}a}}+n_{_{\mathbf{G}}}^{a})}]$, $\check{\mathbf{C}}_{_{Wa}}=\operatorname{diag}(\gamma_{_{\mathbf{H}}}\mathbf{C}_{_{W\mathbf{H}\boldsymbol{y}}},\mathbf{C}_{_{\mathbf{H}\mathbf{I}a}})$,\\
$\mathbf{C}_{_{a1}}^{^{\mathbf{I}}}=[\boldsymbol{0}_{m_{_{\boldsymbol{e}}}\times n_{_{W\mathbf{H}\boldsymbol{y}}}}\ \vdots\ \mathbf{C}_{_{\mathbf{H}\mathbf{I}a}}\ \vdots\ \boldsymbol{0}_{m_{_{\boldsymbol{e}}}\times n_{_{\mathrm{o}a}}}]$.\\
$\mathbf{D}_{_{a}}^{^{\mathrm{T}}}=[\boldsymbol{\sigma}_{_{\boldsymbol{w}}}^{^{1/2}}\gamma_{_{\mathbf{H}}}(\mathbf{D}_{_{W\mathbf{H}\boldsymbol{y}}}^{a})^{^{\mathrm{T}}}\ \vdots\ \mathbf{D}_{_{\mathbf{H}\mathbf{I}a}}^{^{\mathrm{T}}}]$,\ $\mathbf{D}_{_{W\mathbf{H}\boldsymbol{y}}}^{a}=\mathbf{D}_{_{W\mathbf{H}\boldsymbol{y}}}[\mathbf{I}_{m_{_{\boldsymbol{y}}}}\ \vdots\ \boldsymbol{0}_{m_{_{\boldsymbol{y}}}\times 2m_{_{\boldsymbol{v}}}}]$,\ $\mathbf{D}_{_{a\boldsymbol{z}}}^{^{\mathbf{I}}}=[\mathbf{D}_{_{\mathbf{H}\mathbf{I}}}\mathbf{D}_{_{W\boldsymbol{y}}}^{^{-1}}\ \vdots\ \boldsymbol{0}_{m_{_{\boldsymbol{e}}}\times m_{_{\boldsymbol{v}}}}]$,\\ $\mathbf{E}_{_{\mathrm{o}}}=[\mathbf{C}_{_{a1}}^{^{\mathbf{I}}}\ \vdots\mathbf{D}_{_{a1}}^{^{\mathbf{I}}}]$, and\ $\mathbf{E}_{_{\boldsymbol{s}}}=[\mathbf{A}_{_{a1}}\ \vdots\ \mathbf{B}_{_{a1}}^{^{\boldsymbol{z}}}]$.

\subsection*{MSE and the $\mathcal{H}_{2}$ Norm of the Error System}

\vspace*{3mm}

The connection between the mean-squared
estimation error and the $\mathcal{H}_{2}-$norm of the ``error system"
frequency-response is now briefly reviewed for both finite time and as time goes to infinity. To this effect, note first that the assumptions on\ 
$\boldsymbol{y}$\ and\ $\boldsymbol{v}$\ amount to taking them to be filtered 
``versions" of independent, zero-mean sequences\ $\check{\boldsymbol{y}}$\ and\
$\check{\boldsymbol{v}}$\ such that\ $\forall\ k, \ell$,
$\mathrm{E}[\check{\boldsymbol{y}}(k)\check{\boldsymbol{y}}(\ell)^{^{\mathrm{T}}}]
=\delta_{_{k\ell}}\mathbf{I}_{m_{\boldsymbol{y}}}$, $\mathrm{E}[\check{\boldsymbol{v}}(k)\check{\boldsymbol{v}}(\ell)^{^{\mathrm{T}}}]
=\delta_{_{k\ell}}\mathbf{I}_{m_{\boldsymbol{v}}}$, where 
$\delta_{_{k\ell}}=1$\ if\ $k\neq\ell$, $\delta_{_{kk}}=1$.

Thus, denoting by\ $\boldsymbol{F}_{_{\mathbf{e}}}$\ the frequency-response from\ 
$\check{\boldsymbol{z}}=\begin{bmatrix}\check{\boldsymbol{y}}^{^{\mathrm{T}}}&\vdots&\check{\boldsymbol{v}}^{^{\mathrm{T}}} \end{bmatrix}$\ to\ $\mathbf{e}$, \emph{i.e.}, 
$\mathbf{e}=(\mathbf{H}_{_{\mathbf{I}}}-\mathbf{G}\mathbf{H})
\boldsymbol{\phi}_{_{\boldsymbol{y}}}\check{\boldsymbol{y}}
+\mathbf{G}\boldsymbol{\phi}_{_{\boldsymbol{v}}}\check{\boldsymbol{v}}=
\boldsymbol{F}_{_{\mathbf{e}}}\check{\boldsymbol{z}}$\ (by a slight abuse of notation
both signal sequences and their Fourier transforms are denoted by the same
letter), the response from zero initial conditions would be given by\ 
$\mathbf{e}(K)=\displaystyle\sum_{k=0}^{K}\boldsymbol{F}_{_{\boldsymbol{e}K}}
\check{\boldsymbol{z}}(K-k)$, where\ $\boldsymbol{F}_{_{\boldsymbol{e}}}(e^{^{j\phi}})
=\displaystyle \sum_{k=0}^{\infty}\boldsymbol{F}_{_{\mathbf{e}k}}e^{^{-jk\phi}}$\ 
(non-zero initial conditions would fade-out with time due to asymptotic
stability) so that\ $\mathrm{E}\left\{\|\mathbf{e}(K)\|_{_{E}}^{^{2}}
\right\}=\operatorname{tr}\left\{\displaystyle\sum_{k=0}^{K}
\sum_{\ell=0}^{K}\boldsymbol{F}_{_{\mathbf{e}k}}\mathrm{E}[\check{\boldsymbol{z}}
(K-k)\check{\boldsymbol{z}}(K-\ell)^{^{\mathrm{T}}}]
\boldsymbol{F}_{_{\mathbf{e}\ell}}^{^{\mathrm{T}}}\right\}$\ \ 
$\Rightarrow$\ \ $\mathrm{E}\left\{\|\mathbf{e}(K)\|_{_{E}}^{^{2}}
\right\}=\operatorname{tr}\left\{\displaystyle\sum_{\ell=0}^{K}
\boldsymbol{F}_{_{\mathbf{e}k}}\boldsymbol{F}_{_{\mathbf{e}k}}^{^{\mathrm{T}}}\right\}$\ \ 
$\Rightarrow$\ \ $\displaystyle\lim_{K\to\infty}\|\mathbf{e}(K)\|_{_{E}}^{^{2}}=\|\boldsymbol{F}_{_{\mathbf{e}}}\|_{_{2}}^{^{2}}$\ and that\ $\forall\ K$, 
$\mathrm{E}\left\{\|\mathbf{e}(K)\|_{_{E}}^{^{2}}\right\}\leq
 \|\boldsymbol{F}_{_{\mathbf{e}}}\|_{_{2}}^{^{2}}$.
 
If stationarity assumptions are relaxed to the effect that\ $\forall\ k,
\ell$, $\mathrm{E}[\check{\boldsymbol{z}}(k)\check{\boldsymbol{z}}(\ell)]
=\delta_{_{k\ell}}Q_{_{\boldsymbol{z}}}(k)$\ and\ $\forall\ k$, $ 
\lambda_{_{\max}}(Q_{_{\boldsymbol{z}}}(k))\leq \mu_{_{\boldsymbol{z}}}$, 
then\ $\forall\ K$, $\mathrm{E}\left\{\|\mathbf{e}(K)\|_{_{E}}^{^{2}}
\right\}\leq \mu_{_{\boldsymbol{z}}}\|\boldsymbol{F}_{_{\mathbf{e}}}\|_{_{2}}^{^{2}}$ --
\emph{i.e.}, even in this case\ $\|\boldsymbol{F}_{_{\mathbf{e}}}\|_{_{2}}^{^{2}}$\
yields a uniform upper bound on the MS estimation error.
 
To consider sample means of the squared estimation error, note that\ \
$\|\mathbf{e}(K)\|_{_{E}}^{^{2}}=\operatorname{tr}
\left\{\underline{\boldsymbol{F}}_{_{\mathbf{e}N}}\mathbf{R}_{_{K,N}}
\underline{\boldsymbol{F}}_{_{\mathbf{e}N}}^{^{\mathrm{T}}}\right\}$,\ where\ 
$\underline{\boldsymbol{F}}_{_{\mathbf{e}N}}=\begin{bmatrix}
{\boldsymbol{F}_{\mathbf{e}0}}& \cdots& \boldsymbol{F}_{_{\mathbf{e}N}}\end{bmatrix}$\ and\ $\mathbf{R}_{_{K, N}}\triangleq \operatorname{diag}\left(\mathbf{R}_{_{K}},\boldsymbol{0}_{_{N-K,N-K}}\right)$,\
$\mathbf{R}_{_{K}}\triangleq \begin{bmatrix}
                             \check{\boldsymbol{z}}(K)^{^{T}}\\
                             \vdots\\
                             \check{\boldsymbol{z}}(0)\\                             
                             \end{bmatrix}
                             \begin{bmatrix}\check{\boldsymbol{z}}            
                             (K)^{^{\mathrm{T}}}&\cdots&
                             \check{\boldsymbol{z}}(0)^{^{\mathrm{T}}} 
                             \end{bmatrix}$.
\noindent
Thus, 
\begin{equation}\label{remark:eq01}
\frac{1}{N+1}\sum_{K=0}^{N}\|\mathbf{e}(K)\|_{_{E}}^{^{2}}
=\operatorname{tr}\left\{\underline{\boldsymbol{F}}_{_{\mathbf{e}N}}\bar{\mathbf{R}}_{_{N}}
\underline{\boldsymbol{F}}_{_{\mathbf{e}N}}^{^{\mathrm{T}}}\right\},\tag{R1}
\end{equation} 
where\ $\bar{\mathbf{R}}_{_{N}}\triangleq \dfrac{1}{N+1}\displaystyle\sum_{K=0}^{N}
\mathbf{R}_{_{K,N}}$\ is a symmetric matrix of sample correlations with $(p, p+q)$ 
blocks qiven by\ $\forall\ p=0,1,\mathellipsis, N$, 
$\forall\ q=0,1, \mathellipsis, N-p$,\ $\left\{\bar{\mathbf{R}}_{_{N}}
\right\}_{_{p, p+q}}=\dfrac{1}{N+1}\displaystyle\sum_{K=q}^{N-p}
\check{\boldsymbol{z}}(K)
\check{\boldsymbol{z}}(K-q)^{^{\mathrm{T}}}$.

It follows from (\ref{remark:eq01}) that
$$\forall\ N,\ \ \frac{1}{N+1}\sum_{k=0}^{N}\|\mathbf{e}(K)\|_{_{E}}^{^{2}}
\leq \lambda_{_{\max}}(\bar{\mathbf{R}}_{_{N}})\|\boldsymbol{F}_{_{\mathbf{e}}}\|_{_{2}}^{^{2}}$$
so that\ $\|\boldsymbol{F}_{_{\mathbf{e}}}\|_{_{2}}^{^{2}}$\ also yields a uniform upper
bound in the sample-averages of the squared estimation error
which is based on correlation estimates.

\subsection*{MIMO Example Data}

\subsubsection*{Data for the First MIMO Example:}

   $\mathbf{A}_{_{\mathbf{H}\boldsymbol{0}a}}=\left[\begin{smallmatrix}
                                           0&0&0\\
                                           1&0&0\\
                                           0&1&0\\
                                           \end{smallmatrix}\right]$,\ 
                                           $\mathbf{A}_{_{\mathbf{H}\boldsymbol{0}}}=\mathbf{I}_{_{2}}\otimes
  \mathbf{A}_{_{\mathbf{H}\boldsymbol{0}a}}$,\ 
  $\mathbf{b}_{_{\mathbf{H}\boldsymbol{0}a}}=\left[\begin{smallmatrix} 
        1\\
        0\\
        0\\
 \end{smallmatrix}\right]$,\ $\mathbf{B}_{_{\mathbf{H}\boldsymbol{0}}}=\mathbf{I}_{_{2}}\otimes \mathbf{b}_{_{\mathbf{H}\boldsymbol{0}a}}$,\\
 $\mathbf{C}_{_{\mathbf{H}a1}} = \begin{bmatrix}-1.3963 &0.9638& -0.8713& 0.5593& -0.1389& 0.0815\end{bmatrix}$,\\ 
   $\mathbf{C}_{_{\mathbf{H}a2}} = \begin{bmatrix}-0.1770& -0.5893& 0.1982& 0.1795& -0.1013& 0.0543\end{bmatrix}$, 
   $\mathbf{C}_{_{\mathbf{H}\boldsymbol{0}}} = [\mathbf{C}_{_{\mathbf{H}a1}};\mathbf{C}_{_{\mathbf{H}a2}}]$,\\
   or, equivalently, the non-zero Markov parameters are\\
$\boldsymbol{F}_{_{0}} = \left[\begin{smallmatrix}
                                             1.0 & 0.8540\\
                                             0.0 & 1.0\\ 
                                            \end{smallmatrix}\right]$,\ $\boldsymbol{F}_{_{1}} = \left[\begin{smallmatrix}
                                             -1.3963 & 0.5593\\
                                             -0.1770 & 0.1795\\ 
                                            \end{smallmatrix}\right]$,\
$\boldsymbol{F}_{_{2}} = \left[\begin{smallmatrix}
                                             0.9638 & -0.1389\\
                                             -0.5893 & -0.1013\\ 
                                            \end{smallmatrix}\right]$,\ $\boldsymbol{F}_{_{3}} = \left[\begin{smallmatrix}
                                             -0.8713 & 0.0815\\
                                             0.1982 & 0.0543\\ 
                                            \end{smallmatrix}\right]$.\\
The state-space realization for $\boldsymbol{\phi}_{_{\boldsymbol{y}}}$ is given by\\
$\mathbf{A}_{_{\boldsymbol{\phi}_{_{\boldsymbol{y}}}}}=\left[\begin{smallmatrix}
      0.6 & 0.0\\
      0.0 & 0.4\\      
     \end{smallmatrix}\right]$,\ $\mathbf{B}_{_{\boldsymbol{\phi}_{_{\boldsymbol{y}}}}}=\mathbf{I}_{_{2}}$,\ $\mathbf{C}_{_{\boldsymbol{\phi}_{_{\boldsymbol{y}}}}}=\left[\begin{smallmatrix}
      1.0 & 0.2\\
      0.0 & 1.3\\           
     \end{smallmatrix}\right]$,\ $\mathbf{D}_{_{\boldsymbol{\phi}_{_{\boldsymbol{y}}}}}=\left[\begin{smallmatrix}
      1.0 & 0.3\\
      0 & 0.8\\         
     \end{smallmatrix}\right]$
  
\subsubsection*{Data for the Second MIMO Example:}

$\boldsymbol{\lambda}_{_{\mathbf{A}_{_{\mathbf{H}\boldsymbol{0}}}}}=
\begin{bmatrix}0.65 & 0.43 & 0.32 & 0.27 & 0.12\end{bmatrix}$,\
$\boldsymbol{\Lambda}_{_{\mathbf{A}_{_{\mathbf{H}\boldsymbol{0}}}}} = \operatorname{diag}(\boldsymbol{\lambda}_{_{\mathbf{A}_{_{\mathbf{H}\boldsymbol{0}}}}})$,\  $\mathbf{V}_{_{\mathbf{H}\boldsymbol{0}}}
= \left[\begin{smallmatrix}
            1.0 & 1.3 & 0.7 & 1.8 & 0.8\\
            0.0 & 1.2 & 0.4 & 1.6 & 0.6\\
            0.0 & 0.0 & 1.4 & 1.1 & 1.5\\
            0.0 & 0.0 & 0.0 & 2.0 & 1.9\\
            0.0 & 0.0 & 0.0 & 0.0 & 1.7\\
            \end{smallmatrix}\right]$,\\
$\mathbf{A}_{_{\mathbf{H}\boldsymbol{0}}} =
   \mathbf{V}_{_{\mathbf{H}\boldsymbol{0}}}*\boldsymbol{\Lambda}_{_{\mathbf{A}_{_{\mathbf{H}\boldsymbol{0}}}}}*[\mathbf{V}_{_{\mathbf{H}\boldsymbol{0}}}]^{^{-1}}$,\ 
   $\mathbf{B}_{_{\mathbf{H}\boldsymbol{0}a}} = 
   \left[\begin{smallmatrix}
          1.0 & 0.7\\
          0.8 & 1.5\\
          1.3 & 1.1\\ 
          2.1 & 0.7\\
          0.9 & 1.2\\ 
         \end{smallmatrix}\right]$,\  
         $\mathbf{B}_{_{\mathbf{H}\boldsymbol{0}}} = 
         \mathbf{V}_{_{\mathbf{H}\boldsymbol{0}}}
         \mathbf{B}_{_{\mathbf{H}\boldsymbol{0}a}}$,\\
   $\mathbf{C}_{_{\mathbf{H}\boldsymbol{0}a}} =   
              \left[\begin{smallmatrix}
                 0.9638 & -0.8713 & 0.5593 & -0.4389 & 0.4015\\
                -0.5893 &  0.1982 & 0.6795 & -0.4413 & 0.5043\\
              \end{smallmatrix}\right]$,\ $\mathbf{C}_{_{\mathbf{H}\boldsymbol{0}}} = 0.45\times\mathbf{C}_{_{\mathbf{H}\boldsymbol{0}a}}[\mathbf{V}_{_{\mathbf{H}\boldsymbol{0}}}]^{^{-1}}$,\
    $\mathbf{D}_{_{\mathbf{H}\boldsymbol{0}}} = 
        \left[\begin{smallmatrix}
                  1.0 & 0.8540\\
                  0.0 & 0.8\\
              \end{smallmatrix}\right]$.\\
$\boldsymbol{\phi}_{_{\boldsymbol{v}}}$ is as in the first MIMO example.

\subsection*{Proof Outlines}

\noindent
\underline{\textbf{Proof of Proposition 3.1:}}\ To compute\ $ \nu_{_{N}}=\displaystyle\int_{\mathcal{S}_{_{\boldsymbol{\theta}}}^{^{N}}}d\boldsymbol{\theta}$ \ and\ $\nu_{_{\mathcal{J}N}}\triangleq\displaystyle\int_{\mathcal{S}_{_{\boldsymbol{\theta}}}^{^{N}}}\boldsymbol{\theta}^{^{\mathrm{T}}}P_{_{\mathcal{J}}}^{^{N}}(\mathbf{G})\boldsymbol{\theta} d\boldsymbol{\theta}$, define\break  $\widehat{\boldsymbol{\theta}}=\left(P_{_{\boldsymbol{\theta}}}^{^{N}}\right)^{^{1/2}}\boldsymbol{\theta}$\ and rely on polar coordinates to obtain\
$$
 \nu_{_{N}}=\operatorname{det}\left[\left(P_{_{\boldsymbol{\theta}}}^{^{N}}\right)^{^{-1/2}}\right]\frac{\gamma^{^{n}}}{n}\left\{\int_{0}^{\pi}[\sin(\alpha_{_{_{1}}})]^{^{n-2}}d\alpha_{_{_{1}}}\right\}\check{\prod}_{2, n-2}\ \ \ \text{and}
$$
$$\nu_{_{\mathcal{J}N}}=\left\{\sum_{i=1}^{n}[\widehat{P}_{_{\mathcal{J}}}^{^{N}}]_{_{ii}}\right\}\operatorname{det}\left[\left(P_{_{\boldsymbol{\theta}}}^{^{N}}\right)^{^{-1/2}}\right] \left(\frac{\gamma^{n+2}}{n+2}\right)\left\{(1/n)\displaystyle\int_{0}^{\pi}[\sin(\alpha_{_{_{1}}})]^{^{n-2}}d\alpha_{_{_{1}}}\right\}\check{\prod}_{2, n-2}(2\pi),$$
\noindent
where\  $\displaystyle\check{\prod}_{2,n-2}\triangleq\prod_{k=2}^{n-2}\int_{0}^{\pi}[\sin(\alpha_{_{k}})]^{^{n-1-k}}d\alpha_{_{k}}$. Thus,\ $\nu_{_{N}}^{^{-1}}\nu_{_{\mathcal{J}N}}=\displaystyle\frac{\gamma^{^{2}}}{n+2}\left\{\sum_{i=1}^{n}[\widehat{P}_{_{\mathcal{J}}}^{^{N}}(\mathbf{G})]_{_{ii}}\right\}$\ from which Proposition 3.1 follows.\hfill$\blacksquare$ 


\vspace*{3mm}
\noindent
\underline{\textbf{Proof of Proposition 3.2:}}\ It turns out that\ $P_{_{\boldsymbol{\beta}}}^{^{N}}=\mathbf{I}$ and, hence, $P_{_{\boldsymbol{\theta}}}^{^{N}}=\mathbf{I}$\ and\ $\widehat{P}_{_{\mathcal{J}}}^{^{N}}(\mathbf{G})=P_{_{\mathcal{J}}}^{^{N}}(\mathbf{G})$. As $F_{_{N}}(\mathbf{G})=(\mathbf{G}\otimes\boldsymbol{\phi}_{_{\boldsymbol{y}1}}^{^{\mathrm{T}}})(\mathbf{I}_{m_{_{\boldsymbol{v}}}}\otimes(\check{\mathbf{Y}}_{_{\mathbf{X}}}^{^{N}})^{^{\mathrm{T}}})$,\ $[\mathbf{I}_{m_{_{\boldsymbol{v}}}}\otimes(\check{\mathbf{Y}}_{_{\mathbf{X}}}^{^{N}})^{^{\mathrm{T}}}][\mathbf{I}\otimes(\check{\mathbf{Y}}_{_{\mathbf{X}}}^{^{N}})^{c}]
=\mathbf{I}_{m_{_{\boldsymbol{v}}}}\otimes[(\check{\mathbf{Y}}_{_{\mathbf{X}}}^{^{N}})^{^{\mathrm{T}}}(\check{\mathbf{Y}}_{_{\mathbf{X}}}^{^{N}})^{c}]$\
and\break $(\check{\mathbf{Y}}_{_{\mathbf{X}}}^{^{N}})^{^{\mathrm{T}}}(\check{\mathbf{Y}}_{_{\mathbf{X}}}^{^{N}})^{c}=(N+1)\mathbf{I}_{m_{_{\boldsymbol{y}}}}$, it follows that\ $\operatorname{tr}\left\{\widehat{P}_{_{\mathcal{J}}}^{^{N}}(\mathbf{G})\right\}=(N+1) \left\langle 
 \mathbf{G}\otimes \boldsymbol{\phi}_{_{\boldsymbol{y}1}}^{^{\mathrm{T}}}, \mathbf{G}\otimes \boldsymbol{\phi}_{_{\boldsymbol{y}1}}^{^{\mathrm{T}}}\right\rangle$. Thus (in the light of Proposition III.1)\ \ $\boldsymbol{\eta}_{_{N}}(\mathbf{G})=\mathcal{J}(\mathbf{G};\mathbf{H}_{_{\boldsymbol{0}}})+\left\{\frac{\gamma^{^{2}}(N+1)}{m_{_{\boldsymbol{v}}}m_{_{\boldsymbol{y}}}(N+1)+2}\right\}
  \left\langle \mathbf{G}\otimes\boldsymbol{\phi}_{_{\boldsymbol{y}1}}^{^{\mathrm{T}}}, \mathbf{G}\otimes \boldsymbol{\phi}_{_{\boldsymbol{y}1}}^{^{\mathrm{T}}}\right\rangle$\ from which Proposition 3.2 follows.\hfill$\blacksquare$
  
\vspace*{3mm}
\noindent
\underline{\textbf{Proof of Proposition 3.3:}}\ Note that $\boldsymbol{\eta}_{_{\boldsymbol{\alpha}N}}^{a}(\mathbf{G};\mathbf{H})=(\boldsymbol{\mu}_{_{\boldsymbol{\alpha}N}}^{a})^{^{-1}}\boldsymbol{\mu}_{_{\boldsymbol{\alpha}N}}^{^{\boldsymbol{\Gamma}}}(\mathbf{G};\mathbf{H})$, where\break $\boldsymbol{\mu}_{_{\boldsymbol{\alpha}N}}^{^{\boldsymbol{\Gamma}}}=\displaystyle\int_{_{\mathcal{S}_{_{N}}^{\boldsymbol{\alpha}}}}\boldsymbol{\beta}_{_{\boldsymbol{\alpha}}}^{^{\mathrm{T}}}\boldsymbol{\Gamma}_{_{N\boldsymbol{\alpha}}}^{^{\boldsymbol{e}}}\boldsymbol{\beta}_{_{\boldsymbol{\alpha}}}d\boldsymbol{\beta}_{_{\boldsymbol{\alpha}}}$, so that\ $\boldsymbol{\mu}_{_{\boldsymbol{\alpha}N}}^{a}$\ and\ 
$\boldsymbol{\mu}_{_{\boldsymbol{\alpha}N}}^{^{\boldsymbol{\Gamma}}}$\ are formally identical to\ 
$\nu_{_{N}}$\ and\ $\nu_{_{\mathcal{J}N}}$\ above and, hence, the derivation of\ 
$\boldsymbol{\eta}_{_{N}}(\cdot)$\ applies mutatis mutandis to\ $\boldsymbol{\eta}_{_{\boldsymbol{\alpha}N}}^{a}$ -- note that in this case the ``nominal term'' is zero and it is not necessary to rely on Kroenecker products. \hfill$\blacksquare$  


\vspace*{3mm}
\noindent
\underline{\textbf{Derivation of $\boldsymbol{\eta}^{b}(\mathbf{G};
\mathcal{S}_{_{\mathbf{H}\infty}})$ (Subsection 3.3):}} First, the set\ $\bar{\mathcal{S}}_{_{\mathbf{H}\infty}}\supset\mathcal{S}_{_{\mathbf{H}\infty}}$, is introduced where\break
$\bar{\mathcal{S}}_{_{\mathbf{H}\infty}}=\{\mathbf{H}\in\mathcal{R}_{c}^{^{m_{_{\boldsymbol{v}}} \times m_{_{\boldsymbol{y}}} }}:\|(\mathbf{H}-\mathbf{H}_{_{\boldsymbol{0}}})\bar{W}\|_{_{2}}\leq \gamma_{_{\mathbf{H}}}\}$,\ \ $\bar{W}=\boldsymbol{\phi}_{_{\boldsymbol{y}}}^{a}\boldsymbol{\phi}_{_{\boldsymbol{y}W}}^{^{-1}}$\ \ and\ \ 
$\boldsymbol{\phi}_{_{\boldsymbol{y}W}}$\ is a spectral factor of\break 
$\boldsymbol{\phi}_{_{\boldsymbol{y}W}}(e^{j\alpha})\boldsymbol{\phi}_{_{\boldsymbol{y}W}}(e^{j\alpha})^{*}
=\operatorname{tr}\{[W_{_{\mathbf{H}}}^{^{-1}}\boldsymbol{\phi}_{_{\boldsymbol{y}}}^{a}](e^{j\alpha})^{*}[W_{_{\mathbf{H}}}^{^{-1}}\boldsymbol{\phi}_{_{\boldsymbol{y}}}^{a}](e^{j\alpha})\}.$

Then, replacing\ $\gamma$,\ $\boldsymbol{\Gamma}_{_{\boldsymbol{y}}}$,\ $\boldsymbol{\Gamma}_{_{\boldsymbol{v}}}$,\ and\  
$\boldsymbol{\phi}_{_{\boldsymbol{y}1}}$\ respectively by $\gamma_{_{\mathbf{H}}}$,\ $\boldsymbol{\Gamma}_{_{\boldsymbol{y}}}^{a}$,\ 
$\boldsymbol{\Gamma}_{_{\boldsymbol{v}}}^{a}$,\ and\ 
$\bar{\boldsymbol{\phi}}_{_{\boldsymbol{y}1}}=\bar{W}^{^{-1}}\boldsymbol{\phi}_{_{\boldsymbol{y}}}^{a}$\ in the expression of\
$\boldsymbol{\eta}_{_{a\boldsymbol{v}}}(\mathbf{G})$\ leads to (since 
$\bar{\boldsymbol{\phi}}_{_{\boldsymbol{y}1}}=\boldsymbol{\phi}_{_{\boldsymbol{y}W}}\mathbf{I}_{_{m_{_{\boldsymbol{y}}}}}$)
$$\boldsymbol{\eta}^{b}(\mathbf{G};\mathcal{S}_{_{\mathbf{H}}\infty})=\boldsymbol{\eta}^{a}(\mathbf{G};\mathbf{H}_{_{\boldsymbol{0}}})+
(\gamma_{_{\mathbf{H}}}^{^{2}}/m_{_{\boldsymbol{y}}}m_{_{\boldsymbol{v}}})
\left\langle\mathbf{G}\otimes(\boldsymbol{\phi}_{_{\boldsymbol{y}W}}\mathbf{I}_{_{m_{_{\boldsymbol{y}}}}}),
\mathbf{G}\otimes(\boldsymbol{\phi}_{_{\boldsymbol{y}W}}\mathbf{I}_{_{m_{_{\boldsymbol{y}}}}})\right\rangle$$
or, equivalently,\  $\boldsymbol{\eta}^{b}(\mathbf{G};\mathcal{S}_{_{\mathbf{H}}\infty})
=\boldsymbol{\eta}^{a}(\mathbf{G};\mathbf{H}_{_{\boldsymbol{0}}})+
(\gamma_{_{\mathbf{H}}}^{^{2}}/m_{_{\boldsymbol{v}}})
\left\langle\mathbf{G}\boldsymbol{\phi}_{_{\boldsymbol{y}W}},
\mathbf{G}\boldsymbol{\phi}_{_{\boldsymbol{y}W}}\right\rangle$.


\vspace*{3mm}
\noindent
\underline{\textbf{Proof of Proposition 4.1:}}\\
\noindent
\textbf{(a)} Letting\ $\lambda_{_{\mathrm{o}}}\triangleq\|\mathbf{F}_{_{\mathbf{G}\boldsymbol{y}}}\|_{_{\infty}}^{^{2}}$, where\ \ $\mathbf{F}_{_{\mathbf{G}\boldsymbol{y}}}=\mathbf{G}\otimes \mathbf{F}_{_{\boldsymbol{y}}}^{^{\mathrm{T}}}$, and noting that\ $\lambda\in\mathcal{S}_{_{\lambda}}$\ \ $\Leftrightarrow$\ \ $\lambda>\lambda_{_{\mathrm{o}}}$,   the proof consists of the following intermediate statements:\\
\indent\textbf{(1)} $\forall\ \lambda<\lambda_{_{\mathrm{o}}}$, 
$\boldsymbol{\varphi}_{_{\mathbf{D}}}(\lambda;\mathbf{D})=+\infty$.\\
\indent\textbf{(2)} $\inf\{\boldsymbol{\varphi}_{_{\mathbf{D}}}(\lambda;\mathbf{G}):
\lambda\geq\lambda_{_{\mathrm{o}}}\}=\inf\{\boldsymbol{\varphi}_{_{\mathbf{D}}}(\lambda;\mathbf{G}):\lambda\in\mathcal{S}_{_{\lambda}}\}$.\\
\noindent
\textbf{(b)} It is a direct consequence of Lemma A1 ([13]).\\
\noindent
\textbf{(c)} The proof hinges on invoking the (so-called) discrete-time, bounded-real lemma ([27]) 
and using elementary congruence transformations (row and column permutations), say\ $\mathbf{T}$\
and\ $\mathbf{T}^{^{\mathrm{T}}}$, to convert the condition 
``$\lambda\in\mathcal{S}_{_{\lambda}}$'' into the positive-definiteness of a diagonal block of\ \
$\mathbf{T}Q_{_{LQ}}(\mathbf{P};\boldsymbol{\Sigma}_{_{a}}, \mathbf{M}(\lambda))\mathbf{T}^{^{\mathrm{T}}}$.\hfill$\blacksquare$

\vspace*{3mm}
\noindent
\underline{\textbf{Proof of Proposition 4.5:}}\ Writing\
$$- Q_{_{\mathbf{B}\mathbf{R}}}(\mathbf{P},
\boldsymbol{\Sigma}_{_{\mathbf{F}\mathbf{G}}}(\boldsymbol{\theta}), 
\mathbf{M}(\boldsymbol{\sigma}))=\left[\begin{smallmatrix}
                                        \mathbf{P}&\boldsymbol{0}\\
                                        \boldsymbol{0}&\mathbf{M}_{_{\boldsymbol{\sigma}}}\\
                                       \end{smallmatrix}\right]-
                                       \left[\begin{smallmatrix}
                                        \mathbf{A}_{_{\mathbf{F}\mathbf{G}}}^{^{\mathrm{T}}}& \mathbf{C}_{_{\mathbf{F}\mathbf{G}}}^{^{\mathrm{T}}}\\
                                         \mathbf{B}_{_{\mathbf{F}\mathbf{G}}}^{^{\mathrm{T}}}&
                                          \mathbf{D}_{_{\mathbf{F}\mathbf{G}}}^{^{\mathrm{T}}}\\
                                        \end{smallmatrix}\right]\left[\begin{smallmatrix}
                                        \mathbf{P}&\boldsymbol{0}\\
                                        \boldsymbol{0}&\mathbf{I}\\
                                       \end{smallmatrix}\right]\left[\begin{smallmatrix}
                                        \mathbf{A}_{_{\mathbf{F}\mathbf{G}}}& \mathbf{C}_{_{\mathbf{F}\mathbf{G}}}\\
                                         \mathbf{B}_{_{\mathbf{F}\mathbf{G}}}&
                                          \mathbf{D}_{_{\mathbf{F}\mathbf{G}}}\\
                                        \end{smallmatrix}\right]$$
and invoking the Schur complement formula, the condition\ ``$Q_{_{\mathbf{B}\mathbf{R}}}(\cdot)<0$''\ can be recast as\ ``$\bar{\boldsymbol{\psi}}(\mathbf{P},\boldsymbol{\sigma},\boldsymbol{\theta})>0$''\ \ where\ \
 $\bar{\boldsymbol{\psi}}(\mathbf{P},\boldsymbol{\sigma},\boldsymbol{0})=\operatorname{diag}(\mathbf{P},\mathbf{M}_{_{\boldsymbol{\sigma}}},\mathbf{P}^{^{-1}},\mathbf{I}_{_{m_{_{\boldsymbol{e}}}}})+\bar{\boldsymbol{\psi}}_{_{a}}(\boldsymbol{\theta})+\bar{\boldsymbol{\psi}}_{_{a}}(\boldsymbol{\theta})^{^{\mathrm{T}}}$,\
 $\bar{\boldsymbol{\psi}}_{_{a}}(\boldsymbol{\theta})=\left[\begin{smallmatrix}
                                                      \boldsymbol{0}\\
                                                      \mathbf{I}\\
                                                      \end{smallmatrix}\right]\left[\begin{smallmatrix}
                                        \mathbf{A}_{_{\mathbf{F}\mathbf{G}}}& \mathbf{B}_{_{\mathbf{F}\mathbf{G}}}\\
                                         \mathbf{C}_{_{\mathbf{F}\mathbf{G}}}&
                                          \mathbf{D}_{_{\mathbf{F}\mathbf{G}}}\\
                                        \end{smallmatrix}\right][\mathbf{I}\ \boldsymbol{0}]$.
                                        
 The proof is concluded by bringing elementary congruence transformations to bear on\ 
 ``$\bar{\boldsymbol{\psi}}(\mathbf{P},\boldsymbol{\sigma},\boldsymbol{\theta})$''\ and writing
 $\mathbf{A}_{_{\mathbf{FG}}}=\mathbf{A}_{_{\mathrm{o}}}+\mathbf{A}_{_{L}}(\boldsymbol{\theta}),\ 
\ \mathbf{B}_{_{\mathbf{FG}}}=\mathbf{B}_{_{\mathrm{o}}}+\mathbf{B}_{_{L}}(\boldsymbol{\theta}),\ \ 
\mathbf{C}_{_{\mathbf{FG}}}=\mathbf{C}_{_{\mathrm{o}}}+ \mathbf{C}_{_{L}}(\boldsymbol{\theta}),\ \ 
\mathbf{D}_{_{\mathbf{FG}}}=\mathbf{D}_{_{\mathbf{I}\boldsymbol{y}}}+\mathbf{D}_{_{L}}(\boldsymbol{\theta}),$
where\break $\mathbf{A}_{_{\mathrm{o}}}=\operatorname{diag}(\mathbf{A}_{_{a\boldsymbol{z}}}, \boldsymbol{0}_{n_{_{\mathbf{G}}}\times n_{_{\mathbf{G}}}})$,\ \
$\mathbf{B}_{_{\mathrm{o}}}=\left[\begin{smallmatrix}
                 \mathbf{B}_{_{a\boldsymbol{z}}}\\
                 \boldsymbol{0}_{n_{_{\mathbf{G}}}\times m_{\boldsymbol{yv}}}
                \end{smallmatrix}\right]$,\ \ 
$\mathbf{C}_{_{\mathrm{o}}}=[\mathbf{C}_{_{\mathbf{I}\boldsymbol{y}}}\ \vdots\ \boldsymbol{0}_{m_{_{\boldsymbol{e}}}\times (n_{_{a\boldsymbol{z}}}+n_{_{\mathbf{G}}})}]$,\
$\mathbf{D}_{_{\mathrm{o}}}=\mathbf{D}_{_{\mathbf{I}\boldsymbol{y}}}$,\break $\mathbf{A}_{_{L}}(\boldsymbol{\theta})=\left[\begin{smallmatrix}
                                      \boldsymbol{0}_{n_{_{a\boldsymbol{z}}}\times n_{_{a\boldsymbol{z}}}}& \boldsymbol{0}_{n_{_{a\boldsymbol{z}}}\times n_{_{\mathbf{G}}}}\\
                                      \mathbf{B}_{_{\mathbf{G}}}\mathbf{C}_{_{a\boldsymbol{z}}} & \mathbf{A}_{_{\mathbf{G}}}
                                     \end{smallmatrix}\right]$,\ 
                                     $\mathbf{B}_{_{L}}(\boldsymbol{\theta})=\left[\begin{smallmatrix}
                                                                                             \boldsymbol{0}_{n_{_{a\boldsymbol{z}}}\times m_{\boldsymbol{yv}}}\\
                                                                                             \mathbf{B}_{_{\mathbf{G}}}\mathbf{D}_{_{\mathrm{o}\boldsymbol{z}}}
                                                                                             \end{smallmatrix}\right]$,\
$\mathbf{C}_{_{L}}(\boldsymbol{\theta})=[-\mathbf{D}_{_{\mathbf{G}}}\mathbf{C}_{_{a\boldsymbol{z}}}\ \vdots\ -\mathbf{C}_{_{\mathbf{G}}}]$,\
$\mathbf{D}_{_{L}}(\boldsymbol{\theta})=-\mathbf{D}_{_{\mathbf{G}}}\mathbf{D}_{_{\mathrm{o}\boldsymbol{z}}}.\ \ \ \ \ \blacksquare$

\pagebreak
\noindent
\underline{\textbf{Proof of Proposition 4.6:}} \textbf{(a)} follows directly from, Propositions 4.4 and 4.5, \textbf{(\emph{i})} and \textbf{(\emph{ii})}.\\
\noindent
 \textbf{(b)} follows from Proposition 4.5, \textbf{(\emph{i})} and \textbf{(\emph{ii})} with
 ``$\rho(\mathbf{A}_{_{\mathbf{G}}})<1$'' being a consequence of Proposition 4.5 and the fact
 that\ 
 $Q_{_{\mathbf{B}\mathbf{R}}}(\mathbf{P}^{^{\mathrm{o}}},
 \boldsymbol{\Sigma}_{_{\mathbf{F}\mathbf{G}}}(\boldsymbol{\theta}),\mathbf{M}(\boldsymbol{\sigma}^{^{\mathrm{o}}}))<0\ \Rightarrow\  \rho(\mathbf{A}_{_{\mathbf{G}}}) < 1$\
(as, in this case, $\mathbf{A}_{_{\mathbf{F}\mathbf{G}}}^{^{\mathrm{T}}}\mathbf{P}^{^{\mathrm{o}}}
\mathbf{A}_{_{\mathbf{F}\mathbf{G}}}-\mathbf{P}^{^{\mathrm{o}}}<0$ with $\mathbf{P}^{^{\mathrm{o}}}>0$) since\ $\mathbf{A}_{\mathbf{G}\mathbf{F}}$\ is taken to be lower triangular with\ $\mathbf{A}_{_{\mathbf{G}}}$ as one of its diagonal $\text{blocks}.\blacksquare$  

\vspace*{3mm}
\noindent
\underline{\textbf{Proof of Proposition 4.7:}} \textbf{(a)} As $W_{_{a}}^{^{\mathrm{T}}}$\ and\
$W_{_{a}}^{^{\mathrm{T}}}\boldsymbol{\psi}_{_{\mathrm{o}}}(\mathbf{P})W_{_{a}}$\ are given by
$$\mathbf{W}_{_{a}}^{^{\mathrm{T}}}=
\left[\begin{smallmatrix}
\mathbf{I}& \boldsymbol{0}&\vdots&\boldsymbol{0}&\boldsymbol{0}&\vdots&\boldsymbol{0}&\boldsymbol{0}\\
\boldsymbol{0}&\boldsymbol{0}&\vdots&\mathbf{I}&\boldsymbol{0}&\vdots&\boldsymbol{0}&\boldsymbol{0}\\ 
\boldsymbol{0}&\boldsymbol{0}&\vdots&\boldsymbol{0}&\mathbf{I}&\vdots&\boldsymbol{0}&\boldsymbol{0}\\ 
\boldsymbol{0}&\boldsymbol{0}&\vdots&\boldsymbol{0}&\boldsymbol{0}&\vdots&\mathbf{I}&\boldsymbol{0}\\ 
\end{smallmatrix}\right],\ \ \mathbf{W}_{_{a}}^{^{\mathrm{T}}}\boldsymbol{\psi}_{_{\mathrm{o}}}(\mathbf{P})\mathbf{W}_{_{a}}=
\left[\begin{smallmatrix}
 \mathbf{R}&\mathbf{A}_{_{a\boldsymbol{z}}}&\boldsymbol{0}&\mathbf{B}_{_{a\boldsymbol{z}}}\\
 \mathbf{A}_{_{a\boldsymbol{z}}}^{^{\mathrm{T}}}&\mathbf{S}&\mathbf{N}&\boldsymbol{0}\\
 \boldsymbol{0}&\mathbf{N}^{^{\mathrm{T}}}&\mathbf{X}&\boldsymbol{0}\\
 \mathbf{B}_{_{a\boldsymbol{z}}}^{^{\mathrm{T}}}&\boldsymbol{0}&\boldsymbol{0}&\mathbf{M}_{_{\boldsymbol{\sigma}}}\\
\end{smallmatrix}\right].$$
\noindent
 Pre and post-multipliying\ $\mathbf{W}_{_{a}}^{^{\mathrm{T}}}\boldsymbol{\psi}_{_{\mathrm{o}}}(\mathbf{P})\mathbf{W}_{_{a}}$\ by\ $\mathbf{I}_{_{c}}^{^{\mathrm{T}}}$\ and\ $\mathbf{I}_{_{c}}$\ (say), where\ $\mathbf{I}_{_{c}}$\ is a column permutation matrix,  leads to\ 
$ \mathbf{W}_{_{a}}^{^{\mathrm{T}}}\boldsymbol{\psi}_{_{\mathrm{o}}}(\mathbf{P})\mathbf{W}_{_{a}}>0\ 
\Leftrightarrow\ \left[\begin{smallmatrix}
   \mathbf{R}&\mathbf{B}_{_{a\boldsymbol{z}}}&\mathbf{A}_{_{a\boldsymbol{z}}}&\boldsymbol{0}\\
   \mathbf{B}_{_{a\boldsymbol{z}}}^{^{\mathrm{T}}}&\mathbf{M}_{_{\boldsymbol{\sigma}}}&\boldsymbol{0}&\boldsymbol{0}\\
   \mathbf{A}_{_{a\boldsymbol{z}}}^{^{\mathrm{T}}}& \boldsymbol{0}& \mathbf{S}&\mathbf{N}\\
   \boldsymbol{0}&\boldsymbol{0}&\mathbf{N}^{^{\mathrm{T}}}&\mathbf{X}\\
   \end{smallmatrix}\right]>0\ \ \ \Leftrightarrow\ \ (\text{invoking the Schur complement formula}) 
$
$$\left[\begin{smallmatrix}\mathbf{S}&\mathbf{N}\\
   \mathbf{N}^{^{\mathrm{T}}}&\mathbf{X}\\
  \end{smallmatrix}\right]>0\ \ \text{and}\ \ \left[\begin{smallmatrix}
   \mathbf{R}&\mathbf{B}_{_{a\boldsymbol{z}}}\\
   \mathbf{B}_{_{a\boldsymbol{z}}}^{^{\mathrm{T}}}&\mathbf{M}_{_{\boldsymbol{\sigma}}}\\
  \end{smallmatrix}\right]-\left[\begin{smallmatrix}
                \mathbf{A}_{_{a\boldsymbol{z}}}&\boldsymbol{0}\\
                \boldsymbol{0}&\boldsymbol{0}\\
                \end{smallmatrix}\right]\left[\begin{smallmatrix}
                             \mathbf{S}&\mathbf{N}\\
                             \mathbf{N}^{^{\mathrm{T}}}&\mathbf{X}\\
                             \end{smallmatrix}\right]^{^{-1}}\left[\begin{smallmatrix}
                                               \mathbf{A}_{_{a\boldsymbol{z}}}^{^{\mathrm{T}}}&\boldsymbol{0}\\
                                               \boldsymbol{0}&\boldsymbol{0}\\
                                               \end{smallmatrix}\right]>0 
$$
Moreover, since\ \ $\left[\begin{smallmatrix}
                                 \mathbf{S}&\mathbf{N}\\
                                 \mathbf{N}^{^{\mathrm{T}}}&\mathbf{X}\\
                                \end{smallmatrix}\right]^{^{-1}}=\mathbf{P}^{^{-1}}=\left[\begin{smallmatrix}
                                                                                 \mathbf{R}&\mathbf{M}\\
                                                                                 \mathbf{M}^{^{\mathrm{T}}}&\mathbf{Z}\\
                                                                                \end{smallmatrix}\right]$, the last LMI
                                                                                above can be rewritten as
$$
 \left[\begin{smallmatrix}
   \mathbf{R}&\mathbf{B}_{_{a\boldsymbol{z}}}\\
   \mathbf{B}_{_{a\boldsymbol{z}}}^{^{\mathrm{T}}}&\mathbf{M}_{_{\boldsymbol{\sigma}}}\\
  \end{smallmatrix}\right]-\left[\begin{smallmatrix}
                \mathbf{I}\\
                \boldsymbol{0}\\
                \end{smallmatrix}\right]\mathbf{A}_{_{a\boldsymbol{z}}}\mathbf{R}\mathbf{A}_{_{a\boldsymbol{z}}}^{^{\mathrm{T}}}[\mathbf{I}\ \boldsymbol{0}]>0\ \ \ \ \Leftrightarrow\ \ \ \ \left[\begin{smallmatrix}
 \mathbf{R}-\mathbf{A}_{_{a\boldsymbol{z}}}\mathbf{R}\mathbf{A}_{_{a\boldsymbol{z}}}^{^{\mathrm{T}}}& \mathbf{B}_{_{a\boldsymbol{z}}}\\
 \mathbf{B}_{_{a\boldsymbol{z}}}^{^{\mathrm{T}}}&\mathbf{M}_{_{\boldsymbol{\sigma}}}
 \end{smallmatrix}\right]>0.
$$
\noindent
\textbf{(b)} Bringing a similar argument to bear on $W_{_{b}}^{^{\mathrm{T}}}
\boldsymbol{\psi}_{_{\mathrm{o}}}(\mathbf{P}, \boldsymbol{\sigma})W_{_{b}}$\ provides a proof of
Proposition IV.5\textbf{(b)} with\ 
$\mathbf{W}_{_{b}}^{^{\mathrm{T}}}=\left[\begin{smallmatrix}
                  \mathbf{I}&\boldsymbol{0}&\boldsymbol{0}&\boldsymbol{0}&\boldsymbol{0}&\boldsymbol{0}&\boldsymbol{0}\\
                  \boldsymbol{0}&\mathbf{I}&\boldsymbol{0}&\boldsymbol{0}&\boldsymbol{0}&\boldsymbol{0}&\boldsymbol{0}\\
                  \boldsymbol{0}&\boldsymbol{0}&\mathbf{I}&\boldsymbol{0}&\boldsymbol{0}&\mathbf{T}_{_{63}}^{^{\mathrm{T}}}&\boldsymbol{0}\\
                  \boldsymbol{0}&\boldsymbol{0}&\boldsymbol{0}&\boldsymbol{0}&\mathbf{I}&\mathbf{T}_{_{65}}^{^{\mathrm{T}}}&\boldsymbol{0}\\
                  \boldsymbol{0}&\boldsymbol{0}&\boldsymbol{0}&\boldsymbol{0}&\boldsymbol{0}&\boldsymbol{0}&\mathbf{I}\\
                  \end{smallmatrix}\right],\ \ \text{where}\ \mathbf{T}_{_{63}}=-\mathbf{D}_{_{\mathrm{o}\boldsymbol{z}2}}^{^{-1}}\mathbf{C}_{_{a\boldsymbol{z}}},\ \ 
                 \mathbf{T}_{_{65}}=-\mathbf{D}_{_{\mathrm{o}\boldsymbol{z}2}}^{^{-1}}\mathbf{D}_{_{\mathrm{o}\boldsymbol{z}1}},\ \ \mathbf{D}_{_{\mathrm{o}\boldsymbol{z}1}}=\mathbf{D}_{_{\mathbf{H}\mathrm{o}}}\mathbf{D}_{_{\mathbf{W}\boldsymbol{y}}}^{^{-1}}$\ $\mathbf{D}_{_{\mathrm{o}\boldsymbol{z}2}}=\mathbf{D}_{_{\mathbf{W}\boldsymbol{v}}}^{^{-1}}$\ and\ $\mathbf{C}_{_{a\boldsymbol{z}}}$\
                 is as above.\hfill$\blacksquare$

\vspace*{3mm}
\noindent
\underline{\textbf{Proof of Proposition 4.8:}} Proposition 4.8\textbf{(a)} is a direct
consequence of Propositions 4.6 and 4.7 and the fact that\ \
$\mathbf{S}\geq\mathbf{R}^{^{-1}}\ \ \Leftrightarrow\ \ \left[\begin{smallmatrix}
                                                               \mathbf{S}&\mathbf{I}\\
                                                               \mathbf{I}&\mathbf{R}\\
                                                              \end{smallmatrix}\right]\geq0.$\\  
\noindent
\textbf{(b)} The fact that\ $\mathbf{P}^{^{\mathrm{o}}}>0$\ follows from the following logical sequence based on
Schur Complements: as\ $\mathbf{X}>0$, $\mathbf{P}^{^{\mathrm{o}}}>0$\ \ $\Leftrightarrow$\ \
$(\mathbf{S}_{_{\mathrm{o}}}-Q_{_{\mathbf{S}\mathbf{R}}}\mathbf{V}\mathbf{X}^{^{1/2}}\mathbf{X}^{^{-1}}\mathbf{X}^{^{1/2}}\mathbf{V}^{^{\mathrm{T}}}Q_{_{\mathbf{S}\mathbf{R}}})>0$\ \ $\Leftrightarrow$\ \ $(\mathbf{S}_{_{\mathrm{o}}}-Q_{_{\mathbf{S}\mathbf{R}}}^{^{2}})>0$\ \ $\Leftrightarrow$\ $(\mathbf{S}_{_{\mathrm{o}}}-(\mathbf{S}_{_{\mathrm{o}}}-\mathbf{R}_{_{\mathrm{o}}}^{^{-1}}))>0$\ \ $\Leftrightarrow$\ \ $\mathbf{R}_{_{\mathrm{o}}}^{^{-1}}>0$\ \
$\Leftrightarrow$\ \ $\mathbf{R}_{_{\mathrm{o}}}>0$. The statements about\ $\boldsymbol{\theta}$\
at the end of Proposition 4.8\textbf{(b)} follow from Proposition 4.6\textbf{(b)} and 
Proposition 4.7. \hfill$\blacksquare$

\vspace*{3mm}
\noindent
\underline{\textbf{Equations (5.1) and (5.2):}}  (5.1) and (5.2) follow from the equalities
(for\ $Q=Q^{^{\mathrm{T}}}\geq0$)\break
$\operatorname{tr}(\mathbf{M}Q\mathbf{M}^{^{\mathrm{T}}})
=\inf\{\operatorname{tr}(\mathbf{P}):\mathbf{P}=\mathbf{P}^{^{\mathrm{T}}},\ \mathbf{P}\geq\mathbf{M}Q\mathbf{M}^{^{\mathrm{T}}}\}
=\inf\left\{\operatorname{tr}(\mathbf{P}):\mathbf{P}=\mathbf{P}^{^{\mathrm{T}}},\left[\begin{smallmatrix}
                                                 \mathbf{P}&\mathbf{M}Q^{^{1/2}}\\
                                                 Q^{^{1/2}}\mathbf{M}^{^{\mathrm{T}}}&\mathbf{I}\\
                                                \end{smallmatrix}\right]\geq0\right\},$  the
definition of $\langle\cdot,\cdot\rangle$, the parametrizations\ $\mathbf{G}(\boldsymbol{\beta})=
\boldsymbol{\beta}\mathbf{Y}_{_{a}}^{^{i}}$, $i=1,2,3$,\ and the identity\ 
$\mathbf{M}_{_{1}}\mathbf{M}_{_{2}}\otimes\mathbf{M}_{_{3}}^{^{\mathrm{T}}}
=[\mathbf{M}_{_{1}}\otimes\mathbf{I}][\mathbf{M}_{_{2}}\otimes\mathbf{M}_{_{3}}^{^{\mathrm{T}}}].$            

\vspace*{3mm}
\noindent
\underline{\textbf{Proof of Proposition 5.1:}} The first part is an immediate consequence of 
(4.8). The second part follows from Proposition A.1 below.\hfill$\blacksquare$

\subsection*{Proofs}

\setcounter{equation}{0}
\renewcommand{\theequation}{A.\arabic{equation}}

\vspace*{3mm}
\noindent
\underline{\textbf{Upper bounds on\ $\check{\mathcal{J}}_{\infty}(\mathbf{G};\mathbf{H})$\ and\ 
$\check{\mathcal{J}}_{\infty}^{a}(\mathbf{G})$:}} Notice first that  ``$\bar{\boldsymbol{y}}\in \bar{\mathcal{S}}_{_{\boldsymbol{y}}}$\ 
and\ $\bar{\boldsymbol{v}}\in \bar{\mathcal{S}}_{_{\boldsymbol{v}}}$''\ \ \ $\Leftrightarrow$\ \ \ 
``$\|\bar{\boldsymbol{y}}\|_{_{2}}^{^{2}}-\gamma_{_{\boldsymbol{y}}}^{^{2}}\leq 0$\ \ and\ \
$\|\bar{\boldsymbol{v}}\|_{_{2}}^{^{2}}-\gamma_{_{\boldsymbol{v}}}^{^{2}}\leq 0$''.

Thus, $\forall \bar{\boldsymbol{y}}\in \bar{\mathcal{S}}_{_{\boldsymbol{y}}}$, \ \
$\forall \bar{\boldsymbol{v}}\in \bar{\mathcal{S}}_{_{\boldsymbol{v}}}$,\ \ $\forall \boldsymbol{\sigma}_{_{\boldsymbol{y}}}>0$,\ \
 $\forall \boldsymbol{\sigma}_{_{\boldsymbol{v}}}>0$,\\ 
 $Lag_{\infty}(\bar{\boldsymbol{z}}, \boldsymbol{\sigma};\mathbf{G},\mathbf{H})
 \geq \|\bar{\boldsymbol{e}}(\bar{\boldsymbol{z}};\mathbf{G}, \mathbf{H})\|_{_{2}}^{^{2}}$\ \ \ $\Rightarrow$
 \begin{eqnarray*}
 \check{\mathcal{J}}_{\infty}(\mathbf{G};\mathbf{H})
 &\leq& \sup\{Lag_{\infty}(\bar{\boldsymbol{z}}, \boldsymbol{\sigma};\mathbf{G}, \mathbf{H}_{_{\boldsymbol{0}}})
 :\bar{\boldsymbol{y}}\in\mathcal{S}_{_{\boldsymbol{y}}},\ \bar{\boldsymbol{v}}\in\mathcal{S}_{_{\boldsymbol{v}}}\}\\
 &\leq&\sup\{Lag_{\infty}(\bar{\boldsymbol{z}}, \boldsymbol{\sigma};\mathbf{G}, \mathbf{H}_{_{\boldsymbol{0}}})
 :\bar{\boldsymbol{z}}\in \mathcal{R}_{c}^{m_{\boldsymbol{z}}}\}\ \ \ \ \ \ \ \ \ \ \ \ \Rightarrow
 \end{eqnarray*}
 $\forall \boldsymbol{\sigma}_{_{\boldsymbol{y}}}>0$,\ $\forall \boldsymbol{\sigma}_{_{\boldsymbol{v}}}>0$,\ 
 $\check{\mathcal{J}}_{\infty}(\mathbf{G};\mathbf{H}_{_{\boldsymbol{0}}})
 \leq \boldsymbol{\varphi}_{_{\mathbf{D}\infty}}(\boldsymbol{\sigma}; \mathbf{G}, \mathbf{H}_{_{\boldsymbol{0}}})$\ \ \ $\Rightarrow$
$$\check{\mathcal{J}}_{\infty}(\mathbf{G};\mathbf{H}_{_{\boldsymbol{0}}})\leq 
 \mathcal{J}_{\infty}(\mathbf{G};\mathbf{H}_{_{\boldsymbol{0}}})
 \triangleq\inf\{\boldsymbol{\varphi}_{_{\mathbf{D}\infty}}(\boldsymbol{\sigma}; \mathbf{G}, \mathbf{H}_{_{\boldsymbol{0}}})
 :\boldsymbol{\sigma}_{_{\boldsymbol{y}}}>0,\ \boldsymbol{\sigma}_{_{\boldsymbol{v}}}>0\}.$$
 
In an entirely similar way, it can be shown that\\ 
$\forall \boldsymbol{\sigma}_{_{\boldsymbol{y}}}>0$,\ $\forall \boldsymbol{\sigma}_{_{\boldsymbol{v}}}>0$,\  
$\forall \boldsymbol{\sigma}_{_{\boldsymbol{W}}}>0$,\ 
 $\check{\mathcal{J}}_{\infty}^{a}(\mathbf{G})
 \leq \boldsymbol{\varphi}_{_{\mathbf{D}\infty}}^{a}(\boldsymbol{\sigma}; \mathbf{G})$\ \ \ $\Rightarrow$
$$\check{\mathcal{J}}_{\infty}^{a}(\mathbf{G})\leq 
 \mathcal{J}_{\infty}(\mathbf{G})\triangleq\inf\{\boldsymbol{\varphi}_{_{\mathbf{D}\infty}}(\boldsymbol{\sigma}; \mathbf{G})
 :\boldsymbol{\sigma}=(\boldsymbol{\sigma}_{_{\boldsymbol{y}}}, \boldsymbol{\sigma}_{_{\boldsymbol{v}}},
 \boldsymbol{\sigma}_{_{\boldsymbol{w}}} ),\ \boldsymbol{\sigma}_{_{\boldsymbol{y}}}>0,\ \boldsymbol{\sigma}_{_{\boldsymbol{v}}}>0,\
 \boldsymbol{\sigma}_{_{\boldsymbol{w}}}>0 \}.$$
\hfill$\blacksquare$

\vspace*{3mm}
\noindent
\underline{\textbf{Proof of Proposition \ref{prop:01}:}} \textbf{(a)} Note first that 
\begin{equation}\label{eq:A1}
 \nu_{_{N}}=\int_{\mathcal{S}_{_{\boldsymbol{\theta}}}^{^{N}}}d\boldsymbol{\theta}=\int_{\widehat{\mathcal{S}}_{_{\boldsymbol{\theta}}}^{^{N}}}\left|\operatorname{\det}\left[\left(P_{_{\boldsymbol{\theta}}}^{^{N}}\right)^{^{-1/2}}\right]\right|d\widehat{\boldsymbol{\theta}}
 =\left|\operatorname{\det}\left[\left(P_{_{\boldsymbol{\theta}}}^{^{N}}\right)^{^{-1/2}}\right]\right|\int_{\widehat{\mathcal{S}}_{_{\boldsymbol{\theta}}}^{^{N}}}d\widehat{\boldsymbol{\theta}},
\end{equation}
where\ \ $\widehat{\mathcal{S}}_{_{\boldsymbol{\theta}}}^{^{N}}=\left\{\widehat{\boldsymbol{\theta}}\in \mathbb{R}^{n}:\|\widehat{\boldsymbol{\theta}}^{^{2}}\|_{_{E}}^{^{2}}\leq \gamma^{^{2}}\right\}$\ \ 
(\emph{i.e.},\ $\widehat{\boldsymbol{\theta}}=\left(P_{_{\boldsymbol{\theta}}}^{^{N}}\right)^{^{1/2}}\boldsymbol{\theta}$). Note also that for $n>2$
\begin{eqnarray}
 \int_{\widehat{\mathcal{S}}_{_{\boldsymbol{\theta}}}^{^{N}}}d\widehat{\boldsymbol{\theta}}&=&\int_{0}^{2\pi}\int_{0}^{\pi}\cdots\int_{0}^{\pi}\int_{0}^{\gamma}r^{^{n-1}} 
 \prod_{i=1}^{n-2}\left\{\sin(\alpha_{i})\right\}^{n-1-i}dr\ d\alpha_{_{_{1}}}\ \cdots\ d\alpha_{n-2}\ d \alpha_{_{n-1}}\ \ \ \Leftrightarrow\nonumber\\
 \int_{\widehat{\mathcal{S}}_{_{\boldsymbol{\theta}}}^{^{N}}}d\widehat{\boldsymbol{\theta}}&=&\left\{\int_{0}^{\gamma}r^{^{n-1}}dr\right\}\left\{\prod_{i=1}^{n-2}\int_{0}^{\pi}\left\{[\sin(\alpha_{i})]^{n-1-i}d \alpha_{i}\right\}\right\}\int_{0}^{2\pi}d \alpha_{_{n-1}}\ \ \ \Leftrightarrow\nonumber\\
\int_{\widehat{\mathcal{S}}_{_{\boldsymbol{\theta}}}^{^{N}}}d \widehat{\boldsymbol{\theta}}&=&\frac{\gamma^{n}}{n}\left\{\prod_{i=1}^{n-2}\int_{0}^{\pi}[\sin(\alpha_{i})]^{n-1-i}d\alpha_{i}\right\}(2\pi).\label{eq:A2}
 \end{eqnarray}
Proposition \ref{prop:01}\textbf{(a)} follows immediately from (\ref{eq:A1}) and (\ref{eq:A2}).\\
\noindent 
 \textbf{(b)} Let\ \ $\widehat{\boldsymbol{\theta}}=(P_{_{\boldsymbol{\theta}}}^{^{N}})^{^{1/2}}\boldsymbol{\theta}$. Thus,
\begin{eqnarray*}
 \int_{\mathcal{S}_{_{\boldsymbol{\theta}}}^{^{N}}}\boldsymbol{\theta}^{^{\mathrm{T}}}P_{_{\mathcal{J}}}^{^{N}}(\mathbf{G})\boldsymbol{\theta} d\boldsymbol{\theta}
 &=&\int_{\widehat{\mathcal{S}}_{_{\boldsymbol{\theta}}}^{^{N}}}
 \widehat{\boldsymbol{\theta}}^{^{\mathrm{T}}}\left\{(P_{_{\boldsymbol{\theta}}}^{^{N}})^{^{-1/2}}P_{_{\mathcal{J}}}^{^{N}}(\mathbf{G})(P_{_{\boldsymbol{\theta}}}^{^{N}})^{^{-1/2}}\right\}
 \widehat{\boldsymbol{\theta}}\left|\operatorname{det}\left[(P_{_{\boldsymbol{\theta}}}^{^{N}})^{^{-1/2}}\right]\right|d \widehat{\boldsymbol{\theta}}\ \ \ \Leftrightarrow\\
\int_{\mathcal{S}_{_{\boldsymbol{\theta}}}^{^{N}}}\boldsymbol{\theta}^{^{\mathrm{T}}}P_{_{\mathcal{J}}}^{^{N}}(\mathbf{G})\boldsymbol{\theta}d \boldsymbol{\theta}
&=&\left\{\int_{\widehat{\mathcal{S}}_{_{\boldsymbol{\theta}}}^{^{N}}}
\widehat{\boldsymbol{\theta}}^{^{\mathrm{T}}}\widehat{P}_{_{\mathcal{J}}}^{^{N}}(\mathbf{G})\widehat{\boldsymbol{\theta}}d\widehat{\boldsymbol{\theta}}\right\}\left|\operatorname{det}(P_{_{\boldsymbol{\theta}}}^{^{N}})\right|^{^{-1/2}},
 \end{eqnarray*}
where\ \ $\widehat{P}_{_{\mathcal{J}}}^{^{N}}(\mathbf{G})=
(P_{_{\boldsymbol{\theta}}}^{^{N}})^{^{-1/2}}P_{_{\mathcal{J}}}^{^{N}}(\mathbf{G})(P_{_{\boldsymbol{\theta}}}^{^{N}})^{^{-1/2}}$,\ \ so 
that
$$\int_{\mathcal{S}_{_{\boldsymbol{\theta}}}^{^{N}}}\boldsymbol{\theta}^{^{\mathrm{T}}}P_{_{\mathcal{J}}}^{^{N}}(\mathbf{G})\boldsymbol{\theta}d \boldsymbol{\theta}=
\left\{\sum_{i}\sum_{j}\int_{\widehat{\mathcal{S}}_{_{\boldsymbol{\theta}}}^{^{N}}}[\widehat{P}_{_{\mathcal{J}}}^{^{N}}(\mathbf{G})]_{ij}
\widehat{\theta}_{_{i}}\widehat{\theta}_{j}\ d\widehat{\theta}_{_{_{1}}}\ \cdots\ d\widehat{\theta}_{_{n}}\right\}\left|\operatorname{det}(P_{_{\boldsymbol{\theta}}}^{^{N}})\right|^{^{-1/2}}$$
so that (since $\displaystyle\int_{\widehat{\mathcal{S}}_{_{\boldsymbol{\theta}}}^{^{N}}}\widehat{\theta}_{_{i}}\widehat{\theta}_{j}\ 
d\widehat{\theta}_{_{_{1}}}\ \cdots\ d\widehat{\theta}_{_{n}}=0$\ for \ $i\neq j$)
\begin{equation}\label{eq:A3}
 \int_{\mathcal{S}_{_{\boldsymbol{\theta}}}^{^{N}}}\boldsymbol{\theta}^{^{\mathrm{T}}}P_{_{\mathcal{J}}}^{^{N}}(\mathbf{G})
 \boldsymbol{\theta}d\boldsymbol{\theta}
 =\left\{\sum_{i=1}^{n}[\widehat{P}_{_{\mathcal{J}}}^{^{N}}(\mathbf{G})]_{_{ii}}
 \int_{\widehat{\mathcal{S}}_{\theta}^{N}}\widehat{\theta}_{_{i}}^{^{2}}d\widehat{\theta}_{_{_{1}}}\ \cdots\ d\widehat{\theta}_{_{n}}\right\}
 \left|\operatorname{det}(P_{_{\boldsymbol{\theta}}}^{^{N}})\right|^{^{-1/2}}.
\end{equation}
Note now that
\begin{equation}\label{eq:A4}
 \int_{\widehat{\mathcal{S}}_{_{\boldsymbol{\theta}}}^{^{N}}}\widehat{\theta}_{_{i}}^{^{2}}d\widehat{\theta}_{_{_{1}}}\ \cdots\ d\widehat{\theta}_{_{n}}
 = \int_{\widehat{\mathcal{S}}_{_{\boldsymbol{\theta}}}^{^{N}}}\widehat{\theta}_{_{_{1}}}^{^{2}}d\widehat{\theta}_{_{_{1}}}\ \cdots\ d\widehat{\theta}_{_{n}}.
\end{equation}
Moreover,
$$\displaystyle\int_{\widehat{\mathcal{S}}_{_{\boldsymbol{\theta}}}^{^{N}}}\widehat{\theta}_{_{_{1}}}^{^{2}}d\widehat{\theta}_{_{_{1}}}\ \cdots\ \widehat{\theta}_{_{n}}
=\int_{0}^{2\pi}\int_{0}^{\pi}\cdots\int_{0}^{\pi}\int_{r=0}^{\gamma}
\{r\cos(\alpha_{_{_{1}}})\}^{^{2}}r^{^{n-1}}\prod_{k=1}^{n-2}[\sin(\alpha_{_{k}})]^{^{n-1-k}}drd\alpha_{_{_{1}}}\cdots d\alpha_{_{n-1}},$$
or, equivalently,
\begin{equation}\label{eq:A5}
 \int_{\widehat{\mathcal{S}}_{_{\boldsymbol{\theta}}}^{^{N}}}\widehat{\theta}_{_{_{1}}}^{^{2}}d\widehat{\theta}_{_{_{1}}}\cdots \widehat{\theta}_{_{n}}
= \left\{\int_{0}^{\gamma}r^{^{2}}r^{^{n-1}}dr\right\}\left\{\int_{0}^{\pi}[\cos(\alpha_{_{_{1}}})]^{^{2}}[\sin(\alpha_{_{_{1}}})]^{^{n-2}}d \alpha_{_{_{1}}}\right\}\check{\prod}_{2,n-2}(2\pi),
\end{equation}
where\ \ $\displaystyle\check{\prod}_{2,n-2}\triangleq\prod_{k=2}^{n-2}\int_{0}^{\pi}[\sin(\alpha_{_{k}})]^{^{n-1-k}}d\alpha_{_{k}}$.

It then follows from (\ref{eq:A3}) -- (\ref{eq:A5}) that
\begin{eqnarray}
 \int_{\mathcal{S}_{_{\boldsymbol{\theta}}}^{^{N}}}\boldsymbol{\theta}^{^{\mathrm{T}}}P_{_{\mathcal{J}}}^{^{N}}(\mathbf{G})\boldsymbol{\theta}d \boldsymbol{\theta}
 &=&\left\{\sum_{i=1}^{n}[\widehat{P}_{_{\mathcal{J}}}^{^{N}}]_{_{ii}}\right\}\left|\operatorname{det}\left(P_{_{\boldsymbol{\theta}}}^{^{N}}\right)\right|^{^{-1/2}}
 \left(\frac{\gamma^{n+2}}{n+2}\right)\times\nonumber\\
&&\ \ \ \ \ \ \ \ \ \ \  \ \ \ \ \ \times\left\{\int_{0}^{\pi}[\cos(\alpha_{_{_{1}}})]^{^{2}}[\sin(\alpha_{_{_{1}}})]^{^{n-2}}d \alpha_{_{_{1}}}\right\}
\check{\prod}_{2, n-2}(2\pi).\label{eq:A6}
\end{eqnarray}
\noindent 
 \textbf{(c)} In the light of Proposition \ref{prop:01}\textbf{(a)} and (\ref{eq:A6}),
\begin{eqnarray}
\nu_{_{N}}^{^{-1}}\int_{\mathcal{S}_{_{\boldsymbol{\theta}}}^{^{N}}}\boldsymbol{\theta}^{^{\mathrm{T}}}P_{_{\mathcal{J}}}^{^{N}}(\mathbf{G})\boldsymbol{\theta}d \boldsymbol{\theta}
&=&\gamma^{^{2}}\left(\frac{n}{n+2}\right)\left\{\sum_{i=1}^{n}[\widehat{P}_{_{\mathcal{J}}}^{^{N}}(\mathbf{G})]_{_{ii}}\right\}\times\nonumber\\
&&\times\left\{\int_{0}^{\pi}[\cos(\alpha_{_{_{1}}})]^{^{2}}[\sin(\alpha_{_{_{1}}})]^{^{n-2}}d\alpha_{_{_{1}}}\right\}
\left\{\int_{0}^{\pi}[\sin(\alpha_{_{_{1}}})]^{^{n-2}}d\alpha_{_{_{1}}}\right\}^{^{-1}}.\label{eq:A7}
\end{eqnarray}
Note now that 
\begin{equation}\label{eq:A8}
 \int_{0}^{\pi}[\sin(\alpha_{_{_{1}}})]^{^{n-2}}[\cos(\alpha_{_{_{1}}})]^{^{2}}d\alpha_{_{_{1}}}=(1/n)\int_{0}^{\pi}[\sin(\alpha_{_{_{1}}})]^{^{n-2}}
 d\alpha_{_{_{1}}}.
\end{equation}
It then follows from (\ref{eq:A7}) and (\ref{eq:A8}) that 
$$\nu_{_{N}}^{^{-1}}\int_{\mathcal{S}_{_{\boldsymbol{\theta}}}^{^{N}}}\boldsymbol{\theta}^{^{\mathrm{T}}}P_{_{\mathcal{J}}}^{^{N}}(\mathbf{G})\boldsymbol{\theta}d\boldsymbol{\theta}
=\frac{\gamma^{^{2}}}{n+2}\left\{\sum_{i=1}^{n}[\widehat{P}_{_{\mathcal{J}}}^{^{N}}(\mathbf{G})]_{_{ii}}\right\}.$$
Thus, 
$$\boldsymbol{\eta}_{_{N}}(\mathbf{G})=\mathcal{J}(\mathbf{G};\mathbf{H}_{_{\boldsymbol{0}}})+\frac{\gamma^{^{2}}}{(n+2)}\left\{\sum_{i=1}^{n}[\widehat{P}_{_{\mathcal{J}}}^{^{N}}(\mathbf{G})]_{_{ii}}\right\}.$$
\hfill$\blacksquare$

\vspace*{3mm}
\noindent
\underline{\textbf{Proof of equation (\ref{eq:A8}):}} Note first that
$$\int_{0}^{\pi}[\sin(\alpha)]^{n-2}[\cos(\alpha)]^{2}d\alpha=\int_{0}^{\pi}[\sin(\alpha)]^{n-2}d\alpha
-\int_{0}^{\pi}[\sin(\alpha)]^{n}d\alpha$$
and that (integrating by parts yields to)
\begin{eqnarray*}
 \int_{0}^{\pi}[\sin(\alpha)]^{n}d\alpha&=&\int_{0}^{\pi}[\sin(\alpha)]^{n-1}\sin(\alpha)d\alpha=
 \left.[\sin(\alpha)]^{n-1}(-\cos(\alpha))\right|_{0}^{\pi}\\
 &&-\int_{0}^{\pi}(n-1)[\sin(\alpha)]^{n-2}\cos(\alpha)[-\cos(\alpha)]d\alpha\\
 \Leftrightarrow\ \ \  \int_{0}^{\pi}[\sin(\alpha)]^{n}d\alpha&=&
 -(n-1)\int_{0}^{\pi}[\sin(\alpha)]^{n-2}[-\cos^{2}(\alpha)]d\alpha\\
  \Leftrightarrow\ \ \  \int_{0}^{\pi}[\sin(\alpha)]^{n}d\alpha&=&(n-1)\int_{0}^{\pi}[\sin(\alpha)]^{n-2}[\cos(\alpha)]^{2}d\alpha.
\end{eqnarray*}

Thus,
\begin{eqnarray*}
\int_{0}^{\pi}[\sin(\alpha)]^{n-2}[\cos(\alpha)]^{2}d\alpha&=&
\int_{0}^{\pi}[\sin(\alpha)]^{n-2}d\alpha-(n-1)\int_{0}^{\pi}[\sin(\alpha)]^{n-2}[\cos(\alpha)]^{2}d\alpha\\
\Leftrightarrow\ \ \  n\int_{0}^{\pi}[\sin(\alpha)]^{n-2}[\cos(\alpha)]^{2}d\alpha&=&
\int_{0}^{\pi}[\sin(\alpha)]^{n-2}d\alpha\\
\Leftrightarrow\ \ \  \int_{0}^{\pi}[\sin(\alpha)]^{n-2}[\cos(\alpha)]^{2}d\alpha&=&(1/n)\int_{0}^{\pi}[\sin(\alpha)]^{n-2}d\alpha.
\end{eqnarray*}\hfill$\blacksquare$

\vspace*{3mm}
\noindent
\underline{\textbf{Proof of Proposition \ref{prop:02}:}} Recall that
$$P_{_{\boldsymbol{\theta}}}^{^{N}}=\mathbf{I}\otimes\left(P_{_{\boldsymbol{\beta}}}^{^{N}}\right)^{^{\mathrm{T}}}=
[\mathbf{I}\otimes(P_{_{\boldsymbol{\beta}}}^{^{N}})^{^{1/2}}][\mathbf{I}\otimes(P_{_{\boldsymbol{\beta}}}^{^{N}})^{^{1/2}}],$$
where\ \  $P_{_{\boldsymbol{\beta}}}^{^{N}}=(1/2\pi)\displaystyle\int_{0}^{2\pi}\check{\mathbf{Y}}_{_{\mathbf{X}}}^{^{N}}(e^{j\alpha})\check{\mathbf{Y}}_{_{\mathbf{X}}}^{^{N}}(e^{j\alpha})^{*}d\alpha$.
Note now that (for the family\ \ $\mathcal{S}_{_{\mathbf{X}}}^{^{N}}$\ \ of FIRs of length $N$)
\begin{eqnarray*}
 \mathbf{Y}_{_{\mathbf{X}}}^{^{N}}(e^{j\alpha})&=&(e^{j\alpha}\mathbf{I}-\mathbf{A}_{_{N}})^{^{-1}}
 =\operatorname{diag}\left((e^{j\alpha}\mathbf{I}-\mathbf{A}_{c}^{^{N}})^{^{-1}}, \mathellipsis, (e^{j\alpha}\mathbf{I}-\mathbf{A}_{c}^{^{N}})^{^{-1}}\right)\ \ \Rightarrow\\
\mathbf{Y}_{_{\mathbf{X}}}^{^{N}}(e^{j\alpha})\mathbf{B}_{_{N}}&=&
\operatorname{diag}\left((e^{j\alpha}\mathbf{I}-\mathbf{A}_{c}^{^{N}})^{^{-1}}\boldsymbol{b}_{_{c}}^{^{N}},
\mathellipsis, (e^{j\alpha}\mathbf{I}-\mathbf{A}_{c}^{^{N}})^{^{-1}}\boldsymbol{b}_{_{c}}^{^{N}}\right)
\in\mathbf{C}^{^{m_{_{\boldsymbol{y}}}\times(N-1)\times m_{_{\boldsymbol{y}}}}}.
 \end{eqnarray*}

Now, let\ \ $Z_{_{N}}(e^{j\alpha})\triangleq(e^{j\alpha}\mathbf{I}-\mathbf{A}_{_{c}}^{^{N}})^{^{-1}}\boldsymbol{b}_{_{c}}^{^{N}}$\ \ and note that\bigskip

\noindent
\underline{\textbf{Auxiliary Proposition 1:}} $Z_{_{N}}(e^{j\alpha})=[1\ e^{-j\alpha}\ e^{-j2\alpha}\ \cdots\ e^{-j(N-1)\alpha}]^{^{\mathrm{T}}}$. \hfill$\nabla$

As a result $\check{\mathbf{Y}}_{_{\mathbf{X}}}^{^{N}}= \left[\begin{array}{c}
                                          \operatorname{diag}(Z_{_{N}},\mathellipsis, Z_{_{N}}) \\ 
                                      \mathbf{I}_{m_{_{\boldsymbol{y}}}}
                                       \end{array}\right]$\ \ and\ \ 
                                       $(\check{\mathbf{Y}}_{_{\mathbf{X}}}^{^{N}})^{*}=[\operatorname{diag}(Z_{_{N}}^{*}, \mathellipsis, Z_{_{N}}^{*})\ \vdots\ \mathbf{I}_{m_{_{\boldsymbol{y}}}}]$ 
so that 
 \begin{eqnarray*}
 P_{_{\boldsymbol{\beta}}}^{^{N}}&=&
(1/2\pi)\int_{0}^{2\pi}\left[\begin{array}{ccc}
 \operatorname{diag}(Z_{_{N}}Z_{_{N}}^{*}, \mathellipsis, Z_{_{N}}Z_{_{N}}^{*})  & \vdots&  \operatorname{diag}(Z_{_{N}}, \mathellipsis, Z_{_{N}}) \\ 
\operatorname{diag}(Z_{_{N}}^{*}, \mathellipsis, Z_{_{N}}^{*}) &\vdots& \mathbf{I}_{m_{_{\boldsymbol{y}}}} \\
\end{array}\right](e^{j\alpha})d\alpha,\  \ \Rightarrow\\\\
P_{_{\boldsymbol{\beta}}}^{^{N}}&=&\newcommand*{\temp}{\multicolumn{1}{r|}{}}
\left[\begin{array}{ccc}
 \operatorname{diag}(\mathbf{I}_{N-1}, \mathellipsis, \mathbf{I}_{N-1})  & \temp& \boldsymbol{0} \\ \cline{1-3}
\boldsymbol{0} &\temp& \mathbf{I}_{m_{_{\boldsymbol{y}}}} \\
\end{array}\right]=\mathbf{I}.
\end{eqnarray*}
Thus, in this case, \ \ $P_{_{\boldsymbol{\theta}}}^{^{N}}=\mathbf{I}$\ \ and, hence, \ \ $\widehat{P}_{_{\mathcal{J}}}^{^{N}}(\mathbf{G})
=P_{_{\mathcal{J}}}^{^{N}}(\mathbf{G})$.\\

As for\ \ $\widehat{P}_{_{\mathcal{J}}}^{^{N}}(\mathbf{G})\triangleq(1/2\pi)\displaystyle\int_{0}^{2\pi}\left\{F_{_{N}}(\mathbf{G})^{*}F_{_{N}}(\mathbf{G})\right\}(e^{j\alpha})d\alpha$,\ \ 
note first that\ \ $F_{_{N}}(\mathbf{G})=(\mathbf{G}\otimes\boldsymbol{\phi}_{_{\boldsymbol{y}1}}^{^{\mathrm{T}}})(\mathbf{I}_{m_{_{\boldsymbol{v}}}}\otimes(\check{\mathbf{Y}}_{_{\mathbf{X}}}^{^{N}})^{^{\mathrm{T}}})$\ \
so that
$$F_{_{N}}(\mathbf{G})^{*}F_{_{N}}(\mathbf{G})=(\mathbf{I}_{m_{_{\boldsymbol{v}}}}\otimes(\check{\mathbf{Y}}_{_{\mathbf{X}}}^{^{N}})^{^{\mathrm{T}}})^{*}Q_{y}(\mathbf{G})
(\mathbf{I}_{m_{_{\boldsymbol{v}}}}\otimes(\check{\mathbf{Y}}_{_{\mathbf{X}}}^{^{N}})^{^{\mathrm{T}}}),$$
where\ \ $Q_{y}(\mathbf{G})\triangleq(\mathbf{G}\otimes\boldsymbol{\phi}_{_{\boldsymbol{y}1}}^{^{\mathrm{T}}})^{*}(\mathbf{G}\otimes\boldsymbol{\phi}_{_{\boldsymbol{y}1}}^{^{\mathrm{T}}})$.\\
Thus,
\begin{eqnarray*}
\operatorname{tr}\left\{\widehat{P}_{_{\mathcal{J}}}^{^{N}}(\mathbf{G})\right\}&=&\operatorname{tr}\left\{P_{_{\mathcal{J}}}^{^{N}}(\mathbf{G})\right\}\\
&=&(1/2\pi)\displaystyle\int_{0}^{2\pi}\operatorname{tr}
\left\{\mathbf{Q}_{_{_{\mathbf{G}\boldsymbol{y}}}}(\mathbf{G})[(\mathbf{I}_{m_{_{\boldsymbol{v}}}}\otimes(\check{\mathbf{Y}}_{_{\mathbf{X}}}^{^{N}})^{^{\mathrm{T}}})(\mathbf{I}_{m_{_{\boldsymbol{v}}}}\otimes(\check{\mathbf{Y}}_{_{\mathbf{X}}}^{^{N}})^{^{\mathrm{T}}})^{*}]\right\}(e^{j\alpha})d\alpha.
\end{eqnarray*}                              
                                       
Note now that\ \ 
$[\mathbf{I}_{m_{_{\boldsymbol{v}}}}\otimes (\check{\mathbf{Y}}_{_{\mathbf{X}}}^{^{N}})^{^{\mathrm{T}}}]^{*}=\mathbf{I}\otimes[(\check{\mathbf{Y}}_{_{\mathbf{X}}}^{^{N}})^{^{\mathrm{T}}}]^{*}=\mathbf{I}\otimes(\check{\mathbf{Y}}_{_{\mathbf{X}}}^{^{N}})^{c}$,\\
(and since $\mathbf{I}\otimes(\mathbf{M}_{1}\mathbf{M}_{_{2}})^{^{\mathrm{T}}}=(\mathbf{I}\otimes\mathbf{M}_{_{2}}^{^{\mathrm{T}}})(\mathbf{I}\otimes\mathbf{M}_{1}^{^{\mathrm{T}}})$)
$$[\mathbf{I}_{m_{_{\boldsymbol{v}}}}\otimes(\check{\mathbf{Y}}_{_{\mathbf{X}}}^{^{N}})^{^{\mathrm{T}}}][\mathbf{I}\otimes(\check{\mathbf{Y}}_{_{\mathbf{X}}}^{^{N}})^{c}]
=\mathbf{I}_{m_{_{\boldsymbol{v}}}}\otimes[(\check{\mathbf{Y}}_{_{\mathbf{X}}}^{^{N}})^{^{\mathrm{T}}}(\check{\mathbf{Y}}_{_{\mathbf{X}}}^{^{N}})^{c}],$$
and $(\check{\mathbf{Y}}_{_{\mathbf{X}}}^{^{N}})^{^{\mathrm{T}}}(\check{\mathbf{Y}}_{_{\mathbf{X}}}^{^{N}})^{c}=(N+1)\mathbf{I}_{m_{_{\boldsymbol{y}}}}$.\\

As a result, \ \ 
$\operatorname{tr}\left\{\widehat{P}_{_{\mathcal{J}}}^{^{N}}(\mathbf{G})\right\}
=(1/2\pi)\displaystyle\int_{0}^{2\pi}\operatorname{tr}\left\{Q_{y}(\mathbf{G})(N+1)\mathbf{I}_{m_{_{\boldsymbol{v}}}m_{_{\boldsymbol{y}}}}\right\}(e^{j\alpha})d\alpha$
\begin{eqnarray*}
 &\Leftrightarrow&\ \  \operatorname{tr}\left\{\widehat{P}_{_{\mathcal{J}}}^{^{N}}(\mathbf{G})\right\}
 = (N+1)(1/2\pi)\int_{0}^{2\pi}\operatorname{tr}\left\{(\mathbf{G}\otimes\boldsymbol{\phi}_{_{\boldsymbol{y}1}}^{^{\mathrm{T}}})^{*}(\mathbf{G}\otimes\boldsymbol{\phi}_{_{\boldsymbol{y}1}}^{^{\mathrm{T}}})\right\}(e^{j\alpha})d\alpha\\
 &\Leftrightarrow&\ \  \operatorname{tr}\left\{\widehat{P}_{_{\mathcal{J}}}^{^{N}}(\mathbf{G})\right\}=(N+1) \left\langle 
 \mathbf{G}\otimes \boldsymbol{\phi}_{_{\boldsymbol{y}1}}^{^{\mathrm{T}}}, \mathbf{G}\otimes \boldsymbol{\phi}_{_{\boldsymbol{y}1}}^{^{\mathrm{T}}}\right\rangle.
 \end{eqnarray*}
 It then follows  that
 \begin{eqnarray*}
  \lim_{N\rightarrow\infty}\boldsymbol{\eta}_{_{N}}(\mathbf{G})&=&\mathcal{J}(\mathbf{G};\mathbf{H}_{_{\boldsymbol{0}}})+\lim_{N\rightarrow\infty}\left\{\frac{\gamma^{^{2}}(N+1)}{m_{_{\boldsymbol{v}}}m_{_{\boldsymbol{y}}}(N+1)+2}\right\}
  \left\langle \mathbf{G}\otimes\boldsymbol{\phi}_{_{\boldsymbol{y}1}}^{^{\mathrm{T}}}, \mathbf{G}\otimes \boldsymbol{\phi}_{_{\boldsymbol{y}1}}^{^{\mathrm{T}}}\right\rangle\ \ \Rightarrow\\
    \lim_{N\rightarrow\infty}\boldsymbol{\eta}_{_{N}}(\mathbf{G})&=&\mathcal{J}(\mathbf{G};\mathbf{H}_{_{\boldsymbol{0}}})
    + \frac{\gamma^{^{2}}}{m_{_{\boldsymbol{v}}}m_{_{\boldsymbol{y}}}}\left\langle \mathbf{G} \otimes \boldsymbol{\phi}_{_{\boldsymbol{y}1}}^{^{\mathrm{T}}},\mathbf{G} \otimes \boldsymbol{\phi}_{_{\boldsymbol{y}1}}^{^{\mathrm{T}}}\right\rangle.
  \end{eqnarray*}
  \hfill$\blacksquare$
\newpage

\noindent
\underline{\textbf{Proof of Proposition \ref{prop:03}:}} \textbf{(a)} Let $\lambda_{_{\mathrm{o}}}\triangleq \|\mathbf{F}_{_{\mathbf{G}\boldsymbol{y}}}\|_{\infty}^{^{2}}$ and note that\  
\begin{equation}\label{eq:AA01}
\lambda \in \mathcal{S}_{_{\lambda}}\ \ \Leftrightarrow\ \ \lambda> \lambda_{_{\mathrm{o}}}.
\end{equation}
Note also that if $\lambda<\lambda_{_{\mathrm{o}}}$,\ \ $\inf\{L_{a}(Z, \lambda;\mathbf{G}):Z\in\mathcal{R}_{c}^{m_{_{\boldsymbol{v}}}m_{_{\boldsymbol{y}}}}\}=-\infty$\ and\ hence\ \  $\boldsymbol{\varphi}_{_{\mathbf{D}}}(\lambda, \mathbf{G})=+\infty$. Thus, it follows from (\ref{eq:08}) that 
\begin{equation}\label{eq:AA02}
 \bar{\mathcal{J}}(\mathbf{G};\mathcal{S}_{_{\mathbf{X}}})=\inf\{\boldsymbol{\varphi}_{_{\mathbf{D}}}(\lambda;\mathbf{G}): \lambda\geq\lambda_{_{\mathrm{o}}}\}.
\end{equation}
Now, $\forall \lambda>\lambda_{_{\mathrm{o}}}$,\ $\forall Z\in \mathcal{R}_{c}^{m_{_{\boldsymbol{v}}}m_{_{\boldsymbol{y}}}}$,\\
$L_{a}(X, \lambda;\mathbf{G})=L_{a}(X, \lambda_{_{\mathrm{o}}};\mathbf{G})+(\lambda-\lambda_{_{\mathrm{o}}})\langle X\mathbf{F}_{_{W}},X\mathbf{F}_{_{W}}\rangle\geq L_{a}(X, \lambda_{_{\mathrm{o}}};\mathbf{G})$\ \ $\Rightarrow$\\
$\forall \lambda >\lambda_{_{\mathrm{o}}}$,\ $\inf\{L_{a}(X,\lambda; \mathbf{G}): X\in\mathcal{R}_{c}^{m_{_{\boldsymbol{v}}}m_{_{\boldsymbol{y}}}}\}\geq\inf\{L_{a}(X, \lambda_{_{\mathrm{o}}};\mathbf{G}):X\in\mathcal{R}^{m_{_{\boldsymbol{v}}}m_{_{\boldsymbol{y}}}}\}$.\\
Thus, it follows from the fact that
$$\boldsymbol{\varphi}_{_{\mathbf{D}\mathbf{I}}}(\lambda;\mathbf{G})=\lambda_{_{\mathrm{o}}}\gamma^{^{2}}+(\lambda-\lambda_{_{\mathrm{o}}})\gamma^{^{2}}-\inf\{L_{a}(X,\lambda;\mathbf{G}):X\in \mathcal{R}_{c}^{m_{_{\boldsymbol{v}}}m_{_{\boldsymbol{y}}}}\}$$
that\\
$\boldsymbol{\varphi}_{_{\mathbf{D}\mathbf{I}}}(\lambda;\mathbf{G})
\leq\lambda_{_{\mathrm{o}}}\gamma^{^{2}}+(\lambda-\lambda_{_{\mathrm{o}}})\gamma^{^{2}}
-\inf\{L_{a}(X,\lambda_{_{\mathrm{o}}};\mathbf{G}):X\in\mathcal{R}_{c}^{m_{_{\boldsymbol{v}}}m_{_{\boldsymbol{y}}}}\}$\ \ $\Leftrightarrow$\ \ $\boldsymbol{\varphi}_{_{\mathbf{D}\mathbf{I}}}(\lambda;\mathbf{G})\leq\boldsymbol{\varphi}_{_{\mathbf{D}\mathbf{I}}}(\lambda_{_{\mathrm{o}}};\mathbf{G})+(\lambda-\lambda_{_{\mathrm{o}}})\gamma^{^{2}}$. Thus,\ \ $\inf\{\boldsymbol{\varphi}_{_{\mathbf{D}\mathbf{I}}}(\lambda;\mathbf{G}):\lambda>\lambda_{_{\mathrm{o}}}\}\leq \boldsymbol{\varphi}_{_{\mathbf{D}\mathbf{I}}}(\lambda_{_{\mathrm{o}}};\mathbf{G})$\ \ which implies that
\begin{equation}\label{eq:AA03}
 \inf\{\boldsymbol{\varphi}_{_{\mathbf{D}\mathbf{I}}}(\lambda;\mathbf{G}):\lambda\geq\lambda_{_{\mathrm{o}}}\}
 =\inf\{\boldsymbol{\varphi}_{_{\mathbf{D}\mathbf{I}}}(\lambda;\mathbf{G}):\lambda>\lambda_{_{\mathrm{o}}}\}.
\end{equation}
Combining (\ref{eq:AA01}) -- (\ref{eq:AA03}) leads to 
$$\bar{\mathcal{J}}(\mathbf{G};\mathcal{S}_{_{\mathbf{X}}})=\inf\{\boldsymbol{\varphi}_{_{\mathbf{D}\mathbf{I}}}(\lambda;\mathbf{G}):\lambda\in\mathcal{S}_{_{\lambda}}\}.$$

\noindent
\textbf{(b)} It directly follows from Lemma A.1 ([17]) that $\forall \lambda\in \mathcal{S}_{_{\lambda}}$,\\
$\inf\{L_{a}(Z,\lambda;\mathbf{G}): Z\in \mathcal{R}_{c}^{m_{_{\boldsymbol{v}}}m_{_{\boldsymbol{y}}}}\}$ equals the optimal value of\ \ $\inf\{\boldsymbol{x}_{_{\boldsymbol{0}}}^{^{\mathrm{T}}}\mathbf{P}\boldsymbol{x}_{_{\boldsymbol{0}}}:\mathbf{P}\in\mathcal{S}_{_{\mathbf{P}}}(\displaystyle\boldsymbol{\Sigma}_{a}, \lambda)\}$, where
$\mathcal{S}_{_{\mathbf{P}}}(\displaystyle \boldsymbol{\Sigma}_{a}, \lambda)=
\{\mathbf{P}=\mathbf{P}^{^{\mathrm{T}}}: Q_{_{LQ}}(\mathbf{P}; \boldsymbol{\Sigma}_{a},\mathbf{M}(\lambda))\geq0\}$. The proof is concluded by noting
that\ \  $\forall\lambda\in \mathcal{S}_{_{\lambda}}$,\ \
$\mathcal{S}_{_{\mathbf{P}}}^{a}(\displaystyle \boldsymbol{\Sigma}_{a}, \lambda)=
\{\mathbf{P}=\mathbf{P}^{^{\mathrm{T}}}: Q_{_{LQ}}(\mathbf{P}; \boldsymbol{\Sigma}_{a},\mathbf{M}(\lambda))>0\}$ is non-empty (see Proposition 4.1(\textbf{c})) and that\ \ $\mathcal{S}_{_{\mathbf{P}}}^{a}(\boldsymbol{\Sigma}_{a}, \lambda)$ is dense in $\mathcal{S}_{_{\mathbf{P}}}(\displaystyle\boldsymbol{\Sigma}_{a}, \lambda)$ (since $Q_{_{LQ}}(\cdot, \boldsymbol{\Sigma}_{a}, \mathbf{M}(\lambda))$ is affine).\\

\noindent
\textbf{(c)} Note first that (for $F_{_{\mathbf{I}W}}\triangleq\mathbf{I}_{m_{_{\boldsymbol{v}}}}\otimes \mathbf{F}_{_{W}}^{^{\mathrm{T}}}$, 
$\mathbf{F}_{_{\mathbf{G}\boldsymbol{y}}}\triangleq \mathbf{G}\otimes \mathbf{F}_{\boldsymbol{y}}^{^{\mathrm{T}}}$, $\mathbf{F}=\begin{bmatrix}
                                                                      F_{_{\mathbf{I}W}}\\
                                                                      \mathbf{F}_{_{\mathbf{G}\boldsymbol{y}}}
                                                                     \end{bmatrix}
                                                                        $)\\
$\mathbf{F}^{*}\mathbf{M}(\lambda)\mathbf{F}=[F_{_{\mathbf{I}W}}^{*}\ \mathbf{F}_{_{\mathbf{G}\boldsymbol{y}}}^{*}]\operatorname{diag}(\lambda\mathbf{I}, -\mathbf{I})
\begin{bmatrix}
 F_{_{\mathbf{I}W}}\\
 \mathbf{F}_{_{\mathbf{G}\boldsymbol{y}}}
\end{bmatrix}.
$\\
Noting that\ \  $F_{_{\mathbf{I}W}}=\operatorname{diag}(\mathbf{F}_{_{W}}^{^{\mathrm{T}}})$,\ \  so that \ \
$F_{_{\mathbf{I}W}}^{*}F_{_{\mathbf{I}W}}=\operatorname{diag}(\mathbf{F}_{_{W}}\mathbf{F}_{_{W}}^{^{\mathrm{T}}})$, it follows from\break
$\mathbf{F}_{_{W}}=[\mathbf{I}_{m_{_{\boldsymbol{y}}}}\ \vdots\ \boldsymbol{0}_{m_{_{\boldsymbol{y}}}\times m_{_{\boldsymbol{v}}}}]$\ \  that\ \  $\mathbf{F}_{_{W}}\mathbf{F}_{_{W}}^{^{\mathrm{T}}}=\mathbf{I}_{m_{_{\boldsymbol{y}}}}$ \ \  and, hence,\ \ 
$F_{_{\mathbf{I}W}}^{*}F_{_{\mathbf{I}W}}=\operatorname{diag}(\mathbf{I}_{m_{_{\boldsymbol{y}}}})=\mathbf{I}_{m_{_{\boldsymbol{v}}}m_{_{\boldsymbol{y}}}}$.\\
As a result, $\mathbf{F}^{*}\mathbf{M}(\lambda)\mathbf{F}=\lambda\mathbf{I}_{m_{_{\boldsymbol{v}}}m_{_{\boldsymbol{y}}}}-\mathbf{F}_{_{\mathbf{G}\boldsymbol{y}}}^{*}\mathbf{F}_{_{\mathbf{G}\boldsymbol{y}}}$. 
Thus, $\forall \phi \in [0, 2\pi]$, $$(\mathbf{F}^{*}\mathbf{M}(\lambda)\mathbf{F})(e^{j\theta})>0 \ \ 
\Leftrightarrow\ \ \|\lambda^{^{-1/2}}\mathbf{F}_{_{\mathbf{G}\boldsymbol{y}}}\|_{\infty}<1.$$

Now, let\ \  $(\mathbf{A}_{_{_{\mathbf{G}\boldsymbol{y}}}}, \mathbf{B}_{_{_{\mathbf{G}\boldsymbol{y}}}}, \mathbf{C}_{_{_{\mathbf{G}\boldsymbol{y}}}}, \mathbf{D}_{_{\mathbf{G}\boldsymbol{y}}})$\ \ denote a realization of $\mathbf{F}_{_{\mathbf{G}\boldsymbol{y}}}$, $\rho(\mathbf{A}_{_{\mathbf{G}\boldsymbol{y}}})<1$. It then follows from the discrete-time bounded-real lemma ([27]) that\ 
$\|\lambda^{^{-1/2}}\mathbf{F}_{_{\mathbf{G}\boldsymbol{y}}}\|_{\infty}<1\ \ \Leftrightarrow\ \ \exists\ \mathbf{X}=\mathbf{X}^{^{\mathrm{T}}}<0$\ \ such that
\begin{equation}\label{eq:A9}
 \begin{bmatrix}
  \mathbf{A}_{_{_{\mathbf{G}\boldsymbol{y}}}}^{^{\mathrm{T}}}\\
  \mathbf{B}_{_{_{\mathbf{G}\boldsymbol{y}}}}^{^{\mathrm{T}}}
 \end{bmatrix}\mathbf{X}[\mathbf{A}_{_{_{\mathbf{G}\boldsymbol{y}}}}\ \mathbf{B}_{_{_{\mathbf{G}\boldsymbol{y}}}}]-\begin{bmatrix}
                                                                    \mathbf{X} & \boldsymbol{0}\\
                                                                    \boldsymbol{0} & \boldsymbol{0}
                                                                    \end{bmatrix}- \lambda^{^{-1}}\begin{bmatrix}
                                                                                                  \mathbf{C}_{_{_{\mathbf{G}\boldsymbol{y}}}}^{^{\mathrm{T}}}\\
                                                                                                  \mathbf{D}_{_{\mathbf{G}\boldsymbol{y}}}^{^{\mathrm{T}}}
                                                                                                 \end{bmatrix}[\mathbf{C}_{_{_{\mathbf{G}\boldsymbol{y}}}}\ \mathbf{D}_{_{\mathbf{G}\boldsymbol{y}}}]+\begin{bmatrix}
                                                                                                                                                                 \boldsymbol{0} & \boldsymbol{0}\\
                                                                                                                                                                \boldsymbol{0} & \mathbf{I}
                                                                                                                                                                      \end{bmatrix} > 0. 
\end{equation}

Consider now \ \ $Q_{_{LQ}}(\mathbf{P}; \boldsymbol{\Sigma}_{a}, \mathbf{M}(\lambda))$\ \ given by
$$Q_{_{LQ}}(\mathbf{P}; \boldsymbol{\Sigma}_{a}, \mathbf{M}(\lambda))=Q_{_{\mathcal{J}}}(\mathbf{P};\boldsymbol{\Sigma}_{a})+S(\boldsymbol{\Sigma}_{a}; \mathbf{M}(\lambda)),$$
where\ \ $Q_{_{\mathcal{J}}}(\mathbf{P}; \boldsymbol{\Sigma}_{a})\triangleq\begin{bmatrix}
                                                            \mathbf{A}^{^{\mathrm{T}}}\\
                                                            \mathbf{B}^{^{\mathrm{T}}}
                                                           \end{bmatrix}\mathbf{P}[\mathbf{A}\ \mathbf{B}]- \begin{bmatrix}
                                                                                                                                                                  \mathbf{P}& \boldsymbol{0}\\
                                                                                                                                                                \boldsymbol{0} & \boldsymbol{0}                                                                                                                                                                \end{bmatrix}$, \ \ $S(\boldsymbol{\Sigma}_{a}; \mathbf{M}(\lambda))=\mathbf{R}^{^{\mathrm{T}}}\mathbf{M}(\lambda)\mathbf{R}$,\break
                                                                                                                                                                $\mathbf{A}=\begin{bmatrix}
\mathbf{A}_{a}& \mathbf{B}_{a}\\                                                                                                                                                       
\boldsymbol{0} & \boldsymbol{0}                                                                                                                                                                 \end{bmatrix}$, \ \ $\mathbf{B}=\begin{bmatrix}
                                \mathbf{B}_{a}\\                                                                                                                                                       
                                \boldsymbol{0}                                                                                                                                                                 \end{bmatrix}$, \ \ $\mathbf{R}=[\mathbf{C}_{a}\ \boldsymbol{d}_{a}\ \mathbf{D}_{a}]$\ \ and\ \
                                $\boldsymbol{\Sigma}_{a}=(\mathbf{A}_{a}, [\mathbf{B}_{a}\ \vdots\ \boldsymbol{b}_{a}], \mathbf{C}_{a}, 
                                [\mathbf{D}_{a}\ \vdots\ \boldsymbol{d}_{a}])$\ \ is a realization of 
                                $[\mathbf{F}\ \vdots\ -\boldsymbol{\mathcal{X}}_{_{\boldsymbol{0}}}(\mathbf{G})]$, $\rho(\mathbf{A}_{a})<1$.
                                
Note that 
\begin{eqnarray}\label{eq:A10}
 Q_{_{\mathcal{J}}}(\mathbf{P}; \boldsymbol{\Sigma}_{a})&=&\begin{bmatrix}
                                            \mathbf{A}_{a}^{^{\mathrm{T}}}& \vdots& \boldsymbol{0}\\
                                            \boldsymbol{b}_{a}^{^{\mathrm{T}}}& \vdots& \boldsymbol{0}\\
                                            \mathbf{B}_{a}^{^{\mathrm{T}}}& \vdots& \boldsymbol{0}
                                            \end{bmatrix}\begin{bmatrix} 
                                                         \mathbf{P}_{_{11}}&\mathbf{P}_{_{12}}\\
                                                         \mathbf{P}_{_{12}}^{^{\mathrm{T}}}&\mathbf{P}_{_{22}}
                                                         \end{bmatrix}\begin{bmatrix}
                                                                      \mathbf{A}_{a}& \boldsymbol{b}_{a}&\mathbf{B}_{a}\\
                                                                      \boldsymbol{0}&\boldsymbol{0}&\boldsymbol{0}
                                                                      \end{bmatrix}-\begin{bmatrix}
                                                                                     \mathbf{P}_{_{11}}& \mathbf{P}_{_{12}}& \vdots&\boldsymbol{0}\\
                                                                                     \mathbf{P}_{_{12}}^{^{\mathrm{T}}}& \mathbf{P}_{_{22}}& \vdots&\boldsymbol{0}\\
                                                                                   \boldsymbol{0} &\boldsymbol{0} & \vdots&\boldsymbol{0}
                                                                                   \end{bmatrix}\ \ \Rightarrow\nonumber\\\nonumber\\ \nonumber\\
 Q_{_{\mathcal{J}}}(\mathbf{P}; \boldsymbol{\Sigma}_{a})&=&\begin{bmatrix}
                                            \mathbf{A}_{a}^{^{\mathrm{T}}}\\
                                            \boldsymbol{b}_{a}^{^{\mathrm{T}}}\\
                                            \mathbf{B}_{a}^{^{\mathrm{T}}}
                                           \end{bmatrix}\mathbf{P}_{_{11}}[\mathbf{A}_{a}\ \boldsymbol{b}_{a}\ \mathbf{B}_{a}]-
                                                                                     \begin{bmatrix}
                                                                                     \mathbf{P}_{_{11}}& \mathbf{P}_{_{12}}&\boldsymbol{0}\\
                                                                                     \mathbf{P}_{_{12}}^{^{\mathrm{T}}}& \mathbf{P}_{_{22}}&\boldsymbol{0}\\
                                                                                   \boldsymbol{0} &\boldsymbol{0} &\boldsymbol{0} 
                                                                                   \end{bmatrix}.
\end{eqnarray}
Note also that (writing\ \ $\mathbf{R}=\begin{bmatrix}
                                     \mathbf{R}_{_{a1}}\\
                                     \mathbf{R}_{_{a2}}
                                    \end{bmatrix}$,\ \ $\mathbf{R}_{ai}=[\mathbf{C}_{ai}\ \boldsymbol{d}_{ai}\ \mathbf{D}_{ai}], i=1,2$)   
$$S(\boldsymbol{\Sigma}_{a}; \mathbf{M}(\lambda))=[\mathbf{R}_{_{a1}}^{^{\mathrm{T}}}\ \mathbf{R}_{_{a2}}^{^{\mathrm{T}}}]
\operatorname{diag}(\lambda\mathbf{I}, -\mathbf{I})\begin{bmatrix}
                                                   \mathbf{R}_{_{a1}}\\
                                                   \mathbf{R}_{_{a2}}
                                                  \end{bmatrix}=\lambda\mathbf{R}_{_{a1}}^{^{\mathrm{T}}}\mathbf{R}_{_{a1}}-\mathbf{R}_{_{a2}}^{^{\mathrm{T}}}\mathbf{R}_{_{a2}},$$   
 where \ \ $\mathbf{R}_{_{a1}}=[\mathbf{I}_{m_{_{\boldsymbol{v}}}m_{_{\boldsymbol{y}}}}\ \boldsymbol{0}]\mathbf{R}$\ \ and\ \  $\mathbf{R}_{_{a2}}=[\boldsymbol{0}\ \mathbf{I}_{m_{_{\boldsymbol{e}}}m_{_{\boldsymbol{y}}}}]\mathbf{R}\ \ \Rightarrow$   
 \begin{equation}\label{eq:A11}
  S(\boldsymbol{\Sigma}_{a}; \mathbf{M}(\lambda))=\lambda\begin{bmatrix}
                                             \mathbf{C}_{_{a1}}^{^{\mathrm{T}}}\\
                                             \boldsymbol{d}_{_{a1}}^{^{\mathrm{T}}}\\
                                             \mathbf{D}_{_{a1}}^{^{\mathrm{T}}}
                                            \end{bmatrix}[\mathbf{C}_{_{a1}}\ \boldsymbol{d}_{_{a1}}\ \mathbf{D}_{_{a1}}]-\begin{bmatrix}
                                             \mathbf{C}_{_{a2}}^{^{\mathrm{T}}}\\
                                             \boldsymbol{d}_{_{a2}}^{^{\mathrm{T}}}\\
                                              \mathbf{D}_{_{a2}}^{^{\mathrm{T}}}
                                            \end{bmatrix}[\mathbf{C}_{_{a2}}\ \boldsymbol{d}_{_{a2}}\ \mathbf{D}_{_{a2}}].
 \end{equation}
Permuting rows and columns of\  $Q_{_{\mathcal{J}}}(\cdot)$\ and \ $S(\cdot)$\  (\emph{i.e.}, applying congruence transformations) yields $\bar{Q}_{\mathcal{J}}(\mathbf{P}; \boldsymbol{\Sigma}_{a})=\begin{bmatrix}
                                             \mathbf{A}_{a}^{^{\mathrm{T}}}\\
                                             \mathbf{B}_{a}^{^{\mathrm{T}}}\\
                                              \boldsymbol{b}_{a}^{^{\mathrm{T}}}
                                            \end{bmatrix}\mathbf{P}_{_{11}}[\mathbf{A}_{a}\ \mathbf{B}_{a}\ \boldsymbol{b}_{a}]-\begin{bmatrix}
                                                   \mathbf{P}_{_{11}}& \boldsymbol{0}& \mathbf{P}_{_{12}}\\
                                                   \boldsymbol{0}& \boldsymbol{0}&\boldsymbol{0}\\
                                                   \mathbf{P}_{_{12}}^{^{\mathrm{T}}}&\boldsymbol{0}&\mathbf{P}_{_{22}}
                                                   \end{bmatrix}$\ \ and\ \                                                                                                                                                                     $\bar{S}(\boldsymbol{\Sigma}_{a}; \mathbf{M}(\lambda))=\lambda\begin{bmatrix}
                                             \mathbf{C}_{_{a1}}^{^{\mathrm{T}}}\\
                                              \mathbf{D}_{_{a1}}^{^{\mathrm{T}}}\\
                                             \boldsymbol{d}_{_{a1}}^{^{\mathrm{T}}}
                                            \end{bmatrix}[\mathbf{C}_{_{a1}}\ \mathbf{D}_{_{a1}}\ \boldsymbol{d}_{_{a1}}]-\begin{bmatrix}
                                             \mathbf{C}_{_{a2}}^{^{\mathrm{T}}}\\
                                             \mathbf{D}_{_{a2}}^{^{\mathrm{T}}}\\
                                               \boldsymbol{d}_{_{a2}}^{^{\mathrm{T}}}
                                            \end{bmatrix}[\mathbf{C}_{_{a2}}\ \mathbf{D}_{_{a2}}\ \boldsymbol{d}_{_{a2}}]$. \\ \\
Thus, the condition\ \  $Q_{_{LQ}}(\mathbf{P};\boldsymbol{\Sigma}_{a}, \mathbf{M}(\lambda))>0$\ \ is equivalent to 
 \begin{equation}\label{eq:A12}
  \bar{Q}_{_{\mathcal{J}}}(\mathbf{P}; \boldsymbol{\Sigma}_{a})+\bar{S}(\boldsymbol{\Sigma}_{a}; \mathbf{M}(\lambda))>0
 \end{equation}
so that the top diagonal block in (\ref{eq:A12}) is also non-negative definite, \emph{i.e.},
\begin{equation}\label{eq:A13}
 \begin{bmatrix}
  \mathbf{A}_{a}^{^{\mathrm{T}}}\\
  \mathbf{B}_{a}^{^{\mathrm{T}}}
 \end{bmatrix}\mathbf{P}_{_{11}}[\mathbf{A}_{a}\ \mathbf{B}_{a}]-\begin{bmatrix}
                                                         \mathbf{P}_{_{11}}&\boldsymbol{0}\\
                                                         \boldsymbol{0}& \boldsymbol{0}
                                                               \end{bmatrix}+\lambda\begin{bmatrix}
                                                                                    \mathbf{C}_{_{a1}}^{^{\mathrm{T}}}\\
                                                                                     \mathbf{D}_{_{a1}}^{^{\mathrm{T}}}
                                                                                      \end{bmatrix}[\mathbf{C}_{_{a1}}\ \mathbf{D}_{_{a1}}]-\begin{bmatrix}
                                                                                              \mathbf{C}_{_{a2}}^{^{\mathrm{T}}}\\
                                                                                              \mathbf{D}_{_{a2}}^{^{\mathrm{T}}}
                                                                                                \end{bmatrix}[\mathbf{C}_{_{a2}}\ \mathbf{D}_{_{a2}}]>0.
\end{equation}
Note now that \ \ $[\mathbf{F}\ \vdots\ -\boldsymbol{\mathcal{X}}_{_{\boldsymbol{0}}}(\mathbf{G})]=\newcommand*{\temp}{\multicolumn{1}{r|}{}}
\left[\begin{array}{ccc}
 \mathbf{I}_{m_{_{\boldsymbol{v}}}}\otimes \mathbf{F}_{_{W}}^{^{\mathrm{T}}} & \temp& \boldsymbol{0} \\ \cline{1-3}
\mathbf{F}_{_{\mathbf{G}\boldsymbol{y}}} &\temp& -\bar{\boldsymbol{\mathcal{X}}}_{_{\boldsymbol{0}}}(\mathbf{G}) \\
\end{array}\right]$, where\ $\bar{\boldsymbol{\mathcal{X}}}_{_{\boldsymbol{0}}}(\mathbf{G})=\operatorname{rvec}\left\{\mathbf{X}_{_{\boldsymbol{0}}}(\mathbf{G})\mathbf{F}_{\boldsymbol{y}}+\mathbf{G}\mathbf{F}_{\boldsymbol{v}}\right\}$. Thus, 
taking a realization \ 
$(\mathbf{A}_{_{a2}}, [\mathbf{B}_{_{a2}}\ \vdots\ \boldsymbol{b}_{_{a2}}], \mathbf{C}_{_{a2}}, [\mathbf{D}_{_{a2}}\ \vdots\ \boldsymbol{d}_{_{a2}}])$\ \
 of \ \ $[\mathbf{F}_{_{\mathbf{G}\boldsymbol{y}}}\ \vdots\ -\bar{\boldsymbol{\mathcal{X}}}_{_{\boldsymbol{0}}}(\mathbf{G})]$\ \ it follows that\ \ $\mathbf{A}_{a}=\mathbf{A}_{_{a2}}$, \ 
 $[\mathbf{B}_{a}\ \vdots\ \boldsymbol{b}_{a}]=[\mathbf{B}_{_{a2}}\ \vdots\ \boldsymbol{b}_{_{a2}}]$, \ $\mathbf{C}_{a}=\begin{bmatrix}
                                                                                                                   \boldsymbol{0}\\ 
                                                                                                                   \cdots\\
                                                                                                                   \mathbf{C}_{_{a2}}
                                                                                                                  \end{bmatrix}$,\ 
                                                                                                                 $\mathbf{D}_{a}=\begin{bmatrix}
                                                                                                                                \mathbf{I}_{m_{_{\boldsymbol{v}}}}\otimes \mathbf{F}_{_{W}}^{^{\mathrm{T}}}\\
                                                                                                                                \cdots\\
                                                                                                                                \mathbf{D}_{_{a2}}
                                                                                                                                \end{bmatrix}$, \ $\boldsymbol{d}_{a}=\begin{bmatrix}
                                                                                                                                      \boldsymbol{0}\\
                                                                                                                                      \cdots\\
                                                                                                                                      \boldsymbol{d}_{_{a2}}
                                                                                                                                     \end{bmatrix}$\ \
so that\ \  $\mathbf{C}_{_{a1}}=\boldsymbol{0}$ \ \ and\ \ $\mathbf{D}_{_{a1}}=(\mathbf{I}_{m_{_{\boldsymbol{v}}}}\otimes \mathbf{F}_{_{W}}^{^{\mathrm{T}}})$.\\ \\
\noindent
Thus, $\begin{bmatrix}
        \mathbf{C}_{_{a1}}^{^{\mathrm{T}}}\\
        \mathbf{D}_{_{a1}}^{^{\mathrm{T}}}
       \end{bmatrix}[\mathbf{C}_{_{a1}}\ \mathbf{D}_{_{a1}}]= \begin{bmatrix}
                                                          \boldsymbol{0}& \boldsymbol{0}\\
                                                          \boldsymbol{0}& \mathbf{D}_{_{a1}}^{^{\mathrm{T}}}\mathbf{D}_{_{a1}}                                                          
                                                         \end{bmatrix}= \begin{bmatrix}
                                                          \boldsymbol{0}& \boldsymbol{0}\\
                                                          \boldsymbol{0}& \mathbf{I}                                                          
                                                         \end{bmatrix}$\ \ from which (\ref{eq:A13}) is rewritten as
\begin{equation}\label{eq:A14}
 \bar{Q}_{_{\mathcal{J}1}}(\mathbf{P}_{_{11}};\lambda)\triangleq\begin{bmatrix}
  \mathbf{A}_{_{a2}}^{^{\mathrm{T}}}\\
  \mathbf{B}_{_{a2}}^{^{\mathrm{T}}}
 \end{bmatrix}\mathbf{P}_{_{11}}[\mathbf{A}_{_{a2}}\ \mathbf{B}_{_{a2}}]-\begin{bmatrix}
                                                                \mathbf{P}_{_{11}}& \boldsymbol{0}\\
                                                                \boldsymbol{0}& \boldsymbol{0} 
                                                                \end{bmatrix} + \lambda\begin{bmatrix}
                                                                \boldsymbol{0}& \boldsymbol{0}\\
                                                                \boldsymbol{0}& \mathbf{I} 
                                                                \end{bmatrix}-\begin{bmatrix}
                                                                              \mathbf{C}_{_{a2}}^{^{\mathrm{T}}}\\
                                                                              \mathbf{D}_{_{a2}}^{^{\mathrm{T}}}
                                                                              \end{bmatrix}[\mathbf{C}_{_{a2}}\ \mathbf{D}_{_{a2}}]>0.
\end{equation}

Note that, in the light of (\ref{eq:A14}), there exists\  $\mathbf{P}_{_{\mathcal{J}1}}=\mathbf{P}_{_{\mathcal{J}1}}^{^{\mathrm{T}}}>0$\ such that\ $\mathbf{A}_{_{a2}}^{^{\mathrm{T}}}\mathbf{P}_{_{11}}\mathbf{A}_{_{a2}}-\mathbf{P}_{_{11}}-\mathbf{C}_{_{a2}}^{^{\mathrm{T}}}\mathbf{C}_{_{a2}}=\mathbf{P}_{_{\mathcal{J}1}}$\ \ so that as\ \ $\rho(\mathbf{A}_{_{a2}})<1$,\ \
$\mathbf{P}_{_{11}}=-\displaystyle\sum_{k=0}^{\infty}(\mathbf{A}_{_{a2}}^{^{\mathrm{T}}})^{^{k}}(\mathbf{P}_{_{\mathcal{J}1}}+\mathbf{C}_{_{a2}}^{^{\mathrm{T}}}\mathbf{C}_{_{a2}})\mathbf{A}_{_{a2}}^{^{k}}=\mathbf{P}_{_{11}}^{^{\mathrm{T}}}<0.$

Finally, noting that\ \  $\mathbf{F}_{_{\mathbf{G}\boldsymbol{y}}}=[\mathbf{F}_{_{\mathbf{G}\boldsymbol{y}}}\ \vdots\ -\bar{\boldsymbol{\mathcal{X}}_{_{\boldsymbol{0}}}}(\mathbf{G})]\begin{bmatrix}
                                                                                                      \mathbf{I}\\
                                                                                                      \boldsymbol{0}
                                                                                                     \end{bmatrix}$,\ a realization of \ 
                                                                                                     $\mathbf{F}_{_{\mathbf{G}\boldsymbol{y}}}$\
                                                                                                     is given by\
$(\mathbf{A}_{_{_{\mathbf{G}\boldsymbol{y}}}}, \mathbf{B}_{_{_{\mathbf{G}\boldsymbol{y}}}}, \mathbf{C}_{_{_{\mathbf{G}\boldsymbol{y}}}}, \mathbf{D}_{_{\mathbf{G}\boldsymbol{y}}})$, \ where \ 
$\mathbf{A}_{_{_{\mathbf{G}\boldsymbol{y}}}}=\mathbf{A}_{_{a2}}$, \ $\mathbf{B}_{_{_{\mathbf{G}\boldsymbol{y}}}}=\mathbf{B}_{_{a2}}$, $\mathbf{C}_{_{_{\mathbf{G}\boldsymbol{y}}}}=\mathbf{C}_{_{a2}}$\ \ and \ \ 
$\mathbf{D}_{_{\mathbf{G}\boldsymbol{y}}}=\mathbf{D}_{_{a2}}$. Then, for $\lambda>0$ (\ref{eq:A14}) is equivalent to 
$$\begin{bmatrix}
   \mathbf{A}_{_{_{\mathbf{G}\boldsymbol{y}}}}^{^{\mathrm{T}}}\\
   \mathbf{B}_{_{_{\mathbf{G}\boldsymbol{y}}}}^{^{\mathrm{T}}}
  \end{bmatrix}(\lambda^{^{-1}}\mathbf{P}_{_{11}})[\mathbf{A}_{_{_{\mathbf{G}\boldsymbol{y}}}}\ \mathbf{B}_{_{_{\mathbf{G}\boldsymbol{y}}}}]- \begin{bmatrix}
                                                                                                  (\lambda^{^{-1}}\mathbf{P}_{_{11}})& \boldsymbol{0}\\
                                                                                                  \boldsymbol{0}&\boldsymbol{0}
                                                                                                  \end{bmatrix}+ \begin{bmatrix}
                                                                                                                \boldsymbol{0}&\boldsymbol{0}\\
                                                                                                                \boldsymbol{0}& \mathbf{I}
                                                                                                                \end{bmatrix}-\lambda^{-1}\begin{bmatrix}
                                                                                                           \mathbf{C}_{_{\mathbf{G}\boldsymbol{y}}}^{^{\mathrm{T}}}\\
                                                                                                            \mathbf{D}_{_{\mathbf{G}\boldsymbol{y}}}^{^{\mathrm{T}}}
                                                              \end{bmatrix}[\mathbf{C}_{_{\mathbf{G}\boldsymbol{y}}}\ \mathbf{D}_{_{\mathbf{G}\boldsymbol{y}}}]>0.
$$
Thus, for $\lambda>0$ \ \ and \ \ $\mathbf{P}=\begin{bmatrix}
                                               \mathbf{P}_{_{11}}&\mathbf{P}_{_{12}}\\
                                               \mathbf{P}_{_{12}}^{^{\mathrm{T}}}& \mathbf{P}_{_{22}}
                                              \end{bmatrix}=\mathbf{P}^{^{\mathrm{T}}}$ such that\ \ $Q_{_{LQ}}(\mathbf{P}; \boldsymbol{\Sigma}_{a}, \mathbf{M}(\lambda))>0$,\ \
                                              $\exists \mathbf{X}=\mathbf{X}^{^{\mathrm{T}}}<0$\ $(\mathbf{X} \triangleq \lambda^{^{-1}}\mathbf{P}_{_{11}})$\ \ such that (\ref{eq:A9}) holds (in which case\  
                                              $\|\lambda^{^{-1}}\mathbf{F}_{_{\mathbf{G}\boldsymbol{y}}}\|_{\infty}<1$)\ or, equivalently, \ $\lambda \in \mathcal{S}_{_{\lambda}}$.
                                              
 To show the converse, let\ $\lambda \in \mathcal{S}_{_{\lambda}}$\ and write 
 $$\bar{Q}_{_{LQ}}(\mathbf{P}; \boldsymbol{\Sigma}_{a}, \mathbf{M}(\lambda))=\bar{Q}_{_{\mathcal{J}}}(\mathbf{P};\boldsymbol{\Sigma}_{a})+\lambda\bar{\mathbf{R}}_{_{1}}^{^{\mathrm{T}}}\bar{\mathbf{R}}_{_{1}}-\bar{\mathbf{R}}_{_{2}}^{^{\mathrm{T}}}\bar{\mathbf{R}}_{_{2}},$$
where\ \ $\bar{\mathbf{R}}_{_{1}}\triangleq\begin{bmatrix}\mathbf{C}_{_{a1}}&\mathbf{D}_{_{a1}}&\boldsymbol{d}_{_{a1}} \end{bmatrix}$ \ and\ $\bar{\mathbf{R}}_{_{2}}\triangleq\begin{bmatrix}\mathbf{C}_{_{a2}}&\mathbf{D}_{_{a2}}&\boldsymbol{d}_{_{a1}}\end{bmatrix}$ or, equivalently,  
\begin{equation}\label{eq:A14a}
\bar{Q}_{_{LQ}}(\mathbf{P}; \boldsymbol{\Sigma}_{_{a}}, \mathbf{M}(\lambda))=\operatorname{diag}(\bar{Q}_{_{\mathcal{J}1}}(\mathbf{P}_{_{11}}, \lambda),-\mathbf{P}_{_{22}})-\begin{bmatrix}
                                                                           \boldsymbol{0}&\mathbf{P}_{_{12}}\\
                                                                           \mathbf{P}_{_{12}}^{^{\mathrm{T}}}&\boldsymbol{0}\\
                                                                     \end{bmatrix}+\mathbf{T}_{_{LQ}}+\mathbf{T}_{_{LQ}}^{^{\mathrm{T}}},
\end{equation}
where\ \ 
\begin{eqnarray*}
\mathbf{T}_{_{LQ}}&=&\begin{bmatrix}
                              \boldsymbol{0}\\
                              \cdots\\
                              \boldsymbol{b}_{_{a2}}^{^{\mathrm{T}}}
                             \end{bmatrix}\mathbf{P}_{_{11}}\begin{bmatrix}\mathbf{A}_{_{2}}&\mathbf{B}_{_{2}}&\vdots&\boldsymbol{0}\end{bmatrix}+(1/2)\begin{bmatrix}
                              \boldsymbol{0}\\
                              \cdots\\
                              \boldsymbol{b}_{_{a2}}^{^{\mathrm{T}}}
                             \end{bmatrix}\begin{bmatrix}\boldsymbol{0}&\vdots&\boldsymbol{b}_{_{a2}}\end{bmatrix}\\\\
&&\ \ \ \ \ \ \ \ \ \ \ \ \ \ \ \ \ \ \ \ \ \ \ \ \ \ \ \ \ \ \ \ \ -\left\{\begin{bmatrix}
                              \boldsymbol{0}\\
                              \cdots\\
                              \boldsymbol{d}_{_{a2}}^{^{\mathrm{T}}}
                             \end{bmatrix}\begin{bmatrix}\mathbf{C}_{_{a2}}&\mathbf{D}_{_{a2}}&\vdots&\boldsymbol{0}\end{bmatrix}+(1/2)\begin{bmatrix}
                              \boldsymbol{0}\\
                              \cdots\\
                              \boldsymbol{d}_{_{a2}}^{^{\mathrm{T}}}
                             \end{bmatrix}\begin{bmatrix}\boldsymbol{0}&\vdots&\boldsymbol{d}_{_{a2}}\end{bmatrix}\right\}
\end{eqnarray*}
so that 
\begin{equation}\label{eq:A14b}
  \mathbf{T}_{_{LQ}}+\mathbf{T}_{_{LQ}}^{^{\mathrm{T}}}=\newcommand*{\temp}{\multicolumn{1}{r|}{}}
\left[\begin{array}{ccc}
 \boldsymbol{0} & \temp& \boldsymbol{e}_{_{1}}(\mathbf{P}_{_{11}}) \\ \cline{1-3}
\boldsymbol{e}_{_{1}}(\mathbf{P}_{_{11}})^{^{\mathrm{T}}} &\temp& \boldsymbol{b}_{_{a2}}^{^{\mathrm{T}}}\mathbf{P}_{_{11}}\boldsymbol{b}_{_{a2}}\\
\end{array}\right]-\left[\begin{array}{ccc}
 \boldsymbol{0} & \temp& \boldsymbol{e}_{_{2}} \\ \cline{1-3}
\boldsymbol{e}_{_{2}}^{^{\mathrm{T}}} &\temp& \boldsymbol{d}_{_{a2}}^{^{\mathrm{T}}}\boldsymbol{d}_{_{a2}}\\
\end{array}\right],
\end{equation}
where\ \ $\boldsymbol{e}_{_{1}}(\mathbf{P}_{_{11}})^{^{\mathrm{T}}}=\boldsymbol{b}_{_{a2}}^{^{\mathrm{T}}}\mathbf{P}_{_{11}}\begin{bmatrix}\mathbf{A}_{_{2}}&\mathbf{B}_{_{2}}\end{bmatrix}$\ and\
$\boldsymbol{e}_{_{2}}^{^{\mathrm{T}}}=\boldsymbol{d}_{_{a2}}^{^{\mathrm{T}}}\begin{bmatrix}\mathbf{C}_{_{a2}}&\mathbf{D}_{_{a2}}\end{bmatrix}$.

As $\lambda\in \mathcal{S}_{_{\lambda}}$, it follows from (\ref{eq:A14}) and the discrete-time bounded real lemma that\ $\exists\ \mathbf{P}_{_{11}}^{^{\mathrm{o}}}$\ such that $\bar{Q}_{_{\mathcal{J}1}}(\mathbf{P}_{_{11}}^{^{\mathrm{o}}};\lambda)>0$. Take\ $\mathbf{P}_{_{12}}^{^{\mathrm{o}}}\triangleq\boldsymbol{e}_{_{1}}(\mathbf{P}_{_{11}}^{^{\mathrm{o}}})-\boldsymbol{e}_{_{2}}$,\ $\mathbf{P}_{_{22}}^{^{\mathrm{o}}}$\ such that\
$\bar{\boldsymbol{q}}_{_{\mathcal{J}}}(\mathbf{P}_{_{22}}^{^{\mathrm{o}}})\triangleq-\mathbf{P}_{_{22}}^{^{\mathrm{o}}}+\boldsymbol{b}_{_{a2}}^{^{\mathrm{T}}}\mathbf{P}_{_{11}}^{^{\mathrm{o}}}\boldsymbol{b}_{_{a2}}-\boldsymbol{d}_{_{a2}}^{^{\mathrm{T}}}\boldsymbol{d}_{_{a2}}>0$\ \ and\ \ $\mathbf{P}^{^{\mathrm{o}}}=\begin{bmatrix}
                                 \mathbf{P}_{_{11}}^{^{\mathrm{o}}}&\mathbf{P}_{_{12}}^{^{\mathrm{o}}}\\
                                 (\mathbf{P}_{_{12}}^{^{\mathrm{o}}})^{^{\mathrm{T}}}&\mathbf{P}_{_{22}}^{^{\mathrm{o}}}\\
                                 \end{bmatrix}$. It then follows from (\ref{eq:A14a}) and (\ref{eq:A14b}) that
$$\bar{Q}_{_{LQ}}(\mathbf{P}^{^{\mathrm{o}}};\boldsymbol{\Sigma}_{_{a}}, \mathbf{M}(\lambda))=\operatorname{diag}(\bar{Q}_{_{\mathcal{J}1}}(\mathbf{P}_{_{11}}^{^{\mathrm{o}}}),\bar{\boldsymbol{q}}_{_{\mathcal{J}}}^{^{\mathrm{o}}}(\mathbf{P}_{_{22}}^{^{\mathrm{o}}}))>0\ \ \Rightarrow\ \ \bar{Q}_{_{LQ}}(\mathbf{P}^{^{\mathrm{o}}};\boldsymbol{\Sigma}_{_{a}}, \mathbf{M}(\lambda))>0.$$
                                               \hfill$\blacksquare$

\noindent
\underline{\textbf{Proof of Proposition \ref{prop:08a}:}}  Consider first the following sets
$$\mathcal{S}_{_{\boldsymbol{\sigma 1}}}\triangleq \{\boldsymbol{\sigma}=(\boldsymbol{\sigma}_{_{\boldsymbol{y}}},
\boldsymbol{\sigma}_{_{\boldsymbol{v}}}):\boldsymbol{\sigma}_{_{\boldsymbol{y}}}>0,\boldsymbol{\sigma}_{_{\boldsymbol{v}}}>0\ \ 
\text{and}\ \ \forall\ \phi\in[0,\ 2\pi],(\mathbf{M}_{_{\boldsymbol{\sigma}}}-\mathbf{F}_{_{\mathbf{G}}}^{*}\mathbf{F}_{_{\mathbf{G}}})(e^{j\phi})\geq 0\}$$
and
$$\mathcal{S}_{_{\boldsymbol{\sigma 2}}}\triangleq \{\boldsymbol{\sigma}=(\boldsymbol{\sigma}_{_{\boldsymbol{y}}}, \boldsymbol{\sigma}_{_{\boldsymbol{v}}}):\boldsymbol{\sigma}_{_{\boldsymbol{y}}}>0,\boldsymbol{\sigma}_{_{\boldsymbol{v}}}>0\ \ \text{and}\ \ \forall\ \phi\in[0,\ 2\pi],(\mathbf{M}_{_{\boldsymbol{\sigma}}}-\mathbf{F}_{_{\mathbf{G}}}^{*}\mathbf{F}_{_{\mathbf{G}}})(e^{j\phi})> 0\}$$
and note that\ $\forall\ \boldsymbol{\sigma}\in\mathcal{S}_{_{\boldsymbol{\sigma}1}}$, \
$\check{\boldsymbol{\varphi}}_{_{\mathbf{D}\infty}}(\boldsymbol{\sigma};\mathbf{G},
\mathbf{H})=0$\ $(\Rightarrow\ \boldsymbol{\varphi}_{_{\mathbf{D}\infty}}(\boldsymbol{\sigma};\mathbf{G}, 
\mathbf{H})=\boldsymbol{\sigma}_{_{\boldsymbol{y}}}\gamma_{_{\boldsymbol{y}}}^{^{2}}
+\boldsymbol{\sigma}_{_{\boldsymbol{v}}}\gamma_{_{\boldsymbol{v}}}^{^{2}}$) whereas for other values of\
$\boldsymbol{\sigma}>0$,\ $\check{\boldsymbol{\varphi}}_{_{\mathbf{D}\infty}}(\boldsymbol{\sigma};\mathbf{G},
\mathbf{H})=-\infty$\ $(\Rightarrow\ \boldsymbol{\varphi}_{_{\mathbf{D}\infty}}(\boldsymbol{\sigma};\mathbf{G}, \mathbf{H})=+\infty$). Hence,\\
$\bar{\boldsymbol{\varphi}}_{_{\mathbf{D}\infty}}(\mathbf{G};\mathbf{H})=\inf\{\boldsymbol{\varphi}_{_{\mathbf{D}\infty}}(\boldsymbol{\sigma};\mathbf{G}, \mathbf{H})=\boldsymbol{\sigma}_{_{\boldsymbol{y}}}\gamma_{_{\boldsymbol{y}}}^{^{2}}+\boldsymbol{\sigma}_{_{\boldsymbol{v}}}\gamma_{_{\boldsymbol{v}}}^{^{2}}: \boldsymbol{\sigma}\in\mathcal{S}_{_{\boldsymbol{\sigma}1}}\}$.
  
The proof is concluded by noting that\ $\mathcal{S}_{_{\boldsymbol{\sigma}2}}$\ ($\subset \mathcal{S}_{_{\boldsymbol{\sigma}1}}$) is dense in\ $\mathcal{S}_{_{\boldsymbol{\sigma}1}}$ -- indeed, if\ $(\boldsymbol{\sigma}_{_{\boldsymbol{y}}}^{^{\mathrm{o}}},\boldsymbol{\sigma}_{_{\boldsymbol{v}}}^{^{\mathrm{o}}})\in \mathcal{S}_{_{\boldsymbol{\sigma}1}}$,\ $\forall\ \varepsilon>0$,\ $\boldsymbol{\sigma}_{_{\varepsilon}}\triangleq(\boldsymbol{\sigma}_{_{\boldsymbol{y}}}^{^{\mathrm{o}}}+\varepsilon,\boldsymbol{\sigma}_{_{\boldsymbol{v}}}^{^{\mathrm{o}}}+\varepsilon)\in \mathcal{S}_{_{\boldsymbol{\sigma}2}}$\ as
$$(\mathbf{M}_{_{\boldsymbol{\sigma}_{_{\varepsilon}}}}-\mathbf{F}_{_{\mathbf{G}}}^{*}\mathbf{F}_{_{\mathbf{G}}})(e^{j\phi})=(\mathbf{M}_{_{\boldsymbol{\sigma}^{^{\mathrm{o}}}}}-\mathbf{F}_{_{\mathbf{G}}}^{*}\mathbf{F}_{_{\mathbf{G}}})(e^{j\phi})+\varepsilon\mathbf{I}>0.$$\hfill$\blacksquare$ 

               
\noindent
\underline{\textbf{Proof of Proposition \ref{prop:10}:}}\ Note that\  
$\forall \phi \in [0,2\pi] ,\ \ 
 (\mathbf{M}_{_{\boldsymbol{\sigma}}}-\mathbf{F}_{_{\mathbf{G}}}^{*}\mathbf{F}_{_{\mathbf{G}}})(e^{j\phi})>0$\ \ $\Leftrightarrow\ \
\forall \phi \in [0,2\pi]$,\break
$\mathbf{I}-\mathbf{M}_{_{\boldsymbol{\sigma}}}^{^{-1/2}}\mathbf{F}_{_{\mathbf{G}}}^{*}\mathbf{F}_{_{\mathbf{G}}}\mathbf{M}_{_{\boldsymbol{\sigma}}}^{^{-1/2}}(e^{j\phi})>0$\ \ $\Leftrightarrow\ \ \|\mathbf{F}_{_{\mathbf{G}}}\mathbf{M}_{_{\boldsymbol{\sigma}}}^{^{-1/2}}\|_{\infty}<1$.

As, in the light of the so-called (discrete-time) bounded-real lemma ([27]), for a realization\break 
$(\mathbf{A}_{_{\mathbf{F}\mathbf{G}}},\mathbf{B}_{_{\mathbf{F}\mathbf{G}}}, \mathbf{C}_{_{\mathbf{F}\mathbf{G}}}, \mathbf{D}_{_{\mathbf{F}\mathbf{G}}})$\ \ of \
$\mathbf{F}_{_{\mathbf{G}}}$, $\rho(\mathbf{A}_{_{\mathbf{F}\mathbf{G}}})<1$,\ \ $\|\mathbf{F}_{_{\mathbf{G}}}\mathbf{M}_{_{\boldsymbol{\sigma}}}^{^{-1/2}}\|_{\infty}<1$\ \ $\Leftrightarrow$\ \ $\exists \mathbf{X}=\mathbf{X}^{^{\mathrm{T}}}>0$,\ \ such that
\begin{equation*}
\begin{bmatrix}
  \mathbf{A}_{_{\mathbf{F}\mathbf{G}}}^{^{\mathrm{T}}}\mathbf{X}\mathbf{A}_{_{\mathbf{F}\mathbf{G}}}-\mathbf{X} 
  & \mathbf{A}_{_{\mathbf{F}\mathbf{G}}}^{^{\mathrm{T}}}\mathbf{X}\mathbf{B}_{_{\mathbf{F}\mathbf{G}}}\mathbf{M}_{_{\boldsymbol{\sigma}}}^{^{-1/2}}\\
  \mathbf{M}_{_{\boldsymbol{\sigma}}}^{^{-1/2}}\mathbf{B}_{_{\mathbf{F}\mathbf{G}}}^{^{\mathrm{T}}}\mathbf{X}\mathbf{A}_{_{\mathbf{F}\mathbf{G}}}
  & \mathbf{M}_{_{\boldsymbol{\sigma}}}^{^{-1/2}}\mathbf{B}_{_{\mathbf{F}\mathbf{G}}}^{^{\mathrm{T}}}\mathbf{X}\mathbf{B}_{_{\mathbf{F}\mathbf{G}}}\mathbf{M}_{_{\boldsymbol{\sigma}}}^{^{-1/2}}
 \end{bmatrix}+ \begin{bmatrix}
                \mathbf{C}_{_{\mathbf{F}\mathbf{G}}}^{^{\mathrm{T}}}\\
                \mathbf{M}_{_{\boldsymbol{\sigma}}}^{^{-1/2}}\mathbf{D}_{_{\mathbf{F}\mathbf{G}}}^{^{\mathrm{T}}}
                \end{bmatrix}[\mathbf{C}_{_{\mathbf{F}\mathbf{G}}}\ \mathbf{D}_{_{\mathbf{F}\mathbf{G}}}\mathbf{M}_{_{\boldsymbol{\sigma}}}^{\mathrm{-1/2}}]
                 -\begin{bmatrix}
     \boldsymbol{0}& \boldsymbol{0}\\
     \boldsymbol{0}& \mathbf{I}
    \end{bmatrix}<0,
\end{equation*}
it follows that the condition on\ \ $(\mathbf{M}_{_{\boldsymbol{\sigma}}}-\mathbf{F}_{_{\mathbf{G}}}^{*}\mathbf{F}_{_{\mathbf{G}}})$\ \
in Proposition \ref{prop:08a} is equivalent to the matrix inequality above. Moreover, if its left-handed side is pre and 
post-multiplied by $\operatorname{diag}(\mathbf{I}_{n_{_{\mathbf{F}\mathbf{G}}}}, \mathbf{M}_{_{\boldsymbol{\sigma}}}^{^{1/2}})$, it is 
converted into 
$$``\exists \mathbf{X} = \mathbf{X}^{^{\mathrm{T}}}>0\ \ \text{such that}\ \ Q_{_{\mathbf{B}\mathbf{R}}}(\mathbf{X}; \boldsymbol{\Sigma}_{a}, \mathbf{M}_{_{\boldsymbol{\sigma}}})<0"$$
where
\begin{equation*}
Q_{_{\mathbf{B}\mathbf{R}}}(\mathbf{P}; \boldsymbol{\Sigma}_{a}, 
\mathbf{M}_{_{\boldsymbol{\sigma}}})\triangleq\begin{bmatrix}
                       \mathbf{A}_{_{\mathbf{F}\mathbf{G}}}^{^{\mathrm{T}}}\mathbf{P}\mathbf{A}_{_{\mathbf{F}\mathbf{G}}}-\mathbf{P}& \vdots &
                       \mathbf{A}_{_{\mathbf{F}\mathbf{G}}}^{^{\mathrm{T}}}\mathbf{P}\mathbf{B}_{_{\mathbf{F}\mathbf{G}}}\\
                       \mathbf{B}_{_{\mathbf{F}\mathbf{G}}}^{^{\mathrm{T}}}\mathbf{P}\mathbf{A}_{_{\mathbf{F}\mathbf{G}}}& \vdots &
                       \mathbf{B}_{_{\mathbf{F}\mathbf{G}}}^{^{\mathrm{T}}}\mathbf{P}\mathbf{B}_{_{\mathbf{F}\mathbf{G}}}
                       \end{bmatrix}
                      + \begin{bmatrix}
                                       \mathbf{C}_{_{\mathbf{F}\mathbf{G}}}^{^{\mathrm{T}}}\\
                                       \mathbf{D}_{_{\mathbf{F}\mathbf{G}}}^{^{\mathrm{T}}}
                                       \end{bmatrix} [\mathbf{C}_{_{\mathbf{F}\mathbf{G}}}\ \mathbf{D}_{_{\mathbf{F}\mathbf{G}}}]
                                      -\begin{bmatrix}
                                         \boldsymbol{0}& \boldsymbol{0}\\
                                         \boldsymbol{0}& \mathbf{M}_{_{\boldsymbol{\sigma}}}
                                        \end{bmatrix}. 
\end{equation*}
Thus, it follows from Proposition \ref{prop:08a} that\ \ $\mathcal{J}_{\infty}(\mathbf{G}; \mathbf{H})$ can be rewritten as stated above.\hfill$\blacksquare$

\bigskip                  
                  
\noindent
\underline{\textbf{Proof of Proposition \ref{prop:11}(a):}}\ \ Note first that \
$Q_{_{\mathbf{B}\mathbf{R}}}(\mathbf{P}, \boldsymbol{\Sigma}_{a}(\boldsymbol{\theta}), \mathbf{M}_{_{\boldsymbol{\sigma}}})<0$\ \ \ $\Leftrightarrow$
$$\begin{bmatrix}
   \mathbf{P}& \boldsymbol{0}\\
    \boldsymbol{0} & \mathbf{M}_{_{\boldsymbol{\sigma}}}
  \end{bmatrix}-\begin{bmatrix}
                \mathbf{A}_{_{\mathbf{F}\mathbf{G}}}^{^{\mathrm{T}}}\\
                \mathbf{B}_{_{\mathbf{F}\mathbf{G}}}^{^{\mathrm{T}}}
                \end{bmatrix}\mathbf{P} [\mathbf{A}_{_{\mathbf{F}\mathbf{G}}}\ \mathbf{B}_{_{\mathbf{F}\mathbf{G}}}]
                -\begin{bmatrix}
                 \mathbf{C}_{_{\mathbf{F}\mathbf{G}}}^{^{\mathrm{T}}}\\
                 \mathbf{D}_{_{\mathbf{F}\mathbf{G}}}^{^{\mathrm{T}}}
                \end{bmatrix}[\mathbf{C}_{_{\mathbf{F}\mathbf{G}}}\ \mathbf{D}_{_{\mathbf{F}\mathbf{G}}}]>0$$
(which implies that $\mathbf{P}>0$ and, hence, $\rho(\mathbf{A}_{_{\mathbf{F}\mathbf{G}}})<1$) \ \ \ $\Leftrightarrow$
 \begin{equation*}
\newcommand*{\temp}{\multicolumn{1}{r|}{}}
\left[\begin{array}{ccc}
\begin{bmatrix}
   \mathbf{P}& \boldsymbol{0}\\
    \boldsymbol{0} & \mathbf{M}_{_{\boldsymbol{\sigma}}}
  \end{bmatrix}-\begin{bmatrix}
                \mathbf{A}_{_{\mathbf{F}\mathbf{G}}}^{^{\mathrm{T}}}\\
                \mathbf{B}_{_{\mathbf{F}\mathbf{G}}}^{^{\mathrm{T}}}
                \end{bmatrix}\mathbf{P} [\mathbf{A}_{_{\mathbf{F}\mathbf{G}}}\ \mathbf{B}_{_{\mathbf{F}\mathbf{G}}}]
   & \temp& \begin{bmatrix}
             \mathbf{C}_{_{\mathbf{F}\mathbf{G}}}^{^{\mathrm{T}}}\\
             \mathbf{D}_{_{\mathbf{F}\mathbf{G}}}^{^{\mathrm{T}}}
            \end{bmatrix}
 \\ \cline{1-3}
[\mathbf{C}_{_{\mathbf{F}\mathbf{G}}}\ \mathbf{D}_{_{\mathbf{F}\mathbf{G}}}] &\temp& \mathbf{I}_{m_{_{\boldsymbol{e}}}} \\
\end{array}\right]>0\ \ \ \Leftrightarrow
\end{equation*}

 \begin{equation*}
\left[\begin{array}{ccc}
\mathbf{P} &\boldsymbol{0}& \mathbf{C}_{_{\mathbf{F}\mathbf{G}}}^{^{\mathrm{T}}}\\
  \boldsymbol{0} &  \mathbf{M}_{_{\boldsymbol{\sigma}}} & \mathbf{D}_{_{\mathbf{F}\mathbf{G}}}^{^{\mathrm{T}}}\\ 
\mathbf{C}_{_{\mathbf{F}\mathbf{G}}}&  \mathbf{D}_{_{\mathbf{F}\mathbf{G}}} & \mathbf{I}_{m_{_{\boldsymbol{e}}}} \\
\end{array}\right]- \begin{bmatrix}
                     \mathbf{A}_{_{\mathbf{F}\mathbf{G}}}^{^{\mathrm{T}}}\\
                     \mathbf{B}_{_{\mathbf{F}\mathbf{G}}}^{^{\mathrm{T}}}\\
                     \boldsymbol{0}_{m_{_{\boldsymbol{e}}}\times n_{_{\mathbf{F}\mathbf{G}}}}
                    \end{bmatrix}\mathbf{P}
[\mathbf{A}_{_{\mathbf{F}\mathbf{G}}}\ \mathbf{B}_{_{\mathbf{F}\mathbf{G}}}\ \boldsymbol{0}_{n_{_{\mathbf{F}\mathbf{G}}}\times m_{_{\boldsymbol{e}}}}]>0 \ \ \ \ \ \Leftrightarrow
\end{equation*}

 \begin{equation*}
\newcommand*{\temp}{\multicolumn{1}{r|}{}}
\left[\begin{array}{ccccc}
\mathbf{P}& \boldsymbol{0}& \mathbf{C}_{_{\mathbf{F}\mathbf{G}}}^{^{\mathrm{T}}}&\temp& \mathbf{A}_{_{\mathbf{F}\mathbf{G}}}^{^{\mathrm{T}}}\\
  \boldsymbol{0} & \mathbf{M}_{_{\boldsymbol{\sigma}}}&  \mathbf{D}_{_{\mathbf{F}\mathbf{G}}}^{^{\mathrm{T}}} &\temp& \mathbf{B}_{_{\mathbf{F}\mathbf{G}}}^{^{\mathrm{T}}}\\ 
\mathbf{C}_{_{\mathbf{F}\mathbf{G}}}& \mathbf{D}_{_{\mathbf{F}\mathbf{G}}} & \mathbf{I}_{m_{_{\boldsymbol{e}}}} & \temp& \boldsymbol{0}_{m_{_{\boldsymbol{e}}}\times n_{_{\mathbf{F}\mathbf{G}}}} \\ \cline{1-5}
\mathbf{A}_{_{\mathbf{F}\mathbf{G}}}&  \mathbf{B}_{_{\mathbf{F}\mathbf{G}}} & \boldsymbol{0}_{n_{\mathbf{F}}\times m_{_{\boldsymbol{e}}}}& \temp &\mathbf{P}^{^{-1}} 
\end{array}\right]>0
\end{equation*}
$\Leftrightarrow$\ \ (pre and post-multiplying the matrix above by $\mathbf{I}_{\mathbf{p}}^{^{\mathrm{T}}}$ and $\mathbf{I}_{\mathbf{p}}$ 
where, $\mathbf{I}_{\mathbf{p}}$ is the column-block permutation matrix given by 
$\mathbf{I}_{\mathbf{p}}=\begin{bmatrix}
                         \boldsymbol{0}& \mathbf{I}&\boldsymbol{0}&\boldsymbol{0}\\
                         \boldsymbol{0}&\boldsymbol{0}&\mathbf{I}&\boldsymbol{0}\\
                         \boldsymbol{0}&\boldsymbol{0}&\boldsymbol{0}&\mathbf{I}\\
                         \mathbf{I}&\boldsymbol{0}&\boldsymbol{0}&\boldsymbol{0}
                         \end{bmatrix}$)\\

\noindent
$\boldsymbol{\psi}(\mathbf{P}, \boldsymbol{\sigma}, \boldsymbol{\theta})>0$, where \ \ $\boldsymbol{\theta}=\begin{bmatrix}
                                                                                                     \mathbf{A}_{_{\mathbf{G}}}&\mathbf{B}_{_{\mathbf{G}}}\\ 
                                                                                                     \mathbf{C}_{_{\mathbf{G}}}&\mathbf{D}_{_{\mathbf{G}}}
                                                                                                     \end{bmatrix}$, \ \
$(\mathbf{A}_{_{\mathbf{G}}}, \mathbf{B}_{_{\mathbf{G}}}, \mathbf{C}_{_{\mathbf{G}}}, \mathbf{D}_{_{\mathbf{G}}})$\ \ is a realization of\
$\mathbf{G}$\ and\
$$\boldsymbol{\psi}(\mathbf{P}, \boldsymbol{\sigma}, \boldsymbol{\theta})\triangleq
\begin{bmatrix}
\mathbf{P}^{^{-1}}&\mathbf{A}_{_{\mathbf{F}\mathbf{G}}}&\mathbf{B}_{_{\mathbf{F}\mathbf{G}}}&\boldsymbol{0}_{n_{_{\mathbf{F}\mathbf{G}}}\times m_{_{\boldsymbol{e}}}}\\
 \mathbf{A}_{_{\mathbf{F}\mathbf{G}}}^{^{\mathrm{T}}}&\mathbf{P}&\boldsymbol{0}_{n_{_{\mathbf{FG}}}\times m_{\boldsymbol{yv}}} & \mathbf{C}_{_{\mathbf{F}\mathbf{G}}}^{^{\mathrm{T}}}\\
 \mathbf{B}_{_{\mathbf{F}\mathbf{G}}}^{^{\mathrm{T}}}&\boldsymbol{0}_{m_{\boldsymbol{yv}}\times n_{_{\mathbf{FG}}}}&\mathbf{M}_{_{\boldsymbol{\sigma}}}& \mathbf{D}_{_{\mathbf{FG}}}^{^{\mathrm{T}}}\\
\boldsymbol{0}_{m_{_{\boldsymbol{e}}}\times n_{_{\mathbf{FG}}}}&\mathbf{C}_{_{\mathbf{FG}}}&\mathbf{D}_{_{\mathbf{FG}}}&\mathbf{I}_{m_{_{\boldsymbol{e}}}}
 \end{bmatrix}
$$
Thus, 
\begin{equation}\label{eq:23}
 Q_{_{\mathbf{B}\mathbf{R}}}(\mathbf{P}, \boldsymbol{\Sigma}_{a}(\boldsymbol{\theta}), \mathbf{M}_{_{\boldsymbol{\sigma}}})<0 \ \ \ 
\Leftrightarrow\ \ \ \boldsymbol{\psi}(\mathbf{P}, \boldsymbol{\sigma}, \boldsymbol{\theta})>0.
\end{equation}\hfill$\blacksquare$


\vspace*{3mm}                
\noindent
\underline{\textbf{Proof of Proposition \ref{prop:11}(b):}}\ Note first that, as \  
$\mathbf{F}_{_{\mathbf{G}}}=\mathbf{H}_{_{\mathbf{I}\boldsymbol{y}}}-\mathbf{G}\mathbf{H}_{_{\mathrm{o}\boldsymbol{z}}}$, \  a realization\
$(\mathbf{A}_{_{\mathbf{FG}}}, \mathbf{B}_{_{\mathbf{FG}}}, \mathbf{C}_{_{\mathbf{FG}}}, \mathbf{D}_{_{\mathbf{FG}}})$\ of \ $\mathbf{F}_{_{\mathbf{G}}}$\ is 
obtained from realizations\ $(\mathbf{A}_{_{\mathbf{I}\boldsymbol{y}}}, \mathbf{B}_{_{\mathbf{I}\boldsymbol{y}}},
\mathbf{C}_{_{\mathbf{I}\boldsymbol{y}}}, \mathbf{D}_{_{\mathbf{I}\boldsymbol{y}}})$\ of\ $\mathbf{H}_{_{\mathbf{I}\boldsymbol{y}}}$\  and\
$(\mathbf{A}_{_{\mathbf{GO}}}, \mathbf{B}_{_{\mathbf{GO}}}, \mathbf{C}_{_{\mathbf{GO}}}, \mathbf{D}_{_{\mathbf{GO}}})$\ of\break 
$\mathbf{F}_{_{\mathbf{GO}}}\triangleq\mathbf{G}\mathbf{H}_{_{\mathrm{o}\boldsymbol{z}}}$\  as
$$\mathbf{A}_{_{\mathbf{FG}}}=\begin{bmatrix}
                            \mathbf{A}_{_{\mathbf{I}\boldsymbol{y}}}&\boldsymbol{0}\\
                            \boldsymbol{0}&\mathbf{A}_{_{\mathbf{GO}}}
                           \end{bmatrix},\ \ \mathbf{B}_{_{\mathbf{FG}}}=\begin{bmatrix}
                                             \mathbf{B}_{_{\mathbf{I}\boldsymbol{y}}}\\
                                             \mathbf{B}_{_{\mathbf{GO}}}
                                             \end{bmatrix},\ \
                                             \mathbf{C}_{_{\mathbf{FG}}}=[\mathbf{C}_{_{\mathbf{I}\boldsymbol{y}}}\ \vdots\ -\mathbf{C}_{_{\mathbf{GO}}}],\ \
                                             \mathbf{D}_{_{\mathbf{FG}}}=\mathbf{D}_{_{\mathbf{I}\boldsymbol{y}}}-\mathbf{D}_{_{\mathbf{GO}}}.
$$
In turn, a realization of\ $\mathbf{F}_{_{\mathbf{GO}}}$\ is obtained from realizations of\ $\mathbf{G}$\ and\ $\mathbf{H}_{_{\mathrm{o}\boldsymbol{z}}}$\ as
$$\mathbf{A}_{_{\mathbf{GO}}}=\begin{bmatrix}
                            \mathbf{A}_{_{\mathrm{o}\boldsymbol{z}}}&\boldsymbol{0}_{n_{\mathrm{o}\boldsymbol{z}}\times n_{_{\mathbf{G}}}}\\
                            \mathbf{B}_{_{\mathbf{G}}}\mathbf{C}_{_{\mathrm{o}\boldsymbol{z}}}&\mathbf{A}_{_{\mathbf{G}}}
                           \end{bmatrix},\ \ \mathbf{B}_{_{\mathbf{GO}}}=\begin{bmatrix}
                                             \mathbf{B}_{_{\mathrm{o}\boldsymbol{z}}}\\
                                             \mathbf{B}_{_{\mathbf{G}}}\mathbf{D}_{_{\mathrm{o}\boldsymbol{z}}}
                                             \end{bmatrix},\ \
                                             \mathbf{C}_{_{\mathbf{GO}}}=[\mathbf{D}_{_{\mathbf{G}}}\mathbf{C}_{_{\mathrm{o}\boldsymbol{z}}}\ \vdots\ \mathbf{C}_{_{\mathbf{G}}}],\ \
                                             \mathbf{D}_{_{\mathbf{GO}}}=\mathbf{D}_{_{\mathbf{G}}}\mathbf{D}_{_{\mathrm{o}\boldsymbol{z}}}
$$
so that
$$
\mathbf{A}_{_{\mathbf{FG}}}=\newcommand*{\temp}{\multicolumn{1}{r|}{}}
\left[\begin{array}{cccc}
\mathbf{A}_{_{\mathbf{I}\boldsymbol{y}}}& \boldsymbol{0}_{n_{\mathbf{I}\boldsymbol{y}\times n_{\mathrm{o}\boldsymbol{z}}}}  & \temp& \boldsymbol{0}_{n_{_{\mathbf{I}\boldsymbol{y}}}\times n_{_{\mathbf{G}}}} \\ 
\boldsymbol{0}&\mathbf{A}_{_{\mathrm{o}\boldsymbol{z}}}&\temp&\boldsymbol{0}_{n_{\mathrm{o}\boldsymbol{z}}\times n_{_{\mathbf{G}}}}\\ \cline{1-4}
\boldsymbol{0}& \mathbf{B}_{_{\mathbf{G}}}\mathbf{C}_{_{\mathrm{o}\boldsymbol{z}}} &\temp& \mathbf{A}_{_{\mathbf{G}}} \\
\end{array}\right], \ \ \mathbf{B}_{_{\mathbf{FG}}}=\left[\begin{array}{c}
                                                 \mathbf{B}_{_{\mathbf{I}\boldsymbol{y}}}\\
                                                 \mathbf{B}_{_{\mathrm{o}\boldsymbol{z}}}\\  \hline
                                                 \mathbf{B}_{_{\mathbf{G}}}\mathbf{D}_{_{\mathrm{o}\boldsymbol{z}}}
                                                 \end{array}\right],\ \
   \mathbf{C}_{_{\mathbf{FG}}}=[\mathbf{C}_{_{\mathbf{I}\boldsymbol{y}}}\ \vdots\ 
   -\mathbf{D}_{_{\mathbf{G}}}\mathbf{C}_{_{\mathrm{o}\boldsymbol{z}}} \vdots\ -\mathbf{C}_{_{\mathbf{G}}}],$$
$ \mathbf{D}_{_{\mathbf{FG}}}=\mathbf{D}_{_{\mathbf{I}\boldsymbol{y}}}-\mathbf{D}_{_{\mathbf{G}}}\mathbf{D}_{_{\mathrm{o}\boldsymbol{z}}}$,
or, equivalently,\\\\
$\mathbf{A}_{_{\mathbf{FG}}}=\newcommand*{\temp}{\multicolumn{1}{r|}{}}
\left[\begin{array}{ccc}
\mathbf{A}_{_{a\boldsymbol{z}}}& \temp& \boldsymbol{0}_{n_{_{a\boldsymbol{z}}}\times n_{_{\mathbf{G}}}} \\ 
 \mathbf{B}_{_{\mathbf{G}}}\mathbf{C}_{_{a\boldsymbol{z}}} &\temp& \mathbf{A}_{_{\mathbf{G}}} \\
\end{array}\right]$, \ \ $\mathbf{B}_{_{\mathbf{FG}}}=\begin{bmatrix}
                                                    \mathbf{B}_{_{a\boldsymbol{z}}}\\
                                                    \mathbf{B}_{_{\mathbf{G}}}\mathbf{D}_{_{\mathrm{o}\boldsymbol{z}}}
                                                   \end{bmatrix}$, \ \ where\ \ 
$\mathbf{A}_{_{a\boldsymbol{z}}}\triangleq \operatorname{diag}(\mathbf{A}_{_{\mathbf{I}\boldsymbol{y}}},\mathbf{A}_{_{\mathrm{o}\boldsymbol{z}}})$,\
$\mathbf{B}_{_{a\boldsymbol{z}}}=\begin{bmatrix}
                  \mathbf{B}_{_{\mathbf{I}\boldsymbol{y}}}\\
                  \mathbf{B}_{_{\mathrm{o}\boldsymbol{z}}}
                 \end{bmatrix}$\ \ and\ \ 
$\mathbf{C}_{_{a\boldsymbol{z}}}=[\boldsymbol{0}_{m_{_{\boldsymbol{v}}}\times n_{_{\mathbf{I}\boldsymbol{y}}}}\ \vdots\ \mathbf{C}_{_{\mathrm{o}\boldsymbol{z}}}]$, \ so that
$$\mathbf{A}_{_{\mathbf{FG}}}=\mathbf{A}_{_{\mathrm{o}}}+\mathbf{A}_{_{L}}(\boldsymbol{\theta}),\ 
\ \mathbf{B}_{_{\mathbf{FG}}}=\mathbf{B}_{_{\mathrm{o}}}+\mathbf{B}_{_{L}}(\boldsymbol{\theta}),\ \ 
\mathbf{C}_{_{\mathbf{FG}}}=\mathbf{C}_{_{\mathrm{o}}}+ \mathbf{C}_{_{L}}(\boldsymbol{\theta}),\ \ 
\mathbf{D}_{_{\mathbf{FG}}}=\mathbf{D}_{_{\mathbf{I}\boldsymbol{y}}}+\mathbf{D}_{_{L}}(\boldsymbol{\theta}),$$
where\ \ $\mathbf{A}_{_{\mathrm{o}}}=\operatorname{diag}(\mathbf{A}_{_{a\boldsymbol{z}}}, \boldsymbol{0}_{n_{_{\mathbf{G}}}\times n_{_{\mathbf{G}}}})$,\ \
$\mathbf{B}_{_{\mathrm{o}}}=\begin{bmatrix}
                 \mathbf{B}_{_{a\boldsymbol{z}}}\\
                 \boldsymbol{0}_{n_{_{\mathbf{G}}}\times m_{\boldsymbol{yv}}}
                \end{bmatrix}$,\ \ 
$\mathbf{C}_{_{\mathrm{o}}}=[\mathbf{C}_{_{\mathbf{I}\boldsymbol{y}}}\ \vdots\ \boldsymbol{0}_{m_{_{\boldsymbol{e}}}\times (n_{_{a\boldsymbol{z}}}+n_{_{\mathbf{G}}})}]$,\
$\mathbf{D}_{_{\mathrm{o}}}=\mathbf{D}_{_{\mathbf{I}\boldsymbol{y}}}$,\break $\mathbf{A}_{_{L}}(\boldsymbol{\theta})=\begin{bmatrix}
                                      \boldsymbol{0}_{n_{_{a\boldsymbol{z}}}\times n_{_{a\boldsymbol{z}}}}& \boldsymbol{0}_{n_{_{a\boldsymbol{z}}}\times n_{_{\mathbf{G}}}}\\
                                      \mathbf{B}_{_{\mathbf{G}}}\mathbf{C}_{_{a\boldsymbol{z}}} & \mathbf{A}_{_{\mathbf{G}}}
                                     \end{bmatrix}$,\ \ $\mathbf{B}_{_{L}}(\boldsymbol{\theta})=\begin{bmatrix}
                                                                                             \boldsymbol{0}_{n_{_{a\boldsymbol{z}}}\times m_{\boldsymbol{yv}}}\\
                                                                                             \mathbf{B}_{_{\mathbf{G}}}\mathbf{D}_{_{\mathrm{o}\boldsymbol{z}}}
                                                                                             \end{bmatrix}$,\
$\mathbf{C}_{_{L}}(\boldsymbol{\theta})=[-\mathbf{D}_{_{\mathbf{G}}}\mathbf{C}_{_{a\boldsymbol{z}}}\ \vdots\ -\mathbf{C}_{_{\mathbf{G}}}]$,\ \ 
$\mathbf{D}_{_{L}}(\boldsymbol{\theta})=-\mathbf{D}_{_{\mathbf{G}}}\mathbf{D}_{_{\mathrm{o}\boldsymbol{z}}}$.\\

As a result, \ $\boldsymbol{\psi}(\mathbf{P}, \boldsymbol{\theta})
=\boldsymbol{\psi}_{_{\mathrm{o}}}(\mathbf{P})+\mathbf{T}_{_{a}}^{^{\mathrm{T}}}\boldsymbol{\theta}\mathbf{T}_{_{b}}
+\mathbf{T}_{_{b}}^{^{\mathrm{T}}}\boldsymbol{\theta}^{^{\mathrm{T}}}\mathbf{T}_{_{a}}$,\\
where
$$\boldsymbol{\psi}_{_{\mathrm{o}}}(\mathbf{P})
=\begin{bmatrix}
\mathbf{P}^{^{-1}}&\mathbf{A}_{_{\mathrm{o}}}&\mathbf{B}_{_{\mathrm{o}}}&\boldsymbol{0}_{n_{\mathbf{F}}\times m_{_{\boldsymbol{e}}}}\\
\mathbf{A}_{_{\mathrm{o}}}^{^{\mathrm{T}}}&\mathbf{P}&\boldsymbol{0}_{n_{_{\mathbf{FG}}}\times m_{\boldsymbol{yv}}}&\mathbf{C}_{_{\mathrm{o}}}^{^{\mathrm{T}}}\\
\mathbf{B}_{_{\mathrm{o}}}&\boldsymbol{0}_{m_{\boldsymbol{yv}}\times n_{_{\mathbf{FG}}}}&\mathbf{M}_{_{\boldsymbol{\sigma}}}&\mathbf{D}_{_{\mathbf{I}\boldsymbol{y}}}^{^{\mathrm{T}}}\\
\boldsymbol{0}_{m_{_{\boldsymbol{e}}}\times n_{_{\mathbf{FG}}}}& \mathbf{C}_{_{\mathrm{o}}}&\mathbf{D}_{_{\mathbf{I}\boldsymbol{y}}}&\mathbf{I}_{m_{_{\boldsymbol{e}}}}
\end{bmatrix}
$$
and
\begin{equation}\label{eq:A16}
 \mathbf{T}_{_{a}}^{^{\mathrm{T}}}\boldsymbol{\theta}\mathbf{T}_{_{b}}=
\newcommand*{\temp}{\multicolumn{1}{r|}{}}
\left[\begin{array}{ccccccccc}
\boldsymbol{0}_{n_{_{a\boldsymbol{z}}}\times n_{_{a\boldsymbol{z}}}}  & \boldsymbol{0}_{n_{_{a\boldsymbol{z}}}\times n_{_{\mathbf{G}}}}&\temp& \boldsymbol{0}_{n_{_{a\boldsymbol{z}}}\times n_{_{a\boldsymbol{z}}}}  & \boldsymbol{0}_{n_{_{a\boldsymbol{z}}}\times n_{_{\mathbf{G}}}}&\temp&\boldsymbol{0}_{n_{_{a\boldsymbol{z}}}\times m_{\boldsymbol{yv}}}& \temp&\boldsymbol{0}_{n_{\mathbf{F}}\times m_{_{\boldsymbol{e}}}}\\ 
\boldsymbol{0}_{n_{_{\mathbf{G}}}\times n_{_{a\boldsymbol{z}}}}& \boldsymbol{0}_{n_{_{\mathbf{G}}}\times n_{_{\mathbf{G}}}} &\temp& \mathbf{B}_{_{\mathbf{G}}} \mathbf{C}_{_{a\boldsymbol{z}}} &\mathbf{A}_{_{\mathbf{G}}}& \temp& \mathbf{B}_{_{\mathbf{G}}}\mathbf{D}_{_{a\boldsymbol{z}}}&\temp&\boldsymbol{0}_{n_{_{\mathbf{G}}}\times m_{_{\boldsymbol{e}}}}\\ \cline{1-9}
\boldsymbol{0}_{n_{_{a\boldsymbol{z}}}\times n_{_{a\boldsymbol{z}}}}  & \boldsymbol{0}_{n_{_{a\boldsymbol{z}}}\times n_{_{\mathbf{G}}}}&\temp& \boldsymbol{0}_{n_{_{a\boldsymbol{z}}}\times n_{_{a\boldsymbol{z}}}}  & \boldsymbol{0}&\temp&\boldsymbol{0}_{n_{_{a\boldsymbol{z}}}\times m_{\boldsymbol{yv}}}& \temp&\boldsymbol{0}_{n_{_{a\boldsymbol{z}}}\times m_{_{\boldsymbol{e}}}}\\ 
\boldsymbol{0}_{n_{_{\mathbf{G}}}\times n_{_{a\boldsymbol{z}}}}& \boldsymbol{0}_{n_{_{\mathbf{G}}}\times n_{_{\mathbf{G}}}} &\temp& \boldsymbol{0} &\boldsymbol{0}_{n_{_{\mathbf{G}}}\times n_{_{\mathbf{G}}}}& \temp& \boldsymbol{0}_{n_{_{\mathbf{G}}}\times m_{\boldsymbol{yv}}}&\temp& \boldsymbol{0}_{n_{_{\mathbf{G}}}\times m_{_{\boldsymbol{e}}}} \\ \cline{1-9}
\boldsymbol{0}_{m_{\boldsymbol{yv}}\times n_{_{a\boldsymbol{z}}}}& \boldsymbol{0}_{m_{\boldsymbol{yv}}\times n_{_{\mathbf{G}}}} &\temp& \boldsymbol{0}_{m_{\boldsymbol{yv}}\times n_{_{a\boldsymbol{z}}}} &\boldsymbol{0}_{m_{\boldsymbol{yv}}\times n_{_{\mathbf{G}}}}& \temp& \boldsymbol{0}_{m_{\boldsymbol{yv}}\times m_{\boldsymbol{yv}}}&\temp&\boldsymbol{0}_{m_{\boldsymbol{yv}}\times m_{\boldsymbol{yv}}} \\ \cline{1-9}
\boldsymbol{0}_{m_{_{\boldsymbol{e}}}\times n_{_{a\boldsymbol{z}}}}&\boldsymbol{0}_{m_{_{\boldsymbol{e}}}\times n_{_{\mathbf{G}}}}&\temp&-\mathbf{D}_{_{\mathbf{G}}}\mathbf{C}_{_{a\boldsymbol{z}}}&-\mathbf{C}_{_{\mathbf{G}}}&\temp&-\mathbf{D}_{_{\mathbf{G}}}\mathbf{D}_{_{\mathrm{o}\boldsymbol{z}}}&\temp&\boldsymbol{0}_{m_{_{\boldsymbol{e}}}\times m_{_{\boldsymbol{e}}}}\\
\end{array}\right]
\end{equation}
\begin{eqnarray*}
\Leftrightarrow\  \mathbf{T}_{_{a}}^{^{\mathrm{T}}}\boldsymbol{\theta}\mathbf{T}_{_{b}}&=&
\begin{bmatrix}
 \boldsymbol{0}_{n_{_{a\boldsymbol{z}}}\times n_{_{\mathbf{G}}}}&\boldsymbol{0}_{n_{_{a\boldsymbol{z}}}\times m_{_{\boldsymbol{e}}}}\\
 \mathbf{I}_{n_{_{\mathbf{G}}}} & \boldsymbol{0}_{n_{_{\mathbf{G}}}\times m_{_{\boldsymbol{e}}}}\\
 \boldsymbol{0}_{n_{_{a\boldsymbol{z}}}\times n_{_{\mathbf{G}}}}& \boldsymbol{0}_{n_{_{a\boldsymbol{z}}}\times m_{_{\boldsymbol{e}}}}\\
 \boldsymbol{0}& \boldsymbol{0}\\
  \boldsymbol{0}& \boldsymbol{0}\\
  \boldsymbol{0}& \mathbf{I}\\
 \end{bmatrix}\begin{bmatrix}
               \mathbf{B}_{_{\mathbf{G}}}\mathbf{C}_{_{a\boldsymbol{z}}}& \mathbf{A}_{_{\mathbf{G}}}&\mathbf{B}_{_{\mathbf{G}}}\mathbf{D}_{_{\mathrm{o}\boldsymbol{z}}}\\
              -\mathbf{D}_{_{\mathbf{G}}}\mathbf{C}_{_{a\boldsymbol{z}}}& -\mathbf{C}_{_{\mathbf{G}}}&-\mathbf{D}_{_{\mathbf{G}}}\mathbf{D}_{_{\mathrm{o}\boldsymbol{z}}}\\
               \end{bmatrix} \begin{bmatrix}
                             \boldsymbol{0}&\boldsymbol{0}&\mathbf{I}&\boldsymbol{0}&\boldsymbol{0}&\boldsymbol{0}\\
                             \boldsymbol{0}&\boldsymbol{0}&\boldsymbol{0}&\mathbf{I}&\boldsymbol{0}&\boldsymbol{0}\\
                             \boldsymbol{0}&\boldsymbol{0}&\boldsymbol{0}&\boldsymbol{0}&\mathbf{I}&\boldsymbol{0}\\
                              \end{bmatrix}  \ \ \ \Leftrightarrow
\end{eqnarray*}                              
\begin{equation*}
\mathbf{T}_{_{a}}^{^{\mathrm{T}}}\boldsymbol{\theta}\mathbf{T}_{_{b}}=
\left[\begin{smallmatrix}
 \boldsymbol{0}&\boldsymbol{0}\\
 \mathbf{I} & \boldsymbol{0}\\
 \boldsymbol{0}& \boldsymbol{0}\\
 \boldsymbol{0}& \boldsymbol{0}\\
  \boldsymbol{0}& \boldsymbol{0}\\
  \boldsymbol{0}& -\mathbf{I}\\
 \end{smallmatrix}\right]\begin{bmatrix}
               \mathbf{A}_{_{\mathbf{G}}}&\mathbf{B}_{_{\mathbf{G}}}\\
               \mathbf{C}_{_{\mathbf{G}}}&\mathbf{D}_{_{\mathbf{G}}}\\
               \end{bmatrix}\begin{bmatrix}
                            \boldsymbol{0}&\mathbf{I}& \boldsymbol{0}\\
                            \mathbf{C}_{_{a\boldsymbol{z}}}& \boldsymbol{0}&\mathbf{D}_{_{\mathrm{o}\boldsymbol{z}}}\\
                            \end{bmatrix} 
                             \begin{bmatrix}
                             \boldsymbol{0}&\boldsymbol{0}&\mathbf{I}&\boldsymbol{0}&\boldsymbol{0}&\boldsymbol{0}\\
                             \boldsymbol{0}&\boldsymbol{0}&\boldsymbol{0}&\mathbf{I}&\boldsymbol{0}&\boldsymbol{0}\\
                             \boldsymbol{0}&\boldsymbol{0}&\boldsymbol{0}&\boldsymbol{0}&\mathbf{I}&\boldsymbol{0}\\
                              \end{bmatrix}.
\end{equation*}       
Thus, with $\boldsymbol{\theta}\triangleq\begin{bmatrix}
               \mathbf{A}_{_{\mathbf{G}}}&\mathbf{B}_{_{\mathbf{G}}}\\
               \mathbf{C}_{_{\mathbf{G}}}&\mathbf{D}_{_{\mathbf{G}}}\\
               \end{bmatrix}$,\ \ $\mathbf{T}_{_{a}}$\ and \ $\mathbf{T}_{_{b}}$\ are given by
               
  $$\mathbf{T}_{_{a}}= \begin{bmatrix}
                             \boldsymbol{0}&\mathbf{I}&\vdots&\boldsymbol{0}&\boldsymbol{0}&\vdots&\boldsymbol{0}&\boldsymbol{0}\\
                             \boldsymbol{0}&\boldsymbol{0}&\vdots&\boldsymbol{0}&\boldsymbol{0}&\vdots&\boldsymbol{0}&-\mathbf{I}\\
                             \end{bmatrix},\ \ \ \ \ \       
   \mathbf{T}_{_{b}}=\begin{bmatrix}
                  \boldsymbol{0}&\boldsymbol{0}&\vdots&\boldsymbol{0}&\mathbf{I}&\vdots&\boldsymbol{0}&\boldsymbol{0}\\
                  \boldsymbol{0}&\boldsymbol{0}&\vdots&\mathbf{C}_{_{a\boldsymbol{z}}}&\boldsymbol{0}&\vdots&\mathbf{D}_{_{\mathrm{o}\boldsymbol{z}}}&\boldsymbol{0}\\
                  \end{bmatrix}. $$\hfill$\blacksquare$ 
\pagebreak 
                  
\noindent
\underline{\textbf{Proof of Proposition \ref{prop:12}(a):}}\ \ It follows from the fact that (in the 
light of Proposition \ref{prop:11}\textbf{(a)}) the constraint 
``$Q_{_{\mathbf{B}\mathbf{R}}}(\cdot)<0$'' in \emph{Prob. $2a$} can be replaced by 
``$\boldsymbol{\psi}(\mathbf{P},\boldsymbol{\sigma},\boldsymbol{\theta})$''.\hfill$\blacksquare$ 

\bigskip
                  
\noindent
\underline{\textbf{Proof of Proposition \ref{prop:12}(b):}}\ \ The first part follows directly from
Proposition \ref{prop:11}\textbf{(b)}, \textbf{(\emph{i})} and \textbf{(\emph{ii})} above. The fact
that\ ``$\rho(\mathbf{A}_{_{\mathbf{G}}})<1$''\ follows from (\ref{eq:26})\ $\Rightarrow$\ 
``$Q_{_{\mathbf{B}\mathbf{R}}}(\mathbf{P}^{^{\mathrm{o}}}, \boldsymbol{\Sigma}_{_{\mathbf{F}\mathbf{G}}}(\boldsymbol{\theta}^{^{\mathrm{o}}}),
\mathbf{M}(\boldsymbol{\sigma}^{^{\mathrm{o}}}))<0$'' (due to Proposition \ref{prop:12})\ $\Rightarrow$\ $\rho(\mathbf{A}_{_{\mathbf{F}\mathbf{G}}})<1$\ $\Rightarrow$\  $\rho(\mathbf{A}_{_{\mathbf{G}}})<1$ (see the 
definition of $\mathbf{A}_{_{\mathbf{F}\mathbf{G}}}$ above). The last part follows from 
Proposition \ref{prop:10}.\hfill$\blacksquare$


\bigskip
                  
\noindent
\underline{\textbf{Proof of Proposition \ref{prop:13}:}}\ \ To apply the Elimination Lemma on the condition
\begin{equation}\label{eq:A17}
 \boldsymbol{\psi}_{_{\mathrm{o}}}(\mathbf{P}, \boldsymbol{\sigma})+\mathbf{T}_{_{a}}^{^{\mathrm{T}}}\boldsymbol{\theta}\mathbf{T}_{_{b}}
 +\mathbf{T}_{_{b}}^{^{\mathrm{T}}}\boldsymbol{\theta}\mathbf{T}_{_{a}}>0
\end{equation}
it is necessary to obtain matrices (say $\mathbf{W}_{_{a}}$ and $\mathbf{W}_{_{b}}$) whose columns form bases form the null spaces of $\mathbf{T}_{_{a}}$ and 
$\mathbf{T}_{_{b}}$.

To this effect, note that 
$$\operatorname{Ker}(\mathbf{T}_{_{a}})\triangleq\{\boldsymbol{v}=[\boldsymbol{v}_{_{_{1}}}^{^{\mathrm{T}}}\mathellipsis\boldsymbol{v}_{_{6}}^{^{\mathrm{T}}}]^{^{\mathrm{T}}}
:\mathbf{T}_{_{a}}\boldsymbol{v}=0\}
=\{\boldsymbol{v}=[\boldsymbol{v}_{_{_{1}}}^{^{\mathrm{T}}}\mathellipsis\boldsymbol{v}_{_{6}}^{^{\mathrm{T}}}]^{^{\mathrm{T}}}: \boldsymbol{v}_{_{_{2}}}=\boldsymbol{0}, 
\boldsymbol{v}_{_{6}}=\boldsymbol{0}\}$$
so that $\mathbf{W}_{_{a}}$ is given by

$$
\mathbf{W}_{_{a}}=\newcommand*{\temp}{\multicolumn{1}{r|}{}}
\left[\begin{array}{cccc}
        \mathbf{I}& \boldsymbol{0}&\boldsymbol{0}&\boldsymbol{0}\\
        \boldsymbol{0}& \boldsymbol{0}&\boldsymbol{0}&\boldsymbol{0}\\ \cline{1-4}
        \boldsymbol{0}& \mathbf{I}&\boldsymbol{0}&\boldsymbol{0}\\
        \boldsymbol{0}& \boldsymbol{0}&\mathbf{I}&\boldsymbol{0}\\ \cline{1-4}
        \boldsymbol{0}& \boldsymbol{0}&\boldsymbol{0}&\mathbf{I}\\
        \boldsymbol{0}& \boldsymbol{0}&\boldsymbol{0}&\boldsymbol{0}\\
        \end{array}\right]\ \ \ \ (
\mathbf{W}_{_{a}}^{^{\mathrm{T}}}=
\left[\begin{array}{cccccccc}
\mathbf{I}& \boldsymbol{0}&\vdots&\boldsymbol{0}&\boldsymbol{0}&\vdots&\boldsymbol{0}&\boldsymbol{0}\\
\boldsymbol{0}&\boldsymbol{0}&\vdots&\mathbf{I}&\boldsymbol{0}&\vdots&\boldsymbol{0}&\boldsymbol{0}\\ 
\boldsymbol{0}&\boldsymbol{0}&\vdots&\boldsymbol{0}&\mathbf{I}&\vdots&\boldsymbol{0}&\boldsymbol{0}\\ 
\boldsymbol{0}&\boldsymbol{0}&\vdots&\boldsymbol{0}&\boldsymbol{0}&\vdots&\mathbf{I}&\boldsymbol{0}\\ 
\end{array}\right])$$
and a necessary condition for (\ref{eq:A17}) to hold is that 
\begin{equation}\label{eq:A18}
 \mathbf{W}_{_{a}}^{^{\mathrm{T}}}\boldsymbol{\psi}_{_{\mathrm{o}}}(\mathbf{P})\mathbf{W}_{_{a}}>0.
\end{equation}

To partition\ $\boldsymbol{\psi}_{_{\mathrm{o}}}(\mathbf{P})$\ in conformity with (\ref{eq:A16}) let\\ 
$\mathbf{P}=\begin{bmatrix}
                         \mathbf{S}&\mathbf{N}\\
                         \mathbf{N}&\mathbf{X}\\
                         \end{bmatrix}$ \ and\  $\mathbf{P}^{^{-1}}=\begin{bmatrix}
                         \mathbf{R}&\mathbf{M}\\
                         \mathbf{M}^{^{\mathrm{T}}}&\mathbf{Z}\\
                         \end{bmatrix}$, \ where\ $\mathbf{S}=\mathbf{S}^{^{\mathrm{T}}}$\ and\ $\mathbf{R}=\mathbf{R}^{^{\mathrm{T}}}$\ are \ $n_{_{a\boldsymbol{z}}}\times n_{_{a\boldsymbol{z}}}$\ 
                         matrices and\ $\mathbf{X}=\mathbf{X}^{^{\mathrm{T}}}$\ and \ $\mathbf{Z}=\mathbf{Z}^{^{\mathrm{T}}}$\ are\ $n_{_{\mathbf{G}}}\times n_{_{\mathbf{G}}}$\
                         matrices ($n_{_{\mathbf{G}}}\geq n_{_{a\boldsymbol{z}}}$). Thus,\ $\boldsymbol{\psi}_{_{\mathrm{o}}}(\mathbf{P})$ can be rewritten as 

$$
 \boldsymbol{\psi}_{_{\mathrm{o}}}(\mathbf{P})=
\newcommand*{\temp}{\multicolumn{1}{r|}{}}
\left[\begin{array}{ccccccccc}
\mathbf{R}  & \mathbf{M}&\temp& \mathbf{A}_{_{a\boldsymbol{z}}}  & \boldsymbol{0}&\temp&\mathbf{B}_{_{a\boldsymbol{z}}}& \temp&\boldsymbol{0}\\ 
\mathbf{M}^{^{\mathrm{T}}}& \mathbf{Z} &\temp& \boldsymbol{0} &\boldsymbol{0}& \temp& \boldsymbol{0}&\temp&\boldsymbol{0}\\ \cline{1-9}
\mathbf{A}_{_{a\boldsymbol{z}}}^{^{\mathrm{T}}}  & \boldsymbol{0}&\temp& \mathbf{S}  & \mathbf{N}&\temp&\boldsymbol{0}& \temp&\widehat{\mathbf{C}}_{_{\mathrm{o}}}^{^{\mathrm{T}}}\\ 
\boldsymbol{0}& \boldsymbol{0} &\temp& \mathbf{N}^{^{\mathrm{T}}} &\mathbf{X}& \temp& \boldsymbol{0}&\temp& \boldsymbol{0} \\ \cline{1-9}
\mathbf{B}_{_{a\boldsymbol{z}}}^{^{\mathrm{T}}}& \boldsymbol{0} &\temp& \boldsymbol{0} &\boldsymbol{0}& \temp& \mathbf{M}_{_{\boldsymbol{\sigma}}}&\temp&\mathbf{D}_{_{\mathbf{I}\boldsymbol{y}}}^{^{\mathrm{T}}} \\ \cline{1-9}
\boldsymbol{0}&\boldsymbol{0}&\temp&\widehat{\mathbf{C}}_{_{\mathrm{o}}}&\boldsymbol{0}&\temp&\mathbf{D}_{_{\mathbf{I}\boldsymbol{y}}}&\temp&\mathbf{I}_{m_{_{\boldsymbol{e}}}}\\
\end{array}\right],\ \ \text{where}\ \  \widehat{\mathbf{C}}_{_{\mathrm{o}}}=[\mathbf{C}_{_{\mathbf{I}\boldsymbol{y}}}\ \vdots\ \boldsymbol{0}_{m_{_{\boldsymbol{e}}}\times n_{a\boldsymbol{z}}}].$$  
  
As a result,

$$\mathbf{W}_{_{a}}^{^{\mathrm{T}}}\boldsymbol{\psi}_{_{\mathrm{o}}}(\mathbf{P})\mathbf{W}_{_{a}}=\mathbf{W}_{_{a}}^{^{\mathrm{T}}}
\begin{bmatrix}
 \mathbf{R}& \mathbf{A}_{_{a\boldsymbol{z}}}&\boldsymbol{0}&\mathbf{B}_{_{a\boldsymbol{z}}}\\
 \mathbf{M}^{^{\mathrm{T}}}&\boldsymbol{0}&\boldsymbol{0}&\boldsymbol{0}\\
 \mathbf{A}_{_{a\boldsymbol{z}}}^{^{\mathrm{T}}}&\mathbf{S}& \mathbf{N}&\boldsymbol{0}\\
 \boldsymbol{0}&\mathbf{N}^{^{\mathrm{T}}}&\mathbf{X}&\boldsymbol{0}\\
 \mathbf{B}_{_{a\boldsymbol{z}}}^{^{\mathrm{T}}}&\boldsymbol{0}&\boldsymbol{0}&\mathbf{M}_{_{\boldsymbol{\sigma}}}\\
 \boldsymbol{0}&\widehat{\mathbf{C}}_{_{\mathrm{o}}}&\boldsymbol{0}&\mathbf{D}_{_{\mathbf{I}\boldsymbol{y}}}
\end{bmatrix}=
\begin{bmatrix}
 \mathbf{R}&\mathbf{A}_{_{a\boldsymbol{z}}}&\boldsymbol{0}&\mathbf{B}_{_{a\boldsymbol{z}}}\\
 \mathbf{A}_{_{a\boldsymbol{z}}}^{^{\mathrm{T}}}&\mathbf{S}&\mathbf{N}&\boldsymbol{0}\\
 \boldsymbol{0}&\mathbf{N}^{^{\mathrm{T}}}&\mathbf{X}&\boldsymbol{0}\\
 \mathbf{B}_{_{a\boldsymbol{z}}}^{^{\mathrm{T}}}&\boldsymbol{0}&\boldsymbol{0}&\mathbf{M}_{_{\boldsymbol{\sigma}}}\\
\end{bmatrix}.
$$

Moreover, pre and post-multipliying\ $\mathbf{W}_{_{a}}^{^{\mathrm{T}}}\boldsymbol{\psi}_{_{\mathrm{o}}}(\mathbf{P})\mathbf{W}_{_{a}}$\ by \
$\mathbf{I}_{c}^{^{\mathrm{T}}}$\ and\ $\mathbf{I}_{c}$,\ where  $\mathbf{I}_{c}=\begin{bmatrix}
                                                                                     \mathbf{I}&\boldsymbol{0}&\boldsymbol{0}&\boldsymbol{0}\\
                                                                                     \boldsymbol{0}&\boldsymbol{0}&\mathbf{I}&\boldsymbol{0}\\
                                                                                     \boldsymbol{0}&\boldsymbol{0}&\boldsymbol{0}&\mathbf{I}\\
                                                                                     \boldsymbol{0}&\mathbf{I}&\boldsymbol{0}&\boldsymbol{0}\\
                                                                                     \end{bmatrix}$\ (column permutation matrix)
$$\mathbf{W}_{_{a}}^{^{\mathrm{T}}}\boldsymbol{\psi}_{_{\mathrm{o}}}(\mathbf{P})\mathbf{W}_{_{a}}>0\ \ \ 
\Leftrightarrow\ \ \ \mathbf{I}_{c}^{^{\mathrm{T}}}\mathbf{W}_{_{a}}\boldsymbol{\psi}_{_{\mathrm{o}}}(\mathbf{P})\mathbf{W}_{_{a}}\mathbf{I}_{c}>0\ \ \ \Leftrightarrow$$  
$$\begin{bmatrix}
   \mathbf{R}&\mathbf{B}_{_{a\boldsymbol{z}}}&\mathbf{A}_{_{a\boldsymbol{z}}}&\boldsymbol{0}\\
   \mathbf{B}_{_{a\boldsymbol{z}}}^{^{\mathrm{T}}}&\mathbf{M}_{_{\boldsymbol{\sigma}}}&\boldsymbol{0}&\boldsymbol{0}\\
   \mathbf{A}_{_{a\boldsymbol{z}}}^{^{\mathrm{T}}}& \boldsymbol{0}& \mathbf{S}&\mathbf{N}\\
   \boldsymbol{0}&\boldsymbol{0}&\mathbf{N}^{^{\mathrm{T}}}&\mathbf{X}\\
   \end{bmatrix}>0\ \ \ \Leftrightarrow\ \ (\text{invoking the Schur complement formula}) 
$$
$$\begin{bmatrix}\mathbf{S}&\mathbf{N}\\
   \mathbf{N}^{^{\mathrm{T}}}&\mathbf{X}\\
  \end{bmatrix}>0\ \ \text{and}\ \ \begin{bmatrix}
   \mathbf{R}&\mathbf{B}_{_{a\boldsymbol{z}}}\\
   \mathbf{B}_{_{a\boldsymbol{z}}}^{^{\mathrm{T}}}&\mathbf{M}_{_{\boldsymbol{\sigma}}}\\
  \end{bmatrix}-\begin{bmatrix}
                \mathbf{A}_{_{a\boldsymbol{z}}}&\boldsymbol{0}\\
                \boldsymbol{0}&\boldsymbol{0}\\
                \end{bmatrix}\begin{bmatrix}
                             \mathbf{S}&\mathbf{N}\\
                             \mathbf{N}^{^{\mathrm{T}}}&\mathbf{X}\\
                             \end{bmatrix}^{^{-1}}\begin{bmatrix}
                                               \mathbf{A}_{_{a\boldsymbol{z}}}^{^{\mathrm{T}}}&\boldsymbol{0}\\
                                               \boldsymbol{0}&\boldsymbol{0}\\
                                               \end{bmatrix}>0 
$$
Moreover, since\ \ $\begin{bmatrix}
                                 \mathbf{S}&\mathbf{N}\\
                                 \mathbf{N}^{^{\mathrm{T}}}&\mathbf{X}\\
                                \end{bmatrix}^{^{-1}}=\mathbf{P}^{^{-1}}=\begin{bmatrix}
                                                                                 \mathbf{R}&\mathbf{M}\\
                                                                                 \mathbf{M}^{^{\mathrm{T}}}&\mathbf{Z}\\
                                                                                \end{bmatrix}$, the last LMI above can be rewritten as  
\begin{equation}\label{eq:A19}
 \begin{bmatrix}
   \mathbf{R}&\mathbf{B}_{_{a\boldsymbol{z}}}\\
   \mathbf{B}_{_{a\boldsymbol{z}}}^{^{\mathrm{T}}}&\mathbf{M}_{_{\boldsymbol{\sigma}}}\\
  \end{bmatrix}-\begin{bmatrix}
                \mathbf{I}\\
                \boldsymbol{0}\\
                \end{bmatrix}\mathbf{A}_{_{a\boldsymbol{z}}}\mathbf{R}\mathbf{A}_{_{a\boldsymbol{z}}}^{^{\mathrm{T}}}[\mathbf{I}\ \boldsymbol{0}]>0\ \ \ \ \Leftrightarrow\ \ \ \ \begin{bmatrix}
 \mathbf{R}-\mathbf{A}_{_{a\boldsymbol{z}}}\mathbf{R}\mathbf{A}_{_{a\boldsymbol{z}}}^{^{\mathrm{T}}}& \mathbf{B}_{_{a\boldsymbol{z}}}\\
 \mathbf{B}_{_{a\boldsymbol{z}}}^{^{\mathrm{T}}}&\mathbf{M}_{_{\boldsymbol{\sigma}}}
 \end{bmatrix}>0.
\end{equation}

It has thus been established that (\ref{eq:A19}) is a necessary condition for (\ref{eq:A17}) to hold. To obtain the corresponding necessary condition pertaining to 
$\mathbf{T}_{_{b}}$, note that\ $\mathbf{D}_{_{\mathrm{o}\boldsymbol{z}}}=[\mathbf{D}_{_{\mathrm{o}\boldsymbol{z}1}}\ \vdots\ \mathbf{D}_{_{\mathrm{o}\boldsymbol{z}2}}]$, \  where\
$\mathbf{D}_{_{\mathrm{o}\boldsymbol{z}1}}=\mathbf{D}_{_{\mathbf{H}\mathrm{o}}}\mathbf{D}_{_{\mathbf{W}\boldsymbol{y}}}^{^{-1}}$\ and\ $\mathbf{D}_{_{\mathrm{o}\boldsymbol{z}2}}=\mathbf{D}_{_{\mathbf{W}\boldsymbol{v}}}^{^{-1}}$\\

\noindent
so that\ \
$\mathbf{T}_{_{b}}=\newcommand*{\temp}{\multicolumn{1}{r|}{}}
\left[\begin{array}{cccccccccc}
        \boldsymbol{0}& \boldsymbol{0}&\vdots&\boldsymbol{0}&\mathbf{I}&\vdots&\boldsymbol{0}&\boldsymbol{0}&\vdots&\boldsymbol{0}\\
        \boldsymbol{0}& \boldsymbol{0}&\vdots&\mathbf{C}_{_{a\boldsymbol{z}}}&\boldsymbol{0}&\vdots&\mathbf{D}_{_{\mathrm{o}\boldsymbol{z}1}}&\mathbf{D}_{_{\mathrm{o}\boldsymbol{z}2}}&\vdots&\boldsymbol{0}\\
        \end{array}\right]$\ \ and
\begin{eqnarray*}
 \operatorname{Ker}(\mathbf{T})&=&\{\boldsymbol{w}=[\boldsymbol{w}_{_{_{1}}}^{^{\mathrm{T}}}\mathellipsis\boldsymbol{w}_{_{7}}^{^{\mathrm{T}}}]^{^{\mathrm{T}}}:
 \boldsymbol{w}_{_{4}}=\boldsymbol{0}\ \ \text{and}\ \ \mathbf{C}_{_{a\boldsymbol{z}}}\boldsymbol{w}_{_{3}}+\mathbf{D}_{_{\mathrm{o}\boldsymbol{z}1}}\boldsymbol{w}_{_{5}}+\mathbf{D}_{_{\mathrm{o}\boldsymbol{z}2}}\boldsymbol{w}_{_{6}}=\boldsymbol{0}\}\ \ \ \Leftrightarrow\\
 \operatorname{Ker}(\mathbf{T})&=&\{\boldsymbol{w}=[\boldsymbol{w}_{_{_{1}}}^{^{\mathrm{T}}}\mathellipsis\boldsymbol{w}_{_{7}}^{^{\mathrm{T}}}]^{^{\mathrm{T}}}:
 \boldsymbol{w}_{_{4}}=\boldsymbol{0}\ \ \text{and}\ \ \boldsymbol{w}_{_{6}}=-\mathbf{D}_{_{\mathrm{o}\boldsymbol{z}2}}^{^{-1}}(\mathbf{C}_{_{a\boldsymbol{z}}}\boldsymbol{w}_{_{3}}+\mathbf{D}_{_{\mathrm{o}\boldsymbol{z}1}}\boldsymbol{w}_{_{5}})\}.
\end{eqnarray*}

 Thus, the columns of 
$$\mathbf{W}_{_{b}}=\begin{bmatrix}
                  \mathbf{I}&\boldsymbol{0}&\boldsymbol{0}&\boldsymbol{0}&\boldsymbol{0}\\
                  \boldsymbol{0}&\mathbf{I}&\boldsymbol{0}&\boldsymbol{0}&\boldsymbol{0}\\
                  \boldsymbol{0}&\boldsymbol{0}&\mathbf{I}&\boldsymbol{0}&\boldsymbol{0}\\
                  \boldsymbol{0}&\boldsymbol{0}&\boldsymbol{0}&\boldsymbol{0}&\boldsymbol{0}\\
                  \boldsymbol{0}&\boldsymbol{0}&\boldsymbol{0}&\mathbf{I}&\boldsymbol{0}\\
                  \boldsymbol{0}&\boldsymbol{0}&\mathbf{T}_{_{63}}&\mathbf{T}_{_{65}}&\boldsymbol{0}\\
                  \boldsymbol{0}&\boldsymbol{0}&\boldsymbol{0}&\boldsymbol{0}&\mathbf{I}\\
                 \end{bmatrix},\ \ \text{where}\ \mathbf{T}_{_{63}}=-\mathbf{D}_{_{\mathrm{o}\boldsymbol{z}2}}^{^{-1}}\mathbf{C}_{_{a\boldsymbol{z}}}\ \ \text{and}\ \ 
                 \mathbf{T}_{_{65}}=-\mathbf{D}_{_{\mathrm{o}\boldsymbol{z}2}}^{^{-1}}\mathbf{D}_{_{\mathrm{o}\boldsymbol{z}1}}, 
$$ 
form a basis for $\operatorname{Ker}(\mathbf{T}_{_{b}})$\ \
$(\mathbf{W}_{_{b}}^{^{\mathrm{T}}}=\begin{bmatrix}
                  \mathbf{I}&\boldsymbol{0}&\boldsymbol{0}&\boldsymbol{0}&\boldsymbol{0}&\boldsymbol{0}&\boldsymbol{0}\\
                  \boldsymbol{0}&\mathbf{I}&\boldsymbol{0}&\boldsymbol{0}&\boldsymbol{0}&\boldsymbol{0}&\boldsymbol{0}\\
                  \boldsymbol{0}&\boldsymbol{0}&\mathbf{I}&\boldsymbol{0}&\boldsymbol{0}&\mathbf{T}_{_{63}}^{^{\mathrm{T}}}&\boldsymbol{0}\\
                  \boldsymbol{0}&\boldsymbol{0}&\boldsymbol{0}&\boldsymbol{0}&\mathbf{I}&\mathbf{T}_{_{65}}^{^{\mathrm{T}}}&\boldsymbol{0}\\
                  \boldsymbol{0}&\boldsymbol{0}&\boldsymbol{0}&\boldsymbol{0}&\boldsymbol{0}&\boldsymbol{0}&\mathbf{I}\\
                  \end{bmatrix}).
$\\
  
Again, it follows from the Elimination Lemma that a necessary condition for (\ref{eq:A17}) is that\
\begin{equation}\label{eq:A20}
\mathbf{W}_{_{b}}^{^{\mathrm{T}}}\boldsymbol{\psi}_{_{\mathrm{o}}}(\mathbf{P})\mathbf{W}_{_{b}}>0.  
\end{equation}
  
 To write (\ref{eq:A20}) explicitly,\ $\boldsymbol{\psi}_{_{\mathrm{o}}}(\mathbf{P})$\ is partitioned in conformity with\ $\mathbf{W}_{_{b}}$,\emph{i.e.},
 with\break $\mathbf{B}_{_{a\boldsymbol{z}}}=[\mathbf{B}_{_{a\boldsymbol{z}1}}\ \vdots\ \mathbf{B}_{_{a\boldsymbol{z}2}}]$\ \ and\ \
 $\mathbf{D}_{_{\mathbf{I}\boldsymbol{y}}}=[\mathbf{D}_{_{\mathbf{I}\boldsymbol{y}1}}\ \vdots\ \mathbf{D}_{_{\mathbf{I}\boldsymbol{y}2}}]$, then
 
$$
 \boldsymbol{\psi}_{_{\mathrm{o}}}(\mathbf{P})=
\newcommand*{\temp}{\multicolumn{1}{r|}{}}
\left[\begin{array}{cccccccccc}
\mathbf{R}  & \mathbf{M}&\temp& \mathbf{A}_{_{a\boldsymbol{z}}}  & \boldsymbol{0}&\temp&\mathbf{B}_{_{a\boldsymbol{z}1}}&\mathbf{B}_{_{a\boldsymbol{z}2}}& \temp&\boldsymbol{0}\\ 
\mathbf{M}^{^{\mathrm{T}}}& \mathbf{Z} &\temp& \boldsymbol{0} &\boldsymbol{0}& \temp& \boldsymbol{0}&\boldsymbol{0}&\temp&\boldsymbol{0}\\ \cline{1-10}
\mathbf{A}_{_{a\boldsymbol{z}}}^{^{\mathrm{T}}}  & \boldsymbol{0}&\temp& \mathbf{S}  & \mathbf{N}&\temp&\boldsymbol{0}&\boldsymbol{0}& \temp&\widehat{\mathbf{C}}_{_{\mathrm{o}}}^{^{\mathrm{T}}}\\ 
\boldsymbol{0}& \boldsymbol{0} &\temp& \mathbf{N}^{^{\mathrm{T}}} &\mathbf{X}& \temp& \boldsymbol{0}& \boldsymbol{0}&\temp& \boldsymbol{0} \\ \cline{1-10}
\mathbf{B}_{_{a\boldsymbol{z}1}}^{^{\mathrm{T}}}& \boldsymbol{0} &\temp& \boldsymbol{0} &\boldsymbol{0}& \temp& \mathbf{M}_{_{\boldsymbol{\sigma}\boldsymbol{y}}}&\boldsymbol{0}&\temp&\mathbf{D}_{_{\mathbf{I}\boldsymbol{y}1}}^{^{\mathrm{T}}} \\ 
\mathbf{B}_{_{a\boldsymbol{z}2}}^{^{\mathrm{T}}}& \boldsymbol{0} &\temp& \boldsymbol{0} &\boldsymbol{0}& \temp& \boldsymbol{0}&\mathbf{M}_{_{\boldsymbol{\sigma}\boldsymbol{v}}}&\temp&\mathbf{D}_{_{\mathbf{I}\boldsymbol{y}2}}^{^{\mathrm{T}}} \\ \cline{1-10} 
\boldsymbol{0}&\boldsymbol{0}&\temp&\widehat{\mathbf{C}}_{_{\mathrm{o}}}&\boldsymbol{0}&\temp&\mathbf{D}_{_{\mathbf{I}\boldsymbol{y}1}}&\mathbf{D}_{_{\mathbf{I}\boldsymbol{y}2}}&\temp&\mathbf{I}_{m_{_{\boldsymbol{e}}}}\\
\end{array}\right].$$  
Thus,
$$
 \boldsymbol{\psi}_{_{\mathrm{o}}}(\mathbf{P})\mathbf{W}_{_{b}}=
\newcommand*{\temp}{\multicolumn{1}{r|}{}}
\left[\begin{array}{cccccccc}
\mathbf{R}  & \mathbf{M}&\temp& \mathbf{A}_{_{a\boldsymbol{z}}}  +\mathbf{B}_{_{a\boldsymbol{z}2}}\mathbf{T}_{_{63}}&\temp&\mathbf{B}_{_{a\boldsymbol{z}1}}+\mathbf{B}_{_{a\boldsymbol{z}2}}\mathbf{T}_{_{65}}& \temp&\boldsymbol{0}\\ 
\mathbf{M}^{^{\mathrm{T}}}& \mathbf{Z} &\temp& \boldsymbol{0} & \temp& \boldsymbol{0}&\temp&\boldsymbol{0}\\ \cline{1-8}
\mathbf{A}_{_{a\boldsymbol{z}}}^{^{\mathrm{T}}}  & \boldsymbol{0}&\temp& \mathbf{S} &\temp&\boldsymbol{0}& \temp&\widehat{\mathbf{C}}_{_{\mathrm{o}}}^{^{\mathrm{T}}}\\ 
\boldsymbol{0}& \boldsymbol{0} &\temp& \mathbf{N}^{^{\mathrm{T}}} & \temp& \boldsymbol{0}&\temp& \boldsymbol{0} \\ \cline{1-8}
\mathbf{B}_{_{a\boldsymbol{z}1}}^{^{\mathrm{T}}}& \boldsymbol{0} &\temp& \boldsymbol{0} & \temp& \mathbf{M}_{_{\boldsymbol{\sigma}\boldsymbol{y}}}&\temp&\mathbf{D}_{_{\mathbf{I}\boldsymbol{y}1}}^{^{\mathrm{T}}} \\ 
\mathbf{B}_{_{a\boldsymbol{z}2}}^{^{\mathrm{T}}}& \boldsymbol{0} &\temp& \mathbf{M}_{_{\boldsymbol{\sigma}\boldsymbol{v}}}\mathbf{T}_{_{63}}& \temp& \mathbf{M}_{_{\boldsymbol{\sigma}\boldsymbol{v}}}\mathbf{T}_{_{65}}&\temp&\mathbf{D}_{_{\mathbf{I}\boldsymbol{y}2}}^{^{\mathrm{T}}} \\ \cline{1-8} 
\boldsymbol{0}&\boldsymbol{0}&\temp&\widehat{\mathbf{C}}_{_{\mathrm{o}}}+\mathbf{D}_{_{\mathbf{I}\boldsymbol{y}2}}\mathbf{T}_{_{63}}&\temp&\mathbf{D}_{_{\mathbf{I}\boldsymbol{y}1}}+\mathbf{D}_{_{\mathbf{I}\boldsymbol{y}2}}\mathbf{T}_{_{65}}&\temp&\mathbf{I}_{m_{_{\boldsymbol{e}}}}\\
\end{array}\right]$$ 
and, hence,  
$$\boldsymbol{\psi}_{_{b}}\triangleq\mathbf{W}_{_{b}}^{^{\mathrm{T}}}\boldsymbol{\psi}_{_{\mathrm{o}}}(\mathbf{P})\mathbf{W}_{_{b}}=
\begin{bmatrix}
\mathbf{R}&\mathbf{M}&\boldsymbol{\psi}_{_{b13}}&\boldsymbol{\psi}_{_{b14}}&\boldsymbol{\psi}_{_{_{b15}}}\\
\mathbf{M}^{^{\mathrm{T}}}&\mathbf{Z}&\boldsymbol{0}&\boldsymbol{0}&\boldsymbol{0}\\
\boldsymbol{\psi}_{_{b13}}^{^{\mathrm{T}}}&\boldsymbol{0}&(\mathbf{S} + \mathbf{M}_{_{3\boldsymbol{v}}})&\boldsymbol{\psi}_{_{b34}}&\boldsymbol{\psi}_{_{b35}}\\
\boldsymbol{\psi}_{_{b14}}^{^{\mathrm{T}}}&\boldsymbol{0}&\boldsymbol{\psi}_{_{b34}}^{^{\mathrm{T}}}&(\mathbf{M}_{_{\boldsymbol{\sigma}\boldsymbol{y}}}+\mathbf{M}_{_{5\boldsymbol{v}}})&\boldsymbol{\psi}_{_{b45}}\\
\boldsymbol{\psi}_{_{_{b15}}}^{^{\mathrm{T}}}&\boldsymbol{0}&\boldsymbol{\psi}_{_{b35}}^{^{\mathrm{T}}}&\boldsymbol{\psi}_{_{b45}}^{^{\mathrm{T}}}&\mathbf{I}_{m_{_{\boldsymbol{e}}}}\\
\end{bmatrix},
$$  
where  
$$\begin{array}{lll}
   \boldsymbol{\psi}_{_{b13}}=\mathbf{A}_{_{a\boldsymbol{z}}}+\mathbf{B}_{_{a\boldsymbol{z}2}}\mathbf{T}_{_{63}}, &\boldsymbol{\psi}_{_{b14}}=\mathbf{B}_{_{a\boldsymbol{z}1}}+\mathbf{B}_{_{a\boldsymbol{z}2}}\mathbf{T}_{_{65}},&\boldsymbol{\psi}_{_{_{b15}}}=\boldsymbol{0},\\
   \boldsymbol{\psi}_{_{b34}}=\mathbf{T}_{_{63}}^{^{\mathrm{T}}}\mathbf{M}_{_{\boldsymbol{\sigma}\boldsymbol{v}}}\mathbf{T}_{_{65}},& \boldsymbol{\psi}_{_{b35}}=\widehat{\mathbf{C}}_{_{\mathrm{o}}}^{^{\mathrm{T}}}+\mathbf{T}_{_{63}}^{^{\mathrm{T}}}\mathbf{D}_{_{\mathbf{I}\boldsymbol{y}2}}^{^{\mathrm{T}}},&\boldsymbol{\psi}_{_{b45}}=\mathbf{D}_{_{\mathbf{I}\boldsymbol{y}1}}^{^{\mathrm{T}}}+\mathbf{T}_{_{65}}^{^{\mathrm{T}}}\mathbf{D}_{_{\mathbf{I}\boldsymbol{y}2}}^{^{\mathrm{T}}},\\
    \mathbf{M}_{_{3\boldsymbol{v}}}=\mathbf{T}_{_{63}}^{^{\mathrm{T}}}\mathbf{M}_{_{\boldsymbol{\sigma}\boldsymbol{v}}}\mathbf{T}_{_{63}},&\mathbf{M}_{_{5\boldsymbol{v}}}=\mathbf{T}_{_{65}}^{^{\mathrm{T}}}\mathbf{M}_{_{\boldsymbol{\sigma}\boldsymbol{v}}}\mathbf{T}_{_{65}}.&
  \end{array}
$$ 
  
Pre and post-multiplying \ $\boldsymbol{\psi}_{_{b}}$\ by \ $\mathbf{I}_{_{b}}^{^{\mathrm{T}}}$\ and\ $\mathbf{I}_{_{b}}$, where the column permutations matrix
$\mathbf{I}_{_{b}}$ is given by  $\mathbf{I}_{_{b}}=\begin{bmatrix}
                                                     \boldsymbol{0}& \boldsymbol{0}&\boldsymbol{0}&\mathbf{I}&\boldsymbol{0}\\
                                                     \boldsymbol{0}& \boldsymbol{0}&\boldsymbol{0}&\boldsymbol{0}&\mathbf{I}\\
                                                     \mathbf{I}& \boldsymbol{0}&\boldsymbol{0}&\boldsymbol{0}&\boldsymbol{0}\\
                                                     \boldsymbol{0}& \mathbf{I}&\boldsymbol{0}&\boldsymbol{0}&\boldsymbol{0}\\
                                                     \boldsymbol{0}& \boldsymbol{0}&\mathbf{I}&\boldsymbol{0}&\boldsymbol{0}\\
                                                    \end{bmatrix}$,\ it follows that\ \ 
$\boldsymbol{\psi}_{_{b}}>0$\ \ \ $\Leftrightarrow$\ \ \ $\mathbf{I}_{_{b}}^{^{\mathrm{T}}}\boldsymbol{\psi}_{_{b}}\mathbf{I}_{_{b}}>0$\ \ \ $\Leftrightarrow$
$$
\begin{bmatrix}
(\mathbf{S}+\mathbf{M}_{_{3\boldsymbol{v}}})&\boldsymbol{\psi}_{_{b34}}&\boldsymbol{\psi}_{_{b35}}&\boldsymbol{\psi}_{_{b13}}^{^{\mathrm{T}}}&\boldsymbol{0}\\
\boldsymbol{\psi}_{_{b34}}^{^{\mathrm{T}}}&(\mathbf{M}_{_{\boldsymbol{\sigma}\boldsymbol{y}}}+\mathbf{M}_{_{5\boldsymbol{v}}})&\boldsymbol{\psi}_{_{b45}}&\boldsymbol{\psi}_{_{b14}}^{^{\mathrm{T}}}&\boldsymbol{0}\\
\boldsymbol{\psi}_{_{b35}}^{^{\mathrm{T}}}&\boldsymbol{\psi}_{_{b45}}^{^{\mathrm{T}}}&\mathbf{I}_{m_{_{\boldsymbol{e}}}}&\boldsymbol{\psi}_{_{_{b15}}}^{^{\mathrm{T}}}&\boldsymbol{0}\\
\boldsymbol{\psi}_{_{b13}}&\boldsymbol{\psi}_{_{b14}}&\boldsymbol{\psi}_{_{_{b15}}}&\mathbf{R}&\mathbf{M}\\
\boldsymbol{0}&\boldsymbol{0}&\boldsymbol{0}&\mathbf{M}^{^{\mathrm{T}}}&\mathbf{Z}\\
\end{bmatrix}>0
$$                                                      
$\Leftrightarrow$\ \ \ (in the light of the Schur complement formula) \ \ $\begin{bmatrix}
                                                                                  \mathbf{R}& \mathbf{M}\\
                                                                                  \mathbf{M}^{^{\mathrm{T}}}&\mathbf{Z}\\
\end{bmatrix}>0$ \ and
 $$\begin{bmatrix}
   (\mathbf{S}+\mathbf{M}_{_{3\boldsymbol{v}}})&\boldsymbol{\psi}_{_{b34}}&\boldsymbol{\psi}_{_{b35}}\\
   \boldsymbol{\psi}_{_{b34}}^{^{\mathrm{T}}}&(\mathbf{M}_{_{\boldsymbol{\sigma}\boldsymbol{y}}}+\mathbf{M}_{_{5\boldsymbol{v}}})&\boldsymbol{\psi}_{_{b45}}\\
   \boldsymbol{\psi}_{_{b35}}^{^{\mathrm{T}}}&\boldsymbol{\psi}_{_{b45}}^{^{\mathrm{T}}}&\mathbf{I}_{m_{_{\boldsymbol{e}}}}\\
  \end{bmatrix}-\begin{bmatrix}
                \boldsymbol{\psi}_{_{b13}}^{^{\mathrm{T}}}&\boldsymbol{0}\\   
                \boldsymbol{\psi}_{_{b14}}^{^{\mathrm{T}}}&\boldsymbol{0}\\
                \boldsymbol{\psi}_{_{_{b15}}}^{^{\mathrm{T}}}&\boldsymbol{0}\\
                \end{bmatrix}\begin{bmatrix}
                                    \mathbf{R}&\mathbf{M}\\
                                    \mathbf{M}^{^{\mathrm{T}}}&\mathbf{Z}\\
                                    \end{bmatrix}^{^{-1}} \begin{bmatrix}
                                                               \boldsymbol{\psi}_{_{b13}}&\vdots&\boldsymbol{\psi}_{_{b14}}&\vdots&\boldsymbol{\psi}_{_{_{b15}}}\\
                                                               \boldsymbol{0}&\vdots&\boldsymbol{0}&\vdots&\boldsymbol{0}\\
                                                               \end{bmatrix}>0
$$ 
Moreover, since $\begin{bmatrix}
                                \mathbf{R}&\mathbf{M}\\
                                \mathbf{M}^{^{\mathrm{T}}}&\mathbf{Z}\\
                               \end{bmatrix}^{^{-1}} =\mathbf{P}=\begin{bmatrix}
                                                                                              \mathbf{S}&\mathbf{N}\\
                                                                                              \mathbf{N}^{^{\mathrm{T}}}&\mathbf{X}\\
                                                                                              \end{bmatrix}$, the last LMI above can be rewritten as  
\begin{equation}\label{eq:A21}
\begin{bmatrix}
   (\mathbf{S}+\mathbf{M}_{_{3\boldsymbol{v}}})&\boldsymbol{\psi}_{_{b34}}&\boldsymbol{\psi}_{_{b35}}\\
   \boldsymbol{\psi}_{_{b34}}^{^{\mathrm{T}}}&(\mathbf{M}_{_{\boldsymbol{\sigma}\boldsymbol{y}}}+\mathbf{M}_{_{5\boldsymbol{v}}})&\boldsymbol{\psi}_{_{b45}}\\
   \boldsymbol{\psi}_{_{b35}}^{^{\mathrm{T}}}&\boldsymbol{\psi}_{_{b45}}^{^{\mathrm{T}}}&\mathbf{I}_{m_{_{\boldsymbol{e}}}}\\
  \end{bmatrix}-\begin{bmatrix}
                \boldsymbol{\psi}_{_{b13}}^{^{\mathrm{T}}}\\   
                \boldsymbol{\psi}_{_{b14}}^{^{\mathrm{T}}}\\
                \boldsymbol{\psi}_{_{_{b15}}}^{^{\mathrm{T}}}\\
                \end{bmatrix}\mathbf{S} \begin{bmatrix}
                                        \boldsymbol{\psi}_{_{b13}}&\vdots&\boldsymbol{\psi}_{_{b14}}&\vdots&\boldsymbol{\psi}_{_{_{b15}}}\\
                                        \end{bmatrix}>0.
\end{equation}  

To complete the proof note that $\boldsymbol{\psi}_{_{_{b15}}}=0$ so that the last LMI holds if and only if
$$\begin{bmatrix}
   \mathbf{S}&\boldsymbol{0}\\
   \boldsymbol{0}&\mathbf{M}_{_{\boldsymbol{\sigma}\boldsymbol{y}}}
  \end{bmatrix}+ \begin{bmatrix}
                 \mathbf{M}_{_{3\boldsymbol{v}}}(\boldsymbol{\sigma})&\boldsymbol{\psi}_{_{b34}}\\
                 \boldsymbol{\psi}_{_{b34}}^{^{\mathrm{T}}}& \mathbf{M}_{_{5\boldsymbol{v}}}(\boldsymbol{\sigma})
                 \end{bmatrix}-\begin{bmatrix}
                 \boldsymbol{\psi}_{_{b35}}\\
                 \boldsymbol{\psi}_{_{b45}}
                 \end{bmatrix}\begin{bmatrix}\boldsymbol{\psi}_{_{b35}}^{^{\mathrm{T}}}&\vdots&\boldsymbol{\psi}_{_{b45}}^{^{\mathrm{T}}}\end{bmatrix}-\begin{bmatrix}
                 \boldsymbol{\psi}_{_{b13}}\\
                 \boldsymbol{\psi}_{_{b14}}
                 \end{bmatrix}\mathbf{S}\begin{bmatrix}\boldsymbol{\psi}_{_{b13}}^{^{\mathrm{T}}}&\vdots&\boldsymbol{\psi}_{_{b14}}^{^{\mathrm{T}}}\end{bmatrix}>0
$$
or, equivalently,
$$\begin{bmatrix}
   \mathbf{S}&\boldsymbol{0}\\
   \boldsymbol{0}&\boldsymbol{\sigma}_{_{\boldsymbol{y}}}\mathbf{I}_{m_{_{\boldsymbol{y}}}}
   \end{bmatrix}+\boldsymbol{\sigma}_{_{\boldsymbol{v}}}\mathbf{E}_{_{b}}^{^{\mathrm{T}}}\mathbf{E}_{_{b}}-\boldsymbol{\psi}_{_{b}}^{^{\mathrm{T}}}\mathbf{S}\boldsymbol{\psi}_{_{b}}-\mathbf{E}_{_{b\mathrm{o}}}^{^{\mathrm{T}}}\mathbf{E}_{_{b\mathrm{o}}}>0,$$

where\ \ $\mathbf{E}_{_{b}}=\begin{bmatrix}\mathbf{T}_{_{63}}&\vdots&\mathbf{T}_{_{65}}\end{bmatrix}= -\mathbf{D}_{_{\mathrm{o}\boldsymbol{z}2}}^{^{-1}}\begin{bmatrix}\mathbf{C}_{_{a\boldsymbol{z}}}&\vdots&\mathbf{D}_{_{\mathrm{o}\boldsymbol{z}1}}\end{bmatrix}=-\mathbf{D}_{_{W\boldsymbol{v}}}\begin{bmatrix}\mathbf{C}_{_{a\boldsymbol{z}}}&\vdots&\mathbf{D}_{_{\mathbf{H}\mathrm{o}}}\mathbf{D}_{_{W\boldsymbol{y}}}^{^{-1}}\end{bmatrix}$,\break  $\mathbf{E}_{_{b\mathrm{o}}}=\begin{bmatrix}\widehat{\mathbf{C}}_{_{\mathrm{o}}}&\vdots&\mathbf{D}_{_{\mathbf{I}\boldsymbol{y}1}}\end{bmatrix}$ (since $\mathbf{D}_{_{\mathbf{I}\boldsymbol{y}2}}=0$)\ and\ $\boldsymbol{\psi}_{_{b}}=\begin{bmatrix}\mathbf{A}_{_{a\boldsymbol{z}}}&\vdots&\mathbf{B}_{_{a\boldsymbol{z}1}}\end{bmatrix}+\mathbf{B}_{_{a\boldsymbol{z}2}}\mathbf{E}_{_{b}}$.\\  

Thus,\ \ \  $\widehat{Q}_{_{b}}(\mathbf{S},\boldsymbol{\sigma})>0\ \ \ \Leftrightarrow\ \ \ Q_{_{b}}(\mathbf{S},\boldsymbol{\sigma})>0$,\\
where\ \ $Q_{_{b}}(\mathbf{S},\boldsymbol{\sigma})=Q_{_{b1}}(\mathbf{S}, \boldsymbol{\sigma})-\mathbf{E}_{_{b\mathrm{o}}}^{^{\mathrm{T}}}\mathbf{E}_{_{b\mathrm{o}}}$,\ $\check{Q}_{_{b1}}(\mathbf{S}, \boldsymbol{\sigma})=\operatorname{diag}(\mathbf{S},\boldsymbol{\sigma}_{_{\boldsymbol{y}}}\mathbf{I}_{m_{_{\boldsymbol{y}}}})+\boldsymbol{\sigma}_{_{\boldsymbol{v}}}\mathbf{E}_{_{b}}^{^{\mathrm{T}}}\mathbf{E}_{_{b}}-\boldsymbol{\psi}_{_{b}}^{^{\mathrm{T}}}\mathbf{S}\boldsymbol{\psi}_{_{b}}$.\hfill$\blacksquare$

\bigskip
                  
\noindent
\underline{\textbf{Proof of Proposition \ref{prop:14}:}}\ \ It has been established that
$$\boldsymbol{\psi}(\mathbf{P},\boldsymbol{\sigma}, \boldsymbol{\theta})>0 \ \ \Leftrightarrow \ \ 
\boldsymbol{\psi}_{_{\mathrm{o}}}(\mathbf{P},\boldsymbol{\sigma})+\mathbf{T}_{_{a}}^{^{\mathrm{T}}}\boldsymbol{\theta}\mathbf{T}_{_{b}}
+\mathbf{T}_{_{b}}^{^{\mathrm{T}}}\boldsymbol{\theta}^{^{\mathrm{T}}}\mathbf{T}_{_{a}}>0,$$
Thus, it follows from Proposition \ref{prop:13} and the Elimination Lemma
that if \ $Q_{_{a}}(\mathbf{R}_{_{\mathrm{o}}})>0$\ and\break $Q_{_{b}}(\mathbf{S}_{_{\mathrm{o}}})>0$,\  then for any\  $\mathbf{P}=\begin{bmatrix}
                                                                                                                   \mathbf{S}_{_{\mathrm{o}}}&\mathbf{N}\\
                                                                                                                   \mathbf{N}^{^{\mathrm{T}}}&\mathbf{X}\\
                                                                                                                  \end{bmatrix}>0$ \ with\                                                                                                               
$\mathbf{P}^{^{-1}}=\begin{bmatrix}
                                    \mathbf{R}_{_{\mathrm{o}}}&\mathbf{M}\\
                                    \mathbf{M}^{^{\mathrm{T}}}&\mathbf{Z}\\
                                    \end{bmatrix}$\ there exists \ $\boldsymbol{\theta}$\ such that\ \
                                    $\boldsymbol{\psi}_{_{\mathrm{o}}}(\mathbf{P})+\mathbf{T}_{_{a}}^{^{\mathrm{T}}}\boldsymbol{\theta}\mathbf{T}_{_{b}}+\mathbf{T}_{_{b}}^{^{\mathrm{T}}}\boldsymbol{\theta}^{^{\mathrm{T}}}\mathbf{T}_{_{a}}>0.$                                                                                                             
                                                                                                                  
To get\ $\mathbf{P}$\ as above from a given pair\ $(\mathbf{S}_{_{\mathrm{o}}}, \mathbf{R}_{_{\mathrm{o}}})$ as above,\ note that\\ (with $\operatorname{dim}(\mathbf{X})=\operatorname{dim}(\mathbf{S}_{_{\mathrm{o}}})$)
$$\mathbf{R}_{_{\mathrm{o}}}=(\mathbf{S}_{_{\mathrm{o}}}-\mathbf{N}\mathbf{X}^{^{-1}}\mathbf{N}^{^{\mathrm{T}}})^{^{-1}}\ \ \ \Leftrightarrow\ \ \ 
\mathbf{S}_{_{\mathrm{o}}}-\mathbf{N}\mathbf{X}^{^{-1}}\mathbf{N}^{^{\mathrm{T}}}=\mathbf{R}_{_{\mathrm{o}}}^{^{-1}}\ \ \ \Leftrightarrow\ \ \ 
\mathbf{S}_{_{\mathrm{o}}}-\mathbf{R}_{_{\mathrm{o}}}^{^{-1}}=\mathbf{N}\mathbf{X}^{^{-1}}\mathbf{N}^{^{\mathrm{T}}}
$$
so that with\ \ $\mathbf{S}_{_{\mathrm{o}}}-\mathbf{R}_{_{\mathrm{o}}}^{^{-1}}\geq 0$ \ \ \ ($\Leftrightarrow$\ \ \ $\begin{bmatrix}
                                                                                       \mathbf{S}_{_{\mathrm{o}}}&\mathbf{I}\\
                                                                                       \mathbf{I}&\mathbf{R}_{_{\mathrm{o}}}
                                                                                       \end{bmatrix}\geq 0$) \ 
                                                                                       \ 
one might take\ \ $\mathbf{Q}_{_{\mathbf{SR}}}=(\mathbf{S}_{_{\mathrm{o}}}-\mathbf{R}_{_{\mathrm{o}}}^{^{-1}})^{^{1/2}}$,\ \ \ $\mathbf{X}=\mathbf{X}^{^{\mathrm{T}}}>0$\ \ and put\ \
$\mathbf{N}\mathbf{X}^{^{-1/2}}=\mathbf{Q}_{_{\mathbf{SR}}}\mathbf{V}$\ \ \ $\Leftrightarrow$\ \ \ $\mathbf{N}=\mathbf{Q}_{_{\mathbf{SR}}}\mathbf{V}\mathbf{X}^{^{1/2}}$, for any unitary $\mathbf{V}$.\\

The fact that\ $\mathbf{P}^{^{\mathrm{o}}}>0$\ follows from the following logical sequence based on
Schur Complements: as\ $\mathbf{X}>0$, $\mathbf{P}^{^{\mathrm{o}}}>0$\ \ $\Leftrightarrow$\ \
$(\mathbf{S}_{_{\mathrm{o}}}-Q_{_{\mathbf{S}\mathbf{R}}}\mathbf{V}\mathbf{X}^{^{1/2}}\mathbf{X}^{^{-1}}\mathbf{X}^{^{1/2}}\mathbf{V}^{^{\mathrm{T}}}Q_{_{\mathbf{S}\mathbf{R}}})>0$\ \ $\Leftrightarrow$\ \ $(\mathbf{S}_{_{\mathrm{o}}}-Q_{_{\mathbf{S}\mathbf{R}}}^{^{2}})>0$\ \ $\Leftrightarrow$\break $(\mathbf{S}_{_{\mathrm{o}}}-(\mathbf{S}_{_{\mathrm{o}}}-\mathbf{R}_{_{\mathrm{o}}}^{^{-1}}))>0$\ \ $\Leftrightarrow$\ \ $\mathbf{R}_{_{\mathrm{o}}}^{^{-1}}>0$\ \
$\Leftrightarrow$\ \ $\mathbf{R}_{_{\mathrm{o}}}>0$.\medskip

\noindent
As a result in the light of the Proposition \ref{prop:12}, \emph{Prob.\ $2$} can be recast as \emph{Prob.\ $2$d}.

With respect to Proposition \ref{prop:14}\textbf{(b)}, note that it follows from Proposition \ref{prop:10} and \ref{prop:11} that
\begin{eqnarray*}
\mathcal{J}_{\infty}(\mathbf{G}(\boldsymbol{\theta}; \mathbf{H}))
&=&\inf\left\{\boldsymbol{\sigma}_{_{\boldsymbol{y}}}\gamma_{_{\boldsymbol{y}}}^{^{2}}+\boldsymbol{\sigma}_{_{\boldsymbol{v}}}\gamma_{_{\boldsymbol{v}}}^{^{2}}:\boldsymbol{\sigma}_{_{\boldsymbol{y}}}>0, \boldsymbol{\sigma}_{_{\boldsymbol{v}}}>0, \mathbf{P}=\mathbf{P}^{^{\mathrm{T}}}>0\right.\\ 
&&\ \ \ \ \ \ \ \ \ \ \ \ \ \ \ \ \ \ \ \ \ \ \ \ \ \ \ \ \left.\text{and}\ \ \boldsymbol{\psi}_{_{\mathrm{o}}}(\mathbf{P}, \boldsymbol{\sigma})+\mathbf{T}_{_{a}}^{^{\mathrm{T}}}\boldsymbol{\theta}\mathbf{T}_{_{b}}+\mathbf{T}_{_{b}}^{^{\mathrm{T}}}\boldsymbol{\theta}^{^{\mathrm{T}}}\mathbf{T}_{_{a}}^{^{\mathrm{T}}}>0\right\}.
\end{eqnarray*}

Therefore, for\  $\mathbf{P}^{^{\mathrm{o}}}$,\ $\boldsymbol{\sigma}^{^{\mathrm{o}}}$\ and\ $\boldsymbol{\theta}$\ as above,\ $\mathcal{J}_{\infty}(\mathbf{G}(\boldsymbol{\theta}; \mathbf{H}))
\leq\boldsymbol{\sigma}_{_{\boldsymbol{y}}}^{^{\mathrm{o}}}\gamma_{_{\boldsymbol{y}}}^{^{2}}+\boldsymbol{\sigma}_{_{\boldsymbol{v}}}^{^{\mathrm{o}}}\gamma_{_{\boldsymbol{v}}}^{^{2}}=\mathcal{J}_{\infty}^{^{\mathrm{o}}}(\mathbf{H})+\varepsilon$ (the proof that 
$\rho(\mathbf{A}_{_{\mathbf{G}}})<1$\ is as in Proposition \ref{prop:12}\textbf{(b)}). \hfill$\blacksquare$

\bigskip

\noindent
\underline{\textbf{Proof of equation (4.20}):} Proceeding as in the \textbf{Proof of Proposition 4.3}, consider first the following sets
\begin{eqnarray*}\mathcal{S}_{_{\boldsymbol{\sigma a}}}(\mathbf{G})&\triangleq& \{\boldsymbol{\sigma}=(\boldsymbol{\sigma}_{_{\boldsymbol{y}}}, \boldsymbol{\sigma}_{_{\boldsymbol{v}}}, \boldsymbol{\sigma}_{_{\boldsymbol{w}}}):\boldsymbol{\sigma}_{_{\boldsymbol{y}}}>0,\boldsymbol{\sigma}_{_{\boldsymbol{v}}}>0,  \boldsymbol{\sigma}_{_{\boldsymbol{w}}}>0\\
&&\ \ \ \ \ \ \ \ \ \ \ \ \ \ \ \ \ \ \ \ \ \ \text{and}\ \ \forall\ \phi\in[0,\ 2\pi],(\mathbf{F}_{_{\mathbf{G}a}}^{*}\mathbf{F}_{_{\mathbf{G}a}}+\boldsymbol{\sigma}_{_{\boldsymbol{w}}}\gamma_{_{\mathbf{H}}}^{^{2}}\boldsymbol{\Gamma}_{_{\mathbf{H}\boldsymbol{y}}}^{a}-\mathbf{M}_{_{\boldsymbol{\sigma}}}^{a})(e^{j\phi})\leq 0\}
\end{eqnarray*}
where\ \ $\boldsymbol{\Gamma}_{_{\mathbf{H}\boldsymbol{y}}}^{a}\triangleq\operatorname{diag}(W_{_{\mathbf{H}\boldsymbol{y}}}^{*}W_{_{\mathbf{H}\boldsymbol{y}}},\boldsymbol{0}, \boldsymbol{0})$, and
\begin{eqnarray*}
\mathcal{S}_{_{\boldsymbol{\sigma b}}}(\mathbf{G})&\triangleq& \{\boldsymbol{\sigma}=(\boldsymbol{\sigma}_{_{\boldsymbol{y}}}, \boldsymbol{\sigma}_{_{\boldsymbol{v}}}, \boldsymbol{\sigma}_{_{\boldsymbol{w}}}):\boldsymbol{\sigma}_{_{\boldsymbol{y}}}>0,\boldsymbol{\sigma}_{_{\boldsymbol{v}}}>0,  \boldsymbol{\sigma}_{_{\boldsymbol{w}}}>0\\
&&\ \ \ \ \ \ \ \ \ \ \ \ \ \ \ \ \ \ \ \ \ \ \text{and}\ \ \forall\ \phi\in[0,\ 2\pi],(\mathbf{F}_{_{\mathbf{G}a}}^{*}\mathbf{F}_{_{\mathbf{G}a}}+\boldsymbol{\sigma}_{_{\boldsymbol{w}}}\gamma_{_{\mathbf{H}}}^{^{2}}\boldsymbol{\Gamma}_{_{\mathbf{H}\boldsymbol{y}}}^{a}-\mathbf{M}_{_{\boldsymbol{\sigma}}}^{a})(e^{j\phi})<0\}
\end{eqnarray*}
and note that\ \ $\forall\ \boldsymbol{\sigma}\in\mathcal{S}_{_{\boldsymbol{\sigma}a}}(\mathbf{G})$,
$\boldsymbol{\varphi}_{_{\mathbf{D}\infty}}^{a}(\boldsymbol{\sigma};\mathbf{G})
=\boldsymbol{\sigma}_{_{\boldsymbol{y}}}\gamma_{_{\boldsymbol{y}}}^{^{2}}
+\boldsymbol{\sigma}_{_{\boldsymbol{v}}}\gamma_{_{\boldsymbol{v}}}^{^{2}}$,\ \ whereas for the other values of\ $\boldsymbol{\sigma}>0$,\ 
$\boldsymbol{\varphi}_{_{\mathbf{D}\infty}}(\boldsymbol{\sigma};\mathbf{G})=+\infty$ -- hence \ \
$\bar{\boldsymbol{\varphi}}_{_{\mathbf{D}\infty}}^{a}(\mathbf{G})
=\inf\{\boldsymbol{\sigma}_{_{\boldsymbol{y}}}\gamma_{_{\boldsymbol{y}}}^{^{2}}
+\boldsymbol{\sigma}_{_{\boldsymbol{y}}}\gamma_{_{\boldsymbol{v}}}^{^{2}}: 
\boldsymbol{\sigma}\in\mathcal{S}_{_{\boldsymbol{\sigma}a}}(\mathbf{G})\}$.

The proof is concluded by noting that\ $\mathcal{S}_{_{\boldsymbol{\sigma}b}}(\mathbf{G})$\
($\subset \mathcal{S}_{_{\boldsymbol{\sigma}a}}(\mathbf{G})$) is dense in\ $\mathcal{S}_{_{\boldsymbol{\sigma}a}}(\mathbf{G})$ --
indeed, if\break $\boldsymbol{\sigma}^{^{\mathrm{o}}}\triangleq(\boldsymbol{\sigma}_{_{\boldsymbol{y}}}^{^{\mathrm{o}}},\boldsymbol{\sigma}_{_{\boldsymbol{v}}}^{^{\mathrm{o}}},\boldsymbol{\sigma}_{_{\boldsymbol{w}}}^{^{\mathrm{o}}})\in \mathcal{S}_{_{\boldsymbol{\sigma}a}}$,\ $\forall\ \varepsilon>0$,\ $\boldsymbol{\sigma}^{^{\varepsilon}}\triangleq(\boldsymbol{\sigma}_{_{\boldsymbol{y}}}^{^{\mathrm{o}}}+2\varepsilon\gamma_{_{\mathbf{H}}}^{^{2}}\|W_{_{\mathbf{H}\boldsymbol{y}}}\|_{_{\infty}}^{^{2}},\boldsymbol{\sigma}_{_{\boldsymbol{v}}}^{^{\mathrm{o}}}+\varepsilon,\boldsymbol{\sigma}_{_{\boldsymbol{w}}}^{^{\mathrm{o}}}+\varepsilon)\in \mathcal{S}_{_{\boldsymbol{\sigma}b}}$\ as
$$(\mathbf{F}_{_{\mathbf{G}a}}^{*}\mathbf{F}_{_{\mathbf{G}a}}+\boldsymbol{\sigma}_{_{\boldsymbol{w}}}^{^{\mathrm{o}}}\gamma_{_{\mathbf{H}}}^{^{2}}\boldsymbol{\Gamma}_{_{\mathbf{H}\boldsymbol{y}}}^{a}-\mathbf{M}_{_{\boldsymbol{\sigma}^{^{\mathrm{o}}}}}^{a}+\operatorname{diag}(\varepsilon\gamma_{_{\mathbf{H}}}^{^{2}}W_{_{\mathbf{H}\boldsymbol{y}}}^{*}W_{_{\mathbf{H}\boldsymbol{y}}}-2\varepsilon\gamma_{_{\mathbf{H}}}^{^{2}}\|W_{_{\mathbf{H}\boldsymbol{y}}}\|_{_{\infty}}^{^{2}}\mathbf{I}_{m_{_{\boldsymbol{y}}}},-\varepsilon\mathbf{I}_{m_{_{\boldsymbol{v}}}},-\varepsilon\mathbf{I}_{_{W}}))(e^{j\phi})<0.$$\hfill$\blacksquare$

\bigskip

\noindent
\underline{\textbf{Proof of equation (\ref{eq:35}}):} Note that\\
$\mathbf{F}_{_{\mathbf{G}W}}=\begin{bmatrix}
                                           \boldsymbol{\sigma}_{_{\boldsymbol{w}}}^{^{1/2}}\gamma_{\boldsymbol{H}}W_{_{\mathbf{H}\boldsymbol{y}}}^{a}\\
                                           \mathbf{F}_{_{\mathbf{G}a}}\\
                                           \end{bmatrix}= \mathbf{H}_{_{Wa}}-\begin{bmatrix}
                                                                           \boldsymbol{0}\\
                                                                           \mathbf{I}
                                                                           \end{bmatrix}\mathbf{G}\mathbf{H}_{_{\mathrm{o}a}}$\ \ and \ \ 
                                                                           $\mathbf{H}_{_{Wa}}=\begin{bmatrix}
                                                                                             \boldsymbol{\sigma}_{_{\boldsymbol{w}}}^{^{1/2}}\gamma_{_{\mathbf{H}}}W_{_{\mathbf{H}\boldsymbol{y}}}^{a}\\
                                                                                             \mathbf{H}_{_{\mathbf{I}a}}\\
                                                                                            \end{bmatrix}$,\\
                                                                                            
\noindent                                                                                            
so that introducing (minimal) realizations\ $\displaystyle\boldsymbol{\Sigma}_{_{Wa}}$\ of $\mathbf{H}_{_{Wa}}$,\ $\boldsymbol{\Sigma}_{_{\mathbf{G}\mathrm{o}a}}$\ \ of\ \
$\begin{bmatrix}
        \boldsymbol{0}\\
        \mathbf{I}\\
       \end{bmatrix}\mathbf{G}\mathbf{H}_{_{\mathrm{o}a}}$\ and\ $\displaystyle\boldsymbol{\Sigma}_{_{\mathrm{o}a}}$\ \ of\ \ 
       $\mathbf{H}_{_{\mathrm{o}a}}$,\  a
realization\ $\displaystyle\boldsymbol{\Sigma}_{_{\mathbf{G}W}}$\ is obtained as follows:\\

$\mathbf{A}_{_{\mathbf{G}W}}=\begin{bmatrix}
                             \mathbf{A}_{_{Wa}}&\boldsymbol{0}\\
                             \boldsymbol{0}&\mathbf{A}_{_{\mathbf{G}\mathrm{o}a}}\\
                             \end{bmatrix}$, \ \ $\mathbf{B}_{_{\mathbf{G}W}}=\begin{bmatrix}
                                                                         \mathbf{B}_{_{Wa}}\\
                                                                         \mathbf{B}_{_{\mathbf{G}\mathrm{o}a}}\\
                                                                         \end{bmatrix}$, \ \ 
$\mathbf{C}_{_{\mathbf{G}W}}=\begin{bmatrix}\mathbf{C}_{_{Wa}}\ \vdots\ -\mathbf{C}_{_{\mathbf{G}\mathrm{o}a}}\end{bmatrix}$,\
$\mathbf{D}_{_{\mathbf{G}W}}=\mathbf{D}_{_{Wa}}-\mathbf{D}_{_{\mathbf{G}\mathrm{o}a}}$,\break  where\ \ 
$\boldsymbol{\Sigma}_{_{Wa}}=\left(\mathbf{A}_{_{Wa}}, \mathbf{B}_{_{Wa}}, \mathbf{C}_{_{Wa}}, \mathbf{D}_{_{Wa}}\right)$\ \ and\ \
$\boldsymbol{\Sigma}_{_{\mathbf{G}\mathrm{o}a}}=\left(\mathbf{A}_{_{\mathbf{G}\mathrm{o}a}}, \mathbf{B}_{_{\mathbf{G}\mathrm{o}a}},
\mathbf{C}_{_{\mathbf{G}\mathrm{o}a}}, \mathbf{D}_{_{\mathbf{G}\mathrm{o}a}}\right)$ \ \ are given by\\

\noindent
$\mathbf{A}_{_{\mathbf{G}\mathrm{o}a}}=\begin{bmatrix}
                           \mathbf{A}_{_{\mathrm{o}a}}&\boldsymbol{0}_{n_{_{\mathrm{o}a}}\times n_{_{\mathbf{G}}}}\\
                           \mathbf{B}_{_{\mathbf{G}}}\mathbf{C}_{_{\mathrm{o}a}} & \mathbf{A}_{_{\mathbf{G}}}\\
                            \end{bmatrix}$, \ \ $\mathbf{B}_{_{\mathbf{G}\mathrm{o}a}}=\begin{bmatrix}
                                                                           \mathbf{B}_{_{\mathrm{o}a}}\\
                                                                           \mathbf{B}_{_{\mathbf{G}}}\mathbf{D}_{_{\mathrm{o}a}}\\
                                                                           \end{bmatrix}$, \ \ 
$\mathbf{C}_{_{\mathbf{G}\mathrm{o}a}}=\begin{bmatrix}
                            \boldsymbol{0}\\
                            \mathbf{I}_{m_{_{\boldsymbol{e}}}}\\
                            \end{bmatrix}\begin{bmatrix}\mathbf{D}_{_{\mathbf{G}}}\mathbf{C}_{_{\mathrm{o}a}}\ \vdots\ \mathbf{C}_{_{\mathbf{G}}}\end{bmatrix}
$,\ \
$\mathbf{D}_{_{\mathbf{G}\mathrm{o}a}}=
\left[\begin{array}{c}
 \boldsymbol{0} \\ \cline{1-1}
\mathbf{I}_{m_{_{\boldsymbol{e}}}}\\
\end{array}\right]\mathbf{D}_{_{\mathbf{G}}}\mathbf{D}_{_{\mathrm{o}a}}$,\ \ $\mathbf{A}_{_{Wa}}=\begin{bmatrix}
                             \mathbf{A}_{_{W\mathbf{H}\boldsymbol{y}}}&\boldsymbol{0}\\
                             \boldsymbol{0}&\mathbf{A}_{_{\mathbf{H}\mathbf{I}a}}\\
                             \end{bmatrix}$, \ \ $\mathbf{B}_{_{Wa}}=\begin{bmatrix}
                                                                         \mathbf{B}_{_{W\mathbf{H}\boldsymbol{y}}}^{a}\\
                                                                         \mathbf{B}_{_{\mathbf{H}\mathbf{I}a}}\\
                                                                         \end{bmatrix}$, \ \ $\mathbf{C}_{_{Wa}}=\begin{bmatrix}
                                                                         \mathbf{C}_{_{Wa}}^{^{\boldsymbol{\sigma}}}\\
                                                                         \mathbf{C}_{_{Wa}}^{^{\boldsymbol{e}}}\\
                                                                         \end{bmatrix}$,\ \
$\mathbf{D}_{_{Wa}}=\begin{bmatrix}
                                                                         \mathbf{D}_{_{Wa}}^{^{\boldsymbol{\sigma}}}\\
                                                                         \mathbf{D}_{_{Wa}}^{^{\boldsymbol{e}}}\\
                                                                         \end{bmatrix}$,\\ 
$\mathbf{C}_{_{Wa}}^{^{\boldsymbol{\sigma}}}
=\begin{bmatrix}\boldsymbol{\sigma}_{_{\boldsymbol{w}}}^{^{1/2}}\gamma_{_{\mathbf{H}}}\mathbf{C}_{_{W\mathbf{H}\boldsymbol{y}}}& \vdots & 
\boldsymbol{0}_{m_{_{\boldsymbol{y}}}\times n_{_{\mathbf{A}\mathbf{H}\mathbf{I}a}}} \end{bmatrix}$,\ \
$\mathbf{C}_{_{Wa}}^{^{\boldsymbol{e}}}
=\begin{bmatrix}\boldsymbol{0}_{m_{_{\boldsymbol{e}}}\times n_{_{\mathbf{A}W\mathbf{H}\boldsymbol{y}}}}& \vdots & 
\mathbf{C}_{_{\mathbf{H}\mathbf{I}a}}\end{bmatrix}$,\ \ 
$\mathbf{D}_{_{Wa}}^{^{\boldsymbol{\sigma}}}=\boldsymbol{\sigma}_{_{Wa}}^{^{1/2}}\gamma_{_{\mathbf{H}}}\mathbf{D}_{_{W\mathbf{H}\boldsymbol{y}}}^{a}$,\break
$\mathbf{D}_{_{Wa}}^{\boldsymbol{e}}=\mathbf{D}_{_{\mathbf{H}\mathbf{I}a}}$,\ 
--\ \  $\boldsymbol{\Sigma}_{_{Wa}}$\ \ and\ \ $\boldsymbol{\Sigma}_{_{\mathbf{G}\mathrm{o}a}}$\ \ depend on the problem data\ \ $\left(\mathbf{H}_{_{\boldsymbol{0}}}, 
\mathbf{H}_{_{\mathbf{I}}}, W_{\boldsymbol{y}}, W_{\boldsymbol{v}}, W_{_{\mathbf{H}}}\right)$\ \ through the realizations of\ \ 
$\mathbf{H}_{_{oa}}=\begin{bmatrix}\mathbf{H}_{_{\boldsymbol{0}}}W_{\boldsymbol{y}}^{^{-1}}& \vdots & W_{\boldsymbol{v}}^{^{-1}} & \vdots & 
\mathbf{I}_{m_{_{\boldsymbol{v}}}}\end{bmatrix}$,\ \ 
$\mathbf{H}_{_{\mathbf{I}a}}
=\mathbf{H}_{_{\mathbf{I}}}W_{\boldsymbol{y}}^{^{-1}}\begin{bmatrix}\mathbf{I}_{m_{_{\boldsymbol{y}}}}& \vdots & 
\boldsymbol{0}_{m_{_{\boldsymbol{y}}}\times m_{_{\boldsymbol{v}}}} & \vdots & \boldsymbol{0}_{m_{_{\boldsymbol{y}}}\times m_{_{\boldsymbol{v}}}}\end{bmatrix}$,\break
and\ \ $W_{_{\mathbf{H}\boldsymbol{y}}}^{a}
=W_{_{\mathbf{H}\boldsymbol{y}}}\begin{bmatrix}\mathbf{I}_{m_{_{\boldsymbol{y}}}}& \vdots & 
\boldsymbol{0}_{m_{_{\boldsymbol{y}}}\times m_{_{\boldsymbol{v}}}} & \vdots & \boldsymbol{0}_{m_{_{\boldsymbol{y}}}\times m_{_{\boldsymbol{v}}}}\end{bmatrix}$,\ \  
so that\ $\mathbf{D}_{_{W\mathbf{H}\boldsymbol{y}}}^{a}=\begin{bmatrix}\mathbf{D}_{_{W\mathbf{H}\boldsymbol{y}}}& \vdots & 
\boldsymbol{0}_{m_{_{\boldsymbol{y}}}\times m_{_{\boldsymbol{v}}}} & \vdots & \boldsymbol{0}_{m_{_{\boldsymbol{y}}}\times m_{_{\boldsymbol{v}}}}\end{bmatrix}$\ \
and\ \ 
$\mathbf{D}_{_{\mathbf{H}\mathbf{I}a}}=\begin{bmatrix}\mathbf{D}_{_{\mathbf{H}\mathbf{I}}}\mathbf{D}_{_{W_{\boldsymbol{y}}}}^{^{-1}}& \vdots & 
\boldsymbol{0}_{m_{_{\boldsymbol{y}}}\times m_{_{\boldsymbol{v}}}} & \vdots & \boldsymbol{0}_{m_{_{\boldsymbol{y}}}\times m_{_{\boldsymbol{v}}}}\end{bmatrix}$.\

As a result, \ \ $\mathbf{A}_{a}=\operatorname{diag}(\mathbf{A}_{_{a1}}, \boldsymbol{0}_{n_{_{\mathbf{G}}}\times n_{_{\mathbf{G}}}})$, \ \ 
$\mathbf{A}_{_{a1}}=\operatorname{diag}(\mathbf{A}_{_{Wa}}, \mathbf{A}_{_{\mathrm{o}a}})$, \\

$\mathbf{A}_{_{L}}(\boldsymbol{\theta})=\begin{bmatrix}
                                     \boldsymbol{0}&\boldsymbol{0}\\
                                     \mathbf{B}_{_{\mathbf{G}}}\widehat{\mathbf{C}}_{_{\mathrm{o}a}}&\mathbf{A}_{_{\mathbf{G}}}
                                     \end{bmatrix}=\begin{bmatrix}
                                                   \boldsymbol{0}&\boldsymbol{0}\\
                                                   \mathbf{I}&\boldsymbol{0}\\
                                                   \end{bmatrix}\boldsymbol{\theta}\begin{bmatrix}
                                                                                   \boldsymbol{0}&\mathbf{I}\\
                                                                                   \widehat{\mathbf{C}}_{_{\mathrm{o}a}}& \boldsymbol{0}
                                                                                   \end{bmatrix}$, \ \                                                                                    
$\widehat{\mathbf{C}}_{_{\mathrm{o}a}}=\begin{bmatrix}\boldsymbol{0}_{m_{_{\boldsymbol{v}}} \times n_{_{Wa}}}\ \vdots\ \mathbf{C}_{_{\mathrm{o}a}}\end{bmatrix}$,\
$\mathbf{B}_{a}=\begin{bmatrix}
                 \mathbf{B}_{_{a1}}\\
                 \boldsymbol{0}_{n_{_{\mathbf{G}}}\times (m_{_{\boldsymbol{y}}}+2m_{_{\boldsymbol{v}}})}\\
                \end{bmatrix}$,\ \  
                $\mathbf{B}_{_{a1}}=\begin{bmatrix}
                 \mathbf{B}_{_{Wa}}\\
                 \mathbf{B}_{_{\mathrm{o}a}}\\
                \end{bmatrix}$, \ \ 
$\mathbf{B}_{_{L}}(\boldsymbol{\theta})=\begin{bmatrix}
                                                   \boldsymbol{0}&\boldsymbol{0}\\
                                                   \mathbf{I}&\boldsymbol{0}\\
                                                   \end{bmatrix}\boldsymbol{\theta}\begin{bmatrix}
                                                                                   \boldsymbol{0}\\
                                                                                   \mathbf{D}_{_{\mathrm{o}a}}\\
                                                                                   \end{bmatrix}=\left[\begin{array}{c}
 \boldsymbol{0} \\
\mathbf{B}_{_{\mathbf{G}}}\mathbf{D}_{_{\mathrm{o}a}}\\
\end{array}\right]$,\ 
$\mathbf{C}_{a}=\begin{bmatrix}\mathbf{C}_{_{Wa}}\ \vdots\ \boldsymbol{0}_{(m_{_{\boldsymbol{y}}}+m_{_{\boldsymbol{e}}}) \times n_{_{\mathrm{o}a}}}\ \vdots\ 
\boldsymbol{0}_{(m_{_{\boldsymbol{y}}}+m_{_{\boldsymbol{e}}}) \times n_{_{\mathbf{G}}}}\end{bmatrix}$, \ \
$\mathbf{C}_{_{L}}(\boldsymbol{\theta})=-\begin{bmatrix}
                                     \boldsymbol{0}\\
                                     \mathbf{I}_{m_{_{\boldsymbol{e}}}}\\
                                      \end{bmatrix}
                                      \begin{bmatrix}\boldsymbol{0}\ \vdots\ \mathbf{D}_{_{\mathbf{G}}}\mathbf{C}_{_{\mathrm{o}a}}\ \vdots\ \mathbf{C}_{_{\mathbf{G}}}\end{bmatrix}$,
                                    \ \  \emph{i.e.},\                                       
$\mathbf{C}_{_{L}}(\boldsymbol{\theta})=-\begin{bmatrix}
                                     \boldsymbol{0}&\boldsymbol{0}\\
                                     \boldsymbol{0}&\mathbf{I}\\
                                      \end{bmatrix}\boldsymbol{\theta}\left[\begin{array}{ccccc}                                   
                                                                      \boldsymbol{0} & \vdots  & \boldsymbol{0} & \vdots & \mathbf{I}\\
                                                                      \boldsymbol{0}& \vdots & \mathbf{C}_{_{\mathrm{o}a}} &  \vdots & \boldsymbol{0}\\
                                                                      \end{array}\right]$, \ \ 
$\mathbf{D}_{a}=\mathbf{D}_{_{Wa}}$, \ \ $\mathbf{D}_{_{L}}(\boldsymbol{\theta})=-\begin{bmatrix}
                                                                             \boldsymbol{0}&\boldsymbol{0}\\
                                                                             \boldsymbol{0}&\mathbf{I}_{m_{_{\boldsymbol{e}}}}
                                                                            \end{bmatrix}\boldsymbol{\theta}\begin{bmatrix}
                                                                                                            \boldsymbol{0}\\
                                                                                                            \mathbf{D}_{_{\mathrm{o}a}}
                                                                                                            \end{bmatrix}$. \\                                                                     

Thus, partitioning\  $\mathbf{P}$\ and\  $\mathbf{P}^{^{-1}}$\ conformally with\  $\mathbf{A}_{a}$, \emph{i.e.},\ $\mathbf{P}=\begin{bmatrix}
                                                                                                                            \mathbf{S}&\mathbf{N}\\
                                                                                                                            \mathbf{N}^{^{\mathrm{T}}}&\mathbf{X}\\
                                                                                                                           \end{bmatrix}$\ \
and\break 
$\mathbf{P}^{^{-1}}=\begin{bmatrix}
                          \mathbf{R}&\mathbf{M}\\
                          \mathbf{M}^{^{\mathrm{T}}}&\mathbf{Z}\\
                          \end{bmatrix}$,\\

$$
 \boldsymbol{\psi}_{a}^{\mathrm{o}}(\mathbf{P}, \boldsymbol{\sigma})=
\newcommand*{\temp}{\multicolumn{1}{r|}{}}
\left[\begin{array}{ccccccccccc}
\mathbf{R}  & \mathbf{M}&\temp& \mathbf{A}_{_{a1}}  & \boldsymbol{0}&\temp&\mathbf{B}_{_{a1}}^{^{\boldsymbol{z}}}& \boldsymbol{0} & \temp&\boldsymbol{0} &\boldsymbol{0}\\ 
\mathbf{M}^{^{\mathrm{T}}}& \mathbf{Z} &\temp& \boldsymbol{0} &\boldsymbol{0}& \temp& \boldsymbol{0} & \boldsymbol{0} & \temp &\boldsymbol{0} & \boldsymbol{0}\\ \cline{1-11}
\mathbf{A}_{_{a1}}^{^{\mathrm{T}}}  & \boldsymbol{0}&\temp& \mathbf{S}  & \mathbf{N}&\temp&\boldsymbol{0} & \boldsymbol{0}& \temp& (\mathbf{C}_{_{a1}}^{^{\boldsymbol{\sigma}}})^{^{\mathrm{T}}} & (\mathbf{C}_{_{a1}}^{^{\boldsymbol{e}}})^{^{\mathrm{T}}}\\ 
\boldsymbol{0}& \boldsymbol{0} &\temp& \mathbf{N}^{^{\mathrm{T}}} &\mathbf{X}& \temp& \boldsymbol{0} & \boldsymbol{0}&\temp& \boldsymbol{0} & \boldsymbol{0} \\ \cline{1-11}
(\mathbf{B}_{_{a1}}^{\boldsymbol{z}})^{^{\mathrm{T}}}& \boldsymbol{0} &\temp& \boldsymbol{0} &\boldsymbol{0}& \temp& \mathbf{M}_{_{\boldsymbol{\sigma}}}^{\boldsymbol{z}} & \boldsymbol{0} &\temp&(\mathbf{D}_{_{a\boldsymbol{z}}}^{^{\boldsymbol{\sigma}}})^{^{\mathrm{T}}} & (\mathbf{D}_{_{a\boldsymbol{z}}}^{^{\boldsymbol{e}}})^{^{\mathrm{T}}} \\ 
\boldsymbol{0}&\boldsymbol{0}&\temp &\boldsymbol{0} &\boldsymbol{0}&\temp&\boldsymbol{0} & \mathbf{M}_{_{\boldsymbol{\sigma}}}^{^{\boldsymbol{w}}}&\temp&\boldsymbol{0}&\boldsymbol{0}\\ \cline{1-11}
\boldsymbol{0}&\boldsymbol{0}&\temp &\mathbf{C}_{_{a1}}^{^{\boldsymbol{\sigma}}} &\boldsymbol{0}&\temp&\mathbf{D}_{_{a\boldsymbol{z}}}^{^{\boldsymbol{\sigma}}} & \boldsymbol{0}_{m_{_{\boldsymbol{y}}}\times m_{_{\boldsymbol{v}}}}&\temp&\mathbf{I}_{m_{_{\boldsymbol{y}}}}&\boldsymbol{0}\\
\boldsymbol{0}&\boldsymbol{0}&\temp &\mathbf{C}_{_{a1}}^{^{\boldsymbol{e}}} &\boldsymbol{0}&\temp&\mathbf{D}_{_{a\boldsymbol{z}}}^{^{\boldsymbol{e}}} & \boldsymbol{0}_{m_{_{\boldsymbol{e}}}\times m_{_{\boldsymbol{v}}}}&\temp&\boldsymbol{0}&\mathbf{I}_{m_{_{\boldsymbol{e}}}}\\
\end{array}\right]$$  

\noindent
-- note that passing from the $4-$block expression for\ \ $\boldsymbol{\psi}_{a}^{\mathrm{o}}(\mathbf{P}, \boldsymbol{\sigma})$\ \ to the $8-$block one above\ \
$\mathbf{C}_{a}$\ \ and\ \ $\mathbf{D}_{a}$\ \ were written as\\

\noindent
$\mathbf{C}_{a}=\begin{bmatrix}
                \mathbf{C}_{_{a1}}^{^{\boldsymbol{\sigma}}} & \boldsymbol{0}_{m_{_{\boldsymbol{y}}}\times n_{_{\mathbf{G}}}}\\ 
                \mathbf{C}_{_{a1}}^{^{\boldsymbol{e}}} & \boldsymbol{0}_{m_{_{\boldsymbol{e}}}\times n_{_{\mathbf{G}}}}\\
                \end{bmatrix}$, \ \ $\begin{bmatrix}
                                      \mathbf{C}_{_{a1}}^{^{\boldsymbol{\sigma}}}\\
                                      \mathbf{C}_{_{a1}}^{^{\boldsymbol{e}}}\\
                                     \end{bmatrix}=\begin{bmatrix}
                \mathbf{C}_{_{Wa}}^{^{\boldsymbol{\sigma}}} & \boldsymbol{0}_{m_{_{\boldsymbol{y}}}\times n_{_{\mathbf{A}\mathrm{o}a}}}\\ 
                \mathbf{C}_{_{Wa}}^{^{\boldsymbol{e}}} & \boldsymbol{0}_{m_{_{\boldsymbol{e}}}\times n_{_{\mathbf{A}\mathrm{o}a}}}\\
                \end{bmatrix}$\ \ and\ \ $\mathbf{D}_{a}= \mathbf{D}_{_{Wa}}= \begin{bmatrix}
                \mathbf{D}_{_{a\boldsymbol{z}}}^{^{\boldsymbol{\sigma}}} & \vdots&\boldsymbol{0}_{m_{_{\boldsymbol{y}}}\times m_{_{\boldsymbol{v}}}}\\ 
                \mathbf{D}_{_{a\boldsymbol{z}}}^{^{\boldsymbol{e}}} & \vdots & \boldsymbol{0}_{m_{_{\boldsymbol{e}}}\times m_{_{\boldsymbol{v}}}}\\
                \end{bmatrix}$,\break\\

\noindent                
$\begin{bmatrix}
 \mathbf{D}_{_{a\boldsymbol{z}}}^{^{\boldsymbol{\sigma}}}\\
 \mathbf{D}_{_{a\boldsymbol{z}}}^{^{\boldsymbol{e}}}\\
 \end{bmatrix}=\begin{bmatrix}
                \boldsymbol{\sigma}_{_{\boldsymbol{w}}}^{^{1/2}}\gamma_{_{\mathbf{H}}}\mathbf{D}_{_{W\mathbf{H}\boldsymbol{y}}} & \boldsymbol{0}_{m_{_{\boldsymbol{y}}}\times m_{_{\boldsymbol{v}}}}\\ 
                \mathbf{D}_{_{\mathbf{H}\mathbf{I}}}\mathbf{D}_{_{W\boldsymbol{y}}}^{^{-1}} & \boldsymbol{0}_{m_{_{\boldsymbol{e}}}\times m_{_{\boldsymbol{v}}}}\\
                \end{bmatrix}$,\ \  $\mathbf{M}_{_{\boldsymbol{\sigma}}}^{^{\boldsymbol{z}}}\triangleq\operatorname{diag}\left(\boldsymbol{\sigma}_{_{\boldsymbol{y}}}\mathbf{I}_{m_{_{\boldsymbol{y}}}}, \boldsymbol{\sigma}_{_{\boldsymbol{v}}}\mathbf{I}_{m_{_{\boldsymbol{v}}}}\right)$ \ \ and\ \ 
                $\mathbf{M}_{_{\boldsymbol{\sigma}}}^{^{\boldsymbol{w}}}=\boldsymbol{\sigma_{_{W}}}\mathbf{I}_{m_{_{\boldsymbol{v}}}}$.\\
                
Note also that
$$
\mathbf{T}_{_{_{1}}}^{^{\mathrm{T}}}\boldsymbol{\theta}\mathbf{T}_{_{_{2}}}=
\newcommand*{\temp}{\multicolumn{1}{r|}{}}
\left[\begin{array}{ccccccccccc}
\boldsymbol{0}  & \boldsymbol{0}&\temp& \boldsymbol{0}  & \boldsymbol{0}&\temp&\boldsymbol{0}& \boldsymbol{0} & \temp&\boldsymbol{0} &\boldsymbol{0}\\ 
\boldsymbol{0}& \boldsymbol{0} &\temp& \mathbf{B}_{_{\mathbf{G}}}\widehat{\mathbf{C}}_{_{\mathrm{oa}}} &\mathbf{A}_{_{\mathbf{G}}}& \temp& \mathbf{B}_{_{\mathbf{G}}}\mathbf{D}_{_{\mathrm{oa}}}^{^{\boldsymbol{z}}} & \mathbf{B}_{_{\mathbf{G}}} & \temp &\boldsymbol{0} & \boldsymbol{0}\\ \cline{1-11}
\boldsymbol{0} & \boldsymbol{0}&\temp& \boldsymbol{0}  & \boldsymbol{0}&\temp&\boldsymbol{0} & \boldsymbol{0}& \temp& \boldsymbol{0} & \boldsymbol{0}\\ 
\boldsymbol{0}& \boldsymbol{0} &\temp& \boldsymbol{0} &\boldsymbol{0}& \temp& \boldsymbol{0} & \boldsymbol{0}&\temp& \boldsymbol{0} & \boldsymbol{0} \\ \cline{1-11}
\boldsymbol{0}& \boldsymbol{0} &\temp& \boldsymbol{0} &\boldsymbol{0}& \temp& \boldsymbol{0} & \boldsymbol{0} &\temp&\boldsymbol{0} & \boldsymbol{0} \\ 
\boldsymbol{0}&\boldsymbol{0}&\temp &\boldsymbol{0} &\boldsymbol{0}&\temp&\boldsymbol{0} & \boldsymbol{0}&\temp&\boldsymbol{0}&\boldsymbol{0}\\ \cline{1-11}
\boldsymbol{0}&\boldsymbol{0}&\temp &\boldsymbol{0} &\boldsymbol{0}&\temp&\boldsymbol{0} & \boldsymbol{0} &\temp&\boldsymbol{0}&\boldsymbol{0}\\
\boldsymbol{0}&\boldsymbol{0}&\temp &-\mathbf{D}_{_{\mathbf{G}}}\widehat{\mathbf{C}}_{_{\mathrm{o}a}} & -\mathbf{C}_{_{\mathbf{G}}}&\temp&-\mathbf{D}_{_{\mathbf{G}}}\mathbf{D}_{_{\mathrm{o}a}}^{^{\boldsymbol{z}}} &  -\mathbf{D}_{_{\mathbf{G}}}&\temp&\boldsymbol{0}&\boldsymbol{0}\\
\end{array}\right],$$  
where\ \
$\mathbf{D}_{_{\mathrm{o}a}}=\begin{bmatrix}\mathbf{D}_{_{\mathrm{o}a}}^{^{\boldsymbol{z}}}& \vdots & \mathbf{I}_{m_{_{\boldsymbol{v}}}}\end{bmatrix}$\ \ and
\ \ $\mathbf{D}_{_{\mathrm{o}a}}^{^{\boldsymbol{z}}}=\begin{bmatrix}\mathbf{D}_{_{\mathbf{H}\mathrm{o}}}\mathbf{D}_{_{W\boldsymbol{y}}}^{^{-1}}& \vdots
& \mathbf{D}_{_{W\boldsymbol{v}}}^{^{-1}}\end{bmatrix}$\ \ or, equivalently,

$$
\mathbf{T}_{_{_{1}}}^{^{\mathrm{T}}}\boldsymbol{\theta}\mathbf{T}_{_{_{2}}}=\mathbf{T}_{_{_{1}}}^{^{\mathrm{T}}}
\begin{bmatrix}
 \mathbf{B}_{_{\mathbf{G}}}\widehat{\mathbf{C}}_{_{\mathrm{o}a}} & \mathbf{A}_{_{\mathbf{G}}} & \mathbf{B}_{_{\mathbf{G}}}\mathbf{D}_{_{\mathrm{o}a}}^{^{\boldsymbol{z}}}&\mathbf{B}_{_{\mathbf{G}}}\\
 \mathbf{D}_{_{\mathbf{G}}}\widehat{\mathbf{C}}_{_{\mathrm{o}a}} & \mathbf{C}_{_{\mathbf{G}}} & \mathbf{D}_{_{\mathbf{G}}}\mathbf{D}_{_{\mathrm{o}a}}^{^{\boldsymbol{z}}}&\mathbf{D}_{_{\mathbf{G}}}\\
 \end{bmatrix}\begin{bmatrix}
                             \boldsymbol{0}&\boldsymbol{0}&\mathbf{I}&\boldsymbol{0}&\boldsymbol{0}&\boldsymbol{0}&\boldsymbol{0}&\boldsymbol{0}\\
                             \boldsymbol{0}&\boldsymbol{0}&\boldsymbol{0}&\mathbf{I}&\boldsymbol{0}&\boldsymbol{0}&\boldsymbol{0}&\boldsymbol{0}\\
                             \boldsymbol{0}&\boldsymbol{0}&\boldsymbol{0}&\boldsymbol{0}&\mathbf{I}&\boldsymbol{0}&\boldsymbol{0}&\boldsymbol{0}\\
                              \boldsymbol{0}&\boldsymbol{0}&\boldsymbol{0}&\boldsymbol{0}&\boldsymbol{0}&\mathbf{I}&\boldsymbol{0}&\boldsymbol{0}\\
                              \end{bmatrix}\ \ \ \ \Leftrightarrow
$$

$$                              
\mathbf{T}_{_{_{1}}}^{^{\mathrm{T}}}\boldsymbol{\theta}\mathbf{T}_{_{_{2}}}=\mathbf{T}_{_{_{1}}}^{^{\mathrm{T}}}
\begin{bmatrix}
\mathbf{A}_{_{\mathbf{G}}}&\mathbf{B}_{_{\mathbf{G}}}\\
\mathbf{C}_{_{\mathbf{G}}}&\mathbf{D}_{_{\mathbf{G}}}\\
\end{bmatrix}\begin{bmatrix}
                            \boldsymbol{0}&\vdots&\mathbf{I}&\vdots& \boldsymbol{0}&\vdots&\boldsymbol{0}\\
                            \widehat{\mathbf{C}}_{_{\mathrm{o}a}}&\vdots& \boldsymbol{0}&\vdots&\mathbf{D}_{_{\mathrm{o}a}}^{^{\boldsymbol{z}}}&\vdots&\mathbf{I}\\
                            \end{bmatrix} 
                             \begin{bmatrix}
                             \boldsymbol{0}&\boldsymbol{0}&\mathbf{I}&\boldsymbol{0}&\boldsymbol{0}&\boldsymbol{0}&\boldsymbol{0}&\boldsymbol{0}\\
                             \boldsymbol{0}&\boldsymbol{0}&\boldsymbol{0}&\mathbf{I}&\boldsymbol{0}&\boldsymbol{0}&\boldsymbol{0}&\boldsymbol{0}\\
                             \boldsymbol{0}&\boldsymbol{0}&\boldsymbol{0}&\boldsymbol{0}&\mathbf{I}&\boldsymbol{0}&\boldsymbol{0}&\boldsymbol{0}\\
                             \boldsymbol{0}&\boldsymbol{0}&\boldsymbol{0}&\boldsymbol{0}&\boldsymbol{0}&\mathbf{I}&\boldsymbol{0}&\boldsymbol{0}\\
                              \end{bmatrix}.
$$
As a result,
$$\mathbf{T}_{_{_{1}}}=\begin{bmatrix}
                  \boldsymbol{0} & \mathbf{I}     & \vdots & \boldsymbol{0} & \boldsymbol{0} & \vdots & \boldsymbol{0} & \boldsymbol{0} & \vdots & \boldsymbol{0} & \boldsymbol{0}\\
                  \boldsymbol{0} & \boldsymbol{0} & \vdots & \boldsymbol{0} & \boldsymbol{0} & \vdots & \boldsymbol{0} & \boldsymbol{0} & \vdots & \boldsymbol{0} & -\mathbf{I}
                 \end{bmatrix}\ \ \ \text{and}\ \ \ \  \mathbf{T}_{_{2}}=\begin{bmatrix}
                  \boldsymbol{0} & \boldsymbol{0} & \vdots & \boldsymbol{0} & \mathbf{I} & \vdots & \boldsymbol{0} & \boldsymbol{0} & \vdots & \boldsymbol{0} & \boldsymbol{0}\\
                  \boldsymbol{0} & \boldsymbol{0} & \vdots & \widehat{\mathbf{C}}_{_{\mathrm{o}a}} & \boldsymbol{0} & \vdots & \mathbf{D}_{_{\mathrm{o}a}}^{^{\boldsymbol{z}}} & \mathbf{I} & \vdots & \boldsymbol{0} & \boldsymbol{0} \end{bmatrix}.$$
                  \hfill$\blacksquare$\\

\noindent
\underline{\textbf{Proof of Proposition \ref{prop:16a}:}}
It is first necessary to
obtain\ $W_{_{_{1}}}$\ and \ $W_{_{_{2}}}$.\ 
To this effect, note that
$$\operatorname{Ker}(\mathbf{T}_{_{_{1}}})
=\left\{\boldsymbol{v}^{^{\mathrm{T}}}=[\boldsymbol{v}_{_{_{1}}}^{^{\mathrm{T}}}\ \mathellipsis\ \boldsymbol{v}_{_{8}}^{^{\mathrm{T}}}]: 
\mathbf{T}_{_{_{1}}}\boldsymbol{v}=0\right\}=\left\{\boldsymbol{v}:\boldsymbol{v}_{2}=0\ \ \text{and}\ \ \boldsymbol{v}_{8}=0\right\},$$

As a result,                        
$$
W_{_{_{1}}}=
\newcommand*{\temp}{\multicolumn{1}{r|}{}}
\left[\begin{array}{cccccc}
\mathbf{I}      & \boldsymbol{0} & \boldsymbol{0} & \boldsymbol{0} &  \boldsymbol{0}  & \boldsymbol{0}\\ 
\boldsymbol{0}  & \boldsymbol{0} & \boldsymbol{0} & \boldsymbol{0} &  \boldsymbol{0}  & \boldsymbol{0}\\ \cline{1-6}
\boldsymbol{0}  & \mathbf{I}     & \boldsymbol{0} & \boldsymbol{0} &  \boldsymbol{0}  & \boldsymbol{0}\\ 
\boldsymbol{0}  & \boldsymbol{0} & \mathbf{I}     & \boldsymbol{0} &  \boldsymbol{0}  & \boldsymbol{0}\\ \cline{1-6}
\boldsymbol{0}  & \boldsymbol{0} & \boldsymbol{0} & \mathbf{I}     &  \boldsymbol{0}  & \boldsymbol{0}\\ 
\boldsymbol{0}  & \boldsymbol{0} & \boldsymbol{0} & \boldsymbol{0} &  \mathbf{I}      & \boldsymbol{0}\\ \cline{1-6}
\boldsymbol{0}  & \boldsymbol{0} & \boldsymbol{0} & \boldsymbol{0} &  \boldsymbol{0}  & \mathbf{I}    \\ 
\boldsymbol{0}  & \boldsymbol{0} & \boldsymbol{0} & \boldsymbol{0} &  \boldsymbol{0}  & \boldsymbol{0}\\ 
\end{array}\right]
$$
is such that its columns form a basis for the null space of\ $\mathbf{T}_{_{_{1}}}$.


Thus, 
                          
$$
 W_{_{_{1}}}^{^{\mathrm{T}}}\boldsymbol{\psi}_{a}(\mathbf{P}, \boldsymbol{\sigma})W_{_{_{1}}}=
\begin{bmatrix}
\mathbf{R}  & \mathbf{A}_{_{a1}}  & \boldsymbol{0} &\mathbf{B}_{_{a1}}^{^{\boldsymbol{z}}}& \boldsymbol{0} &\boldsymbol{0}\\ 
 \mathbf{A}_{_{a1}}^{^{\mathrm{T}}}& \mathbf{S} &\mathbf{N} &\boldsymbol{0}& \boldsymbol{0} & (\mathbf{C}_{_{a1}}^{^{\boldsymbol{\sigma}}})^{^{\mathrm{T}}}\\
\boldsymbol{0}& \mathbf{N}^{^{\mathrm{T}}} &\mathbf{X}& \boldsymbol{0} & \boldsymbol{0}& \boldsymbol{0}\\ 
(\mathbf{B}_{_{a1}}^{\boldsymbol{z}})^{^{\mathrm{T}}}& \boldsymbol{0} & \boldsymbol{0} & \mathbf{M}_{_{\boldsymbol{\sigma}}}^{\boldsymbol{z}} & \boldsymbol{0} &(\mathbf{D}_{_{a\boldsymbol{z}}}^{^{\boldsymbol{\sigma}}})^{^{\mathrm{T}}}\\ 
\boldsymbol{0}&\boldsymbol{0}&\boldsymbol{0} &\boldsymbol{0}& \mathbf{M}_{_{\boldsymbol{\sigma}}}^{^{\boldsymbol{w}}}&\boldsymbol{0}\\ 
\boldsymbol{0}&\mathbf{C}_{_{a1}}^{^{\boldsymbol{\sigma}}} &\boldsymbol{0}&\mathbf{D}_{_{a\boldsymbol{z}}}^{^{\boldsymbol{\sigma}}} & \boldsymbol{0}&\mathbf{I}_{m_{_{\boldsymbol{y}}}}\\
\end{bmatrix},$$  
so that\ \ $ W_{_{_{1}}}^{^{\mathrm{T}}}\boldsymbol{\psi}_{a}(\mathbf{P}, \boldsymbol{\sigma})W_{_{_{1}}}>0$\ \ \ $\Leftrightarrow$\ \ \ 
$\boldsymbol{\sigma}_{\boldsymbol{w}}>0$\ \  and
$$
\begin{bmatrix}
\mathbf{R}  & \mathbf{A}_{_{a1}}  & \boldsymbol{0} &\mathbf{B}_{_{a1}}^{^{\boldsymbol{z}}}& \boldsymbol{0}\\ 
 \mathbf{A}_{_{a1}}^{^{\mathrm{T}}}& \mathbf{S} &\mathbf{N} &\boldsymbol{0}& (\mathbf{C}_{_{a1}}^{^{\boldsymbol{\sigma}}})^{^{\mathrm{T}}}\\
\boldsymbol{0}& \mathbf{N}^{^{\mathrm{T}}} &\mathbf{X}& \boldsymbol{0} & \boldsymbol{0}\\ 
(\mathbf{B}_{_{a1}}^{\boldsymbol{z}})^{^{\mathrm{T}}}& \boldsymbol{0} & \boldsymbol{0} & \mathbf{M}_{_{\boldsymbol{\sigma}}}^{\boldsymbol{z}} &(\mathbf{D}_{_{a\boldsymbol{z}}}^{^{\boldsymbol{\sigma}}})^{^{\mathrm{T}}}\\ 
\boldsymbol{0}&\mathbf{C}_{_{a1}}^{^{\boldsymbol{\sigma}}} &\boldsymbol{0}&\mathbf{D}_{_{a\boldsymbol{z}}}^{^{\boldsymbol{\sigma}}}&\mathbf{I}_{m_{_{\boldsymbol{y}}}}\\
\end{bmatrix}>0$$  
$\Leftrightarrow$\ \ \ (exchanging rows and columns)\ \ $\boldsymbol{\sigma}_{\boldsymbol{w}}>0$\ \ and
$$
\begin{bmatrix}
\mathbf{R}  &\mathbf{B}_{_{a1}}^{^{\boldsymbol{z}}}   & \boldsymbol{0} &\mathbf{A}_{_{a1}}& \boldsymbol{0}\\ 
(\mathbf{B}_{_{a1}}^{\boldsymbol{z}})^{^{\mathrm{T}}}&\mathbf{M}_{_{\boldsymbol{\sigma}}}^{\boldsymbol{z}}    &(\mathbf{D}_{_{a\boldsymbol{z}}}^{^{\boldsymbol{\sigma}}})^{^{\mathrm{T}}}& \boldsymbol{0} &\boldsymbol{0}\\ 
\boldsymbol{0}&\mathbf{D}_{_{a\boldsymbol{z}}}^{^{\boldsymbol{\sigma}}} &\mathbf{I}_{m_{_{\boldsymbol{y}}}}& \mathbf{C}_{_{a1}}^{^{\boldsymbol{\sigma}}}&\boldsymbol{0}\\
\mathbf{A}_{_{a1}}^{^{\mathrm{T}}}&\boldsymbol{0}& (\mathbf{C}_{_{a1}}^{^{\boldsymbol{\sigma}}})^{^{\mathrm{T}}}& \mathbf{S} &\mathbf{N} \\
\boldsymbol{0}& \boldsymbol{0} & \boldsymbol{0}& \mathbf{N}^{^{\mathrm{T}}} &\mathbf{X}\\ 
\end{bmatrix}>0$$  
$\Leftrightarrow$\ \ $\boldsymbol{\sigma}_{\boldsymbol{w}}>0$\ \ and
$$\boldsymbol{\psi}_{_{a1}}(\mathbf{R}, \boldsymbol{\sigma})-\begin{bmatrix}
                                                             \mathbf{A}_{_{a1}}&\boldsymbol{0}\\
                                                             \boldsymbol{0}&\boldsymbol{0}\\
                                                             \mathbf{C}_{_{a1}}^{^{\boldsymbol{\sigma}}}&\boldsymbol{0}\\
                                                             \end{bmatrix}\begin{bmatrix}
                                                                          \mathbf{R}&\mathbf{M}\\
                                                                          \mathbf{M}^{^{\mathrm{T}}}&\mathbf{Z}\\
                                                                          \end{bmatrix}\begin{bmatrix}
                                                                                       \mathbf{A}_{_{a1}}^{^{\mathrm{T}}}&\boldsymbol{0}&(\mathbf{C}_{_{a1}}^{^{\boldsymbol{\sigma}}})^{^{\mathrm{T}}}\\
                                                                                       \boldsymbol{0}&\boldsymbol{0}&\boldsymbol{0}\\
                                                                                       \end{bmatrix}>0$$
$\Leftrightarrow$\ \ $\boldsymbol{\sigma}_{\boldsymbol{w}}>0$\ \ and
$$\boldsymbol{\psi}_{_{a1}}(\mathbf{R}, \boldsymbol{\sigma})-\begin{bmatrix}
                                                             \mathbf{A}_{_{a1}}\\
                                                             \boldsymbol{0}\\
                                                             \mathbf{C}_{_{a1}}^{^{\boldsymbol{\sigma}}}\\
                                                             \end{bmatrix}\mathbf{R}\begin{bmatrix}
                                                                                       \mathbf{A}_{_{a1}}^{^{\mathrm{T}}}&\boldsymbol{0}&(\mathbf{C}_{_{a1}}^{^{\boldsymbol{\sigma}}})^{^{\mathrm{T}}}
                                                                                    \end{bmatrix}>0,$$
\noindent                                                                         
where\\                                                                                   
$\boldsymbol{\psi}_{_{a1}}(\mathbf{R}, \boldsymbol{\sigma})=\begin{bmatrix}
                                                             \mathbf{R}&\mathbf{B}_{_{a1}}^{^{\boldsymbol{z}}}&\boldsymbol{0}\\
                                                             (\mathbf{B}_{_{a1}}^{^{\boldsymbol{z}}})^{^{\mathrm{T}}}&\mathbf{M}_{_{\boldsymbol{\sigma}}}^{^{\boldsymbol{z}}}&(\mathbf{D}_{_{a\boldsymbol{z}}}^{^{\boldsymbol{\sigma}}})^{^{\mathrm{T}}}\\
                                                             \boldsymbol{0}&\mathbf{D}_{_{a\boldsymbol{z}}}^{^{\boldsymbol{\sigma}}}&\mathbf{I}_{m_{_{\boldsymbol{y}}}}
                                                             \end{bmatrix}$, \ \ 
$\mathbf{C}_{_{a1}}^{^{\boldsymbol{\sigma}}}
=\begin{bmatrix}\boldsymbol{\sigma}_{_{\boldsymbol{w}}}^{^{1/2}}\gamma_{_{\mathbf{H}}}\mathbf{C}_{_{W\mathbf{H}\boldsymbol{y}}}&\vdots&
\boldsymbol{0}_{m_{_{\boldsymbol{y}}}\times n_{_{\mathbf{A}\mathbf{H}\mathbf{I}a}}}&\vdots&\boldsymbol{0}_{m_{_{\boldsymbol{y}}}\times n_{_{\mathbf{A}\mathrm{o}a}}}\end{bmatrix}$,\break\\

$\mathbf{C}_{_{a1}}^{^{\boldsymbol{e}}}
=\begin{bmatrix}\boldsymbol{0}_{m_{_{\boldsymbol{e}}}\times n_{_{\mathbf{A}W\mathbf{H}\boldsymbol{y}}}}&\vdots&\mathbf{C}_{_{\mathbf{H}\mathbf{I}a}}
&\vdots&\boldsymbol{0}_{m_{_{\boldsymbol{e}}}\times n_{_{\mathbf{A}\mathrm{o}a}}}\end{bmatrix}$,\ \ $\mathbf{D}_{_{a\boldsymbol{z}}}^{^{\boldsymbol{\sigma}}}
=\begin{bmatrix}\boldsymbol{\sigma}_{_{\boldsymbol{w}}}^{^{1/2}}\gamma_{_{\mathbf{H}}}\mathbf{D}_{_{W\mathbf{H}\boldsymbol{y}}}&\vdots&
\boldsymbol{0}_{m_{_{\boldsymbol{y}}}\times n_{\boldsymbol{v}}}\end{bmatrix}$\break\\
\noindent
and\ \ \ $\mathbf{D}_{_{a\boldsymbol{z}}}^{^{\boldsymbol{e}}}
=\begin{bmatrix}\mathbf{D}_{_{\mathbf{H}\mathbf{I}}}\mathbf{D}_{_{W\boldsymbol{y}}}^{^{-1}}&\vdots&
\boldsymbol{0}_{m_{_{\boldsymbol{e}}}\times n_{\boldsymbol{v}}}\end{bmatrix}$.\\

Finally, pre and post-multipliying the last matrix inequality above by\ $\operatorname{diag}(\mathbf{I},\mathbf{I},\boldsymbol{\sigma}_{_{\boldsymbol{w}}}^{^{1/2}}\mathbf{I})$\ it follows that\ \ \  $W_{_{1}}^{^{\mathrm{T}}}\boldsymbol{\psi}_{_{a}}^{^{\mathrm{o}}}(\mathbf{P},\boldsymbol{\sigma})W_{_{1}}>0$\ \ $\Leftrightarrow$\ \ $\boldsymbol{\sigma}_{_{\boldsymbol{w}}}>0$,\ $\mathbf{P}>0$\ and\  $Q_{_{1}}(\mathbf{R},\boldsymbol{\sigma})>0$.

Similarly,
$$\operatorname{Ker}(\mathbf{T}_{_{_{2}}})
=\left\{\boldsymbol{v}^{^{\mathrm{T}}}=[\boldsymbol{v}_{_{_{1}}}^{^{\mathrm{T}}}\ \mathellipsis\ \boldsymbol{v}_{_{8}}^{^{\mathrm{T}}}]: 
\boldsymbol{v}_{_{4}}=0\ \ \text{and}\ \ \widehat{\mathbf{C}}_{_{\mathrm{o}a}}\boldsymbol{v}_{_{3}}
+\mathbf{D}_{_{\mathrm{o}a}}^{^{\boldsymbol{z}}}\boldsymbol{v}_{_{5}}+\boldsymbol{v}_{_{6}}=0\right\}.$$  

Thus,
$$\operatorname{Ker}(\mathbf{T}_{_{_{2}}})
=\left\{\boldsymbol{v}^{^{\mathrm{T}}}=[\boldsymbol{v}_{_{_{1}}}^{^{\mathrm{T}}}\ \mathellipsis\  \boldsymbol{v}_{_{8}}]: 
\boldsymbol{v}_{_{4}}=0\ \ \text{and}\ \ \boldsymbol{v}_{_{6}}=\mathbf{T}_{_{63}}\boldsymbol{v}_{_{3}}+\mathbf{T}_{_{65}}\boldsymbol{v}_{_{5}}\right\},$$
where\ \ $\mathbf{T}_{_{63}}=-\widehat{\mathbf{C}}_{_{\mathrm{o}a}}$\ \ and\ \ 
$\mathbf{T}_{_{65}}=-\mathbf{D}_{_{\mathrm{o}a}}^{^{\boldsymbol{z}}}$,\ \ 
$\mathbf{D}_{_{\mathrm{o}a}}^{^{\boldsymbol{z}}}=
\begin{bmatrix}\mathbf{D}_{_{\mathbf{H}\mathrm{o}}}\mathbf{D}_{_{W\boldsymbol{y}}}^{^{-1}}&\vdots&\mathbf{D}_{_{W\boldsymbol{v}}}^{^{-1}} \end{bmatrix}$.

Thus,
$$
W_{_{_{2}}}=
\left[\begin{array}{cccccc}
\mathbf{I}     & \boldsymbol{0} & \boldsymbol{0}  & \boldsymbol{0}  & \boldsymbol{0} & \boldsymbol{0}\\ 
\boldsymbol{0} & \mathbf{I}     & \boldsymbol{0}  & \boldsymbol{0}  & \boldsymbol{0} & \boldsymbol{0}\\ 
\boldsymbol{0} & \boldsymbol{0} & \mathbf{I}      & \boldsymbol{0}  & \boldsymbol{0} & \boldsymbol{0}\\ 
\boldsymbol{0} & \boldsymbol{0} & \boldsymbol{0}  & \boldsymbol{0}  & \boldsymbol{0} & \boldsymbol{0}\\ 
\boldsymbol{0} & \boldsymbol{0} & \boldsymbol{0}  & \mathbf{I}      & \boldsymbol{0}& \boldsymbol{0}\\ 
\boldsymbol{0} & \boldsymbol{0} & \mathbf{T}_{_{63}} & \mathbf{T}_{_{65}} & \boldsymbol{0} & \boldsymbol{0}\\
\boldsymbol{0} & \boldsymbol{0} & \boldsymbol{0}  & \boldsymbol{0}  & \mathbf{I}     & \boldsymbol{0} \\ 
\boldsymbol{0} & \boldsymbol{0} & \boldsymbol{0}  & \boldsymbol{0}  & \boldsymbol{0} & \mathbf{I}\\ 
\end{array}\right], \ \ W_{_{_{2}}}^{^{\mathrm{T}}}=
\left[\begin{array}{cccccccc}
\mathbf{I}     & \boldsymbol{0} & \boldsymbol{0} & \boldsymbol{0} & \boldsymbol{0} & \boldsymbol{0}               & \boldsymbol{0}&\boldsymbol{0}\\ 
\boldsymbol{0} & \mathbf{I}     & \boldsymbol{0} & \boldsymbol{0} & \boldsymbol{0} & \boldsymbol{0}               & \boldsymbol{0} & \boldsymbol{0}\\ 
\boldsymbol{0} & \boldsymbol{0} & \mathbf{I}     & \boldsymbol{0} & \boldsymbol{0} & \mathbf{T}_{_{63}}^{^{\mathrm{T}}} & \boldsymbol{0}& \boldsymbol{0} \\ 
\boldsymbol{0} & \boldsymbol{0} & \boldsymbol{0} & \boldsymbol{0} & \mathbf{I}     & \mathbf{T}_{_{65}}^{^{\mathrm{T}}} & \boldsymbol{0} & \boldsymbol{0} \\ 
\boldsymbol{0} & \boldsymbol{0} & \boldsymbol{0} & \boldsymbol{0} & \boldsymbol{0} & \boldsymbol{0}               & \mathbf{I} & \boldsymbol{0}     \\ 
\boldsymbol{0} & \boldsymbol{0} & \boldsymbol{0} & \boldsymbol{0} & \boldsymbol{0} & \boldsymbol{0}               & \boldsymbol{0} & \mathbf{I} \\ 
\end{array}\right].
$$                   

As a result,
$$
 W_{_{_{2}}}^{^{\mathrm{T}}}\boldsymbol{\psi}_{a}^{\mathrm{o}}(\mathbf{P}, \boldsymbol{\sigma})W_{_{_{2}}}=
\newcommand*{\temp}{\multicolumn{1}{r|}{}}
\left[\begin{array}{ccccccccc}
\mathbf{R}                                 & \mathbf{M}     &\temp& \boldsymbol{\psi}_{_{a13}}                          &\temp& \boldsymbol{\psi}_{_{a14}}                                                                       &\temp& \boldsymbol{0}                  & \boldsymbol{0}                  \\ 
\mathbf{M}^{^{\mathrm{T}}}                 & \mathbf{Z}     &\temp& \boldsymbol{0}                                      &\temp& \boldsymbol{0}                                                                                   &\temp& \boldsymbol{0}                  & \boldsymbol{0}                  \\ \cline{1-9}
\boldsymbol{\psi}_{_{a13}}^{^{\mathrm{T}}} & \boldsymbol{0} &\temp& \mathbf{S} + \mathbf{M}_{_{\boldsymbol{\sigma w}3}} &\temp& \boldsymbol{\psi}_{_{_{a34}}}                                                                    &\temp& \boldsymbol{\psi}_{_{a35}}      & \boldsymbol{\psi}_{_{a36}}      \\ 
\boldsymbol{\psi}_{_{a14}}^{^{\mathrm{T}}} & \boldsymbol{0} &\temp& \boldsymbol{\psi}_{_{_{a34}}}^{^{\mathrm{T}}}       &\temp& \mathbf{M}_{_{\boldsymbol{\sigma}}}^{^{\boldsymbol{z}}} + \mathbf{M}_{_{\boldsymbol{\sigma w}5}} &\temp& \boldsymbol{\psi}_{_{_{a45}}}   & \boldsymbol{\psi}_{_{_{a46}}}   \\ \cline{1-9}
\boldsymbol{0}                             & \boldsymbol{0} &\temp& \boldsymbol{\psi}_{_{a35}}^{^{\mathrm{T}}}          &\temp& \boldsymbol{\psi}_{_{a45}}^{^{\mathrm{T}}}                                                       &\temp& \mathbf{I}_{m_{_{\boldsymbol{y}}}} & \boldsymbol{0}                  \\                                                  
\boldsymbol{0}                             & \boldsymbol{0} &\temp& \boldsymbol{\psi}_{_{a36}}^{^{\mathrm{T}}}          &\temp& \boldsymbol{\psi}_{_{a46}}^{^{\mathrm{T}}}                                                       &\temp& \boldsymbol{0}                  & \mathbf{I}_{m_{_{\boldsymbol{v}}}} \\                                                  
\end{array}\right],$$   
where
\begin{eqnarray}
   \boldsymbol{\psi}_{_{a13}}&=&\mathbf{A}_{_{a1}},\ \ \  \ \ \ \ \  \ \boldsymbol{\psi}_{_{a14}}=\mathbf{B}_{_{a1}}^{^{\boldsymbol{z}}},\label{eq:38}\\
   \boldsymbol{\psi}_{_{_{a34}}}&=&\mathbf{T}_{_{63}}^{^{\mathrm{T}}}\mathbf{M}_{_{\boldsymbol{\sigma}}}^{^{\boldsymbol{w}}}\mathbf{T}_{_{65}}, \boldsymbol{\psi}_{_{a35}}=(\mathbf{C}_{_{a1}}^{^{\boldsymbol{\sigma}}})^{^{\mathrm{T}}}, \boldsymbol{\psi}_{_{a36}}=(\mathbf{C}_{_{a1}}^{^{\boldsymbol{e}}})^{^{\mathrm{T}}}, \boldsymbol{\psi}_{_{_{a45}}}=(\mathbf{D}_{_{a\boldsymbol{z}}}^{^{\boldsymbol{\sigma}}})^{^{\mathrm{T}}}, \boldsymbol{\psi}_{_{_{a46}}}=(\mathbf{D}_{_{a\boldsymbol{z}}}^{^{\boldsymbol{e}}})^{^{\mathrm{T}}},\label{eq:40}\\
   \mathbf{M}_{_{\boldsymbol{\sigma w}3}}&=&\mathbf{T}_{_{63}}^{^{\mathrm{T}}}\mathbf{M}_{_{\boldsymbol{\sigma}}}^{^{\boldsymbol{w}}}\mathbf{T}_{_{63}},\ \mathbf{M}_{_{\boldsymbol{\sigma w}5}}=\mathbf{T}_{_{65}}^{^{\mathrm{T}}}\mathbf{M}_{_{\boldsymbol{\sigma}}}^{^{\boldsymbol{w}}}\mathbf{T}_{_{65}},\  \mathbf{M}_{_{\boldsymbol{\sigma}}}^{^{\boldsymbol{z}}}=\operatorname{diag}(\boldsymbol{\sigma}_{_{\boldsymbol{y}}}\mathbf{I},\ \boldsymbol{\sigma}_{_{\boldsymbol{v}}}\mathbf{I}),\  \mathbf{M}_{_{\boldsymbol{\sigma}}}^{^{\boldsymbol{w}}}=\boldsymbol{\sigma}_{_{\boldsymbol{w}}}\mathbf{I}_{m_{_{\boldsymbol{v}}}}.\nonumber\\\label{eq:41}
\end{eqnarray}

Pre and post-multiplying \ $W_{_{_{2}}}^{^{\mathrm{T}}}\boldsymbol{\psi}_{a}(\mathbf{P}, \boldsymbol{\sigma})W_{_{_{2}}}$\ by \ $\mathbf{I}_{_{a2}}^{^{\mathrm{T}}}$\
and\ $\mathbf{I}_{_{a2}}$, where $\mathbf{I}_{_{a2}}$ is given by\break  $\mathbf{I}_{_{a2}}=\begin{bmatrix}
                                                     \boldsymbol{0}& \boldsymbol{0}&\boldsymbol{0}&\boldsymbol{0}&\mathbf{I}&\boldsymbol{0}\\
                                                     \boldsymbol{0}& \boldsymbol{0}&\boldsymbol{0}&\boldsymbol{0}&\boldsymbol{0}&\mathbf{I}\\
                                                     \mathbf{I}& \boldsymbol{0}&\boldsymbol{0}&\boldsymbol{0}&\boldsymbol{0}&\boldsymbol{0}\\
                                                     \boldsymbol{0}& \mathbf{I}&\boldsymbol{0}&\boldsymbol{0}&\boldsymbol{0}&\boldsymbol{0}\\
                                                     \boldsymbol{0}& \boldsymbol{0}&\mathbf{I}&\boldsymbol{0}&\boldsymbol{0}&\boldsymbol{0}\\
                                                     \boldsymbol{0}&\boldsymbol{0}&\boldsymbol{0}&\mathbf{I}&\boldsymbol{0}&\boldsymbol{0}\\
                                                     \end{bmatrix}$,\ it follows that\ \ 
$W_{_{_{2}}}^{^{\mathrm{T}}}\boldsymbol{\psi}_{a}(\mathbf{P}, \boldsymbol{\sigma})W_{_{_{2}}}>0$\ \ \ $\Leftrightarrow$\ \ \
$\mathbf{I}_{_{a2}}^{^{\mathrm{T}}}W_{_{2}}^{^{\mathrm{T}}}\boldsymbol{\psi}_{_{a}}(\mathbf{P}, \boldsymbol{\sigma})W_{_{2}}\mathbf{I}_{_{a2}}>0$\\\\
$
\newcommand*{\temp}{\multicolumn{1}{r|}{}}
\left[\begin{array}{ccc}
 \boldsymbol{\psi}_{_{Wa}}(\mathbf{S}, \boldsymbol{\sigma}) & \temp& \begin{array}{cc}
                                                                    \boldsymbol{\psi}_{_{a13}}^{^{\mathrm{T}}}&\boldsymbol{0}\\
                                                                    \boldsymbol{\psi}_{_{a14}}^{^{\mathrm{T}}}&\boldsymbol{0}\\
                                                                    \boldsymbol{0}&\boldsymbol{0}\\
                                                                    \boldsymbol{0}&\boldsymbol{0}\\
                                                                    \end{array}               \\ \cline{1-3}
\begin{array}{cccc}
\boldsymbol{\psi}_{_{a13}}& \boldsymbol{\psi}_{_{a1}}&\boldsymbol{0}&\boldsymbol{0}\\
\boldsymbol{0}& \boldsymbol{0}&\boldsymbol{0}&\boldsymbol{0}\\
\end{array} &\temp& \begin{array}{cc}
                     \mathbf{R}&\mathbf{M}\\
                     \mathbf{M}^{^{\mathrm{T}}}&\mathbf{Z}\\
                    \end{array}   \\
\end{array}\right]>0\ \  \Leftrightarrow\  \ \begin{bmatrix}
                                               \mathbf{R}&\mathbf{M}\\
                                               \mathbf{M}^{^{\mathrm{T}}}& \mathbf{Z}\\
                                              \end{bmatrix}>0$\ \ and\break
                                                                         
$\boldsymbol{\psi}_{_{Wa}}(\mathbf{S}, \boldsymbol{\sigma})-\begin{bmatrix}
                                                               \boldsymbol{\psi}_{_{a13}}^{^{\mathrm{T}}}\\
                                                               \boldsymbol{\psi}_{_{a14}}^{^{\mathrm{T}}}\\
                                                               \boldsymbol{0}\\
                                                               \boldsymbol{0}\\
                                                             \end{bmatrix}\mathbf{S}
 \begin{bmatrix}\boldsymbol{\psi}_{_{a13}}&\boldsymbol{\psi}_{_{a14}}& \boldsymbol{0}&\boldsymbol{0} \end{bmatrix}>0$,\ \
where\\\\
$\boldsymbol{\psi}_{_{Wa}}(\mathbf{S}, \boldsymbol{\sigma})=\newcommand*{\temp}{\multicolumn{1}{r|}{}}
\left[\begin{array}{ccc}
 \check{\boldsymbol{\psi}}_{_{a1}}(\mathbf{S}, \boldsymbol{\sigma}) & \temp& \begin{array}{cc}
                                                                    \boldsymbol{\psi}_{_{a35}}& \boldsymbol{\psi}_{_{a36}}\\
                                                                    \boldsymbol{\psi}_{_{a45}}&\boldsymbol{\psi}_{_{a46}}\\
                                                                     \end{array}               \\ \cline{1-3}
\begin{array}{cc}
\boldsymbol{\psi}_{_{a35}}^{^{\mathrm{T}}} & \boldsymbol{\psi}_{_{a45}}^{^{\mathrm{T}}}\\
\boldsymbol{\psi}_{_{a36}}^{^{\mathrm{T}}} & \boldsymbol{\psi}_{_{a46}}^{^{\mathrm{T}}} \\
\end{array} &\temp& \begin{array}{cc}
                     \mathbf{I}_{m_{_{\boldsymbol{y}}}}&\boldsymbol{0}\\
                     \boldsymbol{0}&\mathbf{I}_{m_{_{\boldsymbol{v}}}}\\
                    \end{array}   \\
\end{array}\right]\ \ \  \Leftrightarrow$\ \ $\begin{bmatrix}
                                               \mathbf{R}&\mathbf{M}\\
                                               \mathbf{M}^{^{\mathrm{T}}}& \mathbf{Z}\\
                                              \end{bmatrix}>0$\ \ \  and\break \\\\ 

$\check{\boldsymbol{\psi}}_{_{a1}}(\mathbf{S}, \boldsymbol{\sigma})-\begin{bmatrix}
                                                               \boldsymbol{\psi}_{_{a13}}^{^{\mathrm{T}}}\\
                                                               \boldsymbol{\psi}_{_{a14}}^{^{\mathrm{T}}}\\
                                                            \end{bmatrix}\mathbf{S}
                                                             \begin{bmatrix}\boldsymbol{\psi}_{_{a13}}&\boldsymbol{\psi}_{_{a14}}\end{bmatrix}
                                                             -\left\{\begin{bmatrix}
                                                                       \boldsymbol{\psi}_{_{a35}}&\boldsymbol{\psi}_{_{a36}}\\
                                                                       \boldsymbol{\psi}_{_{a45}}&\boldsymbol{\psi}_{_{a46}}\\
                                                                     \end{bmatrix}\begin{bmatrix}
                                                                       \boldsymbol{\psi}_{_{a35}}^{^{\mathrm{T}}}&\boldsymbol{\psi}_{_{a45}}^{^{\mathrm{T}}}\\
                                                                       \boldsymbol{\psi}_{_{a36}}^{^{\mathrm{T}}}&\boldsymbol{\psi}_{_{a46}}^{^{\mathrm{T}}}\\
                                                                     \end{bmatrix}\right\}>0$,\\\\
                                                                     
where\ \  $\check{\boldsymbol{\psi}}_{_{a1}}(\mathbf{S}, \boldsymbol{\sigma})=\begin{bmatrix} 
                                                                       \mathbf{S}& \boldsymbol{0}\\
                                                                       \boldsymbol{0}&\mathbf{M}_{_{\boldsymbol{\sigma}}}^{^{\boldsymbol{z}}}\\
                                                                      \end{bmatrix}+\begin{bmatrix}
                                                                                   \mathbf{T}_{_{63}}^{^{\mathrm{T}}}\\
                                                                                    \mathbf{T}_{_{65}}^{^{\mathrm{T}}}\\
                                                                                    \end{bmatrix}
                                                                                    \mathbf{M}_{_{\boldsymbol{\sigma}}}^{^{\boldsymbol{w}}}
                                                                                    \begin{bmatrix} \mathbf{T}_{_{63}}& \mathbf{T}_{_{65}}\end{bmatrix}$.\\\\                                                                     
                                                                     
As a result, since\ \ $\begin{bmatrix}
                      \boldsymbol{\psi}_{_{a35}}^{^{\mathrm{T}}}&\boldsymbol{\psi}_{_{a45}}^{^{\mathrm{T}}}\\
                      \boldsymbol{\psi}_{_{a36}}^{^{\mathrm{T}}}&\boldsymbol{\psi}_{_{a46}}^{^{\mathrm{T}}}\\
                      \end{bmatrix} =
                      \mathbf{M}_{_{W\mathbf{I}}}^{^{1/2}}\begin{bmatrix}\widehat{\mathbf{C}}_{_{W\mathbf{H}}}&\widehat{\mathbf{D}}_{_{W\mathbf{H}}}
                      \end{bmatrix}$,\ where\\\\ 

                      $\widehat{\mathbf{C}}_{_{W\mathbf{H}}}=\begin{bmatrix}
                                       \mathbf{C}_{_{W\mathbf{H}\boldsymbol{y}}} & \vdots & \boldsymbol{0}_{m_{_{\boldsymbol{y}}}\times n_{_{\mathbf{A}\mathbf{H}\mathbf{I}a}}} & \vdots & \boldsymbol{0}_{m_{_{\boldsymbol{y}}}\times n_{_{\mathbf{A}\mathrm{o}a}}}\\
                                       \boldsymbol{0}_{m_{_{\boldsymbol{e}}}\times n_{_{\mathbf{A}W\mathbf{H}\boldsymbol{y}}}} & \vdots & \mathbf{C}_{_{\mathbf{H}\mathbf{I}a}} & \vdots & \boldsymbol{0}_{m_{_{\boldsymbol{e}}}\times n_{_{\mathbf{A}\mathrm{o}a}}}\\
                                       \end{bmatrix}$, \ \ 
$\widehat{\mathbf{D}}_{_{a\boldsymbol{z}}}=\begin{bmatrix}
                                    \mathbf{D}_{_{W\mathbf{H}\boldsymbol{y}}} & \vdots & \boldsymbol{0}_{m_{_{\boldsymbol{y}}} \times m_{_{\boldsymbol{v}}}}\\
                                    \mathbf{D}_{_{\mathbf{H}\mathbf{I}}}\mathbf{D}_{_{W\boldsymbol{y}}}^{^{-1}} & \vdots & \boldsymbol{0}_{m_{_{\boldsymbol{e}}} \times m_{_{\boldsymbol{v}}}}\\
                                    \end{bmatrix}$ \break\\\\
\noindent                                    
and \ \ \ $\mathbf{M}_{_{W\mathbf{I}}}=\begin{bmatrix}
                             \boldsymbol{\sigma}_{_{\boldsymbol{w}}}\gamma_{_{\mathbf{H}}}^{^{2}}\mathbf{I}_{m_{_{\boldsymbol{y}}}}& \boldsymbol{0}\\
                             \boldsymbol{0}& \mathbf{I}_{m_{_{\boldsymbol{e}}}}\\
                             \end{bmatrix}$,

\begin{equation}\label{eq:45}
W_{_{_{2}}}^{^{\mathrm{T}}}\boldsymbol{\psi}_{_{a}}^{\mathrm{o}}(\mathbf{P}, \boldsymbol{\sigma})W_{_{_{2}}}>0\ \ \ \Leftrightarrow \ \ \ 
\boldsymbol{\psi}_{_{a\mathbf{S}}}(\mathbf{S}, \boldsymbol{\sigma})-\begin{bmatrix}
                                                                                                                \widehat{\mathbf{C}}_{_{W\mathbf{H}}}^{^{\mathrm{T}}}\\
                                                                                                                \widehat{\mathbf{D}}_{_{a\boldsymbol{z}}}^{^{\mathrm{T}}}\\
                                                                                                                \end{bmatrix}\mathbf{M}_{_{W\mathbf{I}}}(\boldsymbol{\sigma}_{_{\boldsymbol{w}}})\begin{bmatrix}\widehat{\mathbf{C}}_{_{W\mathbf{H}}}&\widehat{\mathbf{D}}_{_{a\boldsymbol{z}}}\end{bmatrix}>0,
\end{equation}
where
\begin{equation}\label{eq:44}
  \boldsymbol{\psi}_{_{a \mathbf{S}}}(\mathbf{S}, \boldsymbol{\sigma})\triangleq \check{\boldsymbol{\psi}}_{_{a1}}(\mathbf{S}, \boldsymbol{\sigma})-\begin{bmatrix}
                                                                                  \boldsymbol{\psi}_{_{a13}}^{^{\mathrm{T}}}\\
                                                                                  \boldsymbol{\psi}_{_{a14}}^{^{\mathrm{T}}}\\
                                                                                  \end{bmatrix}\mathbf{S}
                                                                                  \begin{bmatrix}\boldsymbol{\psi}_{_{a13}}& \vdots& \boldsymbol{\psi}_{_{a14}}\end{bmatrix},
\end{equation}
or, equivalently,
\begin{equation}\label{eq:44a}
  \boldsymbol{\psi}_{_{a \mathbf{S}}}(\mathbf{S}, \boldsymbol{\sigma})\triangleq \check{\boldsymbol{\psi}}_{_{a1}}(\mathbf{S}, \boldsymbol{\sigma})-\begin{bmatrix}
                                                                                  \mathbf{A}_{_{a1}}^{^{\mathrm{T}}}\\
                                                                                  (\mathbf{B}_{_{a1}}^{^{\boldsymbol{z}}})^{^{\mathrm{T}}}\\
                                                                                  \end{bmatrix}\mathbf{S}
                                                                                  \begin{bmatrix}\mathbf{A}_{_{a1}} & \vdots& \mathbf{B}_{_{a1}}^{^{\boldsymbol{z}}}\end{bmatrix}.
\end{equation}

Thus, the condition\ \ $W_{_{_{2}}}^{^{\mathrm{T}}}\boldsymbol{\psi}_{a}^{\mathrm{o}}(\mathbf{P}, \boldsymbol{\sigma})W_{_{_{2}}}>0$\ \ is equivalent to the LMI (\ref{eq:45}) on the variables \ \ $\mathbf{S}$, \ \
$\boldsymbol{\sigma}_{_{\boldsymbol{y}}}$, \ \ $\boldsymbol{\sigma}_{_{\boldsymbol{v}}}$\ \ and \ \ $\boldsymbol{\sigma}_{_{\boldsymbol{w}}}$. The proof is concluded by noting that\ \ $\begin{bmatrix}\mathbf{T}_{_{63}}&\vdots&\mathbf{T}_{_{65}}\end{bmatrix}=-\begin{bmatrix}\widehat{\mathbf{C}}_{_{\mathrm{o}a}}&\vdots&\mathbf{D}_{_{\mathrm{o}a}}^{^{\boldsymbol{z}}}\end{bmatrix}$,\  $\mathbf{M}_{_{\boldsymbol{\sigma}}}^{^{\boldsymbol{w}}}=\boldsymbol{\sigma}_{_{\boldsymbol{w}}}\mathbf{I}$,\ \
$\mathbf{E}_{_{\boldsymbol{s}}}=\begin{bmatrix}\boldsymbol{\psi}_{_{a13}}&\vdots&\boldsymbol{\psi}_{_{a14}}\end{bmatrix}$ ($\mathbf{E}_{_{\boldsymbol{s}}}$\  and\ $\mathbf{E}_{_{\mathrm{o}}}$\ as in the beginning of the Appendix) and\ \
$\begin{bmatrix}
\widehat{\mathbf{C}}_{_{W\mathbf{H}}}^{^{\mathrm{T}}}\\
\widehat{\mathbf{D}}_{_{a\boldsymbol{z}}}^{^{\mathrm{T}}}  
\end{bmatrix}\mathbf{M}_{_{W\mathbf{I}}}(\boldsymbol{\sigma}_{_{w}})\begin{bmatrix}\widehat{\mathbf{C}}_{_{W\mathbf{H}}}&\vdots&\widehat{\mathbf{D}}_{_{a\boldsymbol{z}}}\end{bmatrix}=\boldsymbol{\sigma}_{_{\boldsymbol{w}}}\begin{bmatrix}
                    (\mathbf{C}_{_{a1}}^{^{\boldsymbol{w}}})^{^{\mathrm{T}}}\\
                    (\mathbf{D}_{_{a\boldsymbol{z}}}^{^{\boldsymbol{w}}})^{^{\mathrm{T}}}
                    \end{bmatrix}\begin{bmatrix}
                    \mathbf{C}_{_{a1}}^{^{\boldsymbol{w}}}&\vdots&
                    \mathbf{D}_{_{a\boldsymbol{z}}}^{^{\boldsymbol{w}}}
                    \end{bmatrix}+\mathbf{E}_{_{\mathrm{o}}}^{^{\mathrm{T}}}\mathbf{E}_{_{\mathrm{o}}}.$\hfill$\blacksquare$


\bigskip

\noindent
\underline{\textbf{Proof of equation (5.7):}} Note first that
\begin{eqnarray*}
\mathbf{F}_{_{\mathbf{G}W}}=\begin{bmatrix}
                             \boldsymbol{\sigma}_{_{W}}^{^{1/2}}\gamma_{_{\mathbf{H}}}W_{_{\mathbf{H}\boldsymbol{y}}}^{a}\\
                             \mathbf{F}_{_{\mathbf{G}a}}
                             \end{bmatrix}&=&\begin{bmatrix}
                                           \boldsymbol{\sigma}_{_{W}}^{^{1/2}}\gamma_{_{\mathbf{H}}}\mathbf{I}_{m_{_{\boldsymbol{y}}}}&\boldsymbol{0}\\
                                           \boldsymbol{0}&\mathbf{I}_{_{m_{\boldsymbol{e}}}}
                                           \end{bmatrix}\begin{bmatrix}
                                                        \mathbf{I}_{m_{_{\boldsymbol{y}}}}&\boldsymbol{0}&\boldsymbol{0}\\
                                                        \boldsymbol{0}&\mathbf{I}_{m_{_{\boldsymbol{e}}}}&\boldsymbol{\beta}\\
                                                        \end{bmatrix}\begin{bmatrix}
                                                                     W_{_{\mathbf{H}\boldsymbol{y}}}^{a}\\
                                                                     \mathbf{H}_{_{\mathbf{I}a}}\\
                                                                     -\mathbf{Y}_{_{\mathbf{G}}}^{^{\boldsymbol{z}}}\mathbf{H}_{_{\mathrm{o}a}}
                                                                     \end{bmatrix}\ \ \Rightarrow\\
\begin{bmatrix}\mathbf{C}_{_{\mathbf{G}W}}(\boldsymbol{\sigma}_{_{\boldsymbol{w}}},\boldsymbol{\beta})&\vdots&\mathbf{D}_{_{\mathbf{G}W}}(\boldsymbol{\sigma}_{_{\boldsymbol{w}\boldsymbol{\beta}}})\end{bmatrix}&=&\begin{bmatrix}
                                           \boldsymbol{\sigma}_{_{W}}^{^{1/2}}\gamma_{_{\mathbf{H}}}\mathbf{I}_{m_{_{\boldsymbol{y}}}}&\boldsymbol{0}\\
                                           \boldsymbol{0}&\mathbf{I}_{_{m_{\boldsymbol{e}}}}
                                           \end{bmatrix}\begin{bmatrix}
                                                        \mathbf{I}_{m_{_{\boldsymbol{y}}}}&\boldsymbol{0}&\boldsymbol{0}\\
                                                        \boldsymbol{0}&\mathbf{I}_{m_{_{\boldsymbol{e}}}}&\boldsymbol{\beta}\\
                                                        \end{bmatrix}\begin{bmatrix}\widehat{\mathbf{C}}_{_{\mathbf{G}W}}&\vdots&\widehat{\mathbf{D}}_{_{\mathbf{G}W}}\end{bmatrix}  \ \ \Rightarrow\\\\                                                                   
\begin{bmatrix}\mathbf{C}_{_{\mathbf{G}W}}(\boldsymbol{\sigma}_{_{\boldsymbol{w}}},\boldsymbol{\beta})&\vdots&\mathbf{D}_{_{\mathbf{G}W}}(\boldsymbol{\sigma}_{_{\boldsymbol{w}\boldsymbol{\beta}}})\end{bmatrix}&=& \begin{bmatrix}
                                           \boldsymbol{\sigma}_{_{W}}^{^{1/2}}\gamma_{_{\mathbf{H}}}\mathbf{I}_{m_{_{\boldsymbol{y}}}}&\boldsymbol{0}\\
                                           \boldsymbol{0}&\mathbf{I}_{_{m_{\boldsymbol{e}}}}
                                           \end{bmatrix}\begin{bmatrix}
                                                        \mathbf{T}_{_{\mathbf{G}W}}^{^{\boldsymbol{y}}}\\
                                                        \mathbf{T}_{_{\mathbf{G}W}}^{^{\boldsymbol{e}}}(\boldsymbol{\beta})
                                                        \end{bmatrix},
\end{eqnarray*}
where\ \ \ $ \mathbf{T}_{_{\mathbf{G}W}}^{^{\boldsymbol{y}}}\triangleq\begin{bmatrix}\mathbf{I}_{m_{_{\boldsymbol{y}}}}&\vdots&\boldsymbol{0}_{m_{_{\boldsymbol{y}}}\times m_{_{\boldsymbol{e}}}}&\vdots&\boldsymbol{0}_{m_{_{\boldsymbol{y}}}\times(n_{_{\mathbf{G}}}^{^{\boldsymbol{z}}} m_{_{\boldsymbol{y}}})}\end{bmatrix}\begin{bmatrix}\widehat{\mathbf{C}}_{_{\mathbf{G}W}}&\vdots&\widehat{\mathbf{D}}_{_{\mathbf{G}W}}\end{bmatrix}$\ \ and\\
$ \mathbf{T}_{_{\mathbf{G}W}}^{^{\boldsymbol{e}}}\triangleq\begin{bmatrix}\boldsymbol{0}_{m_{_{\boldsymbol{e}}}\times m_{_{\boldsymbol{y}}}}&\vdots&\mathbf{I}_{m_{_{\boldsymbol{y}}}}&\vdots&\boldsymbol{\beta}\end{bmatrix}\begin{bmatrix}\widehat{\mathbf{C}}_{_{\mathbf{G}W}}&\vdots&\widehat{\mathbf{D}}_{_{\mathbf{G}W}}\end{bmatrix}$.

As a result
$$Q_{_{\mathbf{C}\mathbf{D}}}\triangleq\begin{bmatrix}
                                      \mathbf{C}_{_{\mathbf{G}W}}(\cdot)^{^{\mathrm{T}}}\\
                                      \mathbf{D}_{_{\mathbf{G}W}}(\cdot)^{^{\mathrm{T}}}\\
                                       \end{bmatrix}\begin{bmatrix} \mathbf{C}_{_{\mathbf{G}W}}(\cdot)& \mathbf{D}_{_{\mathbf{G}W}}(\cdot)\end{bmatrix}=\boldsymbol{\sigma}_{_{\boldsymbol{w}}}\gamma_{_{\mathbf{H}}}^{^{2}}(\mathbf{T}_{_{\mathbf{G}W}}^{^{\boldsymbol{y}}})^{^{\mathrm{T}}}(\mathbf{T}_{_{\mathbf{G}W}}^{^{\boldsymbol{y}}})+(\mathbf{T}_{_{\mathbf{G}W}}^{^{\boldsymbol{e}}}(\boldsymbol{\beta}))^{^{\mathrm{T}}}(\mathbf{T}_{_{\mathbf{G}W}}^{^{\boldsymbol{e}}}(\boldsymbol{\beta})).$$

Note now that\ \ \ $Q_{_{\mathbf{B}\mathbf{R}}}(\mathbf{P},\boldsymbol{\Sigma}_{_{\mathbf{G}W}}^{^{b}}(\boldsymbol{\sigma}_{_{\boldsymbol{w}}},\boldsymbol{\beta}),\mathbf{M}_{_{\boldsymbol{\sigma}}}^{a})<0$\ \ \ $\Leftrightarrow$\ \ \ $\check{Q}_{_{b1}}(\mathbf{P},\boldsymbol{\sigma})-Q_{_{\mathbf{C}\mathbf{D}}}>0$\ \ $\Leftrightarrow$ \\
$\check{Q}_{_{b1}}(\mathbf{P},\boldsymbol{\sigma})-\boldsymbol{\sigma}_{_{\boldsymbol{w}}}\gamma_{_{\mathbf{H}}}^{^{2}}(\mathbf{T}_{_{\mathbf{G}W}}^{^{\boldsymbol{y}}})^{^{\mathrm{T}}}(\mathbf{T}_{_{\mathbf{G}W}}^{^{\boldsymbol{y}}})+(\mathbf{T}_{_{\mathbf{G}W}}^{^{\boldsymbol{e}}}(\boldsymbol{\beta}))^{^{\mathrm{T}}}(\mathbf{T}_{_{\mathbf{G}W}}^{^{\boldsymbol{e}}}(\boldsymbol{\beta}))>0\ \ \Leftrightarrow\ \ \check{Q}_{_{b}}(\mathbf{P},\boldsymbol{\sigma},\boldsymbol{\beta})>0.$\hfill$\blacksquare$

\bigskip

\bigskip

\noindent
\underline{\textbf{Rewriting the second constraint of (5.5):}} Note  that\ \ \ $Q_{_{\mathbf{B}\mathbf{R}}}(\mathbf{P};\boldsymbol{\Sigma}_{_{a}}^{^{\mathrm{o}}}(\mathbf{G}(\boldsymbol{\beta})),\mathbf{M}_{_{\boldsymbol{\sigma}}})<0$\ \ \ $\Leftrightarrow$\\\\
$\check{Q}_{_{a1}}(\mathbf{P},\boldsymbol{\sigma})-\begin{bmatrix}
                                                     \mathbf{C}_{_{\mathbf{F}\mathbf{G}}}(\boldsymbol{\beta})^{^{\mathrm{T}}}\\
                                                     \mathbf{D}_{_{\mathbf{F}\mathbf{G}}}(\boldsymbol{\beta})^{^{\mathrm{T}}}
                                                    \end{bmatrix}\begin{bmatrix}\mathbf{C}_{_{\mathbf{F}\mathbf{G}}}(\boldsymbol{\beta})&\mathbf{D}_{_{\mathbf{F}\mathbf{G}}}(\boldsymbol{\beta})\end{bmatrix}>0 \ \ \Leftrightarrow\ \ \check{Q}_{_{a}}(\mathbf{P},  \boldsymbol{\sigma},\boldsymbol{\beta})>0.$ \hfill$\blacksquare$

\bigskip

\bigskip

\vspace*{3mm}
\noindent
\underline{\textbf{Proof of equations (5.1) and (5.2):}}\ Note first that
$$C\left(\mathbf{G}(\boldsymbol{\beta});\widehat{\boldsymbol{\Gamma}}_{_{\boldsymbol{y}}},
\widehat{\boldsymbol{\Gamma}}_{_{\boldsymbol{v}}}\right)
=\left\langle[\mathbf{I}_{m_{_{\boldsymbol{e}}}}\ \vdots\ \boldsymbol{\beta}]
\mathbf{F}_{c}^{^{\boldsymbol{y}}}\widehat{\boldsymbol{\Gamma}}_{_{\boldsymbol{y}}},[\mathbf{I}_{m_{_{\boldsymbol{e}}}}\ \vdots\ \boldsymbol{\beta}]
\mathbf{F}_{c}^{^{\boldsymbol{y}}}\right\rangle+\left\langle[\mathbf{I}_{m_{_{\boldsymbol{e}}}}\ \vdots\ \boldsymbol{\beta}]
\mathbf{F}_{c}^{^{\boldsymbol{v}}}\widehat{\boldsymbol{\Gamma}}_{_{\boldsymbol{v}}},[\mathbf{I}_{m_{_{\boldsymbol{e}}}}\ \vdots\ \boldsymbol{\beta}]
\mathbf{F}_{c}^{^{\boldsymbol{v}}}\right\rangle,$$
where\ \ $\mathbf{F}_{c}^{^{\boldsymbol{y}}}\triangleq\begin{bmatrix}
                                                       \mathbf{H}_{_{\mathbf{I}}}\\
                                                       -\mathbf{Y}_{a}^{^{i}}\mathbf{H}_{_{\boldsymbol{0}}}
                                                      \end{bmatrix}$ \ \ and\ \ $\mathbf{F}_{c}^{^{\boldsymbol{v}}}\triangleq\begin{bmatrix}
                                                       \boldsymbol{0}\\
                                                       \mathbf{Y}_{a}^{^{i}}
                                                      \end{bmatrix}$.\\
Thus,                                                  
\begin{eqnarray*}                        
C\left(\mathbf{G}(\boldsymbol{\beta});\widehat{\boldsymbol{\Gamma}}_{_{\boldsymbol{y}}},
\widehat{\boldsymbol{\Gamma}}_{_{\boldsymbol{v}}}\right)
&=&\left\langle[\mathbf{I}_{m_{_{\boldsymbol{e}}}}\ \vdots\ \boldsymbol{\beta}],[\mathbf{I}_{m_{_{\boldsymbol{e}}}}\ \vdots\ \boldsymbol{\beta}]
\{\mathbf{F}_{c}^{^{\boldsymbol{y}}}\widehat{\boldsymbol{\Gamma}}_{_{\boldsymbol{y}}}(\mathbf{F}_{c}^{^{\boldsymbol{y}}})^{*}
+\mathbf{F}_{c}^{^{\boldsymbol{v}}}\widehat{\boldsymbol{\Gamma}}_{_{\boldsymbol{v}}}(\mathbf{F}_{c}^{^{\boldsymbol{v}}})^{*}\}\right\rangle,\ \ 
\Leftrightarrow\\
C\left(\mathbf{G}(\boldsymbol{\beta});\widehat{\boldsymbol{\Gamma}}_{_{\boldsymbol{y}}},
\widehat{\boldsymbol{\Gamma}}_{_{\boldsymbol{v}}}\right)&=& \operatorname{tr}
\left\{[\mathbf{I}_{m_{_{\boldsymbol{e}}}}\ \vdots\ \boldsymbol{\beta}]Q_{c}
[\mathbf{I}_{m_{_{\boldsymbol{e}}}}\ \vdots\ \boldsymbol{\beta}]^{^{\mathrm{T}}}\right\}.
\end{eqnarray*}

As a result,  $C\left(\mathbf{G}(\boldsymbol{\beta});\widehat{\boldsymbol{\Gamma}}_{_{\boldsymbol{y}}},
\widehat{\boldsymbol{\Gamma}}_{_{\boldsymbol{v}}}\right)= \inf\left\{\operatorname{tr}(\mathbf{P}): \mathbf{P}=\mathbf{P}^{^{\mathrm{T}}},
\mathbf{P}\geq[\mathbf{I}_{m_{_{\boldsymbol{e}}}}\ \vdots\ \boldsymbol{\beta}]Q_{c}
[\mathbf{I}_{m_{_{\boldsymbol{e}}}}\ \vdots\ \boldsymbol{\beta}]^{^{\mathrm{T}}}\right\}.$                       
                        
The proof of (5.1) is concluded by noting that
$$\mathbf{P}\geq[\mathbf{I}_{m_{_{\boldsymbol{e}}}}\ \vdots\ \boldsymbol{\beta}]Q_{c}
[\mathbf{I}_{m_{_{\boldsymbol{e}}}}\ \vdots\ \boldsymbol{\beta}]^{^{\mathrm{T}}}\ \ \Leftrightarrow\ \ 
\mathbf{P}-\left\{\left([\mathbf{I}_{m_{_{\boldsymbol{e}}}}\ \vdots\ \boldsymbol{\beta}]Q_{c}^{^{1/2}}\right)
\left([\mathbf{I}_{m_{_{\boldsymbol{e}}}}\ \vdots\ \boldsymbol{\beta}]Q_{c}^{^{1/2}}\right)^{^{\mathrm{T}}}\right\}\geq0$$
$\Leftrightarrow$\ \ (in the light of the so-called Schur complement formula)\ \ 
$Q_{_{\mathcal{J}q}}(\mathbf{P},[\mathbf{I}_{m_{_{\boldsymbol{e}}}}\ \vdots\ \boldsymbol{\beta}];Q_{c})\geq0$. Equation (5.2) is proved in 
exactly the same way noting at the beginning that
\begin{eqnarray*}
\left\langle\mathbf{G}(\boldsymbol{\beta})\otimes\boldsymbol{\phi}_{_{\boldsymbol{y}}}^{^{\mathrm{T}}},
\mathbf{G}(\boldsymbol{\beta})\otimes\boldsymbol{\phi}_{_{\boldsymbol{y}}}^{^{\mathrm{T}}}\right\rangle
&=&\left\langle\left(\boldsymbol{\beta}\otimes\mathbf{I}\right)
(\mathbf{Y}_{a}^{^{i}}\otimes\boldsymbol{\phi}_{_{\boldsymbol{y}1}}^{^{\mathrm{T}}}),
\left(\boldsymbol{\beta}\otimes\mathbf{I}\right)
(\mathbf{Y}_{a}^{^{i}}\otimes\boldsymbol{\phi}_{_{\boldsymbol{y}1}}^{^{\mathrm{T}}})\right\rangle\ \ \ \ \ \ \  \Leftrightarrow\\
&=&\left\langle\left(\boldsymbol{\beta}\otimes\mathbf{I}\right),\left(\boldsymbol{\beta}\otimes\mathbf{I}\right)
\left\{(\mathbf{Y}_{a}^{^{i}}\otimes\boldsymbol{\phi}_{_{\boldsymbol{y}1}}^{^{\mathrm{T}}})
(\mathbf{Y}_{a}^{^{i}}\otimes\boldsymbol{\phi}_{_{\boldsymbol{y}1}}^{^{\mathrm{T}}})^{*}\right\}\right\rangle\ \ \Leftrightarrow\\
&=&\operatorname{tr}\left\{\left(\boldsymbol{\beta}\otimes\mathbf{I}\right)Q_{_{\mathbf{G}c}}
\left(\boldsymbol{\beta}\otimes\mathbf{I}\right)^{^{\mathrm{T}}}\right\}.
\end{eqnarray*}\hfill$\blacksquare$

\bigskip

\vspace*{3mm}
\noindent
\underline{\textbf{Proof of Proposition \ref{prop:05}:}}\   Proposition \ref{prop:05} follows directly from Proposition \ref{prop:04} and  the
following auxiliary proposition.                                      
                                              
\vspace*{3mm}
\noindent
\underline{\textbf{Proposition A.1:}}\ Let $f:\mathcal{R}_{c}^{m_{_{\boldsymbol{e}}}\times m_{_{\boldsymbol{v}}}}\rightarrow \mathbb{R}$ be $\mathcal{H}_{2}-$continuous,
$$\mathcal{S}_{_{\mathbf{P}r}}\triangleq\{\mathbf{G}\in \mathcal{S}_{_{\mathbf{G}}}:\exists(\lambda, \mathbf{P}), \lambda>0, \mathbf{P}=\mathbf{P}^{^{\mathrm{T}}}\ \text{\textbf{(\emph{i})} and \textbf{(\emph{ii})} hold}\},$$
  where \textbf{(\emph{i})}  $\lambda \gamma^{^{2}}+\boldsymbol{x}_{_{\boldsymbol{0}}}^{^{\mathrm{T}}}\mathbf{P}\boldsymbol{x}_{_{\boldsymbol{0}}}\leq \boldsymbol{\eta}_{_{\mathcal{J}}}$ \ \ and\ \
   \textbf{(\emph{ii})} $Q_{_{\mathcal{J}a}}(\mathbf{P}; \boldsymbol{\Sigma}_{a}, \mathbf{M}(\lambda))<0$, and\\ 
$\mathcal{S}_{\text{eq}}\triangleq\{\mathbf{G}\in \mathcal{S}_{_{\mathbf{G}}}^{^{1}}: \bar{\mathcal{J}}_{_{\mathbf{X}}}(\mathbf{G};
\mathcal{S}_{_{\mathbf{X}}})\leq \boldsymbol{\eta}_{_{\mathcal{J}}}\}$,\ \  $\boldsymbol{\eta}_{_{\mathcal{J}}}\triangleq 
(1+\alpha)\mathcal{J}_{_{o}}^{^{1}},\ \alpha>\varepsilon\ \ \text{and}$\\
$\mathcal{J}_{_{\mathrm{o}}}^{^{1}}=
 \inf\{\bar{\mathcal{J}}_{_{\mathbf{X}}}(\mathbf{G};\mathcal{S}_{_{\mathbf{X}}}):\mathbf{G}\in \mathcal{S}_{_{\mathbf{G}}}\}.
$
 
Then, $\mathcal{S}_{_{\mathbf{P}r}}$ is non-empty, $\mathcal{S}_{_{\mathbf{P}r}}\subset\mathcal{S}_{\text{eq}}$  and 

$$\inf\{f(\mathbf{G}):\mathbf{G}\in\mathcal{S}_{\text{eq}}\}=\inf\{f(\mathbf{G}):\mathbf{G}\in \mathcal{S}_{_{\mathbf{P}r}}\}.$$ \hfill$\nabla$

\vspace*{3mm}
\noindent
\underline{\textbf{Proof of Proposition A.1:}}\ Let 
$\mathcal{S}_{\text{in}}\triangleq\{\mathbf{G}\in\mathcal{S}_{_{\mathbf{G}}}: \bar{\mathcal{J}}_{_{\mathbf{X}}}(\mathbf{G};
\mathcal{S}_{_{\mathbf{X}}})<\boldsymbol{\eta}_{_{\mathcal{J}}}\}$ and note that as\ $\alpha>\varepsilon$,\ $\mathcal{S}_{\text{in}}$\ is 
non-empty. Now, consider the following auxiliary propositions.

\noindent
\underline{\textbf{Auxiliary Proposition 2:}}\ If\ $\mathbf{G}\in \mathcal{S}_{_{\mathbf{P}r}}$,\ then $\mathbf{G}\in \mathcal{S}_{\text{eq}}$\ 
(\emph{i.e.}, $\mathcal{S}_{_{\mathbf{P}r}}\subset \mathcal{S}_{\text{eq}}$). If\ $\mathbf{G} \in \mathcal{S}_{\text{in}}$,\ then $\mathbf{G} \in \mathcal{S}_{_{\mathbf{P}r}}$\ 
(\emph{i.e.}, $\mathcal{S}_{\text{in}}\subset \mathcal{S}_{_{\mathbf{P}r}}$).\hfill$\nabla$
                                       
\noindent
\underline{\textbf{Auxiliary Proposition 3:}}\  For any\ $\mathbf{G}\in \mathcal{S}_{\text{eq}}$\ there exists\ 
$\{\mathbf{G}_{k}\}\subset \mathcal{S}_{\text{in}}$\ such that\break $\mathbf{G}_{k}\stackrel{\mathcal{H}_{2}}{\rightarrow}\mathbf{G}$. \hfill$\nabla$   

\noindent
It follows from \textbf{Auxiliary Proposition 2} that\ $\mathcal{S}_{\text{in}}\subset \mathcal{S}_{_{\mathbf{P}r}}\subset \mathcal{S}_{\text{eq}}$\
(so that\ $\mathcal{S}_{_{\mathbf{P}r}}$\ is non-empty) and, hence,
\begin{equation}\label{eq:A15}
 f_{\text{eq}}\triangleq\inf\{f(\mathbf{G}): \mathbf{G}\in \mathcal{S}_{\text{eq}}\}
 \leq f_{_{\mathbf{P}r}}\triangleq\inf\{f(\mathbf{G}): \mathbf{G}\in \mathcal{S}_{_{\mathbf{P}r}}\}\leq f_{\text{in}}\triangleq\inf\{f(\mathbf{G}): \mathbf{G}\in \mathcal{S}_{\text{in}}\}.
\end{equation}
Now, let\ $\delta>0$\ and take\ $\mathbf{G}_{\delta} \in \mathcal{S}_{\text{eq}}$\  such that\ 
$f(\mathbf{G}_{\delta})\leq f_{\text{eq}}+\delta$. In the light of \textbf{Auxiliary Proposition 3}, 
$\exists\{\mathbf{G}_{\delta_{k}}\}\subset \mathcal{S}_{\text{in}}$\ such that\ 
$\mathbf{G}_{\delta_{k}}\stackrel{\mathcal{H}_{2}}{\rightarrow}\mathbf{G}_{\delta} $\  and, hence (since $f$ in
$\mathcal{H}_{2}-$continuous),\ 
$f(\mathbf{G}_{\delta_{k}})\rightarrow f(\mathbf{G}_{\delta})$.

Thus, as\ $\forall k \in \mathbb{Z}_{+}$,\ $f_{\text{in}}\leq f(\mathbf{G}_{\delta_{k}})$,\ 
$f_{\text{in}}\leq f(\mathbf{G}_{\delta})=f_{\text{eq}}+\delta$. As this holds for any $\delta>0$, it has been established that 
$\forall \delta>0$ \ $f_{\text{in}}\leq f_{\text{eq}}+\delta$. Therefore,\ $f_{\text{in}}\leq f_{\text{eq}}$.

On the other hand, in the light of (\ref{eq:A15}),\ $f_{\text{in}}\geq f_{\text{eq}}$\ \ so that\ $f_{\text{in}}=f_{_{\mathbf{P}r}}=f_{\text{eq}}$.   \textbf{Auxiliary Proposition 2} 
follows
directly from  equation (\ref{eq:13}) and \textbf{Auxiliary Proposition 3} from the fact that\ $\mathcal{S}_{_{\mathbf{G}}}$\ is convex and\ 
$\bar{\mathcal{J}}_{_{\mathbf{X}}}(\cdot; \mathcal{S}_{_{\mathbf{X}}})$\ is convex. \hfill$\blacksquare$
    
\bigskip

\vspace*{3mm}                
\noindent
\underline{\textbf{Recasting \emph{Prob.\ $5$} as (5.5)\ and \emph{Prob.\ $6$} as (5.7):}} These are the counterparts of Proposition 5.1 for
\emph{Prob.\ $5$} and \emph{Prob.\ $6$} and can be proved using the argument (mutatis mutandis) invoked in its $\text{proof.}\hfill\blacksquare$
 
\bigskip 
    
\vspace*{3mm}                
\noindent
\underline{\textbf{Auxiliary Proposition 4:}}\ Let \ $\widehat{\mathbf{X}}\in \mathcal{S}_{_{\mathbf{X}}}$\ be such that\ $\|\widehat{\mathbf{X}}\|_{_{2}}=\gamma$\ and\ 
$\delta_{_{\ell}}(\widehat{\mathbf{X}})\leq 0$. Let\break 
$p(\boldsymbol{\beta}; \widehat{\mathbf{X}})\triangleq \delta_{_{\mathcal{J}}}(\boldsymbol{\beta}\widehat{\mathbf{X}})=\delta_{_{\mathcal{J}\mathrm{o}}}+2\boldsymbol{\beta}|\delta_{_{\ell}}(\widehat{\mathbf{X}})|+\boldsymbol{\beta}^{^{2}}\delta_{_{q}}(\widehat{\mathbf{X}})$,\ 
$S_{_{I\boldsymbol{\beta}}}(\widehat{\mathbf{X}};\mathbf{G})\triangleq\{\boldsymbol{\beta}\in[-1, 1]: p(\boldsymbol{\beta}; \widehat{\mathbf{X}})<0\}$\ and let\ $\mu_{_{I}}(\widehat{\mathbf{X}};\mathbf{G})$\ denote the length of\ $S_{_{I\boldsymbol{\beta}}}(\widehat{\mathbf{X}};\mathbf{G})$.

Let \ $\mathbf{G}$\ and\ $\mathbf{G}_{_{\mathbf{M}}}$\ be such that\
$\delta_{_{\mathcal{J}\mathrm{o}}}\triangleq \mathcal{J}_{_{\mathbf{X}}}(\mathbf{G}; 0)-\mathcal{J}_{_{\mathbf{X}}}(\mathbf{G}_{_{\mathbf{M}}}; 0)<0$. Then,\\
 \textbf{(\emph{i})}\ If\ $\delta_{_{q}}(\widehat{\mathbf{X}})\leq 0$, \ $\forall \varepsilon > 0$,\
$\mu_{_{I}}(\widehat{\mathbf{X}};\mathbf{G})\geq 1+\min(1,\nu_{\varepsilon}(\widehat{\mathbf{X}}))$,
where\  $\nu_{\varepsilon}(\widehat{\mathbf{X}})=(1/2)\dfrac{|\delta_{_{\mathcal{J}\mathrm{o}}}|}{\varepsilon+|\delta_{_{\ell}}(\widehat{\mathbf{X}})|}$.\\
\textbf{(\emph{ii})}\ Let\ $\widehat{\mathbf{X}}$\ be such that\ $\delta_{_{q}}(\widehat{\mathbf{X}})>0$\  and let\
$C(\widehat{\mathbf{X}})\triangleq\dfrac{|\delta_{_{\mathcal{J}\mathrm{o}}}|}{|\delta_{_{q}}(\mathbf{X})|}$,\ $K(\mathbf{X})\triangleq\dfrac{|\delta_{_{\ell}}(\mathbf{X})|}{|\delta_{_{q}}(\mathbf{X})|}$,\break
$$\boldsymbol{\beta}_{_{r1}}\triangleq-K(\widehat{\mathbf{X}})-\{K(\widehat{\mathbf{X}})^{^{2}}+C(\widehat{\mathbf{X}})\}^{^{1/2}}\ \text{and}\ 
\boldsymbol{\beta}_{_{r2}}\triangleq-K(\widehat{\mathbf{X}})+\{K(\widehat{\mathbf{X}})^{^{2}}+C(\widehat{\mathbf{X}})\}^{^{1/2}}.$$
\textbf{(\emph{ii.a})}\ If either\  $\boldsymbol{\beta}_{_{r1}}<-1$\  or  \ $\boldsymbol{\beta}_{_{r2}}>1$,  \ $\mu_{_{I}}(\widehat{\mathbf{X}};\mathbf{G})\geq 1+\min(1, \nu_{_{\beta}}(\widehat{\mathbf{X}}))$,
where\break $\nu_{_{\beta}}(\widehat{\mathbf{X}})\triangleq(1/2)\dfrac{C(\widehat{\mathbf{X}})}{\{K(\widehat{\mathbf{X}})^{^{2}}+C(\widehat{\mathbf{X}})\}^{^{1/2}}}$.\\

\noindent
\textbf{(\emph{ii.b})}\ If $[\boldsymbol{\beta}_{_{r1}}, \boldsymbol{\beta}_{_{r2}}]\subset[-1,1]$, \ \ $\mu_{_{I}}(\widehat{\mathbf{X}};\mathbf{G})=2\{K(\widehat{\mathbf{X}})^{^{2}}+C(\widehat{\mathbf{X}})\}^{^{1/2}}$. \hfill$\nabla$
            
\vspace*{3mm}
\noindent
\underline{\textbf{Proof:}} \textbf{(\emph{i})} If $\delta_{_{q}}(\widehat{\mathbf{X}})\leq 0\ \Rightarrow  \ p(\boldsymbol{\beta}, \widehat{\mathbf{X}})<0\ \forall \boldsymbol{\beta}\in[-1,0]$.
For $\boldsymbol{\beta}\in[0,1]$,\break $p(\boldsymbol{\beta}, \widehat{\mathbf{X}})\leq \delta_{_{\mathcal{J}\mathrm{o}}}+2 \boldsymbol{\beta}|\delta_{_{\ell}}(\mathbf{X})|\leq
\delta_{_{\mathcal{J}\mathrm{o}}}+2\boldsymbol{\beta}(\varepsilon+\delta_{_{\ell}}|\delta_{_{\ell}}(\widehat{\mathbf{X}})|)$ \ for any $\varepsilon>0$ so that for any $\boldsymbol{\beta}\in [0,1]$,\
$p(\boldsymbol{\beta}, \widehat{\mathbf{X}})\leq 0$\ whenever, for any $\varepsilon>0$, 
$$\delta_{_{\mathcal{J}\mathrm{o}}}+2\boldsymbol{\beta}(\varepsilon+|\delta_{_{\ell}}(\widehat{\mathbf{X}})|)\ \ \Leftrightarrow\ \
\boldsymbol{\beta}\leq (1/2)\frac{|\delta_{_{\mathcal{J}\mathrm{o}}}|}{\varepsilon+|\delta_{_{\ell}}(\widehat{\mathbf{X}})|}.$$
Thus\ \ $S_{_{I\boldsymbol{\beta}}}(\widehat{\mathbf{X}};\mathbf{G})\supset[-1, 0]\bigcap[0, \nu_{\varepsilon}]$\ for any $\varepsilon>0\ \Rightarrow\ 
\mu_{_{I}}(\widehat{\mathbf{X}};\mathbf{G})\geq 1+\min(1, \nu_{\varepsilon}(\widehat{\mathbf{X}}))$ for any $\varepsilon>0$. \\

\noindent
\textbf{(\emph{ii})}\  Noting that for $\Delta x>0$\ $(x+\Delta x)^{^{1/2}}= x^{^{1/2}}+\dfrac{1}{2\widehat{x}^{^{1/2}}}\Delta x$\ for some \ 
$\widehat{x}\in (x, x+\Delta x)$, there exists\ \ $\widehat{x}\in (K(\widehat{\mathbf{X}})^{^{2}}, K(\widehat{\mathbf{X}})^{^{2}}+C(\widehat{\mathbf{X}})$\ such that \ 
$\boldsymbol{\beta}_{_{r1}}=-2K(\widehat{\mathbf{X}})-\Delta \boldsymbol{\beta}(\widehat{x})$, \ $\boldsymbol{\beta}_{_{r2}}=\Delta \boldsymbol{\beta}(\widehat{x})$,\ where\
$\Delta\boldsymbol{\beta}(\widehat{\mathbf{X}})=\dfrac{1}{2\widehat{x}^{^{1/2}}}C(\widehat{\mathbf{X}})$. If\ $\boldsymbol{\beta}_{_{r1}}<-1$,\ $S_{_{I\boldsymbol{\beta}}}(\widehat{\mathbf{X}};\mathbf{G})=[-1, \Delta \boldsymbol{\beta}(\widehat{\mathbf{X}})]$. If\ 
$\boldsymbol{\beta}_{_{r2}}>1$, as $\boldsymbol{\beta}_{_{r2}}<-\Delta\boldsymbol{\beta}(\widehat{\mathbf{X}})$, \ $S_{_{I\boldsymbol{\beta}}}(\widehat{\mathbf{X}};\mathbf{G})\supset[-\Delta\boldsymbol{\beta}(\widehat{\mathbf{X}}), 1]$. In both cases,
$$\mu_{_{I}}(\widehat{\mathbf{X}};\mathbf{G})\geq 1+\Delta \boldsymbol{\beta}(\widehat{\mathbf{X}})\geq 1+\frac{C(\widehat{\mathbf{X}})}{2\{K(\widehat{\mathbf{X}})^{^{2}}+C(\widehat{\mathbf{X}})\}^{^{1/2}}}.$$
                
\noindent
\textbf{(\emph{iii})} In this case, \
$\delta_{I}(\widehat{\mathbf{X}})=[\boldsymbol{\beta}_{_{r1}}, \boldsymbol{\beta}_{_{r2}}]\ \Rightarrow\ \mu_{_{I}}(\widehat{\mathbf{X}};\mathbf{G})=2\{K(\widehat{\mathbf{X}})^{^{2}}+C(\widehat{\mathbf{X}})\}^{^{1/2}}$.\hfill$\blacksquare$

\vspace*{3mm}                
\noindent
\underline{\textbf{Proof of Proposition 6.1:}} \textbf{Proposition 6.1} is a straightforward consequence of 
\textbf{Auxiliary Proposition 4}. To see why is this so, note first that $\forall \widehat{\mathbf{X}} \in \mathcal{S}_{_{\mathbf{X}}}$,\ $|\delta_{_{\ell}}(\widehat{\mathbf{X}})|<\bar{\delta}_{_{\ell}}$,\
$|\delta_{_{q}}(\widehat{\mathbf{X}})|<\bar{\delta}_{_{q}}$,

\noindent
\textbf{(a)} $\forall \widehat{\mathbf{X}}\in \mathcal{S}_{_{\mathbf{X}}}$,\ \ $\nu_{\varepsilon}(\widehat{\mathbf{X}})\geq (1/2)\dfrac{|\delta_{_{\mathcal{J}\mathrm{o}}}|}{\varepsilon+\bar{\delta}}_{_{\ell}}$\ \ \ $\Rightarrow$\ \ \ $\sup\{\nu_{_{\varepsilon}}(\widehat{\mathbf{X}}):\varepsilon>0\}\geq \nu_{_{a}}$.

\noindent
\textbf{(b)} $C(\widehat{\mathbf{X}})\geq \nu_{c}$\ \ and\ \ $2\{K(\widehat{\mathbf{X}})^{^{2}}+ C(\widehat{\mathbf{X}})\}^{^{1/2}} \geq 2 C(\widehat{\mathbf{X}})^{^{1/2}}\geq 2\nu_{c}^{^{1/2}}$.

\noindent
\textbf{(c)}
\begin{eqnarray*}
\hat{\nu}_{_{\boldsymbol{\beta}}}(\widehat{\mathbf{X}})&=&(1/2)\dfrac{|\delta_{_{\mathcal{J}\mathrm{o}}}|}{\{|\delta_{_{q}}(\widehat{\mathbf{X}})|^{^{2}}K(\widehat{\mathbf{X}})^{^{2}}+|\delta_{_{q}}(\widehat{\mathbf{X}})^{^{2}}|C(\widehat{\mathbf{X}})\}^{^{1/2}}}=(1/2)\dfrac{\delta_{_{\mathcal{J}\mathrm{o}}}}{\{|\delta_{_{\ell}}(\widehat{\mathbf{X}})^{^{2}}|+|\delta_{_{q}}(\widehat{\mathbf{X}})||\delta_{_{\mathcal{J}\mathrm{o}}}|\}^{^{1/2}}}\\\\
\Rightarrow\ \ \ \hat{\nu}_{_{\boldsymbol{\beta}}}(\widehat{\mathbf{X}})&=&(1/2)\dfrac{|\delta_{_{\mathcal{J}\mathrm{o}}}|}{\bar{\delta_{_{q}}}}\dfrac{1}{\{|\delta_{_{\ell}}(\widehat{\mathbf{X}})|/\bar{\delta}_{_{q}})^{^{2}}+(|\delta_{_{q}}(\widehat{\mathbf{X}})|/\bar{\delta}_{_{q}})(|\delta_{_{\mathcal{J}\mathrm{o}}}|/\bar{\delta}_{_{q}})\}^{^{1/2}}}\\\\
\Rightarrow\ \ \ \hat{\nu}_{_{\boldsymbol{\beta}}}(\widehat{\mathbf{X}})&=&(1/2)\nu_{c}\dfrac{1}{\{|\delta_{_{\ell}}(\widehat{\mathbf{X}})|/\bar{\delta}_{_{q}})^{^{2}}+\nu_{c}(|\delta_{_{q}}(\widehat{\mathbf{X}})|/\bar{\delta}_{_{q}})\}^{^{1/2}}}
\end{eqnarray*}
$\Rightarrow\ \ \ (\text{since} \ \ |\delta_{_{q}}(\widehat{\mathbf{X}})|\leq\bar{\delta}_{_{q}})$\\
$$\hat{\nu}_{_{\boldsymbol{\beta}}}(\widehat{\mathbf{X}})\geq(1/2)\nu_{c}\dfrac{1}{\{|\delta_{_{\ell}}(\widehat{\mathbf{X}})|/\bar{\delta}_{_{q}})^{^{2}}+\nu_{c}\}^{^{1/2}}}\geq\nu_{_{\boldsymbol{\beta}}}.$$                               
In the light of \textbf{(a)}, the lower bound on\ $\mu_{_{\mathcal{J}}}(\widehat{\mathbf{X}};\mathbf{G})$ given in \textbf{(\emph{i})} leads to 
$$\mu_{_{\mathcal{J}}}(\widehat{\mathbf{X}})\geq \min(2, 1+\nu_{a}).$$
Similarly in the light of \textbf{(c)} and \textbf{(b)} the lower bounds on\ \ $\mu_{_{I}}(\widehat{\mathbf{X}};\mathbf{G})$ given by \textbf{(\emph{ii.a})} and  \textbf{(\emph{ii.b})} give rise, respectively, to
$$\mu_{_{I}}(\widehat{\mathbf{X}};\mathbf{G})\geq \min(2, 1+\nu_{_{\beta}})\ \ \ \text{and}\ \ \  \mu_{_{\mathcal{J}}}(\widehat{\mathbf{X}};\mathbf{G})\geq 2\nu_{c}^{^{1/2}}.$$
Thus,\ \ \  $\forall \widehat{\mathbf{X}}\in\mathcal{S}_{_{\mathbf{X}}}$,\ $\mu_{_{I}}(\widehat{\mathbf{X}};\mathbf{G})\geq\min\{\min(2, 1+\nu_{a}), \min(2,1+\nu_{_{\boldsymbol{\beta}}}), 2\nu_{c}^{1/2}\}$\\
$$\Rightarrow\ \ \ \ \ \  \forall \widehat{\mathbf{X}}\in \mathcal{S}_{_{\mathbf{X}}}, \mu_{_{I}}(\widehat{\mathbf{X}};\mathbf{G})\geq\min\{2, 1+\nu_{a},1+\nu_{_{\boldsymbol{\beta}}}, 2\nu_{c}^{^{1/2}}\}.$$\hfill$\blacksquare$

\bigskip

\noindent
\underline{\textbf{Proof of Proposition 6.2:}} \textbf{(a)} Note first that
$$\boldsymbol{\eta}_{_{\mathbf{P}}}^{\infty}(\mathbf{G})=\sup\{\langle\boldsymbol{\Gamma}_{_{\boldsymbol{\delta\gamma}}}\boldsymbol{z}_{_{a}}, \boldsymbol{z}_{_{a}}\rangle:\boldsymbol{z}_{_{a}}\in \mathcal{S}_{_{\boldsymbol{z}_{_{a}}}}\},$$
where \ \ $\boldsymbol{\Gamma}_{_{\boldsymbol{\delta\gamma}}}\triangleq\mathbf{M}_{_{\boldsymbol{\gamma}}}\boldsymbol{\Gamma}_{_{\boldsymbol{\boldsymbol{\delta}}}}\mathbf{M}_{_{\boldsymbol{\gamma}}}$\  \ and
$$\mathcal{S}_{_{\boldsymbol{z}_{_{a}}}}=\left\{\boldsymbol{z}_{_{a}}=\begin{bmatrix}\boldsymbol{y}_{_{a}}^{^{\mathrm{T}}}&\vdots&\boldsymbol{v}_{_{a}}^{^{\mathrm{T}}}\end{bmatrix}: \boldsymbol{y}_{_{a}}\in\mathcal{R}_{_{c}}^{m_{_{\boldsymbol{y}}}}, \boldsymbol{v}_{_{a}}\in\mathcal{R}_{_{c}}^{m_{_{\boldsymbol{v}}}},\|\boldsymbol{y}_{_{a}}\|_{_{2}}\leq1,\|\boldsymbol{v}_{_{a}}\|_{_{2}}\leq1\right\}.$$
Note also that whenever $\|\boldsymbol{z}_{_{a}}\|_{_{2}}\leq1$, $\boldsymbol{z}_{_{a}}\in\mathcal{S}_{_{\boldsymbol{z}_{_{a}}}}$ (Since $1\geq\|\boldsymbol{z}_{_{a}}\|_{_{2}}\geq \max\left\{\|\boldsymbol{y}_{_{a}}\|_{_{2}},\|\boldsymbol{v}_{_{a}}\|_{_{2}}\right\}$). Thus,\break $\boldsymbol{\eta}_{_{\mathbf{P}}}^{\infty}\geq\sup\left\{\langle\boldsymbol{\Gamma}_{_{\boldsymbol{\delta\gamma}}}\boldsymbol{z}_{_{a}}, \boldsymbol{z}_{_{a}}\rangle:\|\boldsymbol{z}_{_{a}}\|_{_{2}}\leq 1\right\}=\bar{\lambda}_{_{\infty}}(\boldsymbol{\Gamma}_{_{\boldsymbol{\delta\gamma}}})$.\medskip

\noindent
 \textbf{(b)} Note first that 
\begin{eqnarray*}
 \boldsymbol{\eta}_{_{\mathbf{R}}}^{\infty}(\mathbf{G})\leq \mu \ \ &\Leftrightarrow&\ \ \
\inf\left\{\langle\boldsymbol{\Gamma}_{_{\mathbf{e}1}}\bar{\boldsymbol{z}},\bar{\boldsymbol{z}}\rangle-\mu\langle\Gamma_{_{\mathbf{e}0}}\bar{\boldsymbol{z}},\bar{\boldsymbol{z}}\rangle:\bar{\boldsymbol{z}}\in\bar{\mathcal{S}}_{_{\boldsymbol{z}}}\right\}\leq 0\\
\ \ &\Leftrightarrow&\ \ \ \sup\left\{\langle\boldsymbol{\Gamma}_{_{\mu}}\boldsymbol{z}_{_{a}},\boldsymbol{z}_{_{a}}\rangle:\boldsymbol{z}_{_{a}}\in\mathcal{S}_{_{\boldsymbol{z}_{_{a}}}}\right\}\geq 0,
 \end{eqnarray*}
where\ \ \ $\boldsymbol{\Gamma}_{_{\mu}}\triangleq\mathbf{M}_{_{\gamma}}(\mu\boldsymbol{\Gamma}_{_{\mathbf{e}0}}-\boldsymbol{\Gamma}_{_{\mathbf{e}1}})\mathbf{M}_{_{\gamma}}$. But
$$\sup\left\{\langle\boldsymbol{\Gamma}_{_{\mu}}\boldsymbol{z}_{_{a}},\boldsymbol{z}_{_{a}}\rangle:\boldsymbol{z}_{_{a}}\in\mathcal{S}_{_{\boldsymbol{z}_{_{a}}}}\right\}\geq \sup\left\{\langle\boldsymbol{\Gamma}_{_{\mu}}\boldsymbol{z}_{_{a}},\boldsymbol{z}_{_{a}}\rangle:\|\boldsymbol{z}_{_{a}}\|_{_{2}}\leq1\right\}=\bar{\lambda}_{\infty}(\boldsymbol{\Gamma}_{_{\mu}}),$$
so that if\ \ \ $\bar{\lambda}_{\infty}(\boldsymbol{\Gamma}_{_{\mu}})\geq0$,\ \ \ $\boldsymbol{\eta}_{_{\mathbf{R}}}^{\infty}\leq \mu$.

Hence,\ \ \ \ $\boldsymbol{\eta}_{_{\mathbf{R}}}^{\infty}(\mathbf{G})\leq \inf\left\{\mu>0:\bar{\lambda}_{\infty}(\boldsymbol{\Gamma}_{_{\mu}})\geq0\right\}$.\hfill$\blacksquare$

 \end{document}